\documentclass[a4paper,12pt,times,numbered,print,index,custommargin]{Classes/PhDThesisPSnPDF}

\input{Preamble/preamble}

\title{Testing gravity on cosmological scales}

\subtitle{Theoretical predictions \\with the COLA method}

\author{Bartolomeo Fiorini}

\dept{Institute of Cosmology and Gravitation}

\university{University of Portsmouth}
\crest{\includegraphics[width=0.5\textwidth]{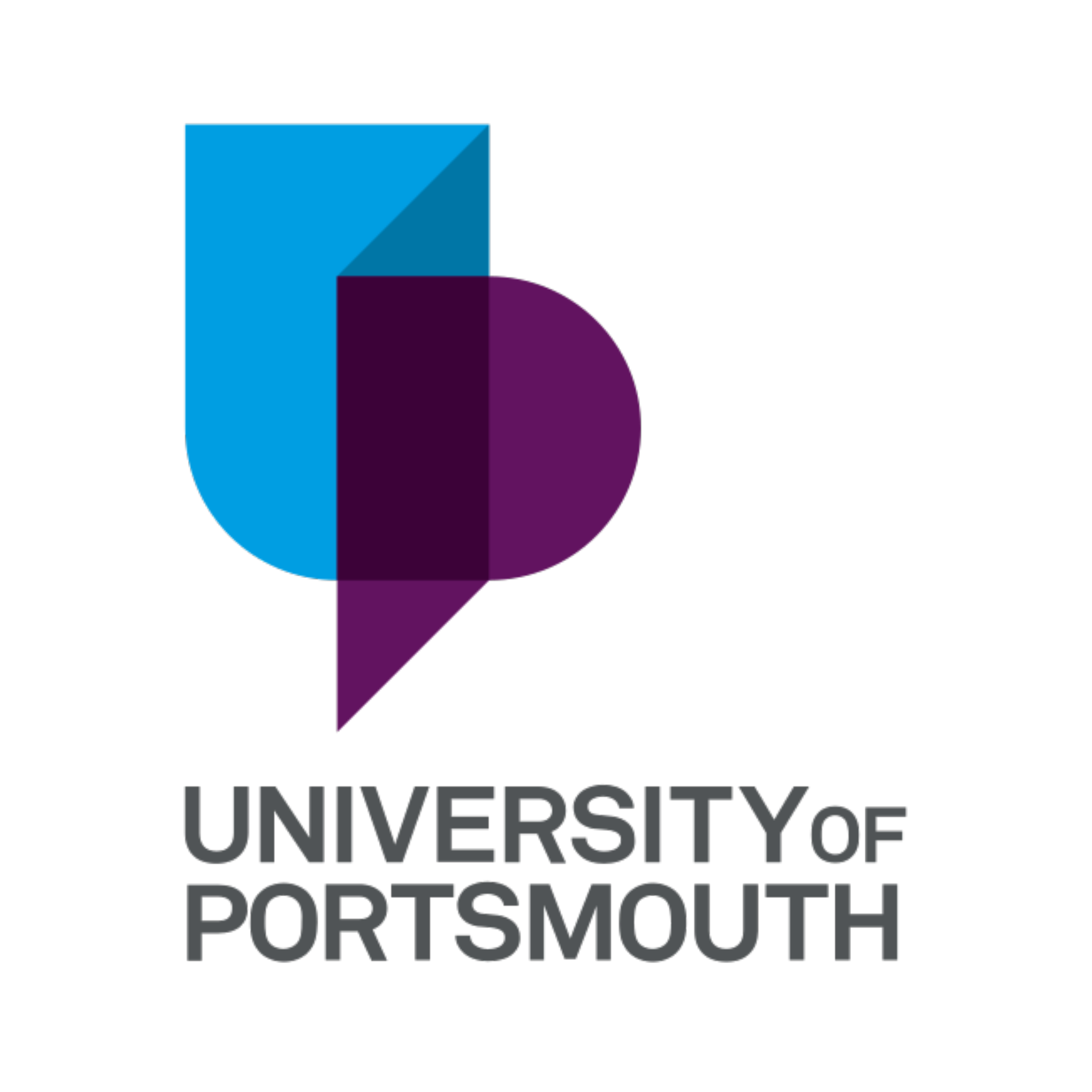}}

\supervisor{Kazuya Koyama\newline
Albert Izard\newline
David Wands}

\supervisorlinewidth{0.25\textwidth}

\degreetitle{Doctor of Philosophy}

\subject{Cosmology} \keywords{{Cosmology} {Modified gravity} {COLA} {Large scale structure} {Tests of gravity} {University of
Portsmouth}}

\pdfoutput=1 

\usepackage[T1]{fontenc} 

\usepackage{graphicx}
\usepackage{amsmath}
\usepackage{mathtools}
\usepackage{nicefrac}
\usepackage[dvipsnames]{xcolor}
\usepackage{booktabs}
\usepackage[normalem]{ulem}
\usepackage{physics}
\usepackage{enumitem}
\usepackage{bibentry}
\usepackage{pdfpages}

\newcommand{\textcode}[1]{{\large\scshape#1}}

\newcommand{\Msun}{\, h^{-1} \,  M_{\odot}}
\newcommand{\hompc}{\,h\,{\rm Mpc}^{-1}}
\newcommand{\mpcoh}{\,h^{-1}\,{\rm Mpc}}

\renewcommand{\vec}{\mathbf}
\newcommand{\vbar}{\Bar{v}}
\NewDocumentCommand{\codeword}{v}{%
\texttt{#1}}
\newcommand{\matr}[1]{\mathbf{#1}}

\newcommand{\NP}[0]{N_{\rm part}^{1/3}}
\newcommand{\NM}[0]{N_{\rm mesh}^{1/3}}
\newcommand{\NS}[0]{N_{\rm step}}
\newcommand{\LF}[0]{\ell_{\rm Force}}
\newcommand{\DispVec}{\xi}

\newcommand{\nbody}[1]{{\it N}-body}

\newcommand{\lcdm}{$\Lambda$CDM}

\newcommand{\CLASS}{\textsc{class}}
\newcommand{\hiclass}{{\tt hi\_class}}

\newcommand{\Pnon}{P_{\mathrm{non}}}
\newcommand{\Plin}{P_{\mathrm{lin}}}
\newcommand{\Pnonref}{P^{\mathrm{ref}}_{\mathrm{non}}}
\newcommand{\Plinref}{P^{\mathrm{ref}}_{\mathrm{lin}}}
\newcommand{\Rnon}{R_{\mathrm{non}}}
\newcommand{\Rlin}{R_{\mathrm{lin}}}
\newcommand{\Bref}{B^{\mathrm{ref}}}

\ifdefineAbstract
\fi

\ifdefineChapter
\fi

\begin{document}

\frontmatter

\maketitle
\cleardoublepage
\includepdf[]{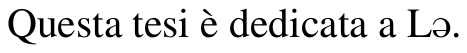}
\cleardoublepage

\begin{declaration}

Whilst registered as a candidate for the above degree, I have not been registered for any other research award. The results and conclusions embodied in this thesis are the work of the named candidate and have not been submitted for any other academic award.



\end{declaration}

\begin{center}
    Approximate word count: 27409
\end{center}
%
%

\chapter*{Dissemination}
This thesis is based on following works
\begin{enumerate}[label={[\arabic*]}] \raggedright
    \item \textbf{Fiorini}, Koyama, Izard, Winther, Wright, Li; \textit{Fast generation of mock galaxy catalogues in modified gravity models with COLA}, \href{https://iopscience.iop.org/article/10.1088/1475-7516/2021/09/021}{\textit{JCAP} \textbf{09} (2021) 021};
    \item \textbf{Fiorini}, Koyama, Izard; \textit{Studying large-scale structure probes of modified gravity with COLA}, \href{https://iopscience.iop.org/article/10.1088/1475-7516/2022/12/028}{\textit{JCAP} \textbf{12} (2022) 028};
    \item Brando, \textbf{Fiorini}, Koyama, Winther; \textit{Enabling matter power spectrum emulation in beyond-$\Lambda$CDM cosmologies with COLA}, \href{https://iopscience.iop.org/article/10.1088/1475-7516/2022/09/051}{\textit{JCAP} \textbf{09} (2022) 051}.
\end{enumerate}

My academic authorship information can be found using either the ORCID record \href{https://orcid.org/0000-0002-0092-4321}{0000-0002-0092-4321} or the arXiv public author identifier \href{http://arxiv.org/a/fiorini_b_1}{fiorini\_b\_1}.
\clearpage

\begin{acknowledgements}     
In case you are reading this acknowledgements to look for your name, I wish to sincerely thank you because of the many reasons that we both know and which I promise I will not forget. 

If, on the other hand, you are just curious about who helped me get to this point, I am sorry to disappoint you: I am very fortunate to have a lot of important people in my life and all have had a part in getting me this far.
Some of these people clearly contributed more, others less. Selecting those who contributed the most would inevitably require leaving out someone who was nonetheless crucial. 

In any case, Numerical computations were done on the Sciama High Performance Compute (HPC) cluster which is supported by the ICG, SEPNet and the University of Portsmouth.

\end{acknowledgements}

\begin{abstract}
The standard model of cosmology based on general relativity needs a dark energy component in the cosmic inventory to successfully describe the accelerated expansion of the universe at late time. This dark energy is not well justified by the standard model of particle physics. An alternative way to explain the accelerated expansion is to consider cosmological models based on modified gravity theories. The simplest extensions to general relativity include an additional scalar field mediating a fifth force. The gravity model is already well constrained with Solar System experiments, which show no sign of departure from general relativity in this environment. However, gravity still needs to be precisely constrained on cosmological scales. Among the modified gravity theories proposed in the literature, we focus on the ones which incorporate screening mechanisms, hiding the fifth force in high-density environments. 

One of the main objectives of stage IV galaxy surveys is to constrain gravity on cosmological scales but to fully take advantage of the constraining power of galaxy surveys it is important to formulate theoretical predictions in the non-linear regime of structure formation. This is possible by means of \nbody{} simulations in combination with models for the galaxy-halo connection. However, full \nbody{} simulations are computationally very expensive and modified gravity further increases their computational cost. In general relativity, approximate simulations method can be used to produce synthetic galaxy catalogues in place of full \nbody{} simulations at the cost of reduced accuracy in the deep non-linear regime. One of these approximate methods, the so-called COLA (COmoving Lagrangian Acceleration) method, has been extended to simulate modified gravity theories. 

In this context, we focus on two modified gravity theories, $f(R)$ and nDGP (the normal branch of the Dvali-Gabadadze-Porrati model), and develop a pipeline based on the COLA method to produce synthetic galaxy catalogues in modified gravity. By performing a comparison of COLA summary statistics with full \nbody{} results, we validate each step of the pipeline and assess the accuracy of the COLA method. Our results show that COLA is able to accurately catch the modified gravity effect on the clustering of galaxies in redshift space, where traces of modified gravity are present in spite of the tuning of the free parameters of the galaxy-halo connection model. We then use the mock galaxy catalogues produced with our pipeline to study the effects of modified gravity and to validate COLA simulations for additional probes of the large-scale structures. These include an estimator of the power spectrum orthogonal to the line of sight $Q_0$, the bispectrum and voids. Comparing $Q_0$ with the real space power spectrum we show that the modified gravity signal contained in the two summary statistics is consistent and that COLA accurately reproduces the \nbody{} results. In the bispectrum of dark matter instead, we find that COLA simulations, due to the screening approximation that they use, lack the modified gravity signal coming from the fifth force non-linearity in $f(R)$ theory. Nonetheless, we show that this is not a problem for the monopole of bispectrum of galaxies in redshift space as this is dominated by non-linearities in the bias model, i.e., the model connecting dark matter with galaxies. We then look at how $f(R)$ and nDGP theories affect the profiles of voids, finding more modified gravity signatures in nDGP than in $f(R)$ theories. By applying a linear model for the redshift space distortions in voids we are able to recover unbiased estimates for the linear growth rate in all gravity theories, even when modified gravity is not taken into account in the redshift space distortion model. We also show that COLA results for voids are consistent with the \nbody{} results.

While much faster than full {\it N}-body simulations, COLA simulations are still too computationally expensive to be directly used for cosmological inference which requires $\sim 10^5$ evaluations of theoretical predictions. Here emulation techniques come in help, requiring as little as $\sim 100$ theoretical predictions to create smooth interpolating functions that cover a wide range of the theory parameter space. To pave the way in this direction, we study the convergence of COLA simulations for predictions of the matter power spectrum by increasing the force, time and mass resolutions employed in the simulations. We find that to achieve convergence it is necessary to increase the three resolutions accordingly. Then we explore the possibility of extending cosmological emulators with COLA simulations using the response function for the matter power spectrum, i.e., the response of the matter power spectrum to changes in cosmological parameters. By comparing COLA predictions of the response function with state-of-the-art cosmological emulators we show that COLA is more accurate in predicting the response function than the power spectrum itself. Finally, we demonstrate the potential of COLA simulations for the extension of cosmological emulators to modified gravity theories producing a suite of simulations in nDGP gravity that we employ to train an emulator for the modified gravity boost factor, i.e., the ratio of the power spectrum in modified gravity with that in general relativity.

The thesis is structured as follows:
\begin{itemize}
    \item in chapter~\ref{chp:introduction} we give a general overview of the main concepts at the base of this work, including the cosmological model, modified gravity theories, the large-scale structure of the universe and \nbody{} simulations;
    \item in chapter~\ref{chp:mocks} we present the pipeline for the efficient production of mock galaxy catalogues with the COLA method and validate it by performing an extensive comparison of summary statistics with full \nbody{} simulations in modified gravity;
    \item in chapter~\ref{chp:statistics} we test the validity of the COLA method for additional probes of the large-scale structure, a real space power spectrum estimator, bispectrum and voids, and investigate if they can help constrain modified gravity theories;
    \item in chapter~\ref{chp:powerspectrum} we study the feasibility of using COLA simulations to accurately extend cosmological emulators of the matter power spectrum to modified gravity theories and give an explicit example in the case of nDGP theory producing an actual emulator;
    \item in chapter~\ref{chp:conclusion} we summarise the main results discussed in this thesis, draw the conclusions and discuss future prospects.
\end{itemize}

\end{abstract}

\chapter*{Notation}
In this work we will use the natural units
\begin{equation*}
c = \hslash \equiv 1
\end{equation*}
so that times and distances will be both of the same dimension of the inverse of energy.

Our metric signature will be $ ( - + + + ) $.

 Greek indices will take the values ${0, 1, 2, 3} $. The latin indices $i,j,k$ will be used for the space dimensions so they will run over ${1,2,3}$. These statements hold unless other definitions are explicitly made. We will adopt the Einstein notation for repeated indices: when an index variable appears twice in a single term, it implies summation of that term over all the values of the index.

 Our Fourier convention will be
\begin{equation*}
\tilde{f}(\vec{k}) = \int d^3x f(\vec{x}) e^{-i \vec{k} \cdot \vec{x}}\,.
\end{equation*}

We will use the subscript index ``$0$'' to denote the present-day values of variables, unless otherwise stated.

The Hubble expansion rate will be indicated through the letter $H$ and its conformal version with $\mathcal{H}$. The dot derivatives $\dot{x}$ will stand for time derivative while the prime derivative $x'$ will be for the conformal time derivative. The partial derivative with respect to the variable $x^{\mu}$ will be written $\partial_{\mu}: \frac{\partial}{\partial x^{\mu}}=\partial_{\mu}$.

\clearpage

\chapter*{Acronyms}

{
\flushleft
\begin{tabular}{p{2cm}l}
     \textbf{2LPT} & second-order Lagrangian Perturbation Theory \\
     \textbf{AMR} & Adaptive mesh refinement  \\
     \textbf{BAO} & Barion Acoustic Oscillations  \\
     \textbf{CCF} & Cross-Correlation Function \\
     \textbf{CMB} & Cosmic Microwave Background  \\
     \textbf{COLA} & COmoving Lagrangian Acceleration \\
     \textbf{DE} & Dark Energy  \\
     \textbf{DGP} & Dvali-Gabadadze-Porrati  \\
     \textbf{DM} & Dark Matter  \\
     \textbf{EE2} & Euclid Emulator 2 \\
     \textbf{F5} & Hu-Sawicki $f(R)$ model with $\left|f_{R 0}\right|=10^{-5}$ \\
     \textbf{FLRW} & Friedmann-Lemaitre-Robertson-Walker  \\
     \textbf{FoF} & Friends-of-Friends \\
     \textbf{GR} & General Relativity  \\
     \textbf{HOD} & Halo Occupation Distribution \\
     \textbf{IC} & Initial Conditions  \\
     \textbf{LHS} & Latin-Hypercube Sampling \\
     \textbf{LSS} & Large Scale Structure  \\
     \textbf{MG} & Modified Gravity  \\
     \textbf{N1} & nDGP model with $H_0r_c = 1$ \\
     \textbf{nDGP} & normal branch of DGP theory   \\
     \textbf{NFW} & Navarro-Frenk-White  \\
     \textbf{PCA} & Principal Components Analysis \\
     \textbf{PM} & Particle Mesh  \\
     \textbf{RSD} & Redshift-Space Distortions  \\     
     \textbf{SHAM} & Sub-Halos Abundance Matching  \\
     \textbf{SO} & Spherical Over-density \\
\end{tabular}
}


\tableofcontents

\listoffigures

\listoftables


\printnomenclature

\mainmatter


\chapter{Introduction}
\label{chp:introduction}

\begin{quote}
    {\it The content of section~\ref{sec:MG} in this chapter is based on the publication \cite{Fiorini:2021dzs}. } 
\end{quote}

\section{Background cosmology}

The standard model of cosmology is based on the assumptions that the universe is homogeneous and isotropic on large-enough scales and that General Relativity (GR) is the correct description of gravity on all scales of cosmological interest. With these assumptions, the metric of space-time can be described using spherical coordinates with the Friedmann-Lemaitre-Robertson-Walker (FLRW) metric given by the line element
\begin{equation}\label{FLRWmetric}
    d s^{2}=-d t^{2}+a(t)^{2}\left(\frac{1}{1-\kappa \chi^{2}} d \chi^{2}+\chi^{2} d \theta^{2}+\chi^{2} \sin ^{2}(\theta) d \phi^{2}\right)
\end{equation}
where $a$ is a time-dependent scale parameter and $\kappa$ describes the spatial curvature of the universe. The coordinate $\chi$ is a {\it comoving} coordinate and it is linked to the physical coordinate $r$ by
\begin{equation}
    r \equiv a(t)\chi \, .
\end{equation}

This definition is useful to describe the expansion of the universe by means of the scale factor. 
The energy-momentum tensor $T_{\mu \nu}$ is related to the metric of space-time $g_{\mu \nu}$ through the Einstein equations
\begin{equation}\label{EinsteinEquation}
    R_{\mu \nu}-\frac{1}{2} R g_{\mu \nu}=8 \pi G T_{\mu \nu} \, ,
\end{equation}
where $R_{\mu \nu}$ is the Ricci tensor and $R$ is the Ricci scalar, describing the curvature of space-time.
Solving Einstein's equations with the metric~\eqref{FLRWmetric} and using the energy-momentum tensor of a perfect fluid
\begin{equation}
    T_{\mu}^{\nu}=\left(\begin{array}{cccc}
    -\rho & 0 & 0 & 0 \\
    0 & P & 0 & 0 \\ 
    0 & 0 & P & 0 \\ 
    0 & 0 & 0 & P\end{array}\right) \, ,
\end{equation}
where $\rho$ is the energy density and $P$ is the pressure of the fluid, gives the Friedmann equations
\begin{gather}
    \left(\frac{\dot{a}}{a}\right)^{2}=\frac{8 \pi G}{3} \rho-\frac{\kappa}{a^{2}}\,, \label{Friedmann1}\\ 
    \frac{\ddot{a}}{a}=-\frac{4 \pi G}{3}(\rho+3 P) \label{Friedmann2} \, .
\end{gather}

In the standard model of cosmology the content of the universe can be described as the sum of three perfect fluids with different equations of state: 
\begin{itemize}
    \item matter, including normal matter (referred to as {\it baryons} in cosmology) and cold Dark Matter (DM), with $P_{\rm m} = 0 $;
    \item radiation, including photons and relativistic neutrinos, with $P_{\rm r} = \frac{1}{3}\rho_{\rm r} $;
    \item Dark Energy (DE), in the form of a cosmological constant $\Lambda$ entering the Einstein equation~\eqref{EinsteinEquation} on the right hand side, with $P_\Lambda = -\rho_\Lambda $ and $\rho_\Lambda = \frac{\Lambda}{8 \pi G}$.
\end{itemize}
Due to the assumptions it makes for the dark components (cosmological constant and cold DM), the standard model of cosmology is also called $\Lambda$CDM model.
Defining the Hubble rate $H\equiv \frac{\dot{a}}{a}$, it is possible to express the Friedmann equations as
\begin{equation}
    \left(\frac{H}{H_{0}}\right)^{2}=\Omega_{\rm m}+\Omega_{\rm r}+\Omega_{\Lambda}+\Omega_{\kappa}
\end{equation}
where $H_{0}$ is the present-day Hubble rate and the dimensionless energy density $\Omega_{i} \equiv \frac{\rho_{i}}{\rho_{c}}$ with $\rho_{c}=\frac{3 H_{0}^{2}}{8 \pi G}$. We have also defined the dimensionless energy density of curvature $\Omega_{\kappa}$ in the same way, with $\rho_{\kappa}=-\frac{3 \kappa}{8 \pi G a^{2}}$.

The joint constraints of the Cosmic Microwave Background (CMB) measurements from the Planck mission \cite{Planck:2018vyg} and the Baryon Acoustic Oscillations (BAO) measurements from galaxy surveys \cite{Beutler:2011hx,Ross:2014qpa,BOSS:2016wmc} are consistent with a null value of $\Omega_{\kappa}$ so in the following, we will assume a spatially flat universe, $\Omega_{\kappa} = 0$.

The supernovae observations show strong evidence for an accelerated expansion of the universe at present time, consistent with the presence of a DE component \cite{SupernovaSearchTeam:1998fmf,SupernovaCosmologyProject:1998vns} as can be understood from the Friedmann equation~\eqref{Friedmann2}. This DE component, in principle, could be explained by the vacuum energy of fields in the standard model of particle physics. However, the amplitude of such vacuum energy would be $\sim 120$ orders of magnitude larger than what is currently observed assuming that the ultraviolet cut-off scale is the Planck scale.

With the assumption of a flat universe, the CMB observations constrain the present-day values of the dimensionless density parameters and Hubble rate to be \cite{Planck:2018vyg}
\begin{equation}
\begin{array}{cc}
    \Omega_{{\rm m}, 0} = 0.315 \pm 0.007 \, , \quad & \Omega_{\Lambda, 0} = 0.685 \pm 0.007 \, , \\ \Omega_{{\rm r}, 0} \simeq 0.0001 \, , \quad &
    H_0 = 67.4 \pm 0.5 \,  {\rm km} \, {\rm Mpc}^{-1} \,  {\rm s}^{-1} \,.
\end{array}
\end{equation}
It is possible to take into account small inhomogeneities on top of the homogeneous background discussed above. These can be described in a GR context with scalar perturbations of the FLRW metric and the energy-momentum tensor in the Newtonian gauge. Having previously set $\kappa=0$, the line element reads
\begin{equation}
    d s^{2}=a^{2}(\tau)\left[-(1+2 \Phi) d \tau^{2}+(1-2 \Psi)\delta_{i j} d x^{i} d x^{j}\right]
\end{equation}
where $\tau$ is the conformal time defined by $\frac{d t}{d \tau} \equiv a $, $\vec{x}$ is the comoving position in cartesian coordinates and $\Phi$ and $\Psi$ are scalar perturbations. 

Considering small perturbations in the energy density due to the inhomogeneous distribution of matter
\begin{equation}
    \rho(\tau, \vec{x}) = \overline{\rho}(\tau) + \overline{\rho}_{\rm m}(\tau)\delta_{\rm m}(\tau, \vec{x})\, ,
\end{equation}
and solving the time-time component of the Einstein equation, we obtain
\begin{equation}\label{PoissonGR}
    \nabla^{2} \Phi=4 \pi G a^{2} \overline{\rho}_{\rm m} \delta_{\rm m} \,,
\end{equation}
which is known as the Poisson equation for the gravitational potential $\Phi$.

\section{Modified Gravity}
\label{sec:MG}

The $\Lambda$CDM model based on GR has been very successful in reproducing various cosmological observations. However, GR still lacks a high energy completion and the $\Lambda$CDM model requires the existence of a highly fine-tuned cosmological constant to explain the current accelerated expansion of the Universe. Furthermore,  the $\Lambda$CDM model has recently been questioned for discrepancies between early-universe and late-universe measurements. In particular, the CMB measurement of $H_0$ \cite{Planck:2018vyg} are in $\sim 5 \sigma$ tension with the value inferred from supernovae \cite{Riess:2020fzl}. Another tension affecting the $\Lambda$CDM model is the $\sim 3 \sigma$ tension in $S_8$ between CMB and stage 3 galaxy surveys \cite{Troster:2019ean, DES:2017qwj, Hildebrandt:2018yau,HSC:2018mrq}. These tensions can be interpreted as hints for new physics beyond the $\Lambda$CDM model that may be due to the underlying gravity theory \cite{Raveri:2019mxg}.

These yet-to-be-solved problems have motivated theorists to formulate alternatives to GR often referred to as modified gravity (MG) theories. A milestone for the systematic study of gravity theories is represented by Lovelock's theorem stating that Einstein’s equations are the only second-order local equations of motion for a single metric derivable from the covariant action in four-dimensional spacetime \cite{Lovelock:1971yv,Lovelock:1972vz,Li:2020uaz}. This implies that MG theories need to break at least one of the assumptions of Lovelock's theorem and this is one of the main approaches guiding the theoretical efforts in the search for models beyond GR. The MG theories of interest to cosmology are the ones showing infrared modifications, which are often associated with an additional force (referred to as the fifth force), mediated by a scalar field. The knowledge of physically motivated gravity models is useful for efficient tests of gravity since it reduces the outcome-space of deviations from GR with respect to model-agnostic approaches.

Solar System tests put very stringent constraints on gravity \cite{Will:2014kxa}.
Because of this, theories that feature deviations from GR relevant to cosmology commonly incorporate a screening mechanism \cite{Joyce:2014kja, Koyama:2015vza} that let the theory evade these constraints.

In the following, we give an overview of well-known screening mechanisms and how they are realised in MG theories. Then we focus on two screening mechanisms of our interest and we introduce two MG theories that naturally incorporate these screening mechanisms. For a comprehensive review that covers the topics of this section see \cite{Koyama:2018som}.
 
\subsection{Screening mechanism}
\label{ScreeningMechanisms}
It is possible to describe the screening mechanisms using the extended Brans-Dicke action
\begin{equation}\label{ExtBransDickeAction}\small
S=\frac{1}{16 \pi G} \int d^{4} x \sqrt{-g}\left(\psi R-\frac{\omega(\psi)}{\psi}(\nabla \psi)^{2}+\mathcal{K}\left[(\nabla \psi)^{2},\left(\nabla^{2} \psi\right)\right]-2 U(\psi)\right)+S_{m}\left[g_{\mu v}\right]
\end{equation}
where the scalar field $\psi$ is linearly coupled with the Ricci scalar $R$ and $S_{m}$ is the action of matter. 
The screening mechanisms can be expressed in the extended Brans-Dicke formalism for specific functional forms of $\omega$, $\mathcal{K}$ and $U$:
\begin{itemize}
\item The {\it chameleon} mechanism \cite{Khoury:2003aq,Khoury:2003rn} is obtained through the scalar field potential $U(\psi)$, which gives a density-dependent effective-mass to the scalar field, thus limiting the range of action of the fifth force;
\item The {\it dilaton} \cite{Brax:2010gi} and {\it symmetron} \cite{Hinterbichler:2010es} mechanisms rely on the function $\omega(\psi)$, which controls the coupling of the scalar field to matter. This coupling can be suppressed in high-density regions for appropriate choices of $\omega(\psi)$;
\item The {\it K-mouflage} \cite{Babichev:2009ee} and {\it Vainshtein} \cite{Vainshtein:1972sx} mechanisms use the non-linear kinetic term $\mathcal{K}\left[(\nabla \psi)^{2},\left(\nabla^{2} \psi\right)\right]$ to suppress the coupling with matter. More specifically, they rely on the non-linearity in the first and second derivatives of the scalar field $\psi$, respectively. 
\end{itemize}

In general, the Poisson equation~\eqref{PoissonGR} is modified by the scalar field as
\begin{equation}\label{PoissonMGgeneral}
\nabla^{2} \Phi=4 \pi G a^{2} \rho-\frac{1}{2} \nabla^{2} \psi \, ,
\end{equation}
where $\psi$ is the perturbation over the background value of the scalar field. To close the system of differential equations, the Poisson equation must be complemented with the scalar field equation (Klein-Gordon) which depends on the specific screening mechanism under consideration.

The Klein-Gordon equation for the scalar field under the quasi-static approximation\footnote{In the quasi-static approximation, time derivatives of the scalar field can be neglected compared with spatial derivatives on sub-horizon scales \cite{Noller:2013wca}.} is, in the chameleon mechanism
\begin{equation}
\left(3+2 w_{\rm B D}\right) \nabla^{2} \psi+\partial_{\psi}U(\psi)=-8 \pi G \rho \, ,
\end{equation}
and in the Vainshtein mechanism
\begin{equation}
\left(3+2 w_{\rm B D}\right) \nabla^{2} \psi+\alpha\left[\left(\nabla^{2} \psi\right)^{2}-\left(\nabla_{i} \nabla_{j} \psi\right)^{2}\right]=-8 \pi G \rho \, .
\end{equation}
In the above expressions, $w_{\rm B D}$ and $\alpha$ are model-dependent quantities. $w_{\rm B D}$ is constrained to be larger than $\approx 40000$ by solar system experiments in the absence of screening mechanisms \cite{Will:2014kxa}.
As we will discuss in section~\ref{sec:Nbody}, full {\it N}-body simulations in MG (i.e., without approximations) solve the Klein-Gordon equations with multi-grid techniques in order to fully capture the dynamics of the fifth force. However, screening approximations based on the spherically symmetric solution have been developed and are currently employed to speed up MG simulations in approximate simulation methods \cite{Winther:2014cia}. We discuss these approximations separately for chameleon and Vainshtein screening in the rest of this section after performing a frame transformation.

In the action~\ref{ExtBransDickeAction} the scalar field $\psi$ is coupled with the Ricci scalar $R$ and matter is minimally coupled with the metric $g_{\mu \nu}$. This is referred to as the {\it Jordan frame}. It is possible to remove the linear coupling between the scalar field and the Ricci scalar through the conformal transformation
\begin{equation}
    g_{\mu \nu}^E = \psi g_{\mu \nu} \, .
\end{equation}
The conformally transformed action can be expressed in terms of a field $\phi$ which is minimally coupled with gravity and satisfies
\begin{equation}
    \left(\frac{d \phi}{d \psi}\right)^{2}=\frac{1}{16 \pi G} \frac{3+2 \omega}{\psi^{2}}, \quad A(\phi)=\psi^{-1 / 2}, \quad V(\phi)=\frac{U(\psi)}{8 \pi G \psi^{2}} \, .
\end{equation}
This frame takes the name of {\it Einstein frame}. In the Einstein frame, the matter is coupled with gravity through the effective metric $A^2(\phi)g_{\mu\nu}^E$ which produces an additional force due to the scalar field with coupling $\beta / M_{\rm Pl}$, where 
\begin{equation}
    \beta \equiv M_{\mathrm{pl}} \frac{d \ln A}{d \phi}\, .
\end{equation}

In the ideal case of a static, spherically symmetric source of radius $r_s$ with density $\rho(r<r_{s}) = \rho_{s}$ embedded in a homogeneous background of density $\rho(r>r_{s}) = \rho_{\infty}$, the Klein-Gordon equation for the Chameleon mechanism in the Einstein frame is 
\begin{equation}
    \frac{d}{d r}\left(r^{2} \frac{d \phi}{d r}\right)=r^{2}\left(V_{, \phi}+\frac{\beta(\phi) \rho_{s}(r)}{M_{\mathrm{Pl}}}\right)
\end{equation}
which has the solution 
\begin{equation}
    \phi(r)=\phi_{\infty}+\frac{\left(\phi_{s}-\phi_{\infty}\right) r_{s}}{r} e^{-m_{\infty} r}, \quad r>r_{s} \, ,
\end{equation}
when the gravitational field of the source $\Phi_{N}$ satisfies
\begin{equation}
    \frac{\Delta r_{s}}{r_{s}} \equiv \frac{\left|\phi_{\infty}-\phi_{s}\right|}{2 \beta_{\infty} M_{\mathrm{Pl}} \Phi_{N}} \ll 1 \, 
\end{equation}
where $\phi_{\infty}$ is the background value of the scalar field satisfying $V_{, \phi_{\infty}}+\frac{\beta_{\infty} \rho_s(r)}{M_{\mathrm{Pl}}} = 0$, $\beta_{\infty} \equiv \beta(\phi_{\infty})$ and $m_{\infty} = \left. \frac{d}{d\phi} \left(V_{, \phi}+\frac{\beta(\phi) \rho_s(r)}{M_{\mathrm{Pl}}}\right) \right|_{\phi_{\infty}}$ is the effective mass of the scalar field away from the source.
The above expression defines the screening factor $\frac{\Delta r_{s}}{r_{s}}$, which can be interpreted as the fraction of the radius that the fifth force is able to penetrate inside the source. Due to this condition, the chameleon mechanism is also described as {\it thin-shell} screening. In the case the gravitational potential is not strong enough, the source is unscreened and the fifth force penetrates the source to the centre, $\frac{\Delta r_{s}}{r_{s}}=1$. We can incorporate this into the theory by re-defining 
\begin{equation}
    \frac{\Delta r_{s}}{r_{s}} \equiv \operatorname{Min}\left[\frac{\left|\phi_{\infty}-\phi_{c}\right|}{2 \beta_{\infty} M_{\mathrm{Pl}} \Phi_{N}}, 1\right] \,.
\end{equation}
The force perceived by a unit-mass test particle outside the source is then
\begin{equation}
    \begin{aligned} 
    F &= \frac{G M}{r^{2}} \left[ 1+ 2 \beta_{\infty}^{2} \left(\frac{\Delta r_{s}}{r_{s}}\right)\left(1+m_{\infty} r\right) e^{-m_{\infty} r} \right] \\
    & \simeq \frac{G M}{r^{2}} \left[ 1+ 2 \beta_{\infty}^{2} \left(\frac{\Delta r_{s}}{r_{s}}\right)\right] \quad \text { for } \quad r \ll m_{\infty}^{-1} \\
    & = F_{\rm N} \left[ 1+ 2 \beta_{\infty}^{2} \left(\frac{\Delta r_{s}}{r_{s}}\right)\right] 
    \end{aligned}
\end{equation}
where $M$ is the total mass of the source. If we neglect the non-linearity of the potential $V(\phi)$, the Klein-Gordon equation for the scalar field becomes
\begin{equation}\label{lin_KG_chameleon}
    \nabla^{2} \phi=a^{2} m^{2}(a) \phi+\frac{\beta(a) a^{2} \bar{\rho}_{m}}{M_{\mathrm{Pl}}} \delta_{m}
\end{equation}
where $m(a) \equiv \left. \frac{\mathrm{d}^{2} \left( V+\rho \ln A\right)}{\mathrm{d} \phi^{2} }\right|_{\phi_{\infty}}$ is the effective mass of the scalar field around the cosmological background. It is possible to artificially implement the screening in eq.~\eqref{lin_KG_chameleon} by replacing $\delta_{m}$ with
\begin{equation}
    \delta_{m}^{\mathrm{eff}}=\delta_{m} \times \operatorname{Min}\left[\frac{\phi(a)}{2 \beta(a) M_{\mathrm{Pl}}\left|\Phi_{N}\right|}, 1\right]
\end{equation}
obtaining the linearised Klein-Gordon equation for the scalar field with screening approximation
\begin{equation}
    \nabla^{2} \phi=a^{2} m^{2}(a) \phi+\frac{\beta(a) a^{2} \bar{\rho}_{m}}{M_{\mathrm{Pl}}} \delta_{m}^{\rm eff}
\end{equation}

In the same spherically symmetric settings as before, the Klein-Gordon equation for the Vainshtein mechanism in the Einstein frame is given by
\begin{equation}
    \frac{1}{r^{2}} \frac{d}{d r}\left(r^{2} \frac{d \phi}{d r}\right)+\frac{2}{\Lambda_{s}^{3}} \frac{d}{d r}\left(r\left(\frac{d \phi}{d r}\right)^{2}\right)=\frac{\beta \rho_{m}}{M_{\mathrm{Pl}}}
\end{equation}
which can be integrated to obtain
\begin{equation}
    \frac{d \phi}{r d r}+\frac{2}{\Lambda_{s}^{3}}\left(\frac{d \phi}{r d r}\right)^{2}=2 \beta M_{\mathrm{Pl}} \frac{G M(r)}{r^{3}}
\end{equation}
This can be solved for $F_{\phi} = -\frac{\beta}{M_{\rm Pl}} \nabla_r\phi$ to get
\begin{equation}\label{vainshtein_screening_rv}
    F_{\phi}=F_{N} \times 2 \beta^{2} \times 2\left(\frac{\sqrt{1+\left(r_{V} / r\right)^{3}}-1}{\left(r_{V} / r\right)^{3}}\right)
\end{equation}

where $r_{V}=\frac{1}{\Lambda_{s}}\left(\frac{2 \beta M}{\pi M_{\mathrm{P} 1}}\right)^{1 / 3}$ is the Vainshtein radius $r_{V}$. The ratio $\left(r_{V} / r\right)^{3}$ can be interpreted as a ratio of densities. Defining $\rho(r)=\rho_{m}(<r) \equiv \frac{M(r)}{4 \pi / 3 r^{3}}$ and $\rho_{\mathrm{crit}}=\frac{3 \Lambda_{s}^{3} M_{\mathrm{Pl}}}{8 \beta}$, the screening factor in eq.~\ref{vainshtein_screening_rv} can be expressed as
\begin{equation}
    2\left(\frac{\sqrt{1+\left(r_{V} / r\right)^{3}}-1}{\left(r_{V} / r\right)^{3}}\right) = \frac{2\left(\sqrt{1+\rho(r) / \rho_{\mathrm{crit}}}-1\right)}{\rho(r) / \rho_{\mathrm{crit}}}
\end{equation}

Neglecting the non-linearities, the Klein-Gordon equation for the scalar field becomes
\begin{equation}
    \nabla^{2} \phi=\frac{\beta a^{2} \bar{\rho}_{m}}{M_{\mathrm{Pl}}} \delta_{m}
\end{equation}
where we again implement the screening by replacing $\delta_{m}$ with
\begin{equation}
    \delta_{m}^{\mathrm{eff}}=\delta_{m} \times \frac{2\left(\sqrt{1+\rho(r) / \rho_{\text {crit }}}-1\right)}{\rho(r) / \rho_{\text {crit }}}
\end{equation}
obtaining the linearised Klein-Gordon equation for the scalar field with screening approximation
\begin{equation}
    \nabla^{2} \phi=a^{2} m^{2}(a) \phi+\frac{\beta(a) a^{2} \bar{\rho}_{m}}{M_{\mathrm{Pl}}} \delta_{m}^{\rm eff} \, .
\end{equation}

Chameleon screening has been tightly constrained by astrophysical tests studying the morphology of dwarf galaxies in void regions \cite{Burrage:2017qrf,Desmond:2020gzn}, at the point that the viable models incorporating the chameleon mechanisms are almost undistinguishable from GR for astrophysical and cosmological interest. 
The Vainshtein mechanism is more difficult to test with astrophysical tests, due to the highly efficient nature of this screening mechanism. The strongest constraints come from testing violations of the strong equivalence principle using supermassive black-holes \cite{Bartlett:2020tjd} and from modelling the redshift space distortions in galaxy surveys \cite{Barreira:2016ovx}.
However, it is still interesting to see whether we can get comparable constraints from cosmological observations on non-linear scales.

\subsection{Hu-Sawicki $f(R)$}
The class of theories where the modifications of gravity can be described by an additional term to the Einstein-Hilbert action in the form of a function of the Ricci scalar is known as $f(R)$ theories \cite{Starobinsky:1980te}:
\begin{equation}
    S=\int \mathrm{d}^{4} x \sqrt{-g}\left[\frac{R+f(R)}{16 \pi G}+\mathcal{L}_{\mathrm{m}}\right] \, ,
\end{equation}
where $\mathcal{L}_{\mathrm{m}}$ is the lagrangian of matter.
A sub-set of $f(R)$ theories, proposed in \cite{Hu:2007nk}, is known as the Hu-Sawicki $f(R)$ model and is described by
\begin{equation}
    f(R)=-M^{2} \frac{c_{1}\left(R / M^{2}\right)^{n}}{c_{2}\left(R / M^{2}\right)^{n}+1} \, ,
\end{equation}
where $c_1$ and $c_2$ are dimensionless parameters, $n>0$ is the power law exponent and $M\equiv H_0 \sqrt{\Omega_{m, 0}}$ is a mass scale defined for convenience. This model evades the Solar System constraints by means of the chameleon mechanism \cite{Khoury:2003aq,Khoury:2003rn}, where the density controls the shape of the scalar field potential, determining the background value of the scalar field and its effective mass (large masses in high-density environments and vice-versa) and the fifth force is consequently screened.
In this work, we focus on the Hu-Sawicki model with $n=1$ and  $\frac{c_1}{c_2}=6 \frac{\Omega_{\Lambda,0}}{\Omega_{m,0}}$ which is described by the following expression in the high-curvature limit ($M^2/R \rightarrow0$) \cite{Hu:2007nk}
\begin{equation}
f(R)=-2 \Lambda+\left| f_{R 0} \right| \frac{\bar{R}_{0}^{2}}{R} \, ,
\label{Hu-Sawicki_fR}
\end{equation}
where $\bar{R}_{0}$ is the background curvature today and $f_{R0}=-n \frac{6\Omega_{\Lambda, 0}}{\Omega_{m,0 }c_2}\left(\frac{\Omega_{m, 0}}{3\left(\Omega_{m, 0}+4\Omega_{\Lambda, 0}\right)}\right)^{n+1}$ is the value of $f_{R} \equiv d f / d R$ today which is used as a model parameter in place of $c_2$.
For a small $|f_{R0}|$, the background cosmology can be approximated as the one given by the $\Lambda$CDM model. In the linear regime, it is described in Fourier space by the modified Poisson equation for the gravitational potential $\Phi$ \cite{Pogosian:2007sw}:
\begin{equation}
k^{2} \Phi=-4 \pi G_{\rm eff} a^{2} \delta \rho, \quad  
G_{\rm eff} = G\left(\frac{4+3 a^{2} m^{2} / k^{2}}{3+3 a^{2} m^{2} / k^{2}}\right),
\label{fR_Poisson}
\end{equation}
where $\delta \rho$ is the fluctuation around the mean energy density and $m^{2}=\frac{1}{6\left| f_{R 0} \right|} \frac{\bar{R}^{3}}{\bar{R}_{0}^{2}}$ is the effective mass of the scalar field. Depending on the scale and cosmic time, the effective gravitational constant $G_{\rm eff}$ in Eq.~\eqref{fR_Poisson} has two limits
\begin{align}
& ma \gg k : G_{\rm eff} \rightarrow G  \, ,\\
& ma \ll k : G_{\rm eff} \rightarrow \frac{4}{3} G\, ,
\end{align}
where in the former case the deviations from GR are suppressed while in the latter gravity is enhanced by the fifth force.

In the non-linear regime, eq.~\eqref{fR_Poisson} for the Poisson equation does not hold and it is necessary to solve the non-linear scalar field equation with a potential. 
The chameleon screening is determined by the gravitational potential of the object as well as the environment. As we have seen in subsection~\ref{ScreeningMechanisms}, the effect of Chameleon screening can be approximated as \cite{winther15}
\begin{equation}
\label{Geff_fR}
G_{\rm eff} = \left(  1 + \frac{1}{3}  
\frac{k^2}{k^2 + a^2 m^2} 
 \text{Min}\left[1, f_s \, \left|\frac{\Phi_{\rm crit}(a)}{\Phi}\right|\right]   \right)
G,
\end{equation}
where we introduced the empirical parameter $f_s$, that we shall call screening efficiency, to tune the strength of the screening and
\begin{equation}
    \Phi_{\rm crit}(a) = \frac{3f_{R0}}{2}\left(\frac{\Omega_{m,0} + 4\Omega_{\Lambda,0}}{\Omega_{m,0} a^{-3} + 4\Omega_{\Lambda,0}} \right)^{2} \, .
\end{equation}
The quantity $\Phi_{\rm crit}$ depends only on background quantities and controls (together with $f_s$) the threshold value of the gravitational potential for which the fifth force is screened in high density regions.


\subsection{Dvali-Gabadadze-Porrati}
\label{dgp}
Another way to modify gravity is by defining a theory in a higher-dimensional space under the assumption that we are living in a 4D brane of the high dimensional space-time. This class of model is often known as braneworld gravity and the simplest of these is the Dvali-Gabadadze-Porrati (DGP) model \cite{Dvali:2000hr}, which is characterised by the action:
\begin{equation}
S=\frac{1}{16 \pi G_{(5)}} \int d^{5} x \sqrt{-\prescript{(5)}{}{g}}\prescript{(5)}{}{R}+\int d^{4} x \sqrt{-g}\left(\frac{1}{16 \pi G} R+\mathcal{L}_{m}\right) \, ,
\end{equation} 
where the index (5) indicates the quantity is the 5-D generalisation of a 4-D quantity.
The normal branch of this model (nDGP) still needs an additional dark energy component to explain the late-time acceleration and the tuning of the cross-over radius $r_{c} \equiv G_{(5)} / 2 G$ to recover the standard cosmology at early times.

Although this model is not theoretically well motivated, it is useful as a toy model to study the effects of its screening mechanism. Indeed focusing on the normal branch, under quasi-static conditions, the theory incorporates the Vainsthein mechanism which screens the fifth force in high curvature environments but allows for deviation from GR in low curvature environments. We assume that the background cosmology is the same as the $\Lambda$CDM model by introducing the appropriate dark energy in the model \cite{Schmidt:2009sg, Bag:2018jle}. Linearising the equation of motion, the resulting modified Poisson equation is \cite{Koyama:2005kd}:
\begin{equation}
\nabla^{2} \Phi=4 \pi G_{\rm eff} a^{2}\rho \delta, 
\quad 
G_{\rm eff}= G \left(1+\frac{1}{3 \beta}\right),
\label{DGP_Poisson}
\end{equation}
where $\beta=1-2 H r_{\mathrm{c}}\left(1+\frac{\dot{H}}{3 H^{2}}\right)$ and the overdot denotes the derivative with respect to the physical time. 
In the case of the Vainshtein mechanism, the scalar field equation satisfies a non-linear equation where non-linearity appears in the second derivative of the scalar field. The Vainshtein screening may be approximated using a density-dependent effective Newton constant. For the DGP model, the approximation discussed in subsection~\ref{ScreeningMechanisms} is given by \cite{winther15}
\begin{equation}
\label{Geff_DGP}
G_{\rm eff} = \left[ 1+ \frac{f_{s}}{3 \beta}\left( \frac{2(\sqrt{1 + x}-1)}{x}  \right)\right] G , \quad 
x = \frac{8(r_cH_0)^2\Omega_{m,0}}{9\beta^2 a^3}\frac{\rho}{\overline{\rho}}, 
\end{equation}
where $\rho$ is the average density within a given radius, $\bar{\rho}$ is the background density and $f_{s}$ is the screening efficiency, an empirical parameter that we introduce to avoid over-screening at early time by setting $f_{s} = a^3$. 
Note that this approximation holds only for spherical mass distribution and can violate the condition that the theory recovers the linear prediction on large scales \cite{Schmidt:2009yj}. Also, the screening depends on the smoothing radius of the density field $\rho$, which needs to be tuned to match the exact result.



\section{Large scale structure}
\label{sec:LSS}

The cosmic structure that we observe today on very large scales ($\sim$ Mpc) in galaxy surveys can be understood as the result of the gravitational evolution of an ``initially'' almost homogenous energy distribution across the visible universe. This very homogenous initial state was characterised by extremely small inhomogeneities ($\sim10^{-5}$) that we can observe today in the CMB. These inhomogeneities have been amplified by the gravitational evolution driven by collisionless cold dark matter. This section aims to introduce the key elements of the LSS of the universe that will form the foundations for the analysis of the following chapters. For more on the topics discussed in this section see \cite{Bertschinger:1998tv,Bernardeau:2001qr,Dodelson:2003ft,Wechsler:2018pic}.



\subsection{Linear theory}
\label{ssec:LinearTheory}
We consider the distribution of cold matter in a patch of the universe large enough that its average density can be approximated with the density of the observable universe for all practical purposes. We focus on a central region of the patch so that the inhomogeneities of the nearby patches are so far apart that they have negligible impact on the gravitational field in the region of interest. We assume that the universe is flat and filled with cold matter and non-clustering DE so that the density perturbations are entirely due to perturbations in the matter distribution.
Let $\overline{\rho}(t) = \overline{\rho}_{\rm m}(t)+ \overline{\rho}_{\rm \Lambda} $ be the time-dependent density of the observable universe. The density in the patch can be described by
\begin{equation}
    \rho (\vec{r},t ) = \overline{\rho}(t) + \delta\rho_{\rm m}(\vec{r},t )\, .
\end{equation}
The gravitational field $\vec{\mathcal{F}}$ due to this density distribution in the Newtonian approximation is given by
\begin{equation}
    \vec{\mathcal{F}}(\mathbf{r},t)=-G \int \rho(\mathbf{s},t) \frac{(\mathbf{r}-\mathbf{s})}{|\mathbf{r}-\mathbf{s}|^3} d^3 s \, .
\end{equation}
Taking the divergence of both sides of the latter equation yields
\begin{equation}\label{eq:GaussLaw}
    \vec{\nabla}_r \cdot \vec{\mathcal{F}}(\mathbf{r},t)=-4 \pi G \rho(\mathbf{r},t) 
\end{equation}
where we have made use of the tensor calculus relation $\vec{\nabla}_r \cdot \frac{(\mathbf{r}-\mathbf{s})}{|\mathbf{r}-\mathbf{s}|^3} = 4 \pi \delta^D(\vec{r}-\vec{s})$, with $\delta^D$ the Dirac delta distribution.
The gravitational field $\vec{\mathcal{F}}$ is irrotational and can therefore be expressed as the gradient of a scalar field which we denote $\Phi_{\rm N}$, the Newtonian potential,  
\begin{equation}\label{eq:NewtPot}
    \vec{\mathcal{F}} \equiv -\vec{\nabla}\Phi_{\rm N} \, .
\end{equation}
Using the definition~\eqref{eq:NewtPot} in eq.~\eqref{eq:GaussLaw} leads to the {\it Poisson equation}
\begin{equation}\label{PoissonTotal}
    \nabla_{\vec{r}}^{2} \Phi_{\rm N}=4 \pi G \rho \, ,
\end{equation}
which, being a differential equation, requires boundary conditions to be solved. Since the universe expands according to the Friedmann equations\footnote{This can be equally thought as descending from GR results or the supernovae observation \cite{SupernovaSearchTeam:1998fmf}.}, an effective gravitational field pulls the matter away from the centre of the reference frame. This gravitational field can be deduced from the Hubble law $\vec{v}_{\rm b g} = H \vec{r}$, by computing the time derivative of the velocity of a comoving object
\begin{equation}
    \vec{\mathcal{F}}_{\rm b g}=\frac{d(H \vec{r})}{d t}=\frac{d(\mathcal{H} \vec{x})}{a\, d \tau} =  \frac{\mathcal{H}'}{a} \, \vec{x} \, .
\end{equation}
To solve the Poisson equation, we impose the boundary condition that the gravitational field computed from the gravitational potential matches the background gravitational field $\vec{\mathcal{F}}_{\rm b g}$ on average on the surface of the patch considered. This fixes the shape of the gravitational potential modulo a constant term that we set to zero. We express the gravitational potential as the sum of the background value due to the boundary conditions and a term due to the inhomogeneities 
\begin{equation}
    \Phi_{\rm N} = \Phi_{\rm b g}(t, r) + \phi(t,\vec{r}) \, 
\end{equation}
where $\Phi_{\rm b g}(t) =  - \mathcal{H}'  \, \frac{x^2}{2} $. Converting the Poisson equation in comoving coordinates and using the first Friedmann equation it is possible to show that  
\begin{equation}
    \nabla_{\vec{x}}^{2} \Phi_{\rm bg}=4 \pi G a^2 \overline{\rho}
\end{equation}
and the Poisson equation becomes
\begin{equation}\label{PoissonNewtonian}
    \nabla_{\vec{x}}^{2} \phi=4 \pi G a^2 \overline{\rho}_{\rm m} \delta_{\rm m}
\end{equation}
which shows that, in the Newtonian description, the gravitational potential $\phi$ is sourced by the matter density perturbation $\delta_{\rm m} \equiv \frac{\delta \rho_{\rm m}}{\overline{\rho}_{\rm m}}$.

The conservation of mass implies that the flux of momentum between a volume $V$ and the surrounding environment through the surface $S$ is the only responsible for the variation in time of the total matter enclosed in the volume $V$
\begin{equation}
    \frac{\partial}{\partial t} \int_{V} \rho  = - \oint_S \rho \vec{u} \cdot dS
\end{equation}
where $\vec{u}$ is the velocity field of matter. Applying the Gauss-Ostrogradsky theorem to the right-hand side, and exploiting the fact that the resulting equation must hold for every arbitrary volume $V$, gives the continuity equation
\begin{equation}
    \frac{\partial \rho}{\partial t}+\vec{\nabla}_{\vec{r}} \cdot(\rho \vec{u})=0 \, .
\end{equation}
The latter can be expressed in comoving coordinates as\footnote{The second term in eq.~\eqref{ComovContEq_exp} is due to the different slicing of physical and conformal time coordinates.}
\begin{equation}\label{ComovContEq_exp}
        \frac{\partial \rho}{\partial \tau} - \mathcal{H}\vec{x}\cdot\vec{\nabla}_{\vec{x}}\rho +\vec{\nabla}_{\vec{x}} \cdot(\rho \vec{v}+ \rho \mathcal{H} \vec{x}) =0 \, .
\end{equation}
where $v$ is the peculiar velocity, and expanding $\rho = \overline{\rho} (1+ \delta)$, gives at first order in the perturbations
\begin{equation}
        \frac{\partial \delta}{\partial \tau}  + \theta = 0 \, .
\end{equation}
where $\theta=\vec{\nabla}_{\vec{x}} \cdot\vec{v}$ is the velocity-divergence potential.
Applying equivalent reasoning to the conservation of momentum in the absence of forces leads to
\begin{equation}
    \left(\partial_{t}+\vec{u} \cdot \vec{\nabla}_{\vec{r}}\right) \vec{u}= 0 \, ,
\end{equation}
which becomes the Euler equation,
\begin{equation}\label{EulerEquation}
    \left(\partial_{t}+\vec{u} \cdot \vec{\nabla}_{\vec{r}}\right) \vec{u}=-\vec{\nabla}_{\vec{r}} \Phi \, ,
\end{equation}
when gravitational forces are taken into account. The linearised Euler equation can be expressed in comoving coordinates as
\begin{equation}
    \partial_{\tau} \vec{v}+\mathcal{H} \vec{v}= -\vec{\nabla}_{\vec{x}} \phi
\end{equation}

Continuity, Euler and Poisson equations form a full set of differential equations uniquely determining the evolution of perturbations. In particular, taking the comoving divergence of the Euler equation and using the results of continuity and Poisson equations on its left and right-hand side respectively, we obtain the second order differential equation in the density contrast
\begin{equation}\label{DensContr_DiffEq}
     \delta''+\mathcal{H}\delta'-  \frac{3}{2}\Omega_{\rm m} H_0^2\delta = 0 \, .
\end{equation}
We decompose the density contrast in two factors, one encapsulating its scale dependence and the other encapsulating its time dependence, $\delta(\tau, \vec{x}) = D(\tau) \delta_{\rm ini}(\vec{x})$, and substitute in the eq.~\eqref{DensContr_DiffEq} to obtain the differential equation
\begin{equation}\label{GrowthFactDiffeq}
    D^{\prime \prime}+\mathcal{H} D^{\prime}-\frac{3}{2} a^2 \Omega_{\mathrm{m}} H_0^{2} D=0 \, .
\end{equation}
The quantity $D(\tau)$ is often referred to as the {\it linear growth factor} as it describes the growth of density perturbation on large scales, where linear theory is accurate. The differential equation~\eqref{GrowthFactDiffeq} has both growing and decaying modes. From the growing one, $D^{(+)}$, we define the {\it linear growth rate}
\begin{equation}\label{LinearGrwothRate}
    f \equiv \frac{\mathrm{d} \ln D^{(+)}}{\mathrm{d} \ln a} \, ,
\end{equation}
that will be useful in subsection~\ref{ssec:RSD_Intro} and in section~\ref{sec:Voids}.

This treatment is valid in a flat universe with cold matter and DE, therefore it is believed to be a good description of our universe during the era of matter and DE domination. The presence of relativistic species requires a GR treatment and affects both the gravitational potential and the background expansion but their effect on the growth function can be neglected. 

\begin{figure}
\centering
\includegraphics[width=.8\textwidth]{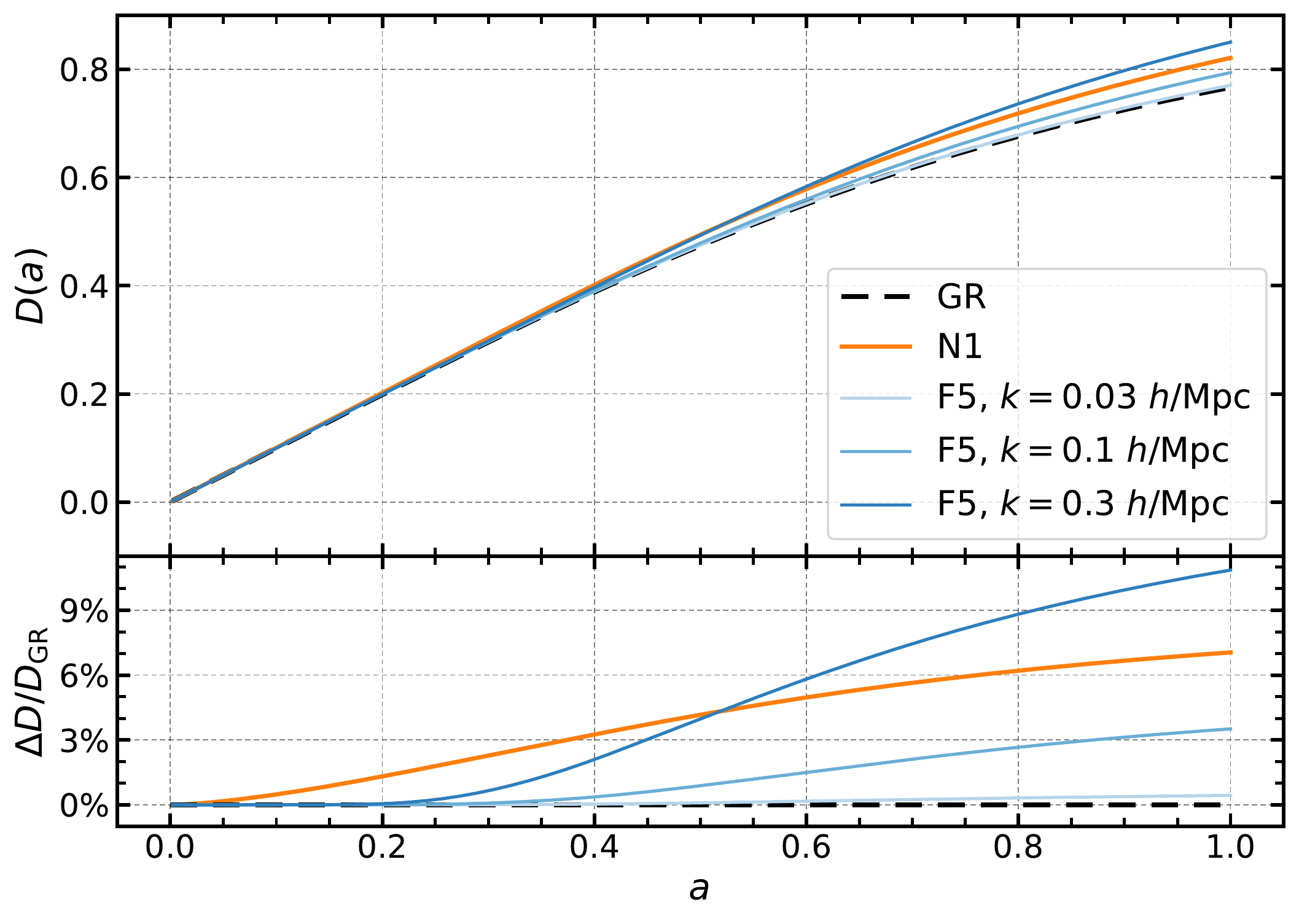}
\caption{\label{fig:nDGP_fR_GR_growth} Comparison of linear growth factors as a function of the scale factor $a$ in GR (black dashed line), in N1 (orange line) and in F5 gravity (blue lines). The growth factor in F5 gravity is scale-dependent so we show it at three different scales, $k= 0.03$, $0.1$ and $0.3 \hompc$ (from light to dark blue lines). The bottom panel shows the fractional difference between the growth factors in MG with that in GR. }
\end{figure}

When MG is taken into account the Poisson equation is modified by the introduction of a (possibly) scale-dependent Newton constant in Fourier space as discussed in section \ref{sec:MG}. The growth equation in MG theories characterised by an effective Newton constant $G_{\rm eff}$ reads
\begin{equation}\label{MG_growth_equation}
    D^{\prime \prime}(\tau, k)+\mathcal{H} D^{\prime}(\tau, k)-\frac{3}{2}\frac{G_{\rm eff}(\tau, k)}{G} a^2 \Omega_{\mathrm{m}} H_0^{2} D(\tau, k)=0 \, .
\end{equation}
This gives rise to a scale-dependent linear growth rate if $G_{\rm eff}$ is scale-dependent. Otherwise, the linear growth rate is scale independent as in the GR case. 
The solution of the growth factor as a function of the scale factor can be computed with numerical methods for a fixed $\Omega_{\rm m}$. In figure~\ref{fig:nDGP_fR_GR_growth} %
we show a comparison of linear growth factors in GR (black dashed line), in the nDGP theory with parameter $H_0 r_c = 1$ (N1, orange line) and in the Hu-Sawicky $f(R)$ theory with $n=1$ and $|f_{R0}| = 10^{-5}$ (F5, blue lines), obtained with $\Omega_{\rm m} = 0.281$. Since $G_{\rm eff}$ in $f(R)$ theories is scale dependent, we show the linear growth factor in F5 at 3 different scales $k= 0.03$, $0.1$ and $0.3 \hompc$ (from light to dark blue lines). At present time ($a=1$) the linear growth factor in N1 is $\approx 7\%$ larger than in GR. Also in F5 the linear growth factor is enhanced at $a=1$ compared to GR, but the enhancement varies with the scale considered, from $\approx 0.5 \%$ at $k = 0.03 \hompc$ to $\approx 11 \%$ at $k = 0.3 \hompc$. This can be understood as the result of the effective mass of the scalar field in F5 that limits the range of interaction of the fifth force leaving the growth on larger scales almost unaffected. 

To derive eq~\ref{MG_growth_equation}, we implicitly used the Fourier transform of the density contrast
\begin{equation}
    \tilde{\delta}(\vec{k},\tau) = \int  \delta(\vec{x},\tau) e^{-i\vec{k}\cdot\vec{x}} \, d^3x \, ,
\end{equation}
and separated again the scale and time dependencies of the density contrast $\tilde{\delta}(\vec{k},\tau) = D(k, \tau) \tilde{\delta}_{\rm ini}(\vec{k})$
The Fourier transform of the density contrast is useful to define the power spectrum 
\begin{equation}
    \left\langle\tilde{\delta}(\vec{k},\tau) \tilde{\delta}\left(\vec{k}^{\prime},\tau \right)\right\rangle \equiv(2 \pi)^3 P(k,\tau) \delta_D\left(\vec{k}+\vec{k}^{\prime}\right) \, ,
\end{equation}
a summary statistics which contains important information of the density field. In the regime of applicability of linear theory, it is possible to show that the power spectrum scales as the square of the growth factor
\begin{equation}
    P(k,\tau) = D^2(k,\tau) P_{\rm ini}(k) \,,
\end{equation}
where we stress again that $D(k, \tau)$ depends on the scale only for theories where $G_{\rm eff}$ is scale-dependent.

\subsection{Lagrangian perturbation theory}
\label{ssec:LPT}
Instead of studying the evolution of the density contrast of given {\it volume elements} using {\it Eulerian} coordinates $\vec{x}$ as done in the previous sub-section, it can be convenient to track the motion of {\it fluid elements} identified with {\it Lagrangian} coordinates $\vec{q}$ using a displacement field $\vec{\DispVec}$ \cite{Bernardeau:2001qr}. 
The two coordinates' systems are related by the mapping
\begin{equation}\label{eq:EulLagMap}
    \vec{x}=\vec{q}+\vec{\DispVec}(\vec{q}, \tau) \, .
\end{equation}
The equation of motion for the fluid elements in the Newtonian approximation is given by
\begin{equation}
    \frac{\mathrm{d}^2 \vec{x}}{\mathrm{d} \tau^2}+\mathcal{H}(\tau) \frac{\mathrm{d} \vec{x}}{\mathrm{d} \tau}=-\vec{\nabla}_x \phi
\end{equation}
which can be expressed in terms of the displacement field as
\begin{equation}
    \frac{\mathrm{d}^2 \vec{\DispVec}}{\mathrm{d} \tau^2}+\mathcal{H}(\tau) \frac{\mathrm{d} \vec{\DispVec}}{\mathrm{d} \tau}=-\vec{\nabla}_x \phi
\end{equation}
or, using the super-comoving time coordinate $\eta$ such that $ d\eta = a d\tau = a^2 dt$,
\begin{equation}\label{eq:LagEqOfMot_SupCom}
    \frac{\mathrm{d}^2 \vec{\DispVec}}{\mathrm{d} \eta^2}=-\vec{\nabla}_x \phi_{\rm sc} \, ,
\end{equation}
where we introduced the super-comoving gravitational potential $\phi_{\rm sc} \equiv a^2\phi$.
Assuming that the mapping~\eqref{eq:EulLagMap} is bijective, it is possible to express the displacement field in terms of the density contrast by means of the mass conservation relation
\begin{equation}\label{eq:LagJacob}
    \int_V d^3 x \rho_{\mathrm{m}}(\vec{x},\eta)=\int_{V} d^3 q \bar{\rho}_{\mathrm{m}}(\eta) \quad \Rightarrow \quad (1+\delta) \mathrm{J}_{\vec{x},\vec{q}}=1
\end{equation}
where $\mathrm{J}_{\vec{x},\vec{q}} \equiv \det (\frac{\partial x_i}{\partial q_j})$. Taking the divergence of eq.~\eqref{eq:LagEqOfMot_SupCom} and using the Poisson equation~\eqref{PoissonNewtonian} we get
\begin{equation}\label{eq:deta2Psi}
    \frac{\mathrm{d}^2 \DispVec_{i,i}}{\mathrm{d} \eta^2} -\DispVec_{j, i} \frac{d^{2}}{d \eta^{2}} \DispVec_{i, j} = - 4 \pi G \bar{\rho}_{\mathrm{m}} \delta \, ,
\end{equation}
 where $\DispVec_{i, j} \equiv \partial \DispVec_i / \partial q_j$. 
Expanding both the density contrast $\delta$ and the displacement field $\vec{\DispVec}$ in a perturbative series in terms of a small parameter $\epsilon$,
\begin{gather}
    \vec{\DispVec} = \epsilon \vec{\DispVec}^{(1)}+ \epsilon^2\vec{\DispVec}^{(2)}+...  \,, \\     
    \delta = \epsilon \delta^{(1)}+ \epsilon^2\delta^{(2)}+...     \,,
\end{gather}
allows us to find the solutions of the set of equations~\eqref{eq:LagJacob} and~\eqref{eq:deta2Psi} order by order in perturbation theory. This framework is known as Lagrangian Perturbation Theory (LPT) \cite{Bernardeau:2001qr}. 
We assume that the curl of the displacement field vanishes so that the displacement field can be expressed as the gradient of a potential field, $\vec{\DispVec} = \vec{\nabla} \phi_{\rm LPT}$. 
Factoring out the time dependence of the displacement field as $\vec{\DispVec}^{(i)}(\eta,\vec{q}) = D_{(i)}(\eta) \vec{\DispVec}^{(i)}(\eta =\eta_{\rm ini},\vec{q})$, the first order solution is given by
\begin{gather}
    \nabla_q^2  \phi_{\rm LPT}^{(1)}=-D_1(\eta) \delta_{\rm ini}(\vec{q})\, , \\
    \frac{d^2 D_1}{d \eta^2}-4 \pi G \bar{\rho}_{\mathrm{m}} {D_1}=0 \, ,
\end{gather}
Truncating the perturbation theory expansion at the first order gives rise to the so-called Zel'dovich approximation \cite{Zeldovich:1969sb}.
The second-order Lagrangian perturbation theory (2LPT) solution is given by
\begin{gather}
    \nabla_q^2 \phi_{\rm LPT}^{(2)}=\frac{1}{2} D_2(\eta) \sum_{i \neq j}\left(\DispVec_{i, i}^{(1)} \DispVec_{j, j}^{(1)}-\DispVec_{i, j}^{(1)} \DispVec_{j, i}^{(1)}\right) \\
    \frac{d^2 D_2}{d \eta^2}-4 \pi G \bar{\rho}_{\mathrm{m}} D_2=-4 \pi G \bar{\rho}_{\mathrm{m}} D_1^2 \, .
\end{gather}
When MG is taken into account, the Poisson equation is in general modified by a time and scale dependent term as in eq.~\eqref{PoissonMGgeneral}. In this case, the LPT solutions can be found mode by mode in Fourier space in terms of modified scale-dependent growth factors \cite{Valogiannis:2016ane,Winther:2017jof,Aviles:2017aor}.

\subsection{Halos}
The growth factor equation~\eqref{GrowthFactDiffeq} describes how the initially small fluctuations of the density field grow over time in the linear regime. However, as more matter falls towards the overdensities, the motion of DM becomes non-linear and bound objects of DM start forming. These are called {\it halos}. 

To understand halos it is useful to study the ideal case of a spherical top-hat overdensity in a homogeneous expanding universe. The magnitude of gravitational energy of the outermost shell due to the overdensity must be larger than its kinetic energy for the shell to stop expanding, revert its motion and collapse reaching the virial equilibrium. In a universe dominated by cold DM, the problem has an analytical (parametric) solution which predicts that the spherical overdensity collapses to a point when its linear evolution would hit the threshold of $\delta_{c} \simeq 1.686$. However, instead of collapsing to a point the matter distribution reaches the virial equilibrium with an overdensity $\Delta_{\rm vir} \approx 180$. 
Including the DE component into the problem affects the results for $\delta_{c}$ and $\Delta_{\rm vir}$ only marginally as the collapsing structures are decoupled from the Hubble flow when DE becomes relevant.
Despite representing an approximation (or better the monopole) of the aspherical halo collapse, this model catches some key properties of halos that are confirmed by computer simulations: 
\begin{itemize}
    \item halo collapse is a local process: the matter belonging to the spherically collapsed halo was originally inside the outer-most shell of the overdensity, with initial radius $R_{\rm i} = \sqrt[3]{\frac{3 M}{4 \pi \rho_{\mathrm{m},0}}}$ for a halo of mass $M$,
    \item the density is (approximately) the same for all virialised objects: in the spherical collapse model it is determined by the virial theorem, $\rho = \Delta_{\rm vir} \times \bar{\rho}_{\mathrm{m}} $.
\end{itemize}

The spherical collapse framework is used as a starting point in the extended Press-Schechter formalism \cite{Press:1973iz, Bond:1990iw}, which applies the excursion-set theory to the Brownian motion of the density contrast as a function of its variance $\sigma^{2}(R)$ computed in spheres of comoving radius $R$. For a single Brownian walk, i.e. studying the density centred in a single point in space, the first crossing of the barrier $\delta_{c}$ determines the current mass of the halo which hosts the element of mass initially located in the point considered. From a different perspective, studying the fraction of walks that are above the threshold $\delta_c(z)$ in terms of the variance $\sigma^{2}$, gives the fraction of halos collapsed as a function of their mass and redshift in the Press-Schechter formalism. This can be used to estimate the number density of halos $n(M, z)$ at a given mass and redshift which satisfies
\begin{equation}\label{PS_HaloAbundance}
    \frac{d n(M, z)}{d M}=\frac{\rho_{\mathrm{m},0}}{M^2} f_{\mathrm{PS}}\left(\frac{\delta_{\mathrm{c}}}{\sigma(M, z)}\right)\left|\frac{d \ln \sigma(M, z)}{d \ln M}\right|, \quad f_{\mathrm{PS}}(v)=\sqrt{\frac{2}{\pi}} \nu e^{-v^{2} / 2} \, ,
\end{equation}
where $M$ is the mass enclosed in the sphere of comoving radius $R$, $M(R)= \frac{4 \pi}{3} R^3 \rho_{{\rm m},0} $.
For large halo masses, which correspond to spheres of large comoving radii, $\sigma(M, z)\ll \delta_{\mathrm{c}}$ and the abundance of halos is exponentially suppressed.

This description is useful to gain intuition about how halos form and why halos of large mass are very rare. However, this is just a toy model and to have reliable halo statistics it is important to fully take into account the non-linear dynamics of halo collapse. This can be done by running high resolution {\it N}-body simulations which simulate the dynamics of a large number of DM particles under the action of gravity forces (see section~\ref{sec:Nbody} for more details). Thanks to the analysis of {\it N}-body simulations it is possible to derive fitting formulae for the average radial profile of halos and for the {\it halo mass function}, $n(>M, z)$, i.e., the abundance of halos with a mass larger than $M$ as a function of $M$. 

A practical way of identifying halos in simulations is to identify the largest spheres of comoving radius $R_{\Delta}$ that contain the same average density
\begin{equation}
    \frac{M\left(<R_\Delta\right)}{4 \pi R_\Delta^{3} / 3}=\Delta_{\rm m} \times \rho_{\mathrm{m},0}= \Delta_{\rm c} \times \rho_{\mathrm{c}}
\end{equation}
where $\Delta$ is the overdensity parameter that can be defined in terms of the mean matter density $\rho_{{\rm m},0}$ or the critical matter density $\rho_{\rm c}$\footnote{The critical overdensity parameter is related to the matter overdensity parameter by $\Delta_{\rm c} = \Omega_{{\rm m},0} \Delta_{\rm m} $.}. The halos found in this way are referred to as {\it spherical overdensity} halos and their properties depend on the overdensity parameter $\Delta$.

The profile of spherical overdensity halos has been shown to be approximately universal and well described by the so-called Navarro-Frenk-White (NFW) profile  \cite{Navarro:1995iw}
\begin{equation}
    \rho_{\mathrm{NFW}}(r)=\frac{\rho_{s}}{\left(r / r_{s}\right)\left(1+r / r_{s}\right)^{2}}
\end{equation}
in terms of two parameters, the scale radius $r_s$ and the density $\rho_s$. Alternatively, the NFW profile can be expressed in terms of the concentration parameter $c_{\Delta}\equiv R_{\Delta}/r_s$ and the mass 
\begin{equation}
    M_{\Delta} \equiv \int_{0}^{R_{\Delta}} 4 \pi r^{2} \rho_{\rm NFW}(r) d r = 4 \pi \rho_{s} r_{s}^{3}\left[\ln (1+c)-\frac{c}{1+c}\right] \, .
\end{equation}

The velocity of DM particles inside halos can be described as the sum of two components: a circular velocity component $v_{\rm circ}$ and a radial velocity component $v_{r}$. 
Under the assumption that the matter inside the sphere of radius $R_{\Delta}$ is virialised, the NFW profile can be used to predict the average velocity profiles of halos.
The average modulus of the circular velocity depends on the mass $M(<r)$ enclosed in the sphere of radius $r$ as
\begin{equation}
    v_{\rm circ}(r) = \sqrt{\frac{G M(<r)}{r}}
\end{equation}
while the radial-velocity dispersion profile can be expressed as \cite{NFW_VelDisp}
\begin{equation}\label{V_disp_NFW_Intro}
    \langle v_r^2(r)\rangle=\frac{1}{\rho_{\text {NFW }}(r)} \int_{r}^{\infty} \mathrm{d} r \rho_{\text {NFW }}(r) \frac{\mathrm{d} \Phi(r)}{\mathrm{d} r} \, , \quad \Phi(r) =\frac{-G M(<r)}{r} .
\end{equation}

The halo mass functions found in {\it N}-body simulations have been used in literature to tune analytical fitting functions (see for instance \cite{Jenkins:2000bv,Sheth:2001dp,Warren:2005ey}) and more recently to train emulators (e.g., \cite{Heitmann:2015xma, McClintock:2018uyf,Bocquet:2020tes}). 
In particular, by estimating the abundance of spherical overdensity halos from a suite of {\it N}-body simulations with different resolutions for a wide range of halo masses, the authors of \cite{Tinker:2008ff} have produced a fitting formula for the halo mass function with $5\%$ accuracy in the mass range $10^{11} \leq M \leq 10^{15} h^{-1} M_{\odot}$, becoming the standard in cluster abundance analysis of the last decade.

\subsection{The bias model}
\label{ssec:BiasModelIntro}
Let us consider a long wavelength density perturbation $\delta_\ell(\vec{x},t)$ in the excursion set theory. The perturbation $\delta_\ell$ adds up to the smaller scales density contrast smoothed on some radius $R$, $\delta_R(\vec{x},t)$, increasing the chances of the density contrast to hit the spherical collapse barrier $\delta_{\rm c}$ in the regions of space where $\delta_\ell(\vec{x},t)>0$. From a different point of view, the perturbation $\delta_\ell(\vec{x},t)$ can be seen as affecting the height of the barrier $\delta_{\rm c}$
\begin{equation}
    \delta_{\mathrm{c}} \longrightarrow \delta_{\mathrm{c}}-\delta_{\ell}^{(1)}(\boldsymbol{x}, t) \, .
\end{equation}
where the apex $(1)$ indicates that the density contrasts
is evolved to the time $t$ using the linear growth factor $D(t)$ discussed in section~\ref{ssec:LinearTheory}.
Now the expected number density of halos of eq.~\eqref{PS_HaloAbundance}, in the presence of this long wavelength overdensity, becomes
\begin{equation}
    \left.\frac{d n(M, z)}{d M}\right|_{\delta_\ell}=\frac{\rho_{\mathrm{m}}\left(t_{0}\right)}{M^{2}} f_{\mathrm{PS}}\left(\frac{\delta_{\mathrm{c}}-\delta_{\ell}^{(1)}(\vec{x}, t)}{\sigma(M, z)}\right)\left|\frac{d \ln \sigma(M, z)}{d \ln M}\right| \, ,
\end{equation}
which defines the halo bias $b(M,z)$ due to the long wavelength perturbation $\delta_{\ell}^{(1)}(\vec{x}, t)$ in terms of the halo number density contrast as
\begin{equation}
    \delta_{\mathrm{h}, \ell}(\vec{x}, t)=\frac{d n /\left.d \ln M\right|_{\delta_{\ell}}}{d n /\left.d \ln M\right|_{0}}-1 \equiv b(M, z) \delta_{\ell}^{(1)}(\vec{x}, t) \, .
\end{equation}
At the first order in $\delta_{\ell}^{(1)}(\boldsymbol{x}, t)$, in the extended Press-Schechter formalism, the linear bias is given by
\begin{equation}
    b_{1}^{\mathrm{PS}}(M, z)=\frac{\delta_{\rm c}^{2}-\sigma^2(M, z)}{\delta_{\mathrm{c}}\sigma^2(M, z) } \approx \frac{\delta_{\rm c}}{\sigma^2(M, z) } \,, \quad \text{for} \quad \sigma(M, z)\ll \delta_{\rm c}
\end{equation}
which shows that the bias is larger for larger halo masses (smaller $\sigma(M, z)$). From a statistical perspective, the halo bias describes how much more clustered the halos are with respect to the underlying matter distribution. This concept of bias can be extended to other tracers, such as galaxies. 

In the current understanding of galaxy formation, galaxies form inside halos \cite{Wechsler:2018pic}. Hydrodynamical simulations let us study the formation of galaxies by incorporating hydrodynamical models of {\it subgrid physics} in high-resolution {\it N}-body simulations\footnote{See section~\ref{sec:Nbody} for more details on {\it N}-body simulations}. However, hydrodynamical simulations are computationally very expensive, so empirical prescriptions are often used to connect halos found in DM-only simulations with galaxies. Among these, we focus on {\it abundance matching } and {\it Halo Occupation Distribution} (HOD) models. 

The abundance matching prescription assumes that the most massive or luminous galaxies live in the most massive halos, so if $N_{\rm g}$ galaxies are to be placed in a catalogue of $N_{\rm h}$ halos, halos are ordered by mass and the $N_{\rm g}$ most massive halos are selected to host a galaxy. This method has the advantage of being non-parametric, but to provide accurate results the halos substructure needs to be well resolved \cite{Klypin:2013rsa}. This is possible in high-resolution {\it N}-body simulations where both halos and sub-halos are identified and galaxies are assigned to them as central and satellite galaxies respectively. When abundance matching is applied to both halos and sub-halos it is often referred to as Sub-Halos Abundance Matching (SHAM) \cite{Kravtsov:2003sg, Vale:2004yt}. One difficulty of accurately employing SHAM is that subhalos and galaxies are stripped of their mass by the host halos in different ways \cite{Nagai:2004ac}, hence galaxies should be matched to sub-halos at the time they are accreted. This requires tracking the evolution of halos with {\it merger-trees} \cite{Conroy:2005aq} further increasing the computational cost of the already expensive high-resolution {\it N}-body simulations. 
Conversely, in simulations with lower resolution, only host halos are resolved \cite{Wechsler:1997fz} and the clustering statistics predicted by this technique are not accurate enough to be employed in modern galaxy surveys \cite{Klypin:1997fb}.

The HOD is a parametric method to populate halos with galaxies. It gives the probability, $P(N_{\rm g} \mid M_{\rm h}, \dots)$, that a number of galaxies, $N_{\rm g}$, are found in a halo depending on its properties, normally on its mass $M_{\rm h}$. The galaxies are often split between central and satellite galaxies and their probability distributions follow the Bernoulli and Poisson distributions respectively. The functional form for the satellite galaxies is studied in high-resolution simulations assuming that the number of satellite galaxies depends on the halo mass similarly to the number of subhalos, hence a power law is often assumed \cite{Kravtsov:2003sg}. The satellites' positions and velocity are assigned based on analytical halo profiles whose functional form is derived from high-resolution simulations, such as NFW or Einasto profile \cite{Navarro:1995iw,Graham:2005xx}. On the one hand, HOD models can be applied to low-resolution simulations since they rely on basic halo properties (e.g., the halo mass). On the other hand, they have 3-5 parameters that need to be tuned to reproduce some observed clustering signal, which may decrease the constraining power of galaxy surveys by introducing degeneracies in the theory parameter space.

From an observational perspective, an operative definition of bias can be given in terms of the power spectrum of matter $P_{\rm m}$ and the one of the biased tracer $P_{\rm t}$ as 
\begin{equation}
    P_{\rm t}(k,\tau) \equiv b_{\rm t}^2(k,\tau)P_{\rm m}(k,\tau) \, ,
\end{equation}
where the tracer's bias $b_{\rm t}(k,\tau)$ is a function of both scale and time.

\subsection{Redshift space distortions}
\label{ssec:RSD_Intro}
Galaxy surveys provide information on galaxy positions in terms of angular coordinates and redshift. The redshift we observe can be interpreted as the result of two effects, the cosmological redshift due to the expansion of the universe and the Doppler shift due to the peculiar velocity of galaxies in the reference frame of the expanding universe:
\begin{equation}\label{ObsRedshift}
    1+z_{\rm obs} = (1+z_{\rm cos}) \cdot \left( 1+v_{||}(\vec{r}) \right) \, ,
\end{equation}
where $v_{||} = \vec{v}\cdot \hat{r}$ and $\vec{r}$ is the vector originating from the observer and pointing towards the source\footnote{Eq.~\ref{ObsRedshift} assumes $v_{||} \ll 1$ and neglects relativistic effects beyond the first order. It also assumes that the observer is at rest in the reference frame of the expanding universe.}. In the FLRW metric, the cosmological redshift is due to the expansion of the universe from the time of emission to that of the observation
\begin{equation}    
1+z_{\rm cos}=\frac{a_{\rm obs}}{a_{\rm emit}} \, .
\end{equation}
This can be used to obtain the instantaneous distance from the observer to the source at the time of the observation
\begin{equation}\label{InstDistance}
    r =a_{0} \int_{0}^{\chi_{s}} d \chi=a_{0} \int_{t_{s}}^{t_{0}}  \frac{d t}{a} = \int_{0}^{z_{\rm cos}} \frac{d z^{\prime}}{H\left(z^{\prime}\right)} \, ,
\end{equation}
where $\chi_s$ is the comoving distance of the source.
While eq.~\eqref{InstDistance} provides a way to map cosmological redshift to instantaneous positions, the cosmological redshift of the sources is not directly observed and the observed redshift is affected also by the peculiar velocities of the sources, as we discussed above. 
Using the observed redshift $z_{\rm obs}$ it is possible to infer the redshift space position
\begin{equation} \label{RSD_map_Intro}
    s \equiv \int_{0}^{z_{\rm obs }} \frac{d z^{\prime}}{H\left(z^{\prime}\right)} \simeq r +  \frac{v_{||}(\vec{r})}{\mathcal{H}\left(z_{\rm cos}\right)} \, .
\end{equation}
On the one hand, this does not allow us to know the exact positions of the sources, on the other hand, it makes cosmological probes of clustering sensitive also to the matter velocity field.

In fact, the Doppler shift produces well-studied distortions in the clustering of galaxies known as Redshift-Space Distortions (RSD). To understand RSD it is convenient to split the analysis into large (linear) scales and small (non-linear) scales. 

On large scales, the coherent infall of matter towards the overdensity produces a squashing of the structures along the line-of-sight that is well described by the Kaiser model in the distant-observer approximation \cite{Kaiser:1987qv}. In this model the velocity field is due to the gradient of the gravitational potential and vorticity can be neglected. This is valid on very large scales where linear theory applies and the galaxy number density contrast $\delta_{\rm g}$ is related to the matter density contrast $\delta_{\rm m}$ by the linear bias relation
\begin{equation}
    \delta_{\mathrm{g}}(\vec{r}, \tau)=b(\tau) \delta_{\mathrm{m}}(\vec{r}, \tau) \, .
\end{equation}
The conservation of galaxy counts under the redshift space remapping determined by eq.~\eqref{RSD_map_Intro} relates the redshift-space galaxy number-density contrast $\delta_{{\rm g}, \vec{s}}$ to the matter density contrast $\delta_{\rm m}$ through
\begin{equation}
    \delta_{\mathrm{g}, \vec{s}}(\vec{r})=b \, \delta_{\mathrm{m}}(\vec{r})-\frac{\partial}{\partial \vec{r}}\left[\frac{v_{||}(\vec{r})}{\mathcal{H}}\right] \, ,
\end{equation}
which, in Fourier space, using the Euler equation~\eqref{EulerEquation} and incorporating time dependency into the linear growth factor, yields
\begin{equation}\label{kaiser_delta_gs}
    \delta_{\mathrm{g}, \vec{s}}(\vec{k})=\left[b+f \mu_{k}^{2}\right] \delta_{\mathrm{m}}(\vec{k}) \, ,
\end{equation}
where $\mu_{\vec{k}}$ is the cosine of the angle between the versor $\hat{k} $ and the line-of-sight, and $f$ is the linear growth rate already defined in eq.~\eqref{LinearGrwothRate}. Eq.~\eqref{kaiser_delta_gs} shows that the clustering of galaxies depends on the cosine with the line-of-sight $\mu_{k}$ making redshift-space galaxies more clustered than real-space galaxies along the line-of-sight. This clearly affects also the galaxy power spectrum in redshift space which can be expressed in terms of the matter power spectrum as
\begin{equation}
    P_{{\rm g},\vec{s}}(k, \mu_k) = \left(b+f \mu_k^2\right)^2 P_{\rm m}(k)\, .
\end{equation}
By expanding the redshift space galaxy power spectrum in its multipoles through the projection on the Legendre polynomials,
\begin{equation}
    P_{\ell}^{g,\vec{s}}(k)=\frac{2 \ell+1}{2} \int_{-1}^1 P^{g,\vec{s}}(k, \mu_k) \mathcal{L}_{\ell}(\mu_k) d \mu_k \, ,
\end{equation}
it is possible to show that, in the Kaiser model, only monopole, quadrupole and hexadecapole moments are sourced by redshift space distortions, with the three multipoles given by:
\begin{equation}
\begin{aligned} P_{0}^{g,\vec{s}}(k) &=\left(b^2+\frac{2}{3} f\,b+\frac{1}{5} f^{2}\right) P_{\text{m}}(k) \, , \\
 P_{2}^{g,\vec{s}}(k) &=\left(\frac{4}{3} f\, b+\frac{4}{7} f^{2}\right) P_{\text{m}}(k) \, , \\
 P_{4}^{g,\vec{s}}(k) &=\frac{8}{35} f^{2} P_{\text{m}}(k) \, . \end{aligned}
\label{eq:Pk_gal_multipoles}
\end{equation}

On small scales, instead, the matter is virialised, hence the motion of galaxies is incoherent and their velocities are larger. The redshift-space displacement due to the peculiar velocity is larger than the separations between galaxies which results in an elongation of structures along the line-of-sight known as the Fingers-of-God effect \cite{Jackson:1971sky}. Due to their different nature, unlike the large-scale squashing of structures, these small-scale distortions require a much more complicated non-linear treatment and, as a result, they are more often regarded as a problem to deal with rather than a cosmological probe.


\subsection{Galaxy surveys}
Extracting information from the large scale structure is a non-trivial task and equally (if not more) difficult is to design and run experiments that are able to accurately map the LSS. 
It is possible to classify the galaxy surveys in two main kinds, the ones devoted to collect imaging data (photometric surveys) and the ones devoted to collect spectrographic data (spectroscopic surveys). To greater interest of this work are the latter which, unlike the former, can provide precise determinations for the redshift of the observed galaxies (necessary to study RSD for example). To achieve this goal in these experiments first it is necessary to identify a population of galaxies by means of imaging data, then, known the angular coordinates of each galaxy, the light collected by the telescope is conveyed to the spectrographs by means of optical fibers and finally the redshift is determined by analysing the shift of the observed spectral lines. Depending on the experimental design several systematic errors need to be taken into account like {\it fiber collisions} \cite{Hahn:2016kiy} and {\it selection effects} \cite{2012MNRAS.424..564R}.

Chronologically, the first spectroscopic galaxy survey with significant cosmological implications is the 2 degrees Field Galaxy Redshift Survey (2dFGRS) \cite{2DFGRS:2001zay} that, from the year 1997 to the 2002, measured the spectra of more than 200.000 galaxies in a region of $\sim 2.000 \,{\rm deg}^2$ around redshift $z=0.11$ \cite{2dFGRS:2005yhx}. The analysis of the large scale galaxy power spectrum in the 2dFGRS provided the first constraints on the matter content of the universe from LSS observations \cite{2dFGRS:2001csf, 2dFGRS:2005yhx}. 
Soon after the 2dFGRS, other galaxy surveys started collecting data.
The 6 degrees Field Galaxy Survey (6dFGS) carried out between 2001 and 2009 measured the spectra of $\sim 130.000$ galaxies across $\sim17.000\,{\rm deg}^2$ (almost half of the sky) becoming the widest spectroscopic survey of the low redshift universe \cite{Jones:2004zy,Beutler:2011hx}.
Between the years 2006 and 2011, the WiggleZ survey \cite{Drinkwater:2009sd} probed a $\sim1 \, {\rm Gpc}^3$ volume of the universe by focusing on a relatively small area of the sky ($\sim1.000\,{\rm deg}^2$) but targeting a higher-redshift galaxy population. With a single robotic spectrograph, this survey collected precise redshift measurements of about $240.000$ galaxies emission line galaxies in the redshift range $0.2 \lesssim z \lesssim 1 $ \cite{Blake:2012pj}.
Across the first 20 years of the 21 century, the Sloan Digital Sky Survey \cite{SDSS:2000hjo,eBOSS:2020yzd} measured the spectroscopic redshift of more than 1 million galaxies in an area of around $\sim14.000\,{\rm deg}^2$, roughly a quarter of the sky. Having targeted several galaxy populations from nearby galaxies to Luminous Red Galaxies, Emission Line Galaxies and quasars (or {\it quasi stellar objects}) this survey covered the redshift range $0.07 < z < 2.2 $. With a volume of more than $10 \, {\rm Gpc}^3$, SDSS is the largest spectroscopic galaxy catalogue to date and has enabled precise inference of cosmological parameters, in particular when combined with imaging surveys like DES and CMB measurements like Planck.

Motivated by the amazing achievements of these surveys and the relentless advance in technology and engineering that our times are witnessing, new surveys have been designed and are now becoming a reality. Among these new exciting experiments, often referred to as Stage IV surveys, the following are worthy of a particular mention:
\begin{itemize}
    \item the Dark Energy Spectroscopic Instrument (DESI) \cite{DESI:2016fyo}, a ground based telescope with robotically-orientable optical fibers fed in an array of 10 spectrographs, started collecting data in April 2021 
    and after only 7 months of observations became the largest spectroscopic survey available\footnote{Source: Berkeley lab website \href{https://www.lbl.gov}{https://www.lbl.gov}.}. The experiment is designed to collect the spectra of magnitude limited bright galaxies at $z\sim0.2$, LRG up to $z<1$, ELG up to $z<1.7$, and quasars covering $14.000\,{\rm deg}^2$ of the sky. With its wide field and improved sensitivity, DESI will observe a number of galaxies and a volume $\sim10$ times larger than SDSS.
    \item the Euclid mission \cite{Euclid:2021icp} of the European Space Agency is a space-based experiment. The launch of the satellite is expected in July 2023 and the main survey (the {\it wide survey}) is planned to start shortly thereafter. Unlike the others, the Euclid telescope will perform slit-less spectroscopy which may lead to different systematic errors. The main target of the wide survey will be the H$\alpha$ emitting galaxies across an area of $15.000\,{\rm deg}^2$ in the redshift range $0.9 < z < 1.8$. Given the estimated number density of these galaxies \cite{Pozzetti:2016cch}, Euclid is expected to measure the spectroscopic redshift of around 30 million galaxies \cite{Euclid:2019clj}.
    \item The Nancy Grace Roman space telescope \cite{Akeson:2019biv}, also known as WFIRST, is scheduled to launch by May 2027\footnote{From the press release: \href{https://www.nasa.gov/feature/goddard/2021/nasa-confirms-roman-missions-flight-design-in-milestone-review}{NASA Confirms Roman Mission's Flight Design in Milestone Review}.}. During the first five year of the mission, WFIRST will devote a significant part of its observing time to map a deep but relatively narrow region of the sky producing the so-called High Latitude Survey (HLS).
    The telescope will measure the spectroscopic redshifts of galaxies emitting the H$\alpha$ (10 million spectra at  z=1-2) and O-III lines (2 million spectra at  z=2-3) in a area of $2.000\,{\rm deg}^2$ \cite{Wang:2021oec}. Similarly to the Euclid mission, WFIRST will perform slit-less spectroscopy.
\end{itemize}
As the experiments progress, the theoretical modelling must keep the pace to maximise the scientific return of Stage IV surveys. Since structure formation is an intrinsically non-linear process, numerical simulations will play a key role in future cosmological inference with the LSS.

\section{{\it N}-body simulations}
\label{sec:Nbody}

Cosmological simulations are based on the {\it N}-body framework, where a large number of particles (typically $\sim 10^8-10^{11}$) is used to discretise the distribution of mass in a comoving portion of the universe referred to as the simulation box, normally a cubic box with periodic boundary conditions. 
They rely on the Newtonian approximation and use comoving coordinates to incorporate the effects of background expansion. It is also possible to include large-scales relativistic effects in Newtonian {\it N}-body simulations thanks to the {\it N}-body gauge formalism \cite{Fidler:2015npa,Tram:2018znz,Dakin:2019vnj,Brando:2020ouk,Brando:2021jga}. The {\it N}-body particles are assumed to be collision-less and to interact only through gravity forces. Their equations of motions are given by the Hamilton equations
\begin{gather}\label{HamiltonEq}
    {\vec{x}}'_{n} =\frac{\vec{p}_{n}}{a(\tau)} \\
    {\vec{p}}'_{n} =a(\tau)\left(-\frac{\partial \phi}{\partial \vec{x}}\right)_{n} 
\end{gather}
where $\vec{x}_{n}$ is the comoving position of the $n^{\rm th}$ particle and $\vec{p}_{n}$ is its conjugate momentum. The gravitational potential $\phi$ is computed from the density contrast using the Poisson equation~\eqref{PoissonNewtonian}. The scale factors in the Hamilton equations take into account the weakening of gravitational interactions due to background expansion. The number of particles $N_{\rm part}$ and the size of the box $L_{\rm box}$ are crucial in determining the mass resolution of the simulation. For particles of equal mass in a cubic box, the mass of each particle is given by
\begin{equation}
    M_{\rm part} = \bar{\rho} \frac{L_{\mathrm{box}}^{3}}{N_{\mathrm{part}}} \, ,
\end{equation}
where $\bar{\rho}$ is the average density of the universe for the given cosmology.

The Initial Conditions (IC) are normally set using (first or second order) Lagrangian Perturbation Theory (LPT) in the matter domination at a stage where nonlinearities are negligible for the scales and accuracy of interest in the particular case. It has been shown that redshift of $z \gtrsim 100$ should be used to set the IC with the Zeldovich approximation (i.e., the first order lagrangian perturbation theory) to achieve percent level accuracy up to $k\sim 1 \hompc$ \cite{Schneider:2015yka}. 

The particles' positions and velocities are evolved from the IC to the desired redshift using {\it N}-body techniques that solve the equations of motion~\eqref{HamiltonEq} for the {\it N}-body particles together with the Poisson equation~\eqref{PoissonNewtonian} at each time-step. The solution of the Poisson equation is the most computationally expensive part in a $\Lambda$CDM simulation, therefore many efforts have been made to adopt the most efficient techniques. Some of the most relevant techniques used to compute the forces are:
\begin{itemize}
    \item {\it Direct-summation method} \cite{Ewald1921}: The equation of motion are integrated once analytically. The forces are computed by adding the contributions from each pair of particles, making the computational complexity of this algorithm scale with $N_{\rm part}^2$.
    \item {\it Particle-Mesh (PM) method}: A regular mesh is constructed and the particles are assigned to the mesh intersections following specific interpolation methods like Cloud-in-Cell (CIC) or Triangular-Shaped-Clouds (TSC) \cite{Sefusatti:2015aex}. This has the advantage that the FFT algorithm can be applied to solve the Poisson equation in Fourier space. The size of the mesh cells determines the force resolution $\ell_{\rm F} = L_{\rm box}/\NM$, where $N_{\rm mesh}$ is the total number of knots in the mesh. The cosmological code \codeword{GLAM} \cite{Klypin:2017iwu} uses the PM method.
    \item {\it Tree method} \cite{1986Natur.324..446B}: The particles are assigned to a hierarchical tree to speed up the force computation. The forces are computed with direct summation between particle pairs on small scales and between particles and cells on larger scales. \codeword{PKDGRAV3} \cite{pkdgrav} implements the tree method.
    \item {\it Adaptive Mesh Refinement (AMR) method}: Based on the PM method, the AMR method achieves a better force resolution while mitigating the loss of efficiency by sampling high-density regions with increasingly finer meshes, until a maximum refinement level is reached. \codeword{RAMSES} \cite{Teyssier:2001cp} is an example of AMR {\it N}-body code.
    \item {\it Tree-PM method}: The forces are calculated with the tree method on small scales and with the PM method on large scales. The codes \codeword{GADGET3} \cite{gadget2,angulo2012} and \codeword{AREPO} \cite{Arepo}) adopt this technique to carry out the force computation.
\end{itemize}

\begin{figure}
\centering
\includegraphics[width=.8\textwidth]{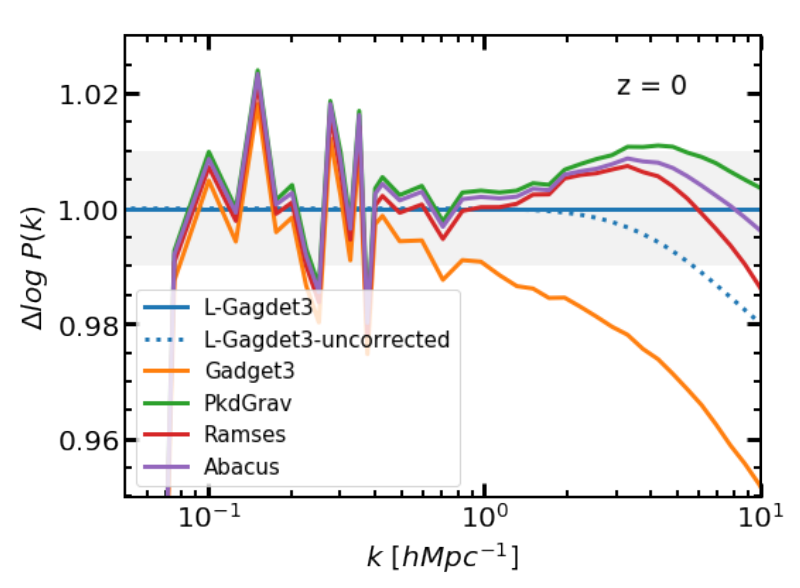}
\caption{\label{fig:NbodyCodeComparison_LCDM} Comparison of matter power spectra at redshift $z=0$ obtained with the {\it N}-body codes listed in the legend. Adapted from \cite{Angulo:2020vky}.}
\end{figure}

The codes \codeword{PKDGRAV3}, \codeword{RAMSES} and \codeword{GADGET3} have been compared in \cite{Schneider:2015yka} for the power spectrum predictions where they have been found to be in $1\%$ agreement up to $k \sim 1 \hompc$ with larger discrepancy arising at smaller scales (in particular at higher redshift). More recently, another comparison has been carried out using different simulations parameters \cite{Angulo:2020vky} finding below percent agreement in the matter power spectrum between these codes up to $k \sim 5 \hompc$ at redshift $z=0$. This is shown in figure~\ref{fig:NbodyCodeComparison_LCDM}, depicting a comparison of the matter power spectra at redshift $z=0$ between different {\it N}-body codes. With the exception of ``Gadget3'', relative to the \codeword{GADGET3} simulation performed in \cite{Schneider:2015yka}, all the other results agree are in sub-percent agreement up to $k \sim 5 \hompc$. The codes involved in this more recent comparison are the three also used in \cite{Schneider:2015yka} (i.e., \codeword{GADGET3}, \codeword{PKDGRAV3} and \codeword{RAMSES}) with the addition of \codeword{ABACUS}, a full {\it N}-body code based on direct-summation for near-field forces and multipole expansion for far-field forces. 


\subsection{{\it N}-body simulations in modified gravity}
When MG is taken into account, the Poisson equation is modified accordingly (see eq.~\eqref{PoissonMGgeneral}) and the Klein-Gordon equation for the scalar field also need to be solved at each time step. The latter is a highly non-linear partial differential equation that needs to be solved with multi-grid techniques, where the equation is discretised on a grid (possibly using AMR) and solved iteratively with the Gauss-Seidel scheme \cite{Llinares:2018maz}. This makes cosmological simulations in MG much more computationally expensive than their $\Lambda$CDM counterparts, being typically $\sim10$ times slower \cite{Winther:2014cia}.

Some of the {\it N}-body codes in MG that solve the full differential equation for the scalar field have been compared in \cite{Winther:2015wla}. The codes included in the comparison are:
\begin{itemize}
    \item \codeword{ECOSMOG} \cite{Li:2011vk,Li:2013nua}, based on the AMR code \codeword{RAMSES}
    \item \codeword{MG-GADGET} \cite{Puchwein:2013lza}, based on the tree-PM code \codeword{GADGET3}, with AMR for the MG solver
    \item \codeword{DGPM} \cite{Schmidt:2009sg,Oyaizu:2008sr}, fixed grid PM code
    \item \codeword{ISIS} \cite{Llinares:2013jza}, also based on the AMR code \codeword{RAMSES}
    \item \codeword{ISIS-NONSTATIC} \cite{Llinares:2013qbh}, version of \codeword{ISIS} that goes beyond the quasi-static approximation
\end{itemize}

\codeword{ECOSMOG}, \codeword{MG-GADGET} and \codeword{ISIS} are used to produce $f(R)$ simulations. \codeword{DGPM} and \codeword{ECOSMOG} are used to produce simulations in DGP theories\footnote{The comparison in DGP theories is performed using a fixed PM grid in ECOSMOG to match the force resolution of DGPM which does not incorporate an AMR solver.}. Finally, \codeword{ISIS} and \codeword{ISIS-NONSTATIC} are used to produce simulations in a Symmetron model. 
The codes are compared at the level of the matter power spectrum, velocity divergence, halo mass function and halo profiles. Since differences in these statistics between the various codes are already present in GR, the comparison focuses on the enhancement of the statistics in MG compared to GR (the {\it MG boost factor}). 
For what concerns the MG boost factor of the power spectrum at redshift $z=0$, all codes are found to be in $1\%$ agreement up to $k \sim 7 \hompc$ in $f(R)$ and DGP theories. Also, the agreement in the velocity divergence power spectrum is found to be excellent, with better than $1 \%$ agreement for $k \lesssim 3 h \mathrm{Mpc}^{-1}$ in $f(R)$ theory and better than $\lesssim 2 \%$ at all scales considered for DGP theories. The halo statistics are found to be in good agreement too in both $f(R)$ and DGP theories, but sample variance limits the validity of the comparison.
Interestingly, \codeword{ISIS} and \codeword{ISIS-NONSTATIC} are found to give almost identical results for all the statistics considered in the symmetron model, suggesting that it is fine to neglect the time derivatives in the Klein-Gordon equation with the quasi-static approximation.

More recently, other simulation codes have been extended to MG. Of interest to this work are the MG extension of \codeword{AREPO} and \codeword{GLAM}, named \codeword{MG-AREPO}~\cite{Arepo_fR,Arepo_nDGP} and \codeword{MG-GLAM}~\cite{Hernandez-Aguayo:2021kuh,Ruan:2021wup}, that we will encounter again in chapter~\ref{chp:powerspectrum}.

An alternative way to incorporate screening in {\it N}-body simulations in MG is thanks to the screening approximation proposed in \cite{Winther:2014cia} and that we have discussed in the case of chameleon and Vainshtein mechanisms in section~\ref{ScreeningMechanisms}. This approximation, which can be obtained by linearising the Klein-Gordon equation, allows reducing the slow-down of {\it N}-body simulations due to MG to as little as a factor $2$-$3$ while still capturing the suppression of the fifth force on small scales. 









\chapter{Fast production of galaxy mock catalogues in modified gravity}\label{chp:mocks}
\begin{quote}
    {\it The content of this chapter is based on the publication \cite{Fiorini:2021dzs}. The particles data from ELEPHANT simulations used in this chapter were provided by Baojiu Li.}
\end{quote}
It is crucial to make accurate theoretical predictions of the properties of the LSS on non-linear scales, both in terms of comparison to measurements and to construct realistic covariance matrices, to successfully constrain the gravity model (see e.g. \cite{Alam:2020jdv}). This can be achieved by means of cosmological simulations, but in excess of ${\cal O}(10^3)$ realisations must be produced to match the volume of Stage IV surveys and compute an accurate estimate of the covariance matrices. This poses a serious challenge for full {\it N}-body simulations in MG models due to their high computational cost. Such models usually have screening mechanisms that hide modified gravity effects on small scales. To describe the screening mechanism, an additional non-linear equation needs to be solved in {\it N}-body simulations, which significantly slows down the MG simulations, as discussed in section~\ref{sec:MG}. 

An alternative to full {\it N}-body simulations is to exploit approximate methods (see \cite{Monaco16} for a comprehensive review and \cite{Chuang:2014toa} for a comparison project in GR), which lower the computational cost at the expenses of accuracy on non-linear scales. Amongst these, the COLA method \cite{Tassev:2013pn, Koda:2015mca, Izard:2015dja, Howlett:2015hfa} with its extension to MG \cite{Valogiannis:2016ane, Winther:2017jof, Wright:2017dkw} offers an interesting compromise between speed-up and accuracy without introducing any additional free parameter and is therefore ideal to access the MG information on (mildly) non-linear scales without losing predictability (see \cite{Moretti:2019bob} for an alternative approach based on \textcode{pinocchio}\footnote{\textcode{pinocchio} uses analytical LPT solutions to evolve the DM distribution and the {\it orbit-crossing} collapse model to produce halo catalogues \cite{Monaco:2001jg}.}).

Having a bridge between observed galaxies and simulated dark matter distribution makes it possible to extract the LSS information encoded in the datasets from galaxy surveys.
As we have seen in section~\ref{ssec:BiasModelIntro}, the HOD prescription allows us to do so starting from the DM halos identified in the density field by a halo-finder algorithm \cite{Knebe:2011rx} and populating them with galaxies with a probability distribution conditioned on some halo properties.
In particular, the HOD model proposed in \cite{Zheng:2007zg} relies on the halo mass and the density: using the analytical NFW model (introduced in sub-section~\ref{sec:Halos}) for the halo profile, it is possible to apply this formalism to COLA simulations that do not resolve the internal halo properties \cite{Koda:2015mca}.
Due to the non-trivial dynamics of the screening mechanisms, the internal halo structure in MG theories can differ from the one in GR and this needs to be considered to produce realistic galaxy mocks \cite{Mitchell:2018qrg, Mitchell:2019qke}.

In this context, we investigate the feasibility of producing mock galaxy catalogues from COLA simulations in MG producing a simulation suite with \textcode{mg-picola}
\cite{Winther:2015wla, Wright:2017dkw} and using a suite of full {\it N}-body simulations performed by the \textcode{ecosmog} code \cite{Li:2011vk} to validate the COLA results and to estimate the accuracy in reproducing galaxy clustering statistics.

This chapter is organised as follows. 
After introducing the MG theories of interest to this work in Section~\ref{sec:MG}, we discuss the simulation techniques and suites in Section~\ref{sec:Sims}. In Section~\ref{sec:Halos} we investigate the production of halo catalogues. We then apply the HOD formalism in Section~\ref{sec:Galaxies} to create mock galaxy catalogues and study the multipole moments of the galaxy spectrum in redshift space.
\section{Simulations}
\label{sec:Sims}

To produce mock catalogues for the modified gravity models previously described, we start by running dark matter simulations of the large-scale structure. In particular, we explore the following modified gravity models: the normal branch DGP model with $H_0 r_c=1$ (N1) and Hu-Sawicki $f(R)$ model with $|f_{R0}| = 10^{-5}$ (F5), plus the vanilla GR model.

Cosmological simulations can be accurate up to non-linear scales, but they are computationally very expensive, as forces between particles need to be calculated for thousands of time steps typically. The cost is even higher for MG models, which in addition also need to solve the non-linear equation for the fifth force. In recent years, the so-called approximate methods have become a popular alternative to full $N$-body simulations, and they are aimed at speeding up the creation of a density field at the expense of not resolving accurately sub-halo scales (see \cite{Chuang:2014toa,Lippich_2018,Blot_2019,Colavincenzo_2018} for studies on their accuracy, and  \cite{Monaco16} for a review). The simulations in this work make use of the COmoving Lagrangian Acceleration (COLA) method \cite{Tassev:2013pn}, in particular, we use \textcode{mg-picola} \cite{Winther:2017jof}, that includes gravity models other than GR, and is an ideal tool to efficiently run simulations for MG.

\subsection{COLA method}
The COLA method \cite{Tassev:2013pn} uses the Particle-Mesh (PM) algorithm (see sec~\ref{sec:Nbody}) to evolve the displacement of particles with respect to their second-order Lagrangian Perturbation Theory (2LPT) positions which are obtained analytically (see sec~\ref{ssec:LPT}).
In practice, the PM technique in COLA solves the following equation of motion for the particles trajectories,
\begin{equation}
    \partial_t^2 \vec{x}_{\mathrm{res}}=-\vec{\nabla} \Phi-\partial_t^2 \vec{x}_{\mathrm{2LPT}} \, ,
\end{equation}
where $\vec{x}_{\mathrm{2LPT}} = q + \Psi^{(1)}+\Psi^{(2)}$ is the 2LPT solution that is computed beforehand. The second time derivative of $\vec{x}_{\mathrm{2LPT}}$ acts as a fictitious force in the (non-inertial) COLA reference frame.
Thanks to this, a small number of time-steps (typically of $\mathcal{O}(10)$) are enough to accurately recover the DM density field up to mildly non-linear scales, as well as the halo positions and masses. This results in a speed-up of a factor $100-1000$ with respect to full {\it N}-body simulations, at the expense of not resolving internal halo properties because of the low force and time resolution that are used for the PM technique. For the production of accurate mock halo catalogues, \cite{Izard:2015dja} proposed an optimal configuration of the parameters in COLA that control the trade-off between accuracy and computational cost, such as the grid resolution used to compute forces and the distribution of time steps, which we adopt for this work.

\subsection{MG extension to COLA}
When MG is incorporated in cosmological simulation, the complexity increases because the motion of particles is governed by both the Newtonian force and the fifth force. This can lead to a slow-down of a factor of $\mathcal{O}(10)$ in theories where computing the fifth force requires the solution of non-linear equations. 
One way to avoid this significant slow-down is to use the screening approximations introduced in section~\ref{sec:MG} where the fifth force dynamics is described by means of an effective mass of the scalar field and a coupling determined by either the value of gravitational potential or its derivatives, depending on the type of screening mechanism. To extend the COLA method to MG, it is also necessary to reformulate the Lagrangian Perturbation Theory to take into account the effect of the fifth force on the dynamics \cite{Valogiannis:2016ane, Winther:2017jof, Aviles:2017aor}. 

\textcode{mg-picola} \cite{Winther:2017jof} is a publicly available code for cosmological simulations with the COLA method in modified gravity and it implements the scale-dependent 2LPT and some screening approximations. Its accuracy has already been tested against full {\it N}-body simulations at the level of the DM density field. In this work, we study the accuracy of \textcode{mg-picola} also for the DM velocity field, the halo abundance and halo clustering, and we show how it can be used in combination with a HOD algorithm to generate mock galaxy catalogues (with some tweaks relative to applications in GR).

\subsection{Simulation suites}
We use a full {\it N}-body simulation suite called \textcode{elephant}, which was introduced and validated in \cite{Cautun:2017tkc}, to benchmark the results of our runs based on COLA.
The \textcode{elephant} suite was produced using \textcode{ecosmog} \cite{Li:2011vk}, a MG extension of the adaptive mesh refinement code \textcode{ramses} \cite{Teyssier:2001cp} that solves the exact equations for the fifth force; these simulations were recently used to create mock galaxy catalogues in \cite{Hernandez-Aguayo:2018yrp,Hernandez-Aguayo:2018oxg, Alam:2020jdv}\footnote{see \cite{Barreira:2016ovx} and \cite{Devi:2019swk} for other attempts to create mock galaxy catalogues in MG theories}. Table~\ref{tab:elephant_details} describes the parameters of the suite such as the box size, the mass resolution and the gravity models implemented, and the cosmological parameters employed are:

\begin{equation}
\begin{array}{ccc}
\Omega_{m,0}=0.281 \, , & \Omega_{\Lambda,0}=0.719 \, , & \Omega_{b,0}=0.046 \, , \\
n_{s}=0.971 \, , & \sigma_{8}=0.842 \, , & h=0.697 \, .
\end{array}
\label{cosmology}
\end{equation}
The initial conditions were generated using the Zel'dovich approximation at $z=49$.
For a given realisation, the same initial seed was used for GR, F5 and N1\footnote{Due to an anomaly with one snapshot, we use only 4 realisations for N1 in this work.}.

We develop a new suite of simulations with \textcode{mg-picola} that matches the cosmology and parameters (such as the mass resolution, the box size and the number of realisations) of the fiducial \textcode{elephant} set. We call this new COLA suite PIpeline TEsting Run (\textcode{piter}). We stop the simulations at redshift $z=0.5057$ after 30 timesteps to match the redshift of the \textcode{elephant} snapshot closest to redshift $z=0.5$. As discussed in the previous section, with the screening approximation it is possible to tune the behaviour of the screening mechanism: in Eq.~\eqref{Geff_fR} we use a screening efficiency $f_s = 4.0$ for F5 and in Eq.~\eqref{Geff_DGP} we use a Gaussian smoothing with scale 1 $\mpcoh$ for N1. We have found these values to provide the best agreement for the dark matter power spectrum with {\it N}-body simulations amongst the values tested.
A summary of the \textcode{piter} simulation details is given in Table~\ref{tab:piter_details}.
\begin{table}
\centering 
\small
\parbox{.47\linewidth}{
\caption{\textcode{elephant} simulations} \label{tab:elephant_details}
\begin{tabular}{cc}
\toprule
Models & GR, F5, N1 \\
Realisations & 5 \\
Box size & 1024 $\mpcoh$ \\
$N_{\mathrm{part}}$ & $1024^3$ \\
$M_{part}$ & $7.7985 \cdot 10^{10}\Msun$ \\
Domain grid & $1024^3$ \\
Refinement criterion & $8$ \\
Initial conditions & Zel'dovich, $z=49$ \\
\bottomrule
\end{tabular}
}
\hfill
\parbox{.47\linewidth}{
\caption{\textcode{piter} simulations} \label{tab:piter_details}
\begin{tabular}{ccccc}
\toprule
Models & GR, F5, N1 \\
Realisations & 5 \\
Box size & 1024 $\mpcoh$ \\
$N_{\mathrm{part}}$ & $1024^3$ \\
$M_{part}$ & $7.7985 \cdot 10^{10}\Msun$ \\
Force grid & $3072^3$ \\
Timesteps & $30$ \\
Initial conditions & 2LPT, $z=49$ \\
\bottomrule
\end{tabular}
}
\end{table}
Initial conditions are generated using the 2LPT at $z=49$. As in \textcode{elephant}, for a given realisation, the same initial seed is used for GR, F5 and N1. We also ran a GR COLA simulation using the same initial condition as one of the realisations in \textcode{elephant} and checked that our conclusions were not affected by the cosmic variance.
For simplicity of notation, in the following we will just use {\it N}-body and COLA to refer to \textcode{elephant} and \textcode{piter} respectively.

\begin{figure}[t]
\centering 
\subfloat[][GR]{
\includegraphics[width=.48\textwidth,clip]{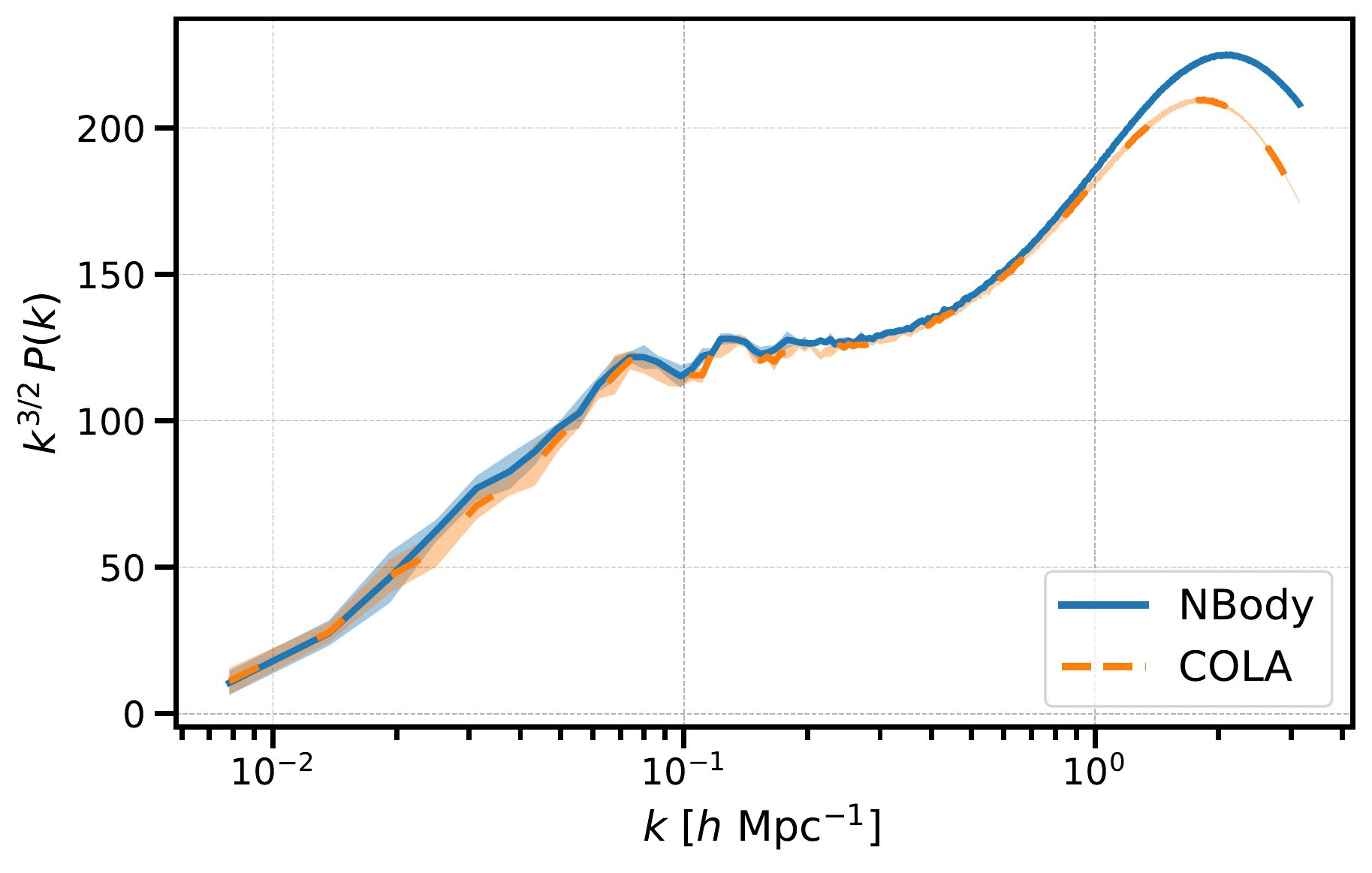}
}
\hfill
\subfloat[][GR ratio]{
\includegraphics[width=.48\textwidth]{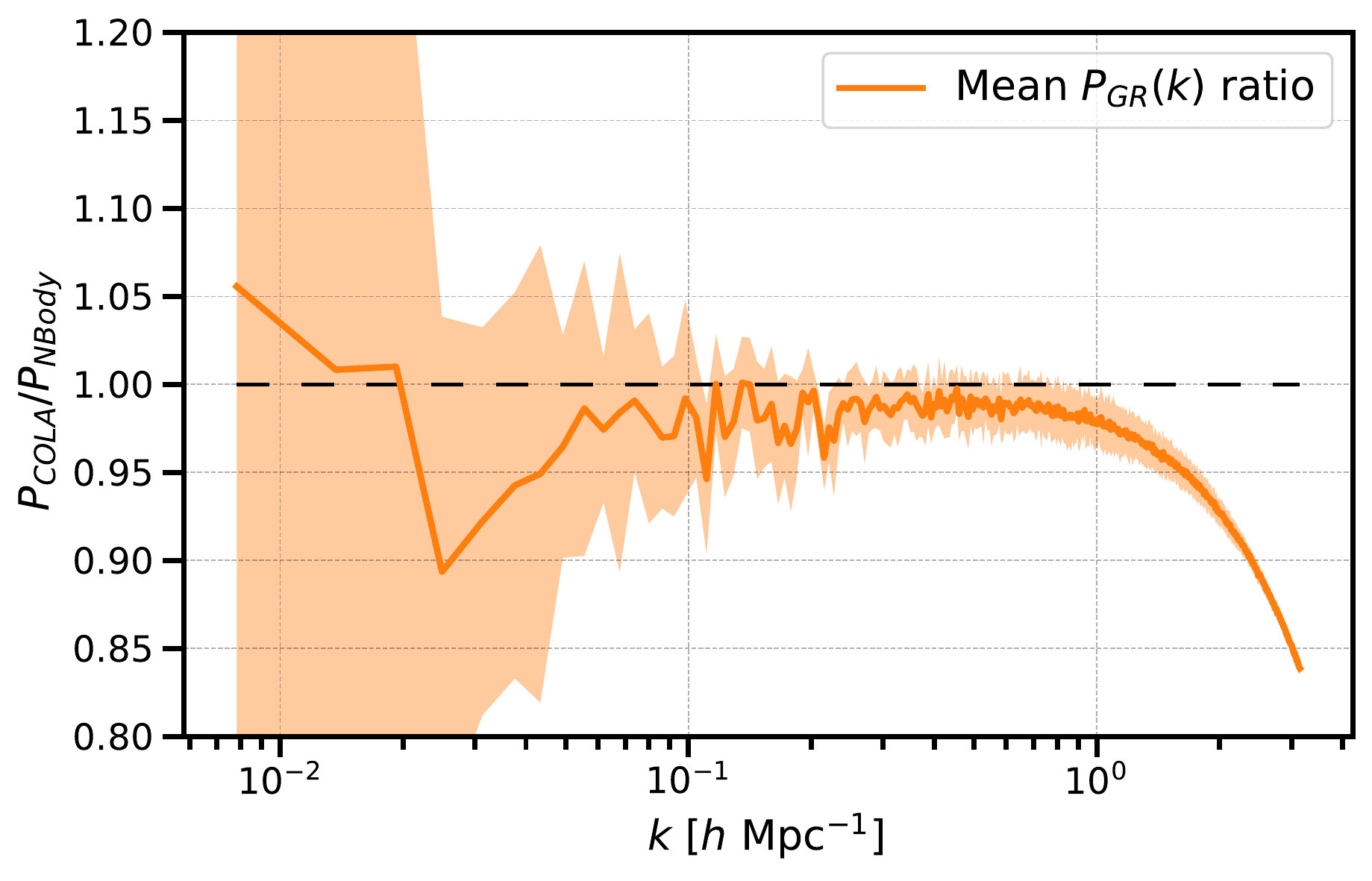}
}
\vfill
\subfloat[][F5 boost factor]{
\includegraphics[width=.48\textwidth]{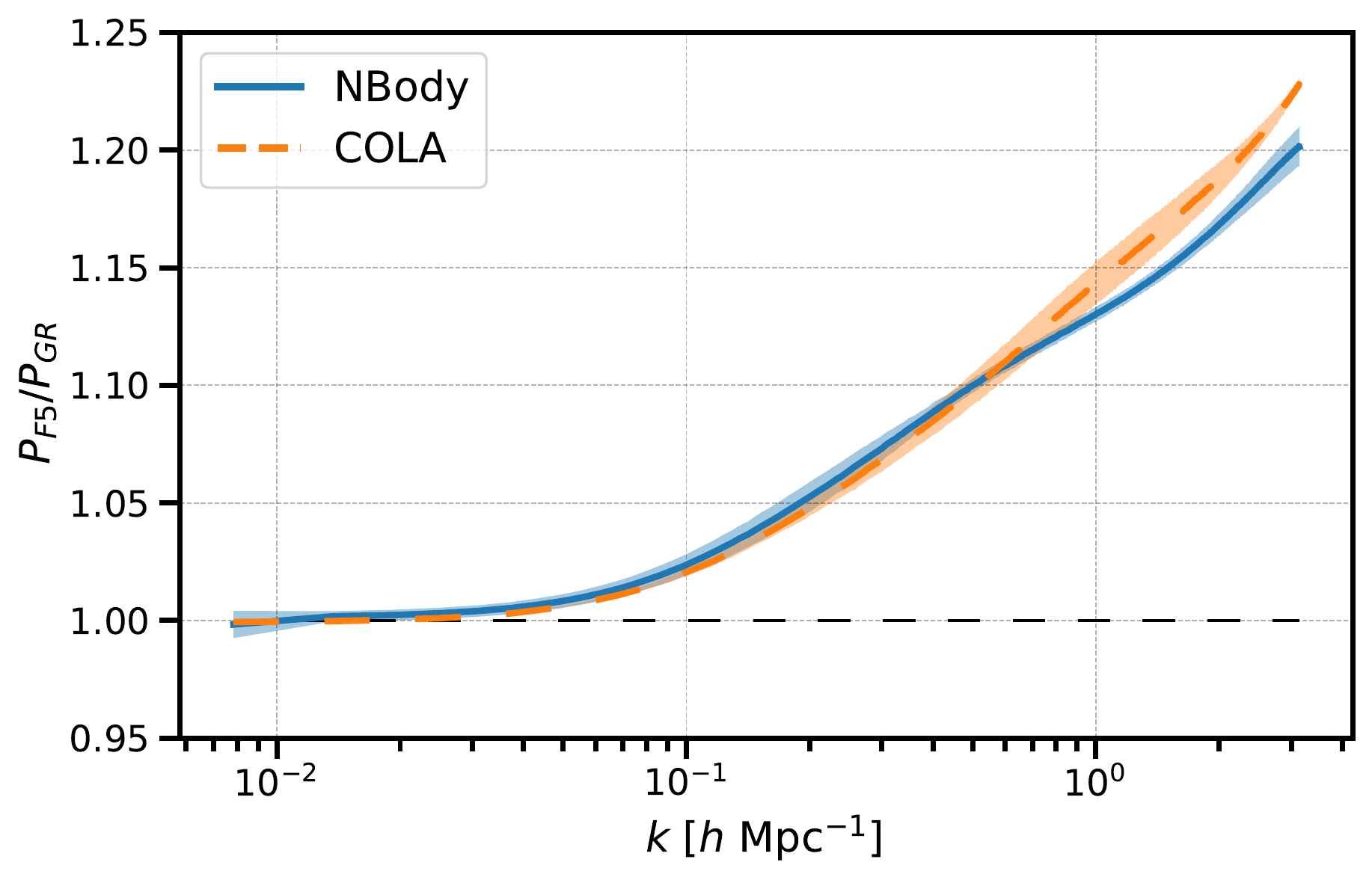}
}
\hfill
\subfloat[][N1 boost factor]{
\includegraphics[width=.48\textwidth]{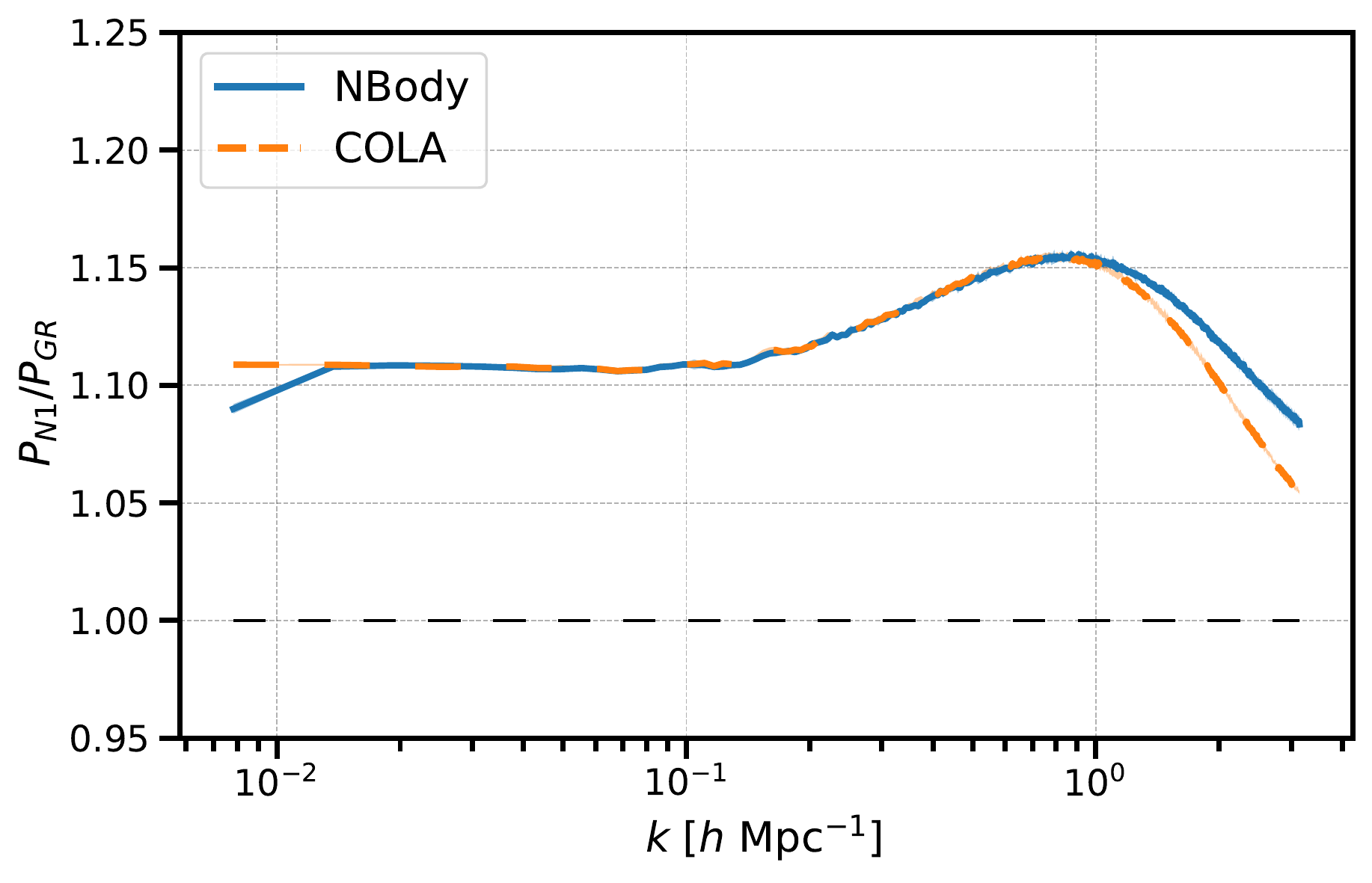}
}
\caption{\label{fig:DM_Pk} Dark Matter power spectrum comparison between COLA (orange dashed lines) and {\it N}-body (blue solid lines) obtained by taking the average over 5 realisations. The shaded regions represent the standard deviation over the 5 realisations. The power spectrum in GR (top panels) and the boost-factors in F5 (bottom left) and N1 (bottom right) show agreement within the variance up to $k \sim 1 \hompc$.}
\end{figure}

\begin{figure}[t]
\centering 
\subfloat[][GR]{
\includegraphics[width=.48\textwidth,clip]{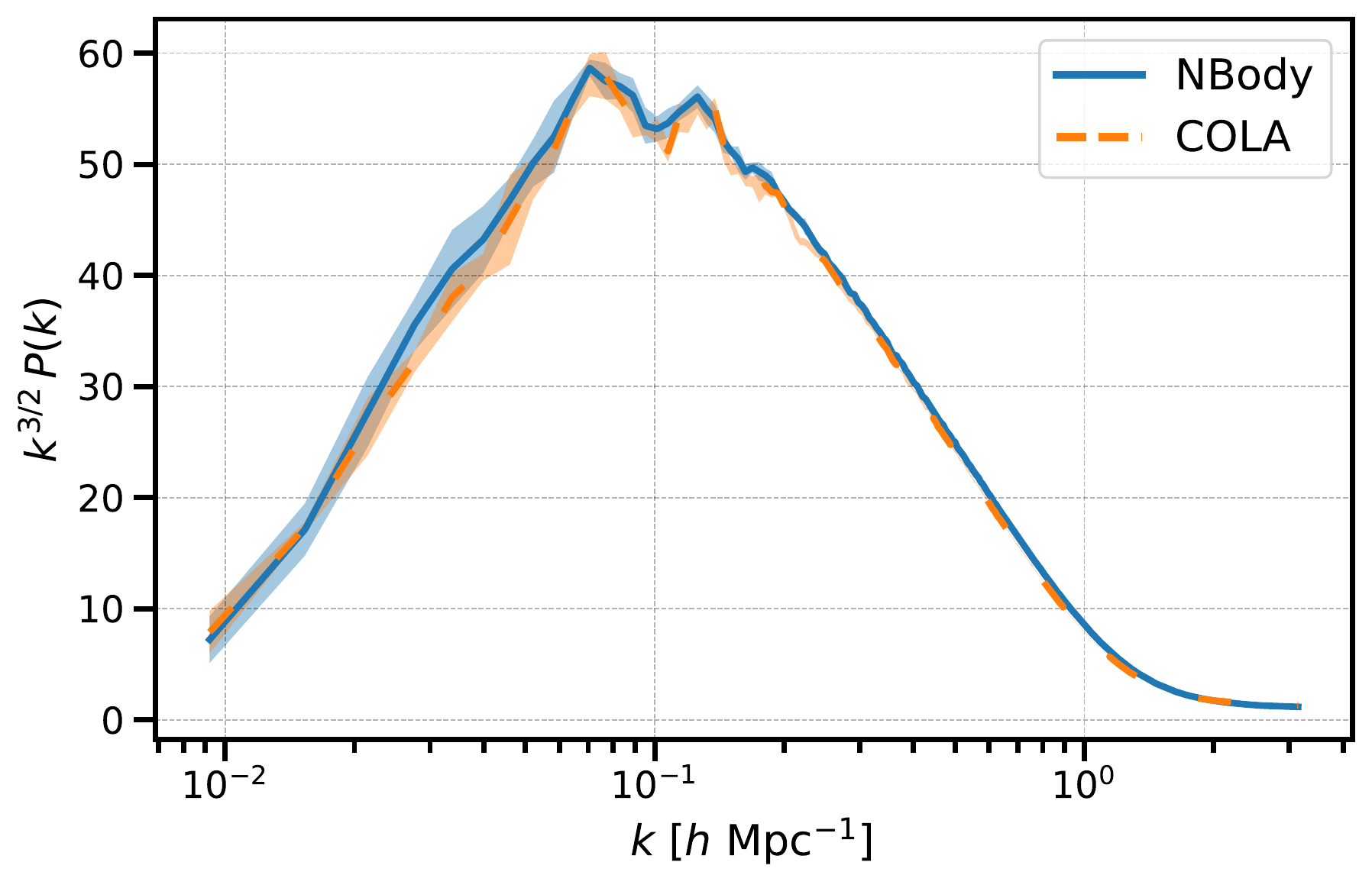}
}
\hfill
\subfloat[][GR ratio]{
\includegraphics[width=.48\textwidth]{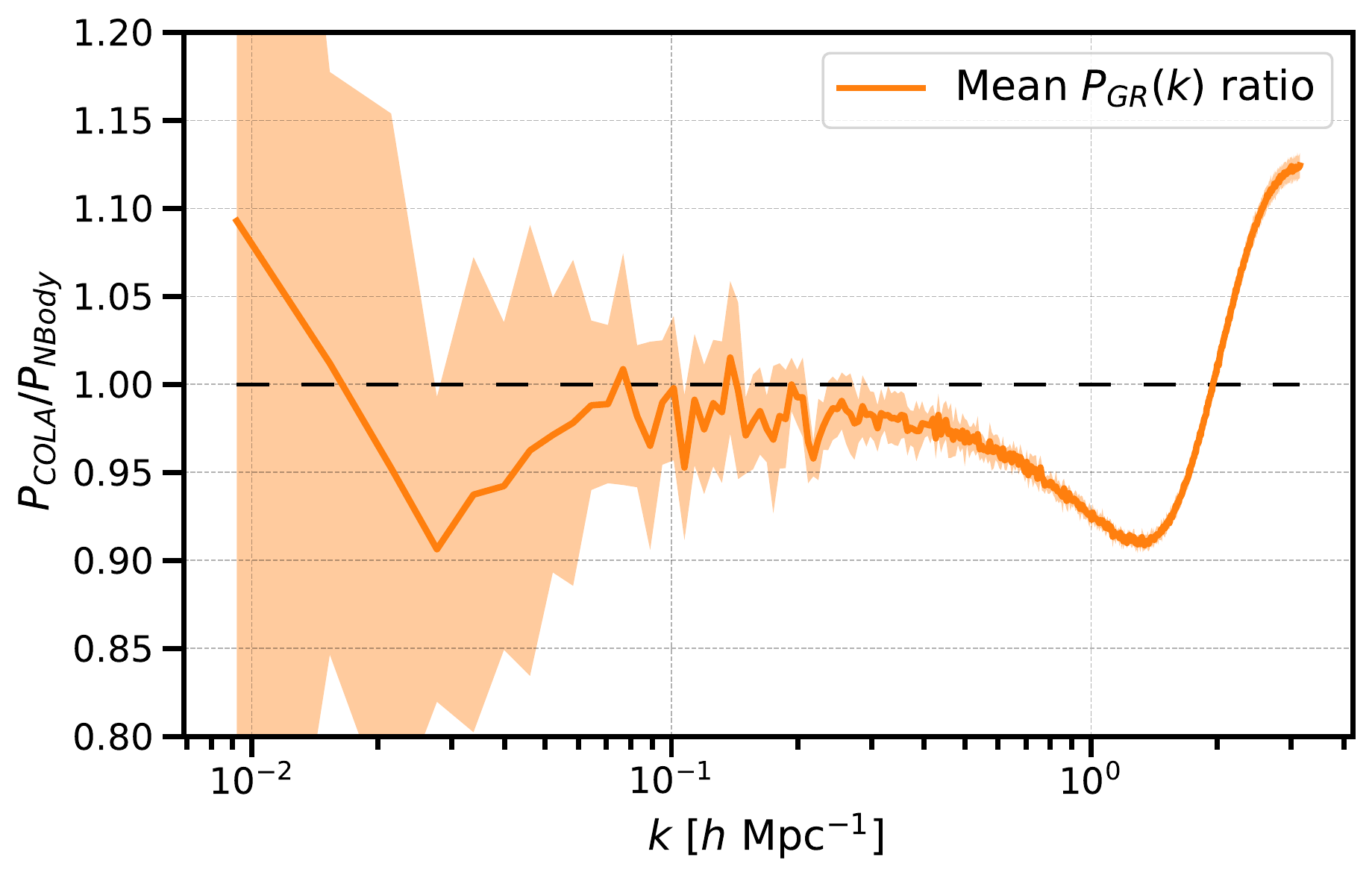}
}
\vfill
\subfloat[][F5 boost factor]{
\includegraphics[width=.48\textwidth]{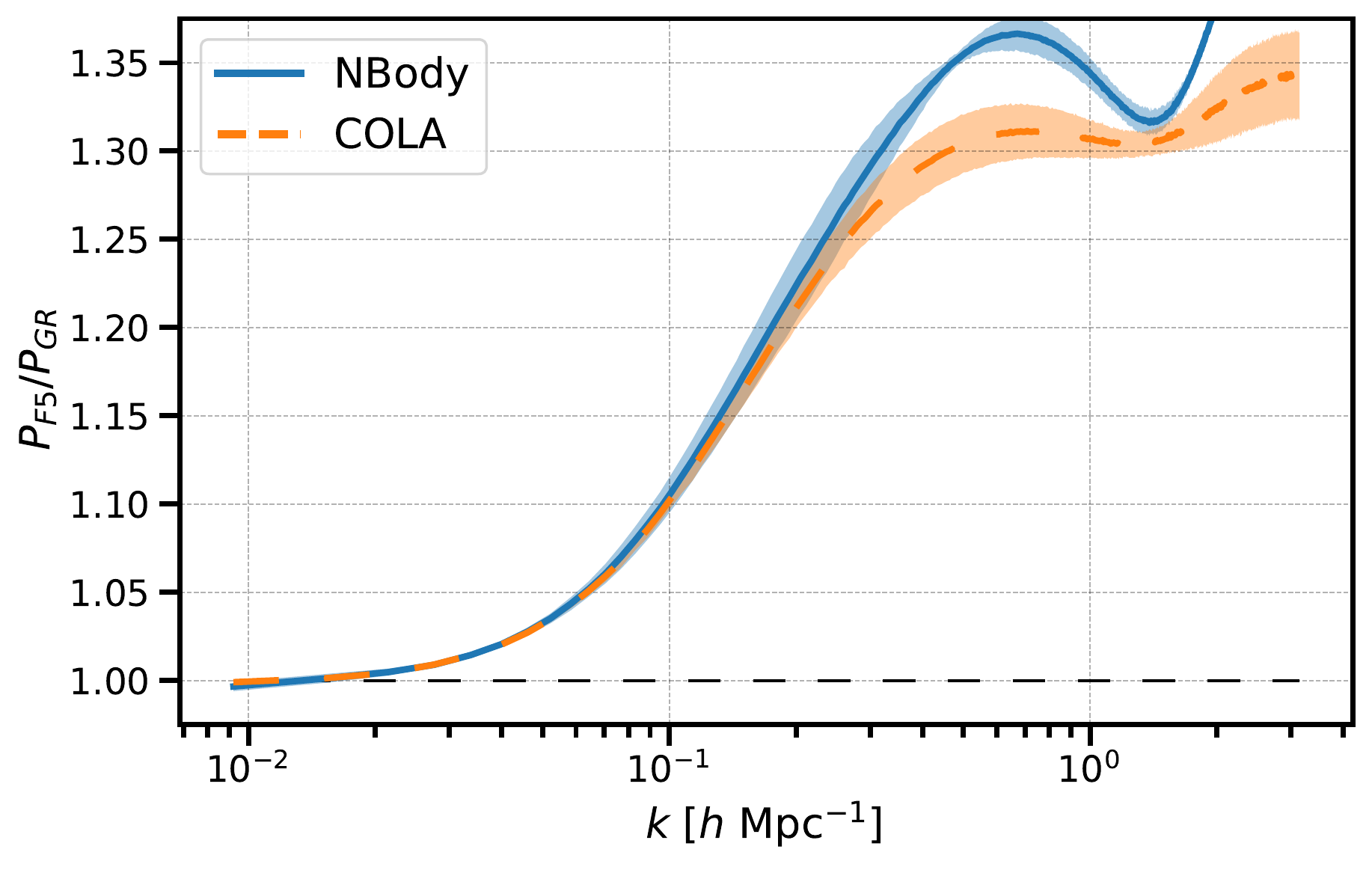}
}
\hfill
\subfloat[][N1 boost factor]{
\includegraphics[width=.48\textwidth]{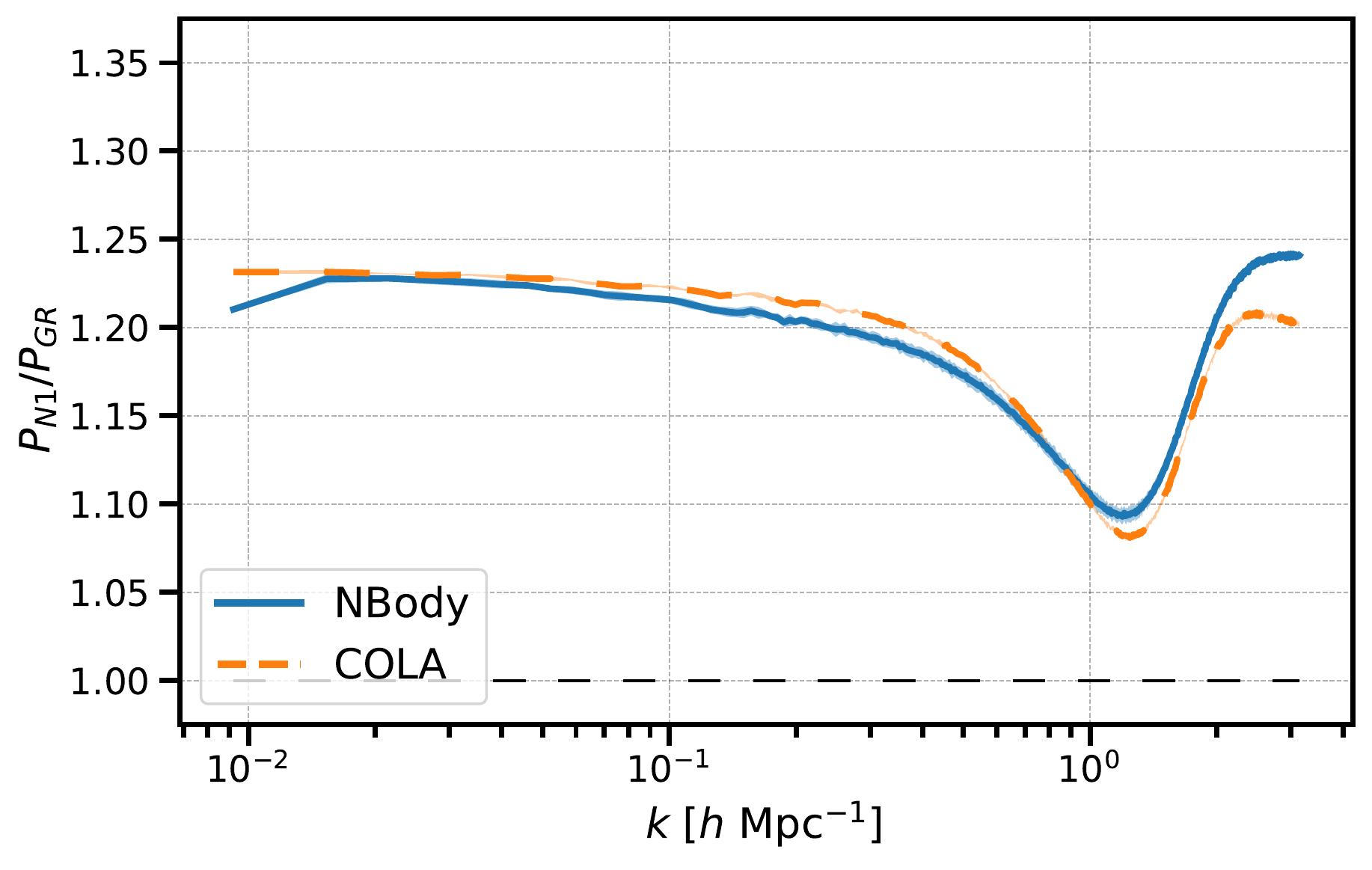}
}
\caption{\label{fig:DM_DTFE_Pk} Dark Matter velocity divergence power spectrum comparison between COLA (orange dashed lines) and {\it N}-body (blue solid lines) obtained by taking the average over 5 realisations. The shaded regions represent the standard deviation over the 5 realisations. The power spectrum in GR (top panels) is in agreement within $3\%$ up to $k \sim 0.4 \hompc$ where COLA starts to lose power. The boost-factor in F5 (bottom left) and N1 (bottom right) show better than $5\%$ accuracy even deeper in the non-linear regime, up to $k \sim 2 \hompc$.}
\end{figure}

We use the public codes PowerI4\footnote{\url{https://github.com/sefusatti/PowerI4}} and DTFE \cite{Cautun:2011gf} to compute and compare the density and velocity-divergence power spectra in the \textcode{elephant} and \textcode{piter} simulations. In Figures~\ref{fig:DM_Pk} and \ref{fig:DM_DTFE_Pk} we compare the clustering signals in GR (top left panel), the ratio between COLA and {\it N}-body (top right panel) and the enhancement in MG with respect to GR, which we refer as the boost-factor (bottom panels, F5 on the left and N1 on the right). The measurements correspond to the mean over 5 realisations and the shaded regions display the standard deviation (just to have a rough estimate of the errors to guide the eye). The errors are smaller in the boost-factor plots because the noise cancels out: the MG and the GR simulations were started from the same initial conditions and therefore have a very similar cosmic variance at late redshifts that vanishes in ratios. 
We find better than $3\%$ agreement between COLA and {\it N}-body for the density power spectra up to modes of $\sim 1 \hompc$ in GR while the boost-factors in F5 and N1 show agreement within the variance up to $k \sim 1 \hompc$. This shows that COLA simulations capture the MG effects very accurately even with the use of approximations for screening. The agreement in the velocity divergence power spectra is within $3\%$ up to $k \sim 0.4 \hompc$ in GR and for the boost factor in F5 while the boost factor in N1 agrees within $2\%$ up to $k \sim 2 \hompc$. In GR, after losing power at $k \sim 1 \hompc$, the velocity divergence power spectrum in COLA shows an enhancement compared with {\it N}-body at very high $k$. This is because COLA has less random motion at small scales that washes out the signal in {\it N}-body more than in COLA.
The better agreement in the density power spectra is expected because the velocity divergence power spectrum is more affected by non-linear physics.

From the boost-factors, we can already appreciate the intrinsic difference between F5 and N1: while F5 has a scale-dependent enhancement, N1 is almost scale-invariant across most scales. This behavior can be explained by comparing the respective Poisson equations: $G_{\rm eff}$ is a function of $k$ in Eq.~\eqref{fR_Poisson} for $f(R)$ gravity while it is scale-independent in  Eq.~\eqref{DGP_Poisson} for DGP.  

\section{Halos}
\label{sec:Halos}
Halos can be found by means of specific algorithms often referred to as halo-finders. Amongst the several halo-finders proposed in literature we focus on two popular codes: \textcode{rockstar} \cite{Behroozi13} and the Friends-of-Friends (FoF) finder \cite{Davis85} (as implemented in \texttt{nbodykit}\footnote{\url{https://nbodykit.readthedocs.io}}). \textcode{rockstar} uses a 6D phase-space FoF finder and is capable of measuring many halo properties like the spherical over-density mass, the concentration parameter and the velocity dispersion just to name a few. 
The FoF finder instead is purely based on particles positions.

\subsection{\textcode{rockstar} halos}
\begin{figure}[t]
        \centering 
        \subfloat[][GR]{
        \includegraphics[width=.48\textwidth,clip]{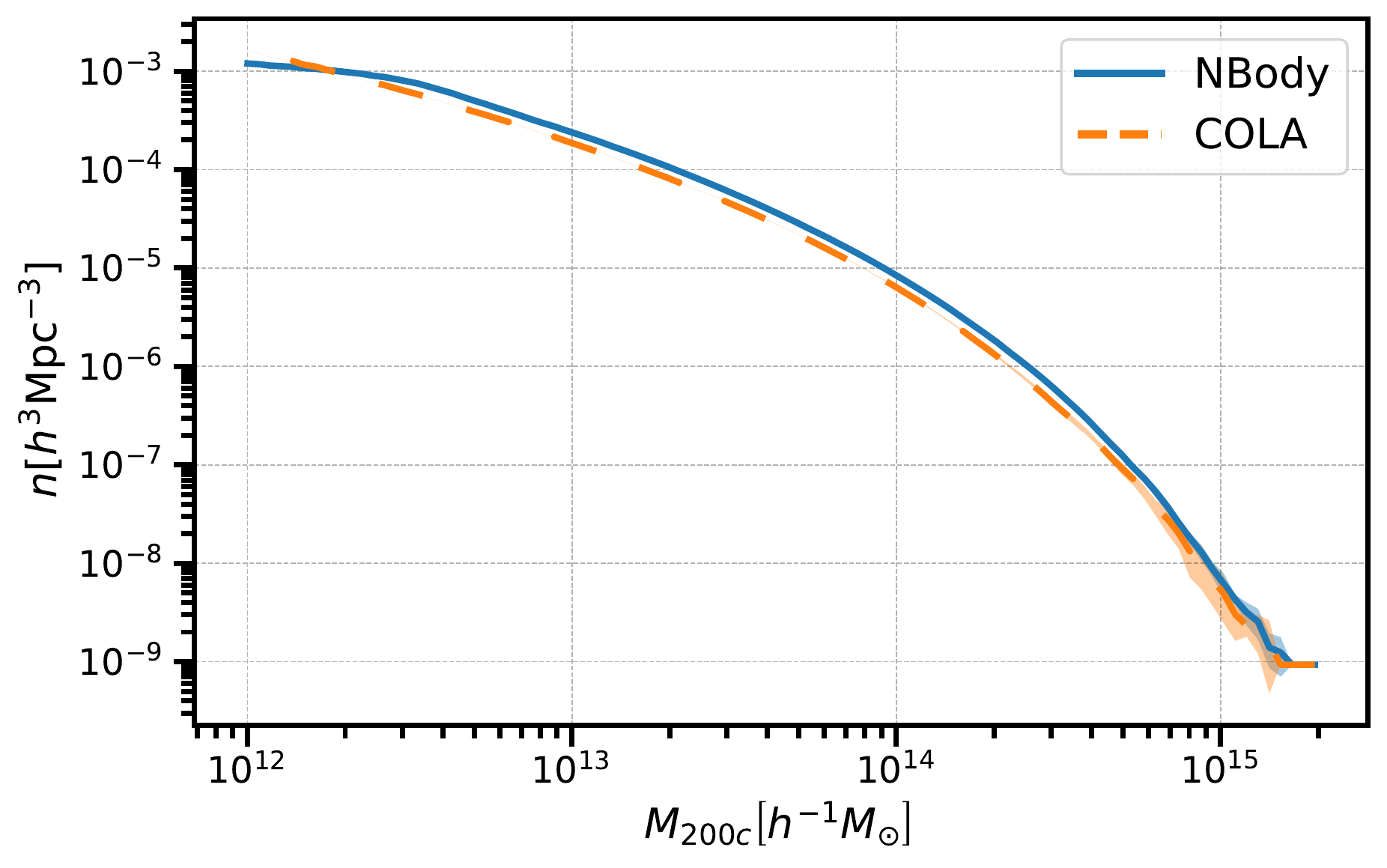}
        }
        \hfill
        \subfloat[][GR ratio]{
        \includegraphics[width=.48\textwidth]{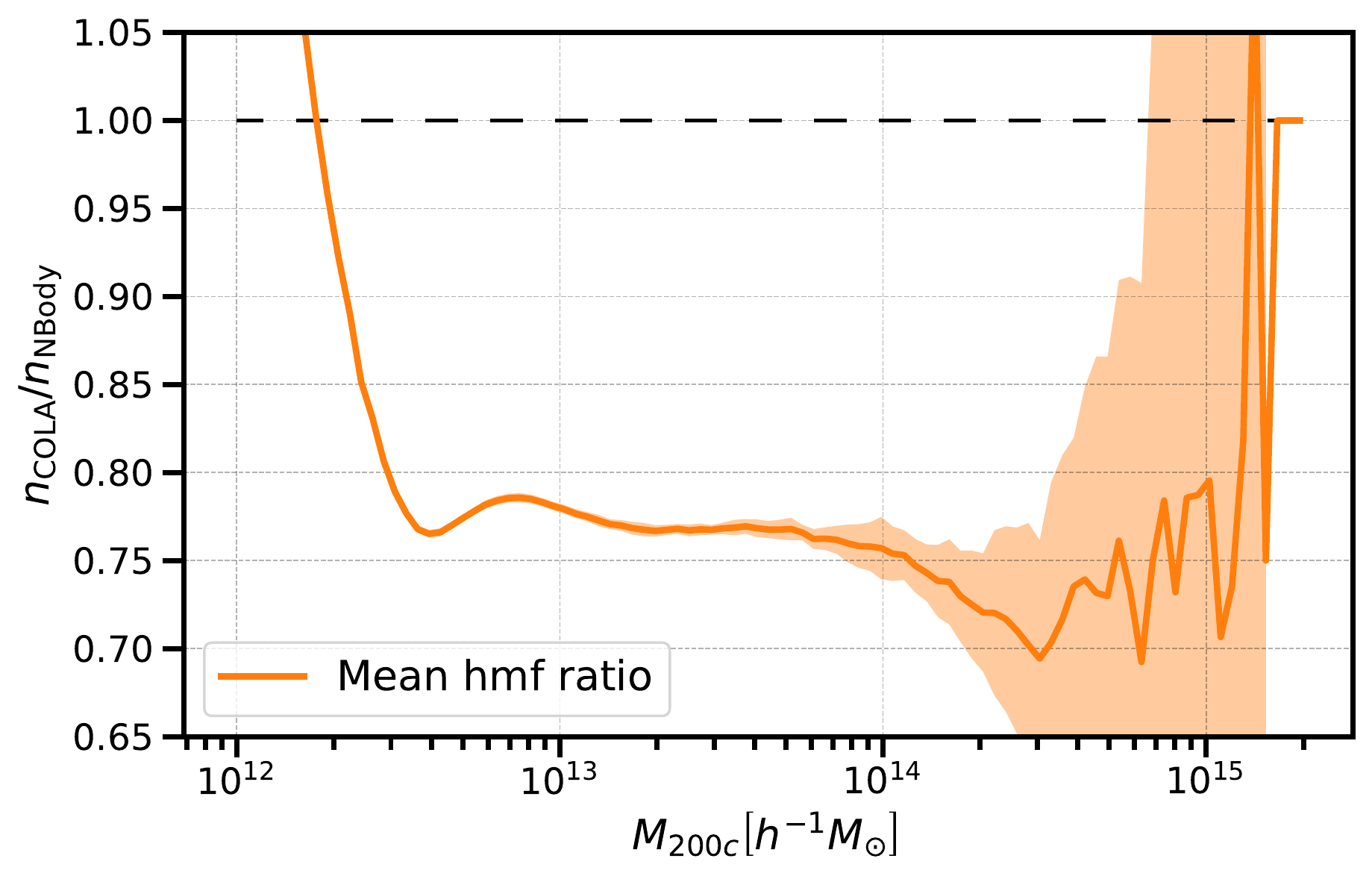}
        }
    \caption{\textcode{rockstar} halo mass function comparison in GR. The hmf in COLA and {\it N}-body (left panel) and their ratio (right panel) show $\sim 25 \%$ discrepancy for masses greater than $10^{13} \Msun$.}
    \label{fig:hmf_Rock}
\end{figure}

\begin{figure}[t]
        \centering 
        \subfloat[][GR]{
        \includegraphics[width=.48\textwidth,clip]{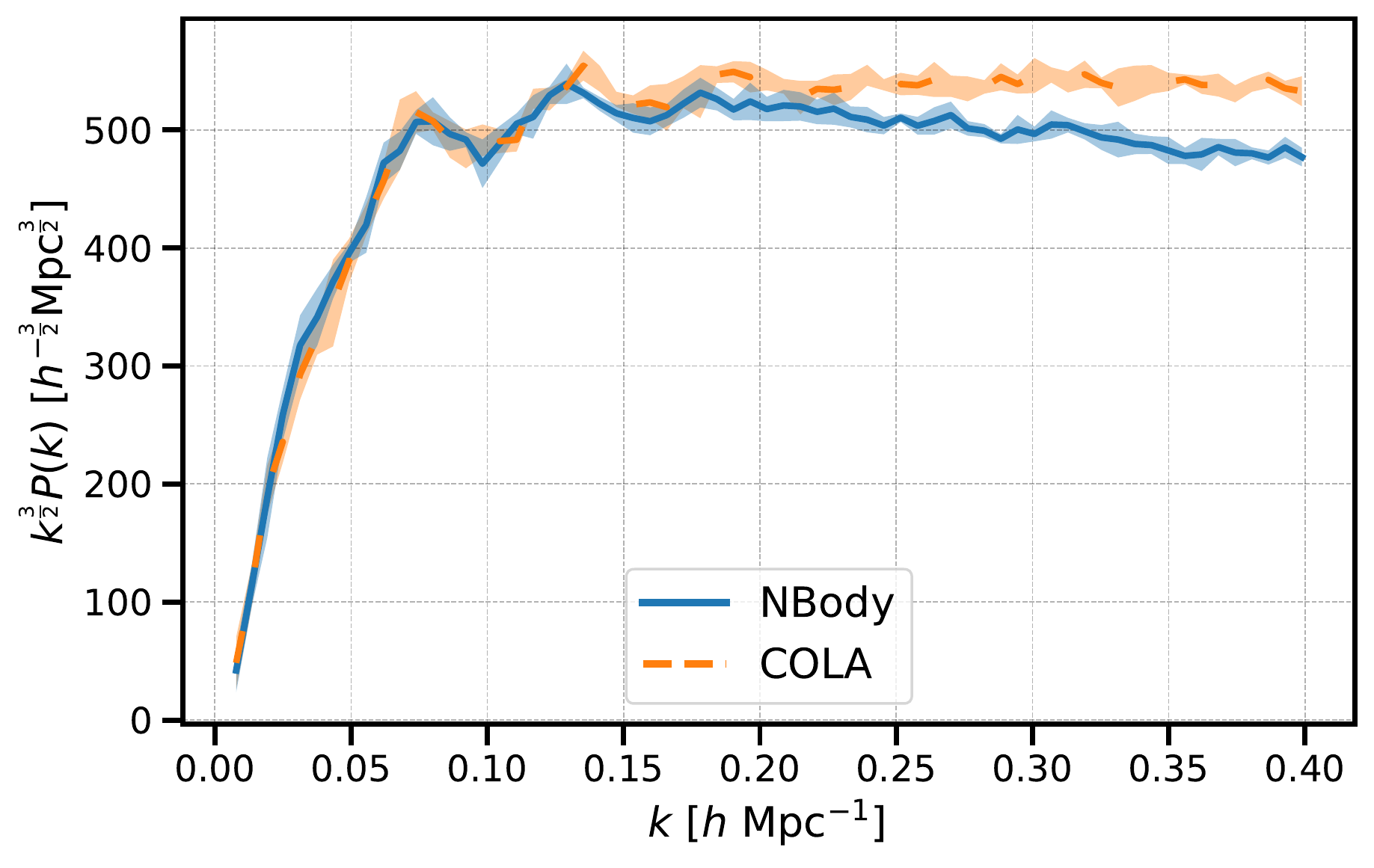}
        }
        \hfill
        \subfloat[][GR ratio]{
        \includegraphics[width=.48\textwidth]{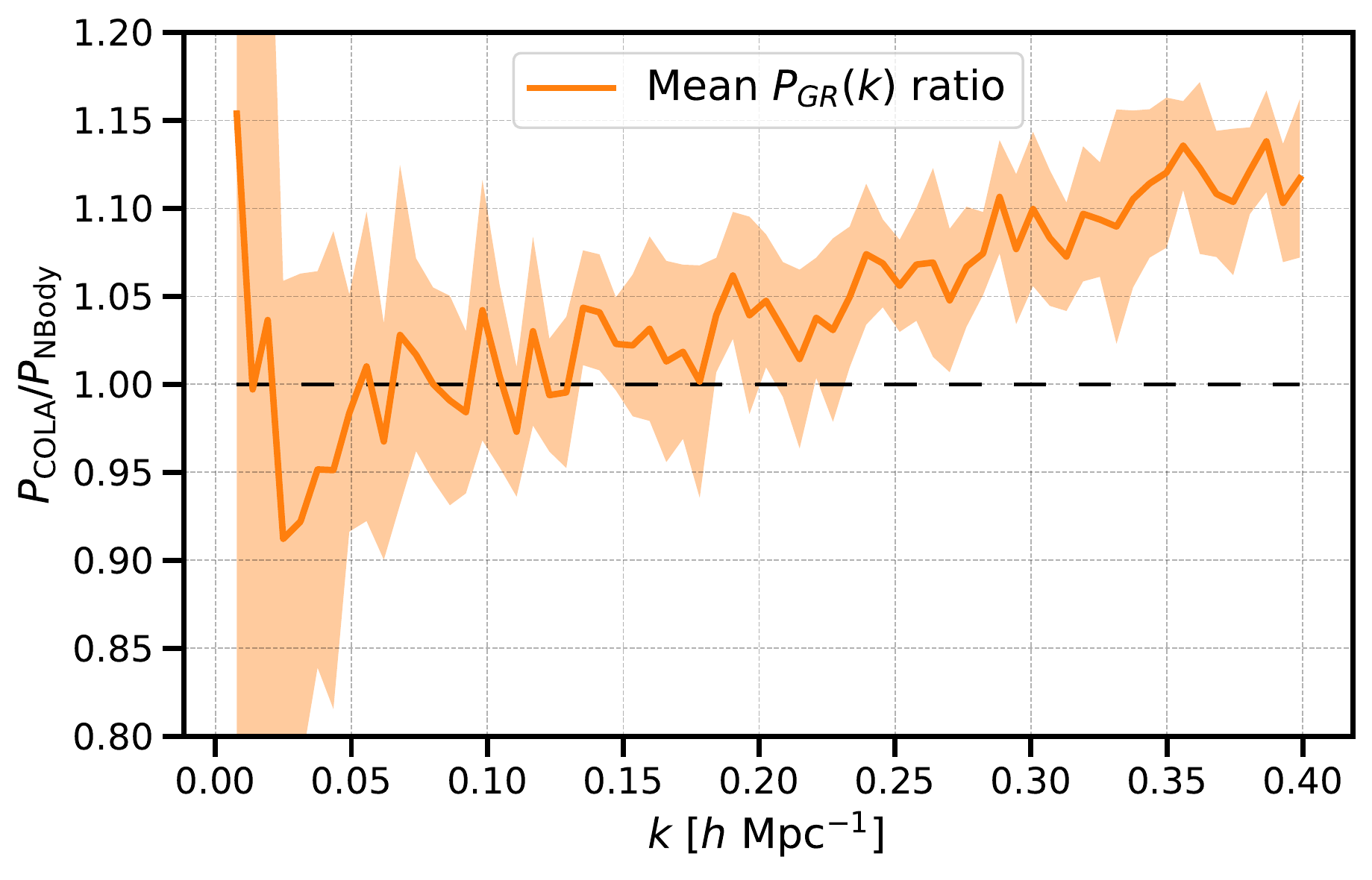}
        }
    \caption{\textcode{rockstar} power spectrum comparison between the power spectra of COLA and {\it N}-body in GR. The discrepancy, which is already clear in the left panel, is shown to reach the $10\%$ with a scale-dependent behaviour in the right panel.}
    \label{fig:halo_Pk_Rock}
\end{figure}

We first produce \textcode{rockstar} halo catalogues in COLA and {\it N}-body and compare the cumulative halo mass function (hmf) and the halo power spectrum. 
We use the setting to remove unbound particles and keep only parent halos\footnote{We found that, if we do not remove unbound particles, the agreement between {\it N}-body and COLA is substantially worse for the halo power spectrum}.
The hmf is simply defined as the number density of halos with mass greater than $M$. Before measuring the halo power spectrum, we draw a halo sample defined by an abundance matching procedure, in which the target density is given by a cut at $10^{13} \Msun$ in the {\it N}-body catalogues for GR.
Figures~\ref{fig:hmf_Rock} and~\ref{fig:halo_Pk_Rock} show the comparisons of the hmf and the halo power spectrum for COLA and {\it N}-body \textcode{rockstar} catalogues respectively, both in GR. The hmf is $25\%$ off for masses greater than $10^{13} \Msun$, and the halo power spectrum ratio is well predicted in the large-scale limit but shows a scale-dependent deviation, up to $10\%$ at $k\sim 0.4 \hompc$. These differences are likely due to the lower force resolution and accuracy in recovering the velocity field in COLA simulations.

Given this result, in the rest of the chapter, we will use only the FoF finder to generate halo catalogues from COLA and {\it N}-body simulations. However, we still use \textcode{rockstar} in the {\it N}-body simulations to quantify the effects of MG on the concentration and velocity dispersion of halos in F5 simulations (see Section~\ref{sec:F5_NFWtweaks}), which we then incorporate in the prescriptions that we use to populate the COLA simulations with galaxies.

\subsection{Friends-of-Friends halos}
The FoF algorithm links together particles that are separated by a distance smaller than a certain linking length $b$, expressed in units of the mean inter-particle distance. We adopt the standard value $b=0.2$. This value, which is commonly used in {\it N}-body simulations, has been shown to also be valid in COLA simulations (see \cite{Koda:2015mca,Howlett:2014opa,Izard:2015dja}), although, in some other approximate methods that use very low-resolution density fields, $b$ has been tweaked, see \cite{Manera:2012sc}.  
At this point, we are interested in converting FoF masses, $M_{\rm FoF}$, to  spherical over-density (SO) masses, $M_{\rm200c}$, which we adopt in the rest of the chapter when expressing halo masses.
$M_{\rm200c}$ denotes the mass enclosed in a spherical region with density 200 times the critical density. We use the fit in \cite{Lukic:2008ds} to convert mass definitions with an accuracy of $5\%$ for most halos in the mass range $10^{12.5}-10^{15.5} \Msun$:
\begin{equation}
    \frac{M_{\rm FoF}}{M_{200 \mathrm{c}}}=\frac{a_{1}}{c_{200}^{2}}+\frac{a_{2}}{c_{200}}+a_{3}
    \label{eq:mfof_m200_conversion} \, ,
\end{equation}
where the coefficients $a_1$, $a_2$ and $a_3$ depend on the number of particles (see their Table 1), and $c_{200}$ is the concentration parameter, which we compute using the empirical formula from \cite{Dutton:2014xda}:

\begin{align}
   \log _{10} c_{200} &=a+b \log _{10}\left( \frac{M_{\rm 200 c}}{10^{12} \Msun}\right), \label{eq:concentration}\\
   a& =0.520+(0.905-0.520) \exp \left(-0.617 z^{1.21}\right), \nonumber \\ 
   b& =-0.101+0.026 z. \nonumber
\end{align}
The empirical relations in eq.~\eqref{eq:mfof_m200_conversion} and eq.~\eqref{eq:concentration} have been calibrated in $\Lambda$CDM simulations but given the small amplitude of MG effects on the halo concentration that we discuss in section~\ref{sec:F5_NFWtweaks}, we argue that the above relations are valid within our target accuracy also in the MG theories we consider in this chapter.

\begin{figure}
        \centering 
        \subfloat[][Mass conversion factor]{
        \includegraphics[width=.48\textwidth,clip]{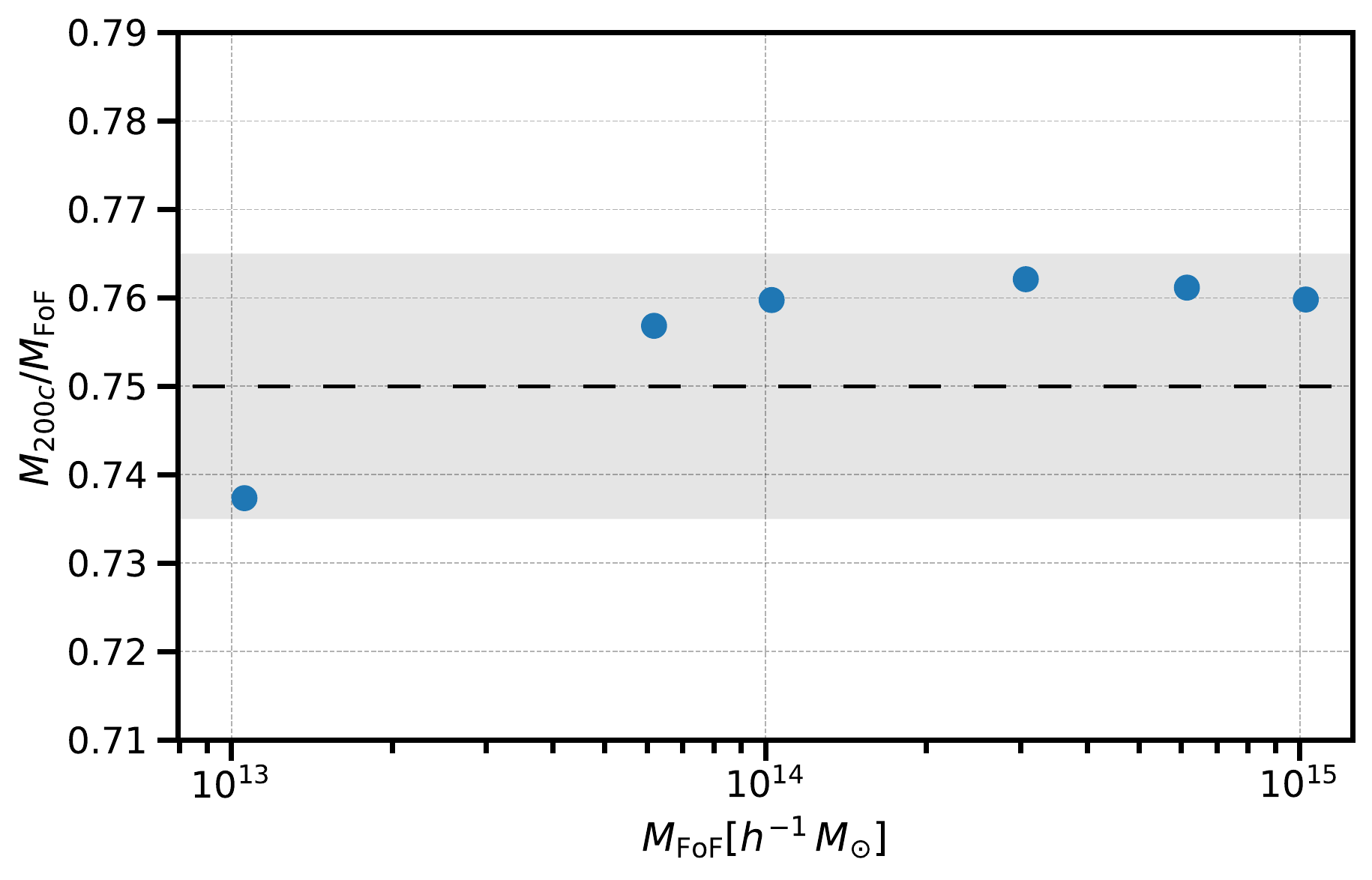}
        }
        \hfill
        \subfloat[][\textcode{rockstar} comparison]{
        \includegraphics[width=.48\textwidth]{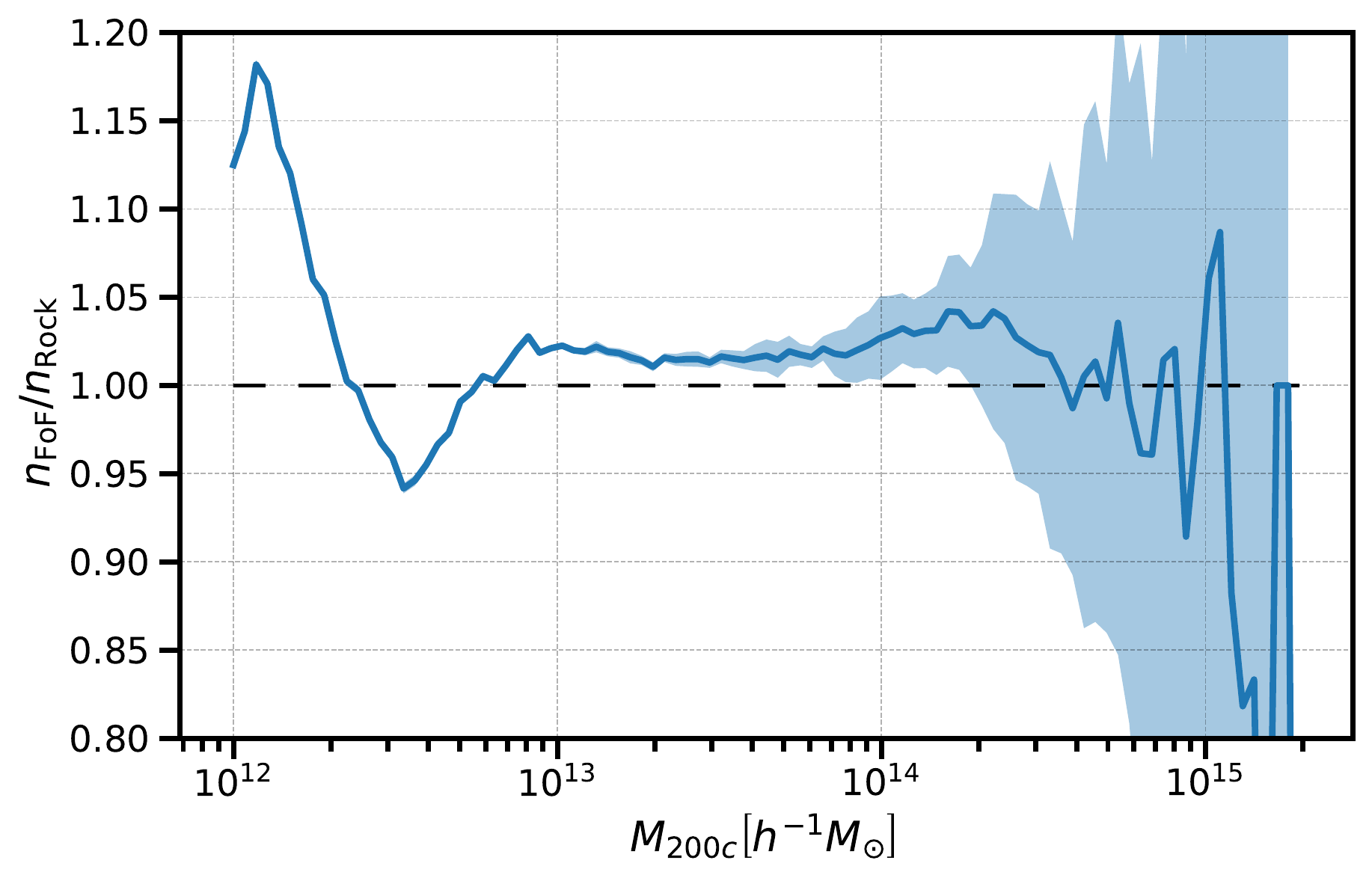}
        }
    \caption{Mass conversion between FoF and $M_{200c}$ spherical over-density masses. The conversion factor is evaluated for 6 mass values (left panel), corresponding to the SO mass of halos composed of 100, 600, 1000, 3000, 6000 and 10000 DM particles. All the six points lie within $2\%$ from the value $0.75$ which we select as a constant to convert the masses of the FoF halos to SO masses. In the right panel, we use the $N$-body simulations in GR to compare the FoF hmf resulting after this conversion with the hmf of \textcode{rockstar} catalogues. The ratio between the two (right panel) shows a better than $3\%$ agreement for masses above $10^{13} \Msun$.}
    \label{fig:mass_conversion}
\end{figure}

The left panel in Figure~\ref{fig:mass_conversion} shows the conversion factor of Eq.~\eqref{eq:mfof_m200_conversion} (for a set of values corresponding to SO masses of 100, 600, 1000, 3000, 6000 and 10000 particles). There is very little dependence on halo mass, with all the values within $2\%$ of 0.75. Hence we use the constant value 0.75 to convert the FoF masses to SO masses. 
To validate this mass conversion, we compare the mass function for the FoF halo catalogues $n_{\rm FoF}$ with rescaled $M_{200 \mathrm{c}}=0.75 M_{\rm FoF}$ to the mass function as measured by \textcode{rockstar} $n_{\rm Rock}$ in terms of $M_{\rm 200c}$. The right panel of Figure~\ref{fig:mass_conversion} shows this comparison for {\it N}-body GR halo catalogues, and the agreement of these two mass functions is within $3\%$ for masses grater than $10^{13} \Msun$. 
We thus produce halo catalogues for both the {\it N}-body and COLA simulations using the FoF method and assign the over-density mass $M_{200 \mathrm{c}}=0.75 M_{\rm FoF}$.

\begin{figure}[t]
\centering 
\subfloat[][GR]{
\includegraphics[width=.48\textwidth,clip]{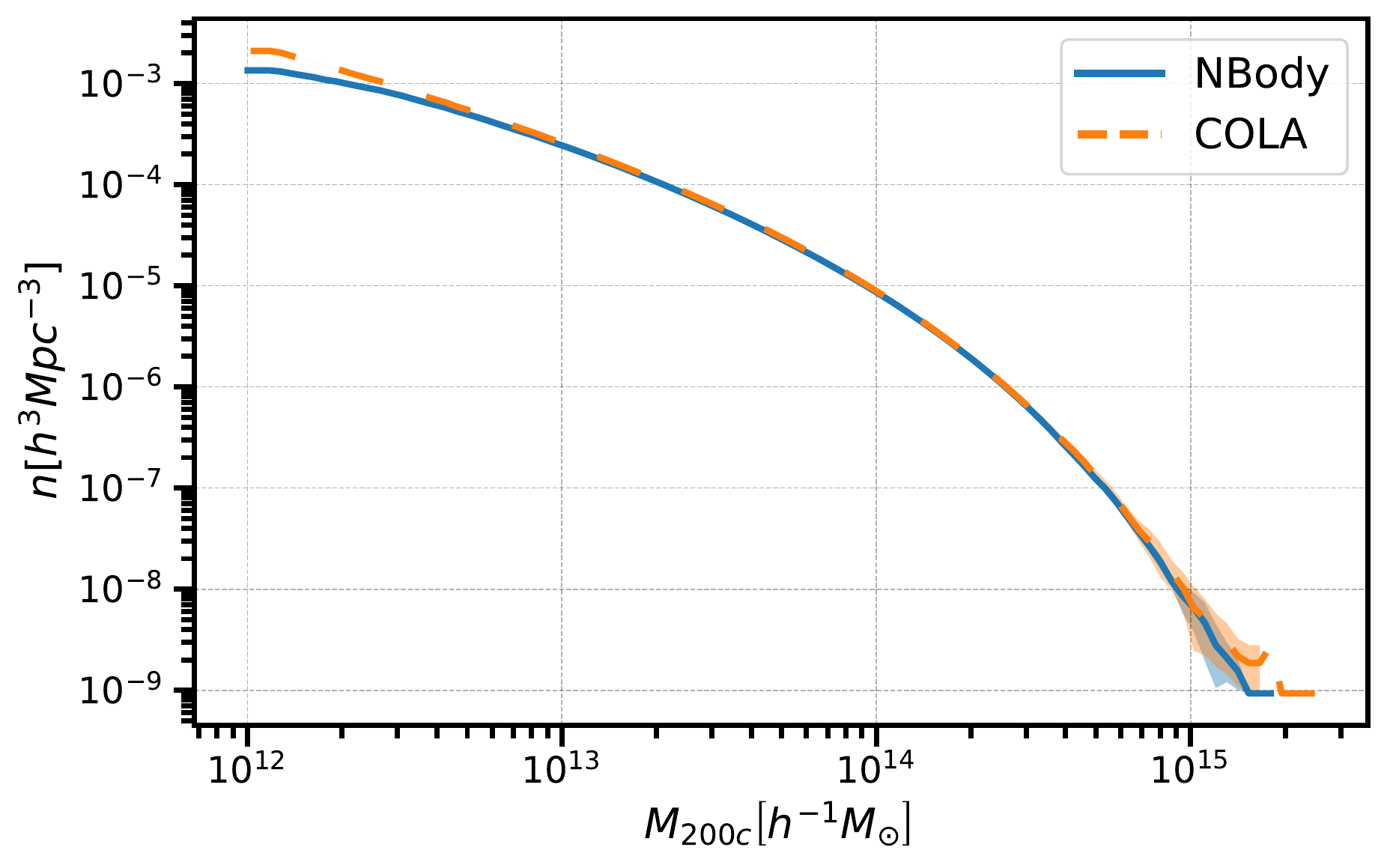}
}
\hfill
\subfloat[][GR ratio]{
\includegraphics[width=.48\textwidth]{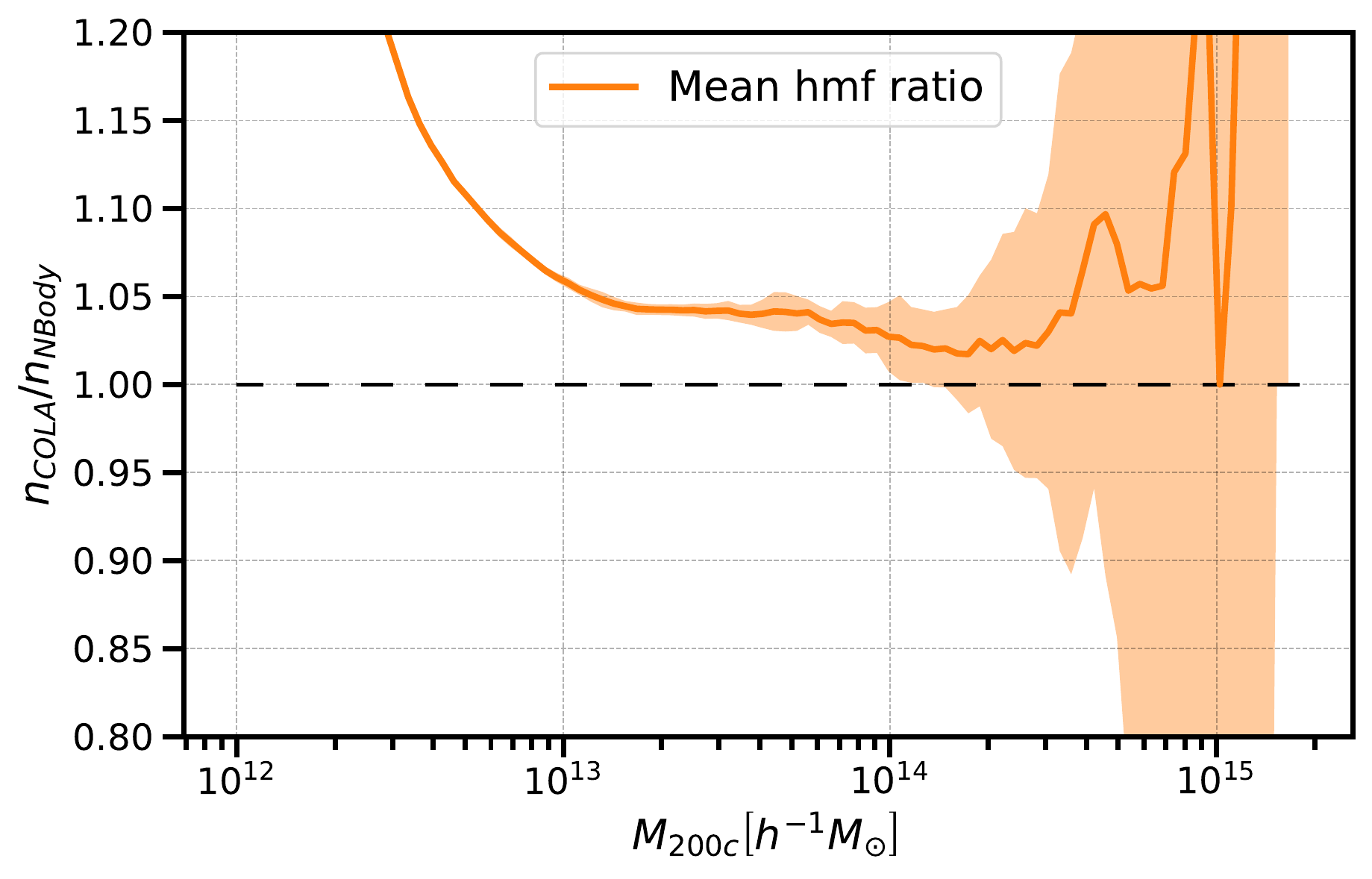}
}
\vfill
\subfloat[][F5 boost factor]{
\includegraphics[width=.48\textwidth]{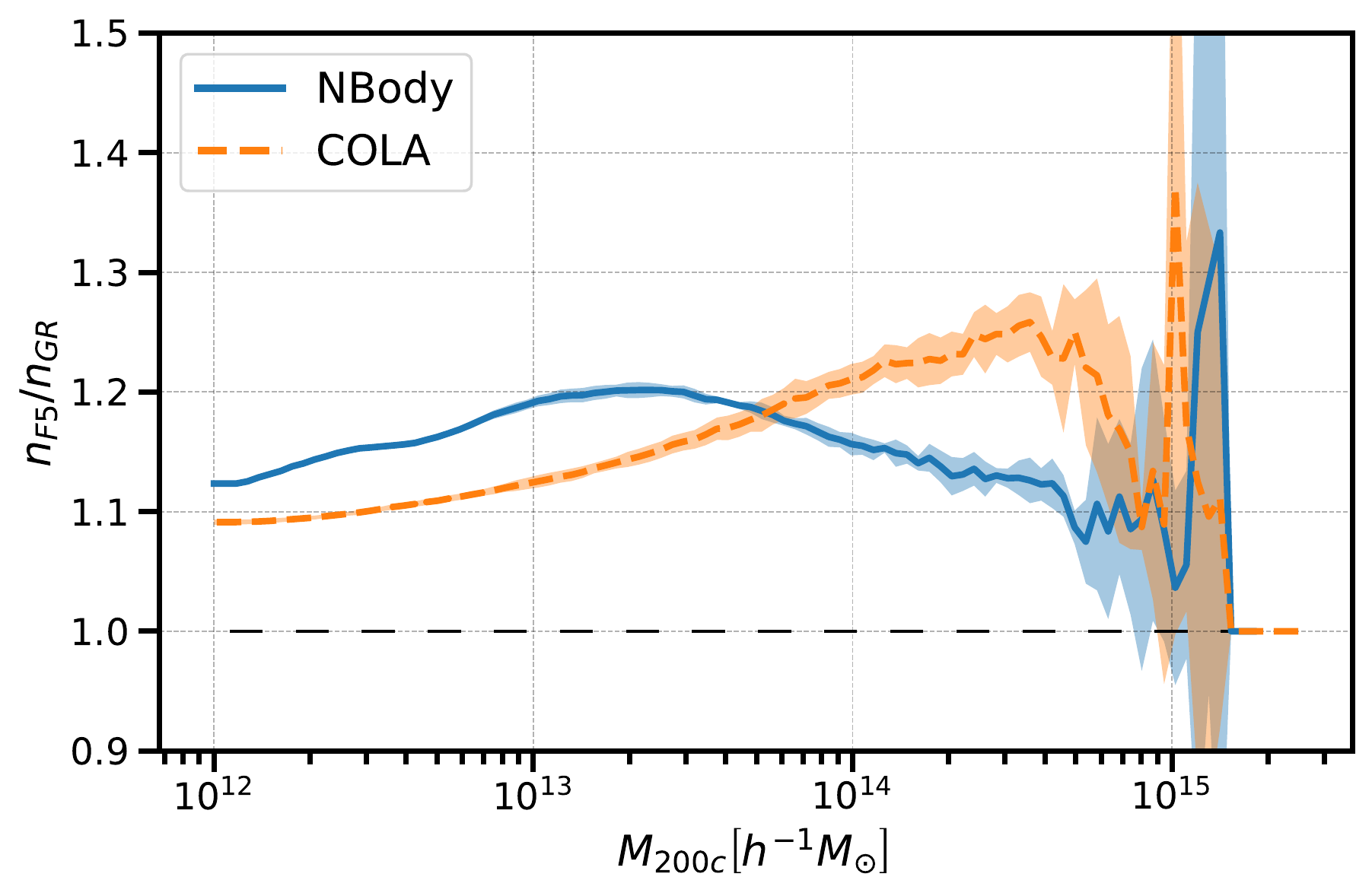}
}
\hfill
\subfloat[][N1 boost factor]{
\includegraphics[width=.48\textwidth]{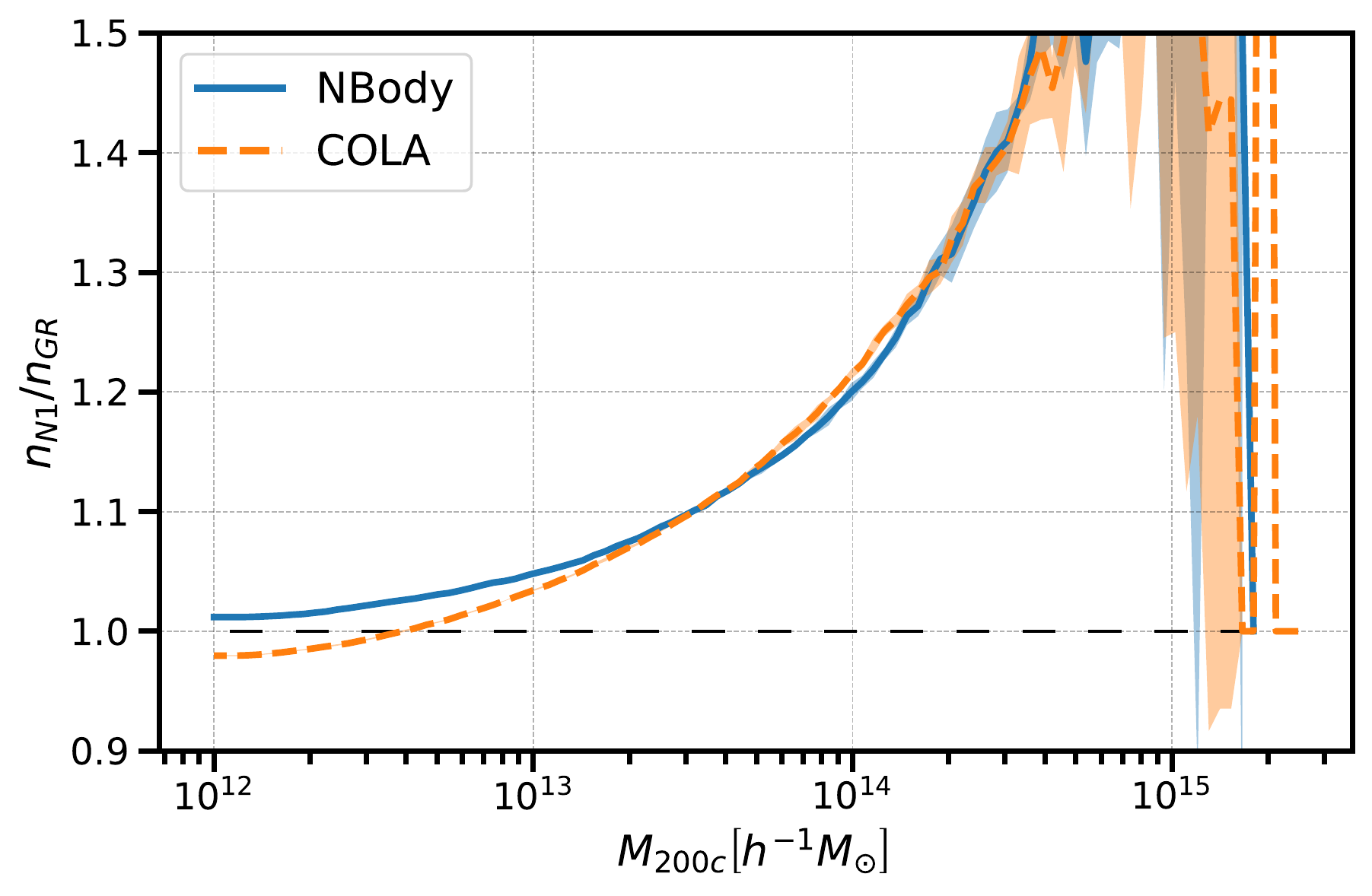}
}
\caption{Halo mass function comparison between COLA (orange dashed lines) and {\it N}-body (blue solid lines) obtained by taking the average over 5 realisations. The shaded regions represent the standard deviation over the 5 realisations. The hmf in GR (top panels) show a $\sim 5 \%$ accuracy for masses above $10^{13} \Msun$. The boost factor in N1 (bottom right) in COLA agree with that in {\it N}-body within the $5 \%$ at all masses, while the boost factor in F5 (bottom left) features an agreement within the $10 \%$.}
\label{fig:Halo_hmf}
\end{figure}

\begin{figure}[t]
\centering 
\subfloat[][GR]{
\includegraphics[width=.48\textwidth,clip]{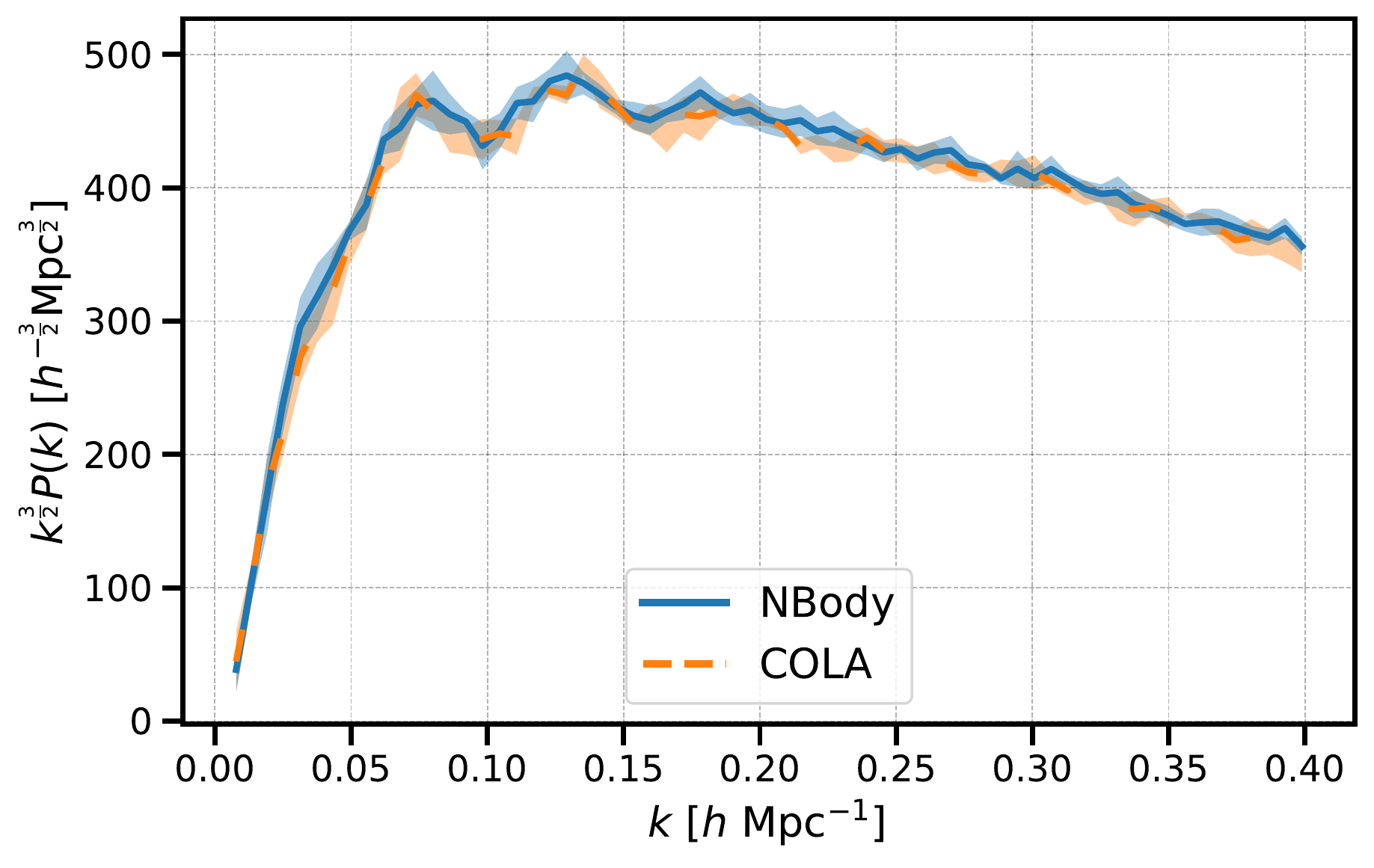}
}
\hfill
\subfloat[][GR ratio]{
\includegraphics[width=.48\textwidth]{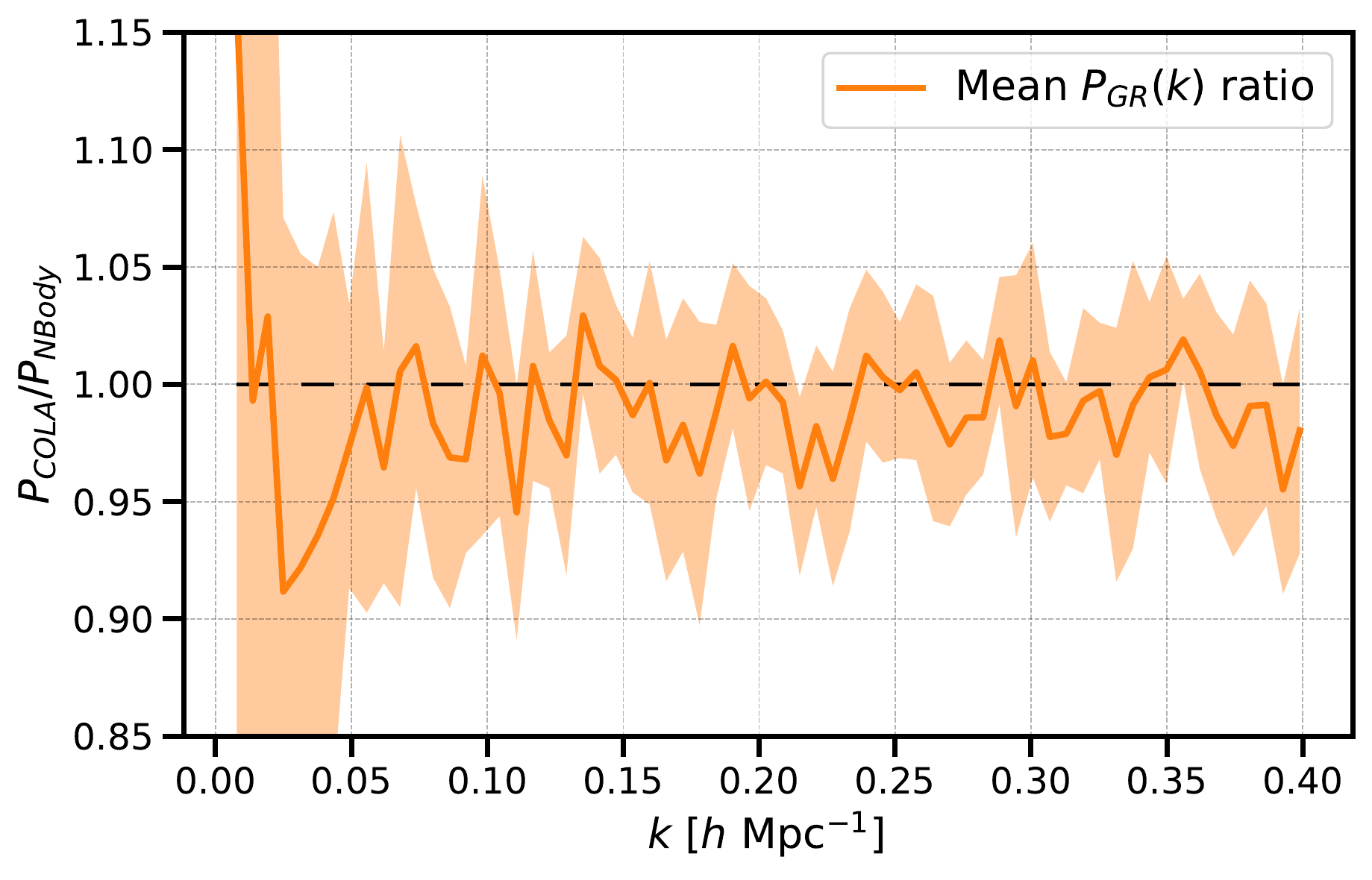}
}
\vfill
\subfloat[][F5 boost factor]{
\includegraphics[width=.48\textwidth]{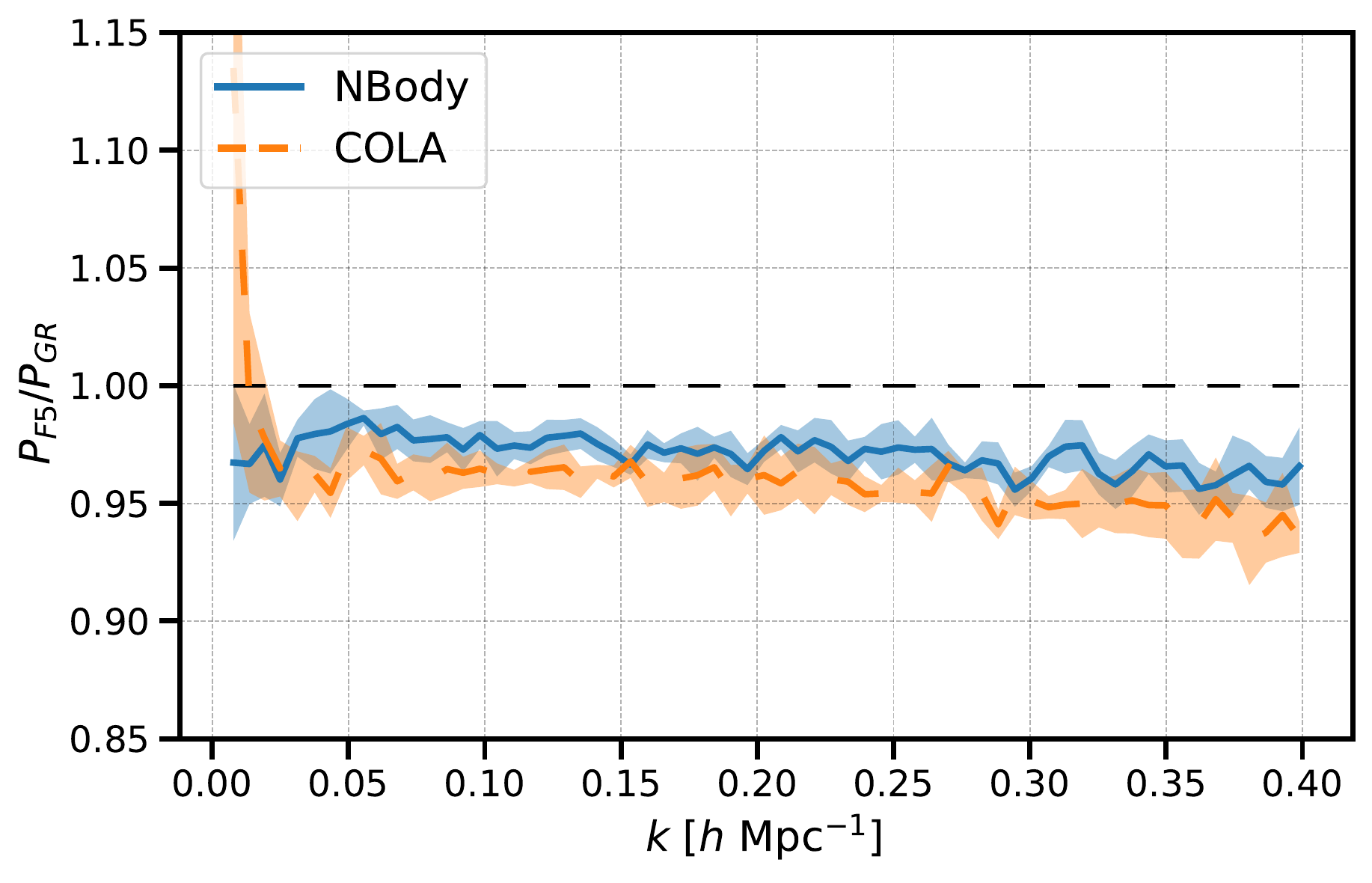}
}
\hfill
\subfloat[][N1 boost factor]{
\includegraphics[width=.48\textwidth]{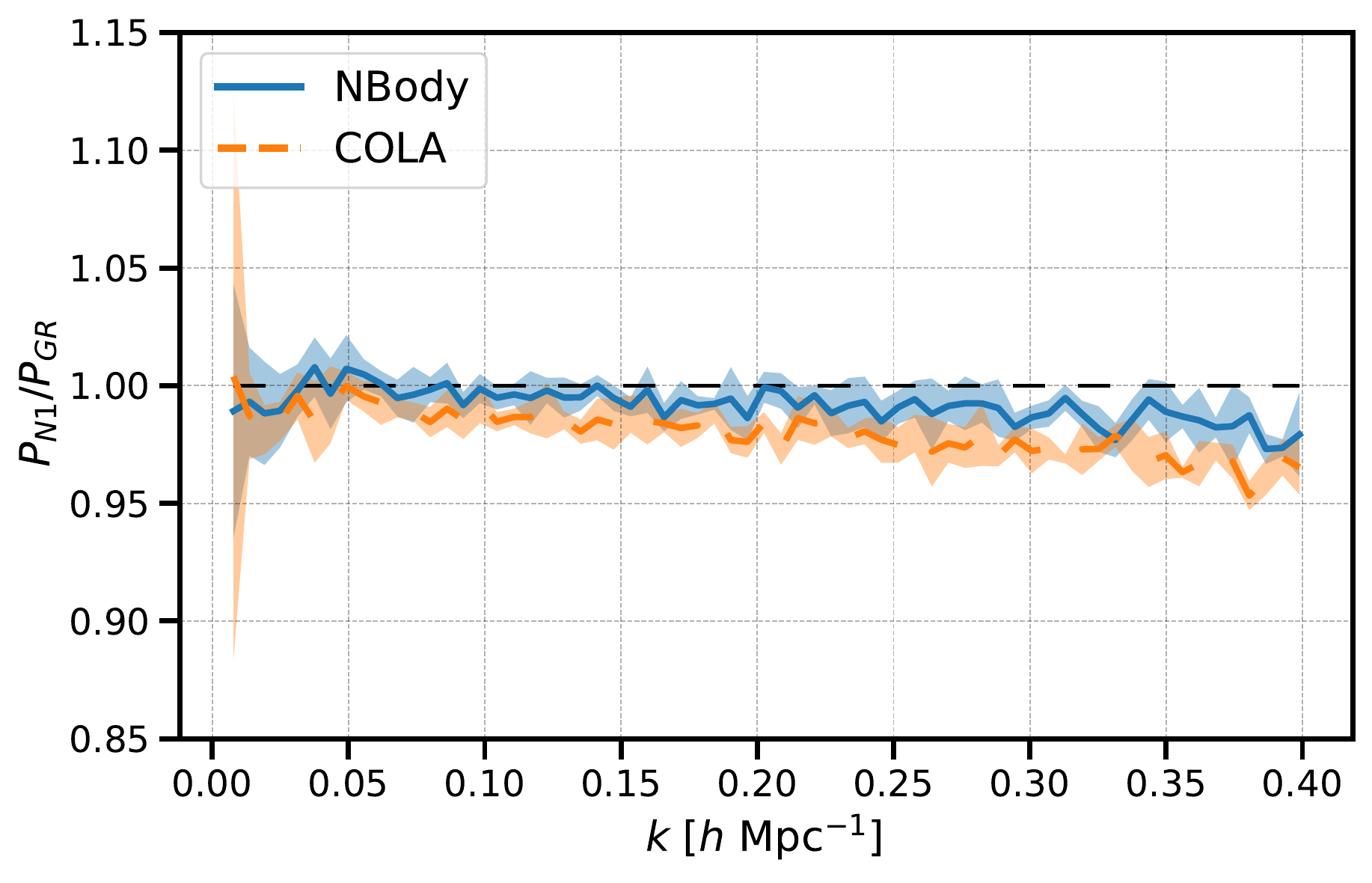}
}
\caption{Halo power spectrum comparison between COLA (orange dashed lines) and {\it N}-body (blue solid lines) obtained by taking the average over 5 realisations. The shaded regions represent the standard deviation over the 5 realisations. The halo power spectrum in GR (top panels) and the boost factors in F5 and N1 (bottom panels) show agreement within the variance up to $k \sim 0.4 \hompc$.}
\label{fig:Halo_Pk}
\end{figure}

To validate the COLA halo catalogues we compare against the {\it N}-body halo catalogues for the following two summary statistics: the halo mass function and the halo power spectrum, which are shown in Figure~\ref{fig:Halo_hmf} and Figure~\ref{fig:Halo_Pk} respectively. 
For the halo power spectrum, we draw a halo sample defined by an abundance matching procedure, in which the target density is given by a cut at $10^{13} \Msun$ in the {\it N}-body catalogues for GR. We see that COLA reproduces the {\it N}-body mass function with an accuracy better than $5\%$ for halos with more than 130 particles (or $10^{13}\Msun$) in GR, while the hmf boost-factors agree to better than $10\%$ for all the halo masses. In regards to the halo power spectrum, Figure~\ref{fig:Halo_Pk} shows that COLA and {\it N}-body are in agreement within the variance both for GR and for the boost-factors. 

\subsection{Internal structure of halos}
\label{sec:F5_NFWtweaks}
Certain halo properties are needed to create galaxy mock catalogues with the HOD model discussed in the next section; the phase space distribution of satellite galaxies is based on the NFW profile, which in turn is completely determined by the concentration parameter and the mass of the halo. 

COLA is purposely designed to accurately determine basic halo quantities (positions, velocities and masses) but not the internal halo structure, which would require simulations with a much higher computational cost. Because of that, in the pipeline developed for this work, we use several prescriptions calibrated from high resolution {\it N}-body simulations in $\Lambda$CDM to assign certain halo properties in COLA (see Eq.~\eqref{eq:mfof_m200_conversion} and Eq.~\eqref{eq:concentration}). Furthermore, MG may slightly modify some of these relationships (e.g. the concentration parameter and the velocity dispersion). We use the \textcode{elephant} simulations to estimate these effects, which we then introduce in the prescription to create galaxy mock catalogues in the MG models.

In F5 the fifth force is unscreened for halos of mass lower than $10^{14} \Msun$, affecting the radial density profile and the virial equilibrium for such halos. We account for this by implementing corrections to the concentration parameter and the velocity dispersion. 
Detailed studies of the effect of the fifth force on these quantities were performed in \cite{Mitchell:2018qrg} and \cite{Mitchell:2019qke} using higher-resolution simulations, which derived the fitting formulae for them as a function of halo mass. However, these studies used a different halo mass definition, thus we use simple hyperbolic functions to fit concentration parameter and velocity dispersion boost-factors measured with the \textcode{rockstar} halos in \textcode{elephant} simulations. Although the concentration parameter measurement for halos with less than $500-1000$ particles, which corresponds to $(4-7) \times 10^{13} \Msun$ in \textcode{elephant}, needs to be treated with care, the study using higher-resolution simulations found a similar increase of the concentration parameter for unscreened halos \cite{Mitchell:2019qke}.
In the case of the concentration parameter we use the fitting formula:
\begin{equation}
    \frac{c_{\rm F5}}{c_{\rm GR}} = 1+ a_c \, (1- \tanh{ ( b_c \, \log_{10}(M_{\rm 200 c}) - M_c^* } ) \, .
    \label{c_fit}
\end{equation}
The velocity dispersion is obtained by solving the equations for the virial equilibrium between kinetic energy and gravitational energy of DM in halos: the solution depends linearly on the square root of the effective gravitational constant, $G_{\rm eff}$, as can be deduced from eq.~\eqref{V_disp_NFW_Intro}. To fit the transition of the velocity dispersion between screened and unscreened halos we use the fitting function:
\begin{equation}
    \frac{V_{\rm rms, F5}}{V_{\rm rms, GR}} = 1+ a_V \, (1- \tanh{ ( b_V \, \log_{10}(M_{\rm 200 c}) - M_V^* } ) \, .
    \label{V_rms_fit}
\end{equation}

\begin{figure}
        \centering 
        \subfloat[][Concentration]{
        \includegraphics[width=.48\textwidth,clip]{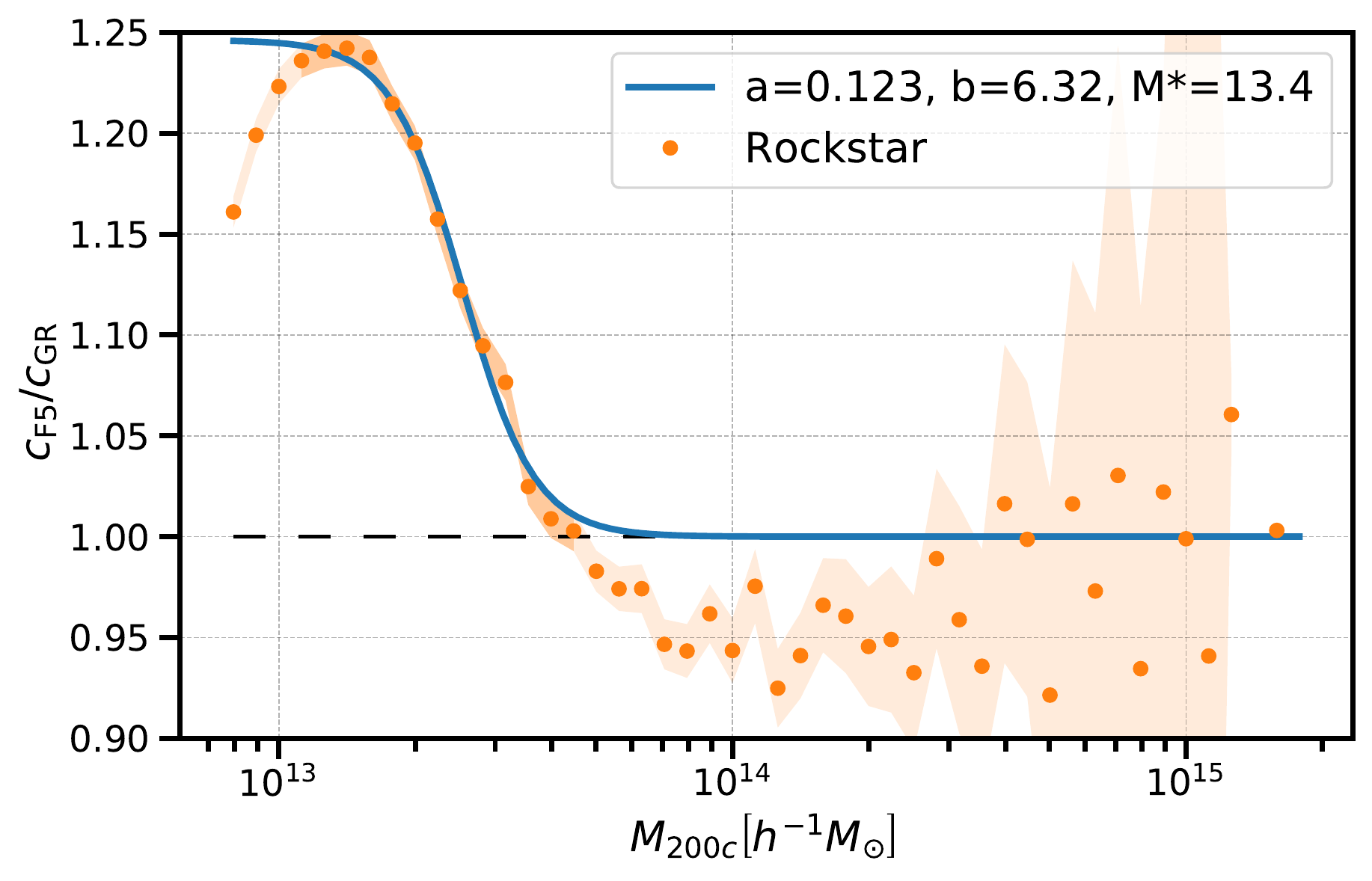}
        }
        \hfill
        \subfloat[][Velocity dispersion]{
        \includegraphics[width=.48\textwidth]{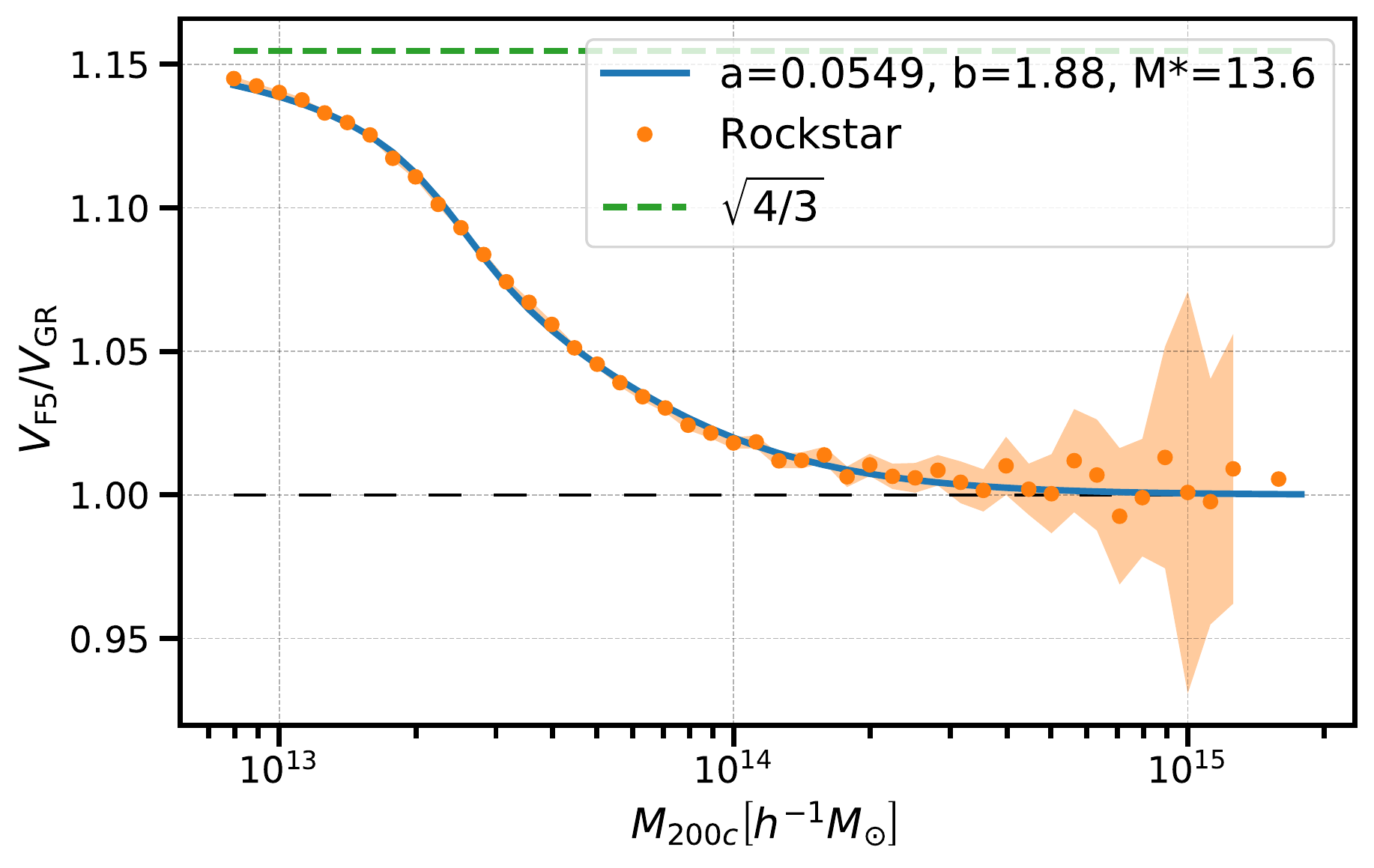}
        }
    \caption{Fit of F5 boost factors for concentration parameter (left panel) and velocity dispersion (right panel). The values measured from \textcode{rockstar} catalogues (orange dots) are used to fit the functions in Eq.~\eqref{c_fit} and Eq.~\eqref{V_rms_fit}. Only halos with masses above $10^{13} \Msun$ are used to perform the fits.}
    \label{fig:NFWfitF5}
\end{figure}

The results of the fit compared with the data from \textcode{rockstar} are shown in Figure~\ref{fig:NFWfitF5}: the hyperbolic functions correctly represents the transition from unscreened to screened halos and provides an effective description of the fifth force effect on these halo properties. The fitted parameters, $a$, $b$ and $M_*$, are shown in Figure~\ref{fig:NFWfitF5}. Note that these fitted parameters are valid only for $|f_{R0}|=10^{-5}$ and they need to be re-fitted for different values of $f_{R0}$.

In the case of N1, the Vainshtein mechanism operates efficiently irrespectively of halo mass and we confirmed that the concentration parameter and the velocity dispersion in N1 are not modified from GR.

\section{Galaxy catalogues}
\label{sec:Galaxies}

The HOD technique allows us to produce galaxy mock catalogues from low-resolution dark matter simulations that do not resolve the internal structure of halos. 

\subsection{HOD model}
Due to the limited resolution of \textcode{elephant} simulations, we create mock galaxies similar to those found in the Baryon Oscillation Spectroscopic Survey CMASS galaxy sample. These galaxies can be modelled by the HOD model proposed in \cite{Zheng:2007zg}\footnote{
We used \texttt{Halotools} \cite{Hearin:2016uxs} (\url{https://halotools.readthedocs.io}) to implement the HOD model in our simulations.}. In this model, galaxies are split between centrals and satellites and the probabilities of a halo of mass $M$ to host either galaxy type are given by:  
\begin{equation}
\begin{split}
   \left\langle N_{\mathrm{cen}}(M)\right\rangle &=\frac{1}{2}\left[1+\operatorname{erf}\left(\frac{\log M-\log M_{\min }}{\sigma_{\log M}}\right)\right], \\
   \left\langle N_{\mathrm{sat}}(M)\right\rangle &= \left\langle N_{\mathrm{cen}}(M)\right\rangle\left(\frac{M-M_{0}}{M_{1}}\right)^{\alpha} \,,
\end{split}
\end{equation}
where each halo can have only one central galaxy but multiple satellite galaxies. The parameters $M_{\min }$ and $\sigma_{\log M}$ control respectively the scale and the width of the transition between probability 0 and 1 to host a central galaxy. The number of satellites is instead modelled as a power law with exponent $\alpha$, scale $M_{1}$ and cut-off $M_{0}$, multiplied by the number of central galaxies. 

Once the halo occupation has been determined, we need to assign the position and velocity for each galaxy. We place central galaxies at the halo centre of mass position and assign them the velocity of the host halo. On the other hand, to place satellites we sample a NFW halo density profile centred at the halo centre of mass and with a concentration parameter determined by Eq.~(\ref{eq:concentration}). We set the velocities to the halo velocity plus a dispersion term that corresponds to the halo virial velocity, as calculated in Eq.~(24) in \cite{More:2008yy}. In the case of F5, we correct the concentration parameter and the velocity dispersion as discussed in Section~\ref{sec:F5_NFWtweaks} employing the fitting formulae in Eq.~\eqref{c_fit} and Eq.~\eqref{V_rms_fit}.   

\subsection{HOD parameters fitting}
The 5 parameters of this HOD model are to be determined by requiring that the galaxy clustering reproduces a reference clustering signal, typically coming from real observations. In this work, instead, we use as reference the clustering of the galaxy catalogues measured in the {\it N}-body halo catalogues in GR with a fiducial set of parameters that were constrained in \cite{Manera:2012sc} to fit the clustering of the CMASS galaxies in BOSS.
Since we can measure our reference signal directly from simulations, we can choose the statistics most convenient for our purpose. The results of \cite{Hernandez-Aguayo:2018oxg} show the importance of RSD in breaking the degeneracy between galaxy bias and MG parameters. Furthermore, it has been shown that the multipoles of the power spectrum can break the degeneracy between different HOD parameters better than the projected correlation function \cite{Hikage:2014bza}. In light of these works, we employ the multipoles of the galaxy power spectrum in redshift space to fit the HOD parameters.

\begin{table}
    \centering
    \begin{tabular}{lrclrcl}
    \toprule
    {} & \multicolumn{3}{c}{COLA} & \multicolumn{3}{c}{{\it N}-body} \\
    {} &      F5 &      GR &      N1 &      F5 &      \textbf{GR} &      N1 \\
    \midrule
    $\log_{10} \left(M_{\min }\right)$    &  13.205 &  13.155 &  13.170 &  13.174 &  \textbf{13.090} &  13.116 \\
    $\sigma_{\log_{10} M}$ &   0.579 &   0.589 &   0.590 &   0.578 &   \textbf{0.596} &   0.600 \\
    $\log_{10} \left(M_{0}\right)$      &  13.070 &  13.093 &  12.809 &  12.832 &  \textbf{13.077} &  12.677 \\
    $\log_{10} \left(M_{1}\right)$      &  13.969 &  13.955 &  14.056 &  14.030 &  \textbf{14.000} &  14.107 \\
    $\alpha$      &   1.059 &   1.044 &   1.043 &   1.040 &   \textbf{1.013} &   1.057 \\
    \midrule
    $n_s [10^{-4}(\hompc)^3]$      &   3.210 &   3.218 &   3.217 &   3.231 &   \textbf{3.226} &   3.230 \\
    \bottomrule
    \end{tabular}
    \caption{HOD parameters and number density of galaxies obtained by minimising the objective function in Eq.~\eqref{RSD_ObjFun} with the simplex search algorithm, except for the {\it N}-body GR column where the HOD parameters are the fiducial parameters taken from \cite{Manera:2012sc}.}
    \label{tab:RSD_HOD_params}
\end{table}

The galaxy power spectrum in redshift space can be decomposed in angular moments using the projection
\begin{equation}
    P_{\ell}(k)=(2 \ell+1) \int_{0}^{1} d \mu P(k, \mu) \mathcal{L}_{\ell}(\mu) \, ,
\end{equation}
where $\mathcal{L}_{\ell}(\mu)$ is the Legendre polynomial of order $\ell$. In the Kaiser approximation \cite{Kaiser:1987qv}, one can see how the quadrupole and the hexadecapole moments carry information on the redshift space distortion, which is key to break the degeneracy between cosmological and modified gravity parameters. 
The hexadecapole, however, is strongly affected by noise if estimated in the cosmological volume covered by our simulations. Thus, the objective function used to fit the HOD parameters includes only the monopole and the quadrupole. We use a Gaussian model \cite{Taruya:2010mx} to estimate the covariance matrix $\text{Cov}_{\ell, \ell^\prime}$ in terms of the volume spanned by the 5 realisations ($\sim5 (h^{-1} \, {\rm Gpc})^3$), and the galaxy number density and bias measured from {\it N}-body simulations in GR.  We add the tuning of the number density of galaxies to define the objective function as:
\begin{equation}
    \chi^2 = \sum_{\ell, \ell^\prime \in 0,2} \sum_{i \in k_{\rm bins}} \left(P_{\ell, i} - P^{\text{ref}}_{\ell, i}\right)\text{Cov}_{\ell, \ell^\prime, i}^{-1} \left(P_{\ell^\prime, i} - P^{\text{ref}}_{\ell^\prime, i} \right) + W_{n_{s}} \left( \frac{n_s - n_{s,\text{ref}}}{n_{s,\text{ref}}} \right)^2	\, .
\label{RSD_ObjFun}
\end{equation}
Here $W_{n_{s}}$  is the weight to control the importance of the number density tuning. We will set $W_{n_{s}}=10^4$ to enforce roughly a one per cent agreement in the number density. The multipoles of the power spectrum and the covariance matrix are evaluated in 25 linearly distributed bins in the range $k=[0.05,0.3] \hompc$.
We use the simplex search algorithm \cite{Nelder:1965zz} to find the set of HOD parameters that minimise the objective function \eqref{RSD_ObjFun}. The reference power spectra are computed from {\it N}-body GR simulations by populating galaxies using the fiducial HOD parameters of \cite{Manera:2012sc}, marked in bold in Table~\ref{tab:RSD_HOD_params}. We average over the 5 realisations of halo catalogues in each model and over 3 orthogonal lines-of-sight to produce the multipole moments that are used in the objective function. The best fit HOD parameters resulting from the simplex search are summarised in Table~\ref{tab:RSD_HOD_params} together with the number density of galaxies $n_s$.

Using the best-fit parameters, we produce 5 HOD realisations from each halo catalogue and we average over them to compute galaxy statistics. We add redshift space distortions along the three different axes, producing 3 redshift space galaxy catalogues for each real space galaxy catalogue, and use the average over these when computing redshift space clustering statistics. 

\begin{figure}
        \centering 
        \subfloat[][Mass Distribution Centrals]{
        \includegraphics[width=.48\textwidth,clip]{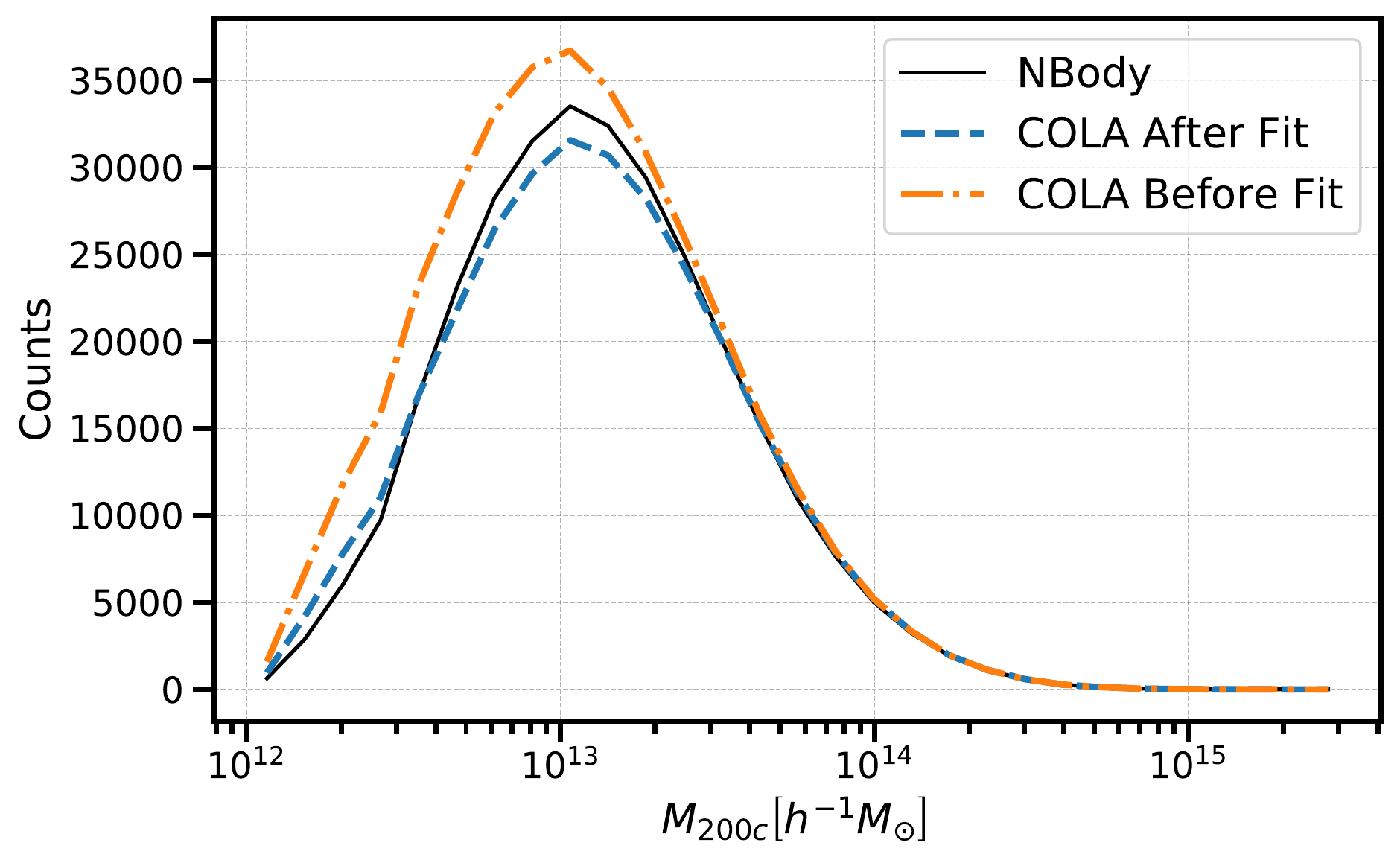}
        }
        \hfill
        \subfloat[][Mass Distribution Satellites]{
        \includegraphics[width=.48\textwidth]{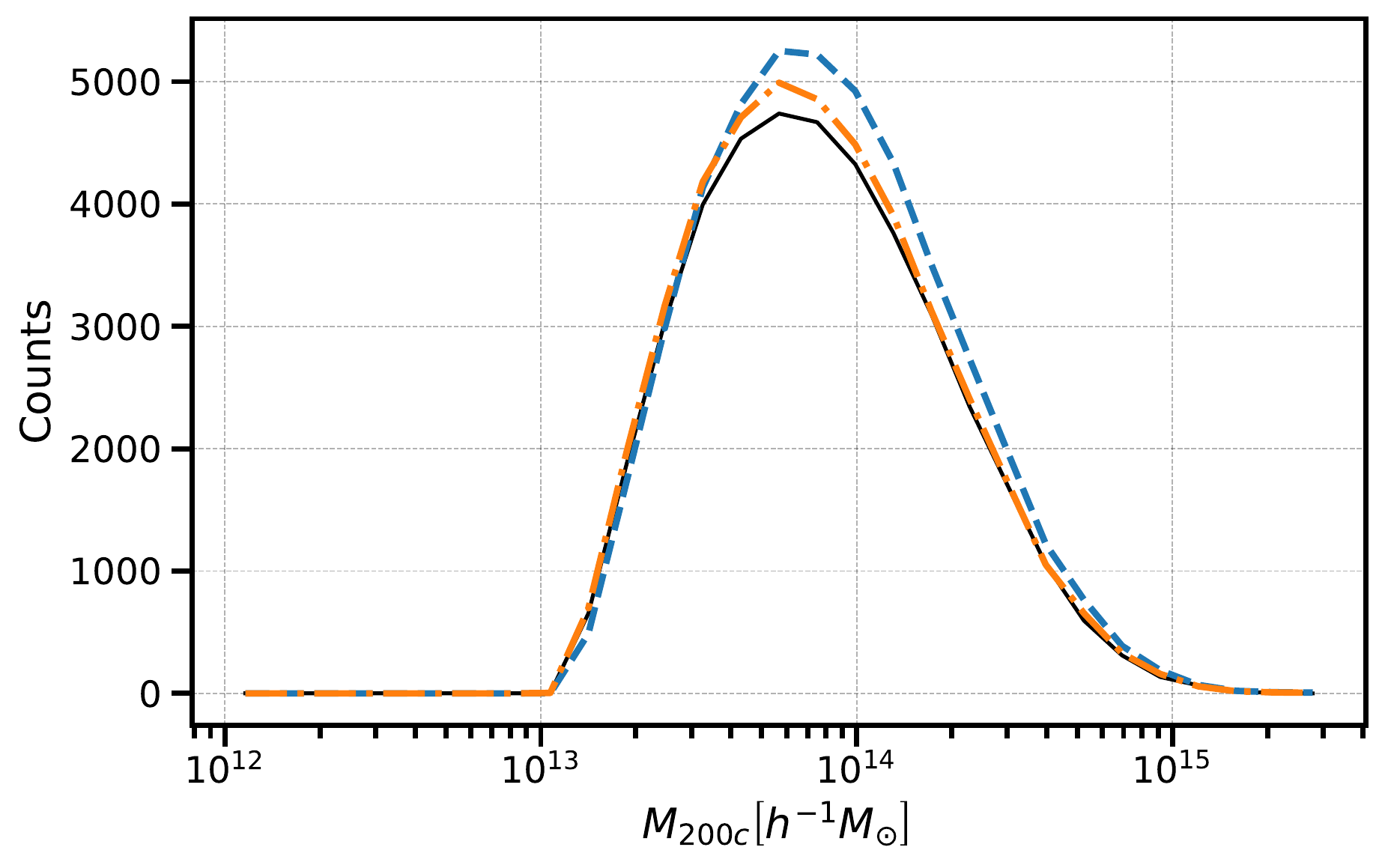}
        }
        \vfill
        \subfloat[][Velocity Distribution Centrals]{
        \includegraphics[width=.48\textwidth,clip]{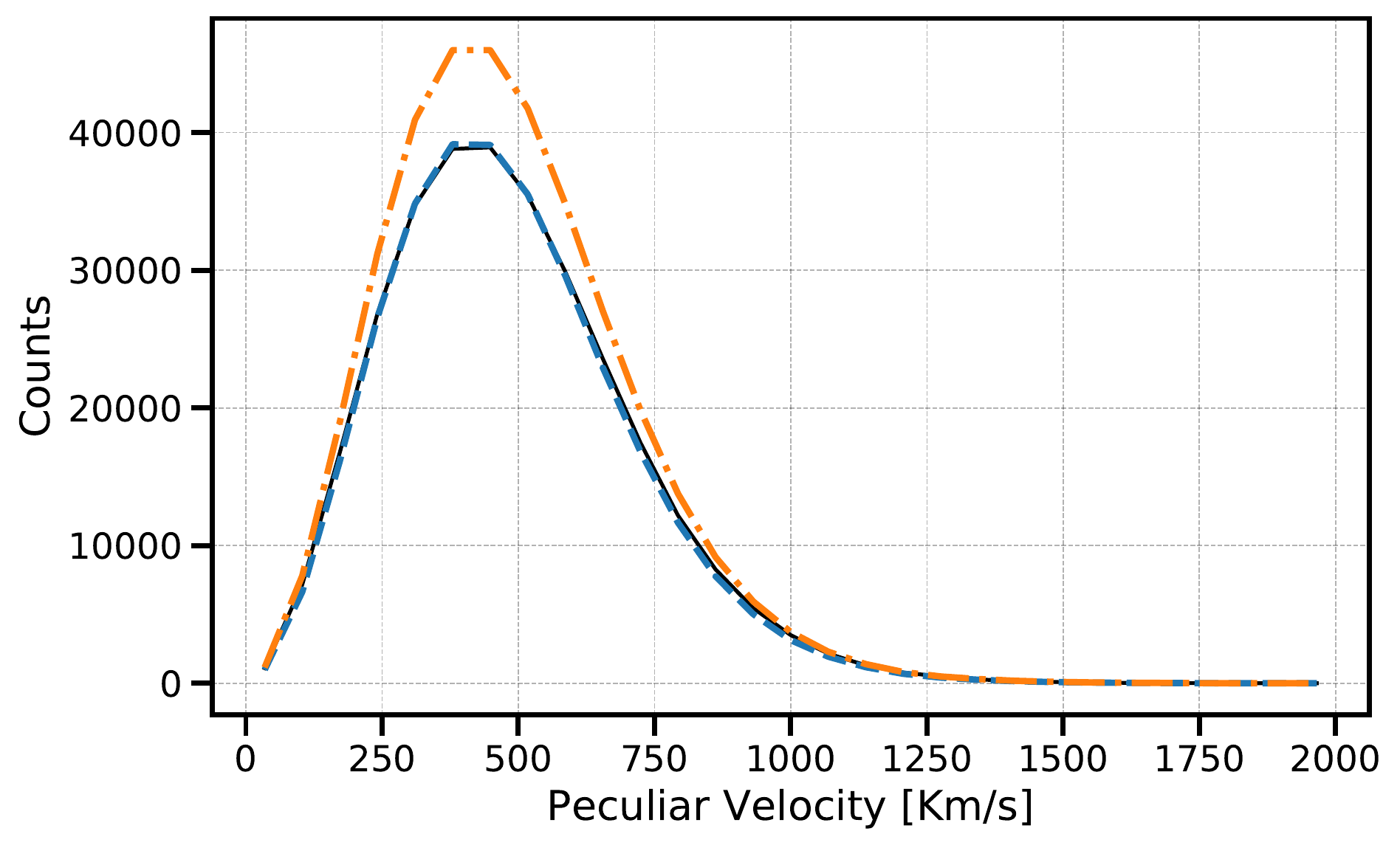}
        }
        \hfill
        \subfloat[][Velocity Distribution Satellites]{
        \includegraphics[width=.48\textwidth]{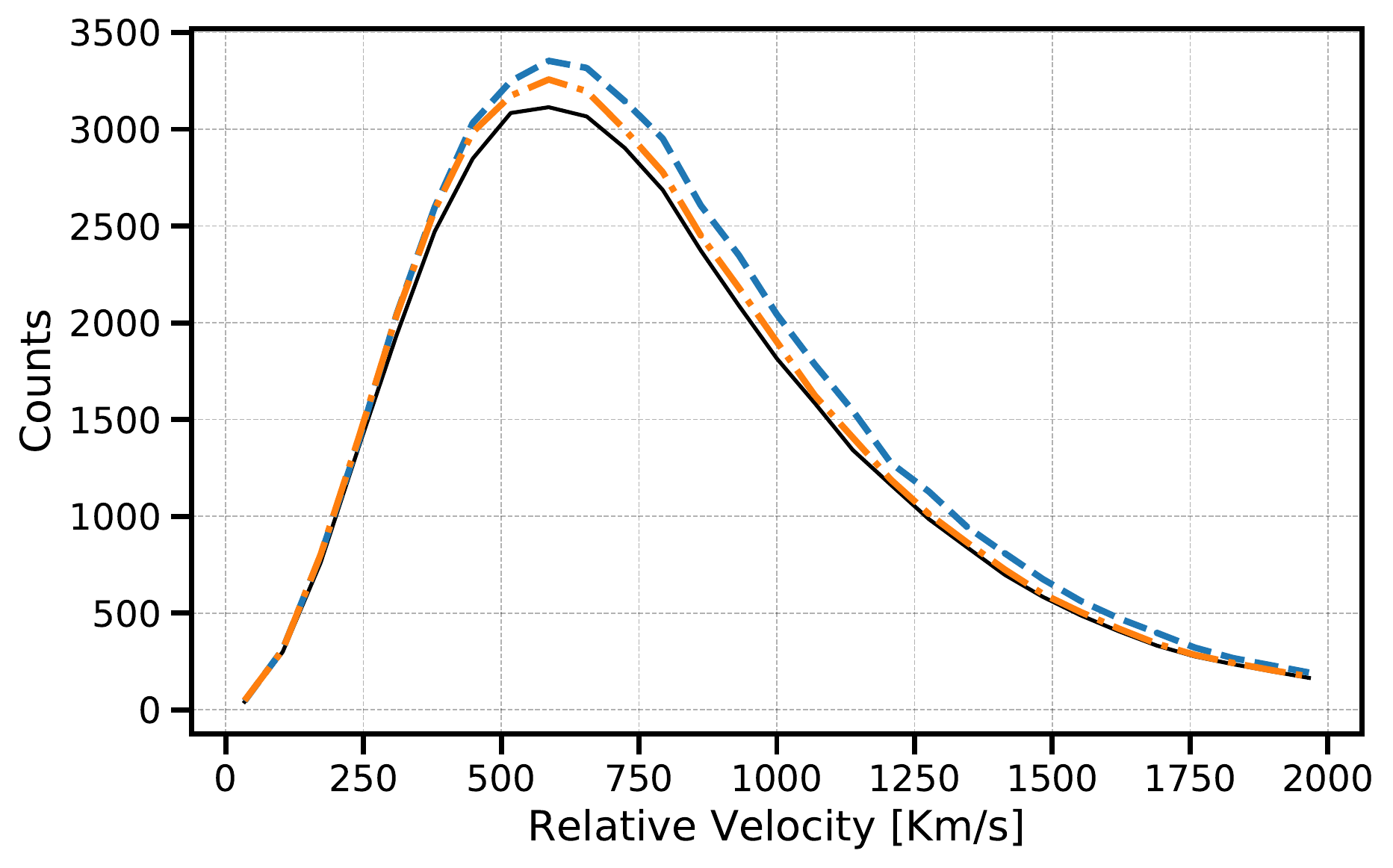}
        }
\caption{\label{fig:GalDistr_NoFit}Halo mass (top panels) and velocity distributions (bottom panels) of central (left panels) and satellite (right panels) galaxies in GR. The velocity distribution shows the number of galaxies in each velocity bin, which corresponds to peculiar velocities for centrals and velocities with respect to the host halo rest frame for satellites. The distributions of galaxies in COLA before (orange dot-dashed lines) and after (blue dashed lines) the fit compared with the distributions of {\it N}-body in GR (black solid lines) show that the fit is mostly driven by central galaxies.}
\end{figure}

\begin{figure}
        \centering 
        \subfloat[][Mass Distribution Centrals]{
        \includegraphics[width=.48\textwidth,clip]{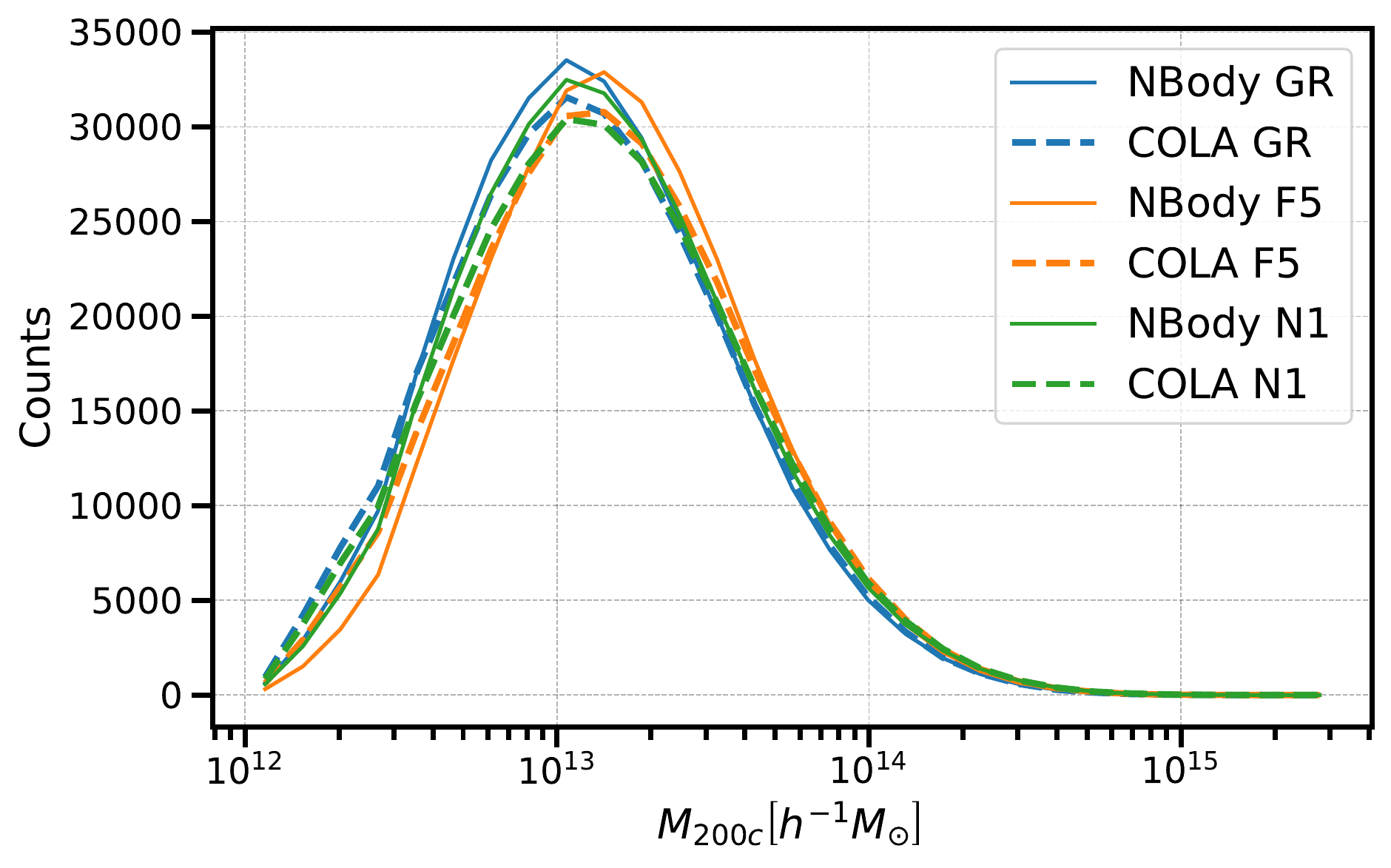}
        }
        \hfill
        \subfloat[][Mass Distribution Satellites]{
        \includegraphics[width=.48\textwidth]{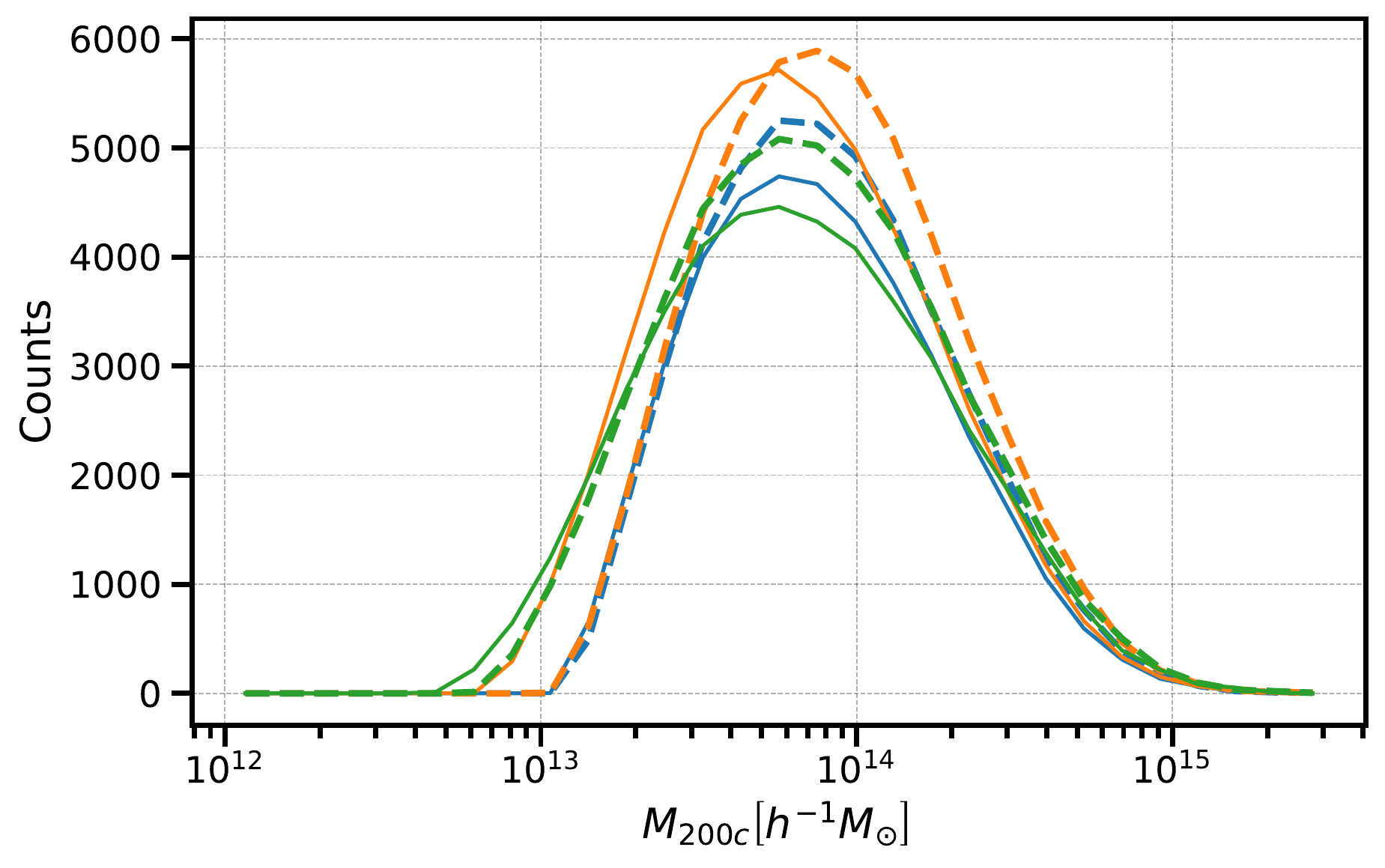}
        }
        \vfill
        \subfloat[][Velocity Distribution Centrals]{
        \includegraphics[width=.48\textwidth,clip]{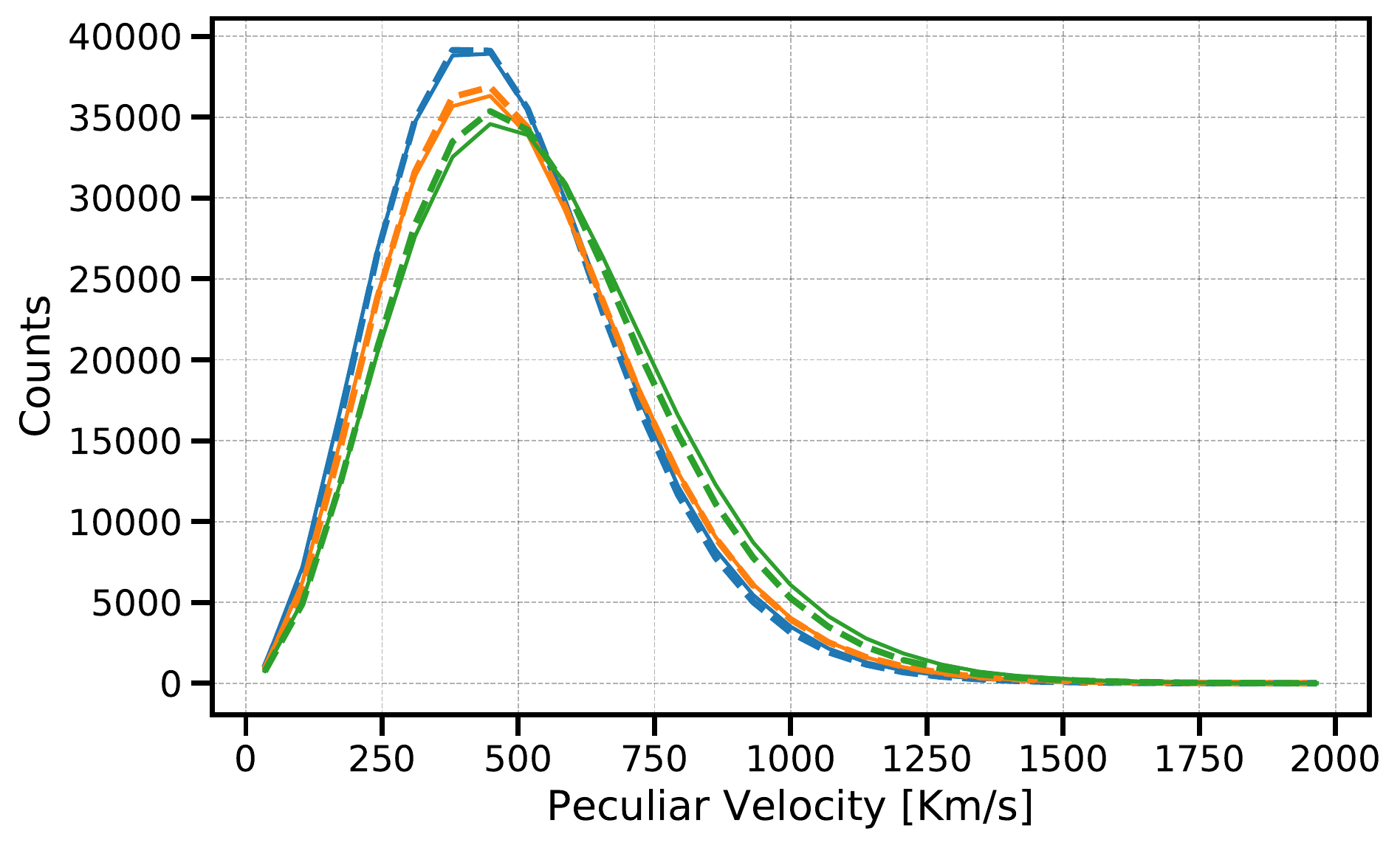}
        }
        \hfill
        \subfloat[][Velocity Distribution Satellites]{
        \includegraphics[width=.48\textwidth]{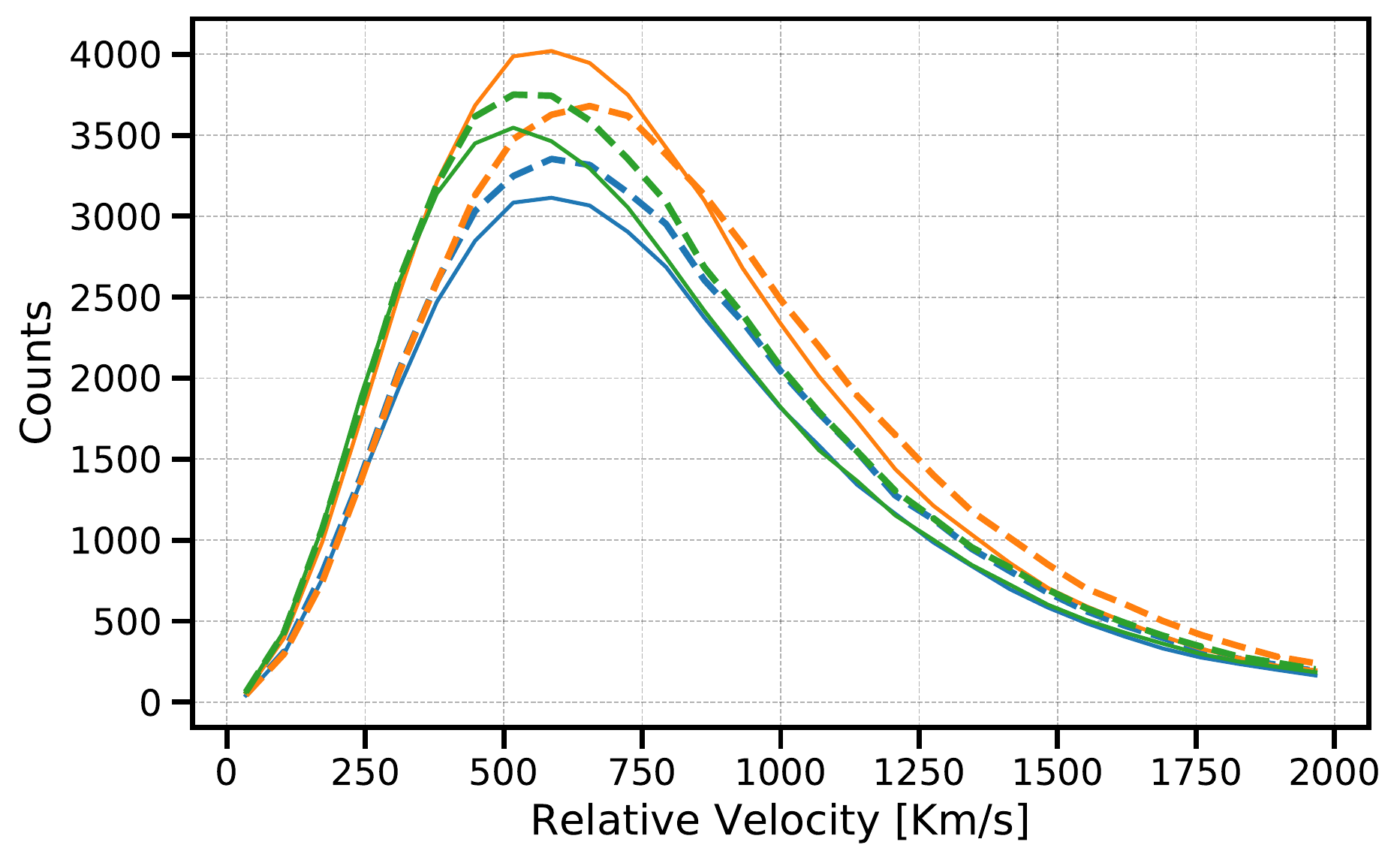}
        }
\caption{\label{fig:GalDistr_MultipolesFit}
Same as Figure~\ref{fig:GalDistr_NoFit} but for all the gravity models (GR, F5 and N1 in blue, orange and green respectively) after fitting the HOD parameters, which are shown in Table~\ref{tab:RSD_HOD_params}.
}
\end{figure}

To interpret the results of the HOD parameters tuning, we measure in the GR mocks the galaxy number count as a function of the halo mass and the peculiar velocity (which for satellites we take as the velocity with respect to the reference frame of the hosting halo).
Figure~\ref{fig:GalDistr_NoFit} compares the counts before\footnote{The HOD parameters before the fit are the same as the default values for the GR case in $N$-body, highlighted in bold in Table~\ref{tab:RSD_HOD_params}.} and after fitting the HOD parameters in COLA and $N$-body in GR.
We can see how the fit is driven by central galaxies, due to their higher abundance and the stronger signal-to-noise ratio in the monopole. The larger value of $M_{\mathrm{min}}$ in COLA places central galaxies in more massive halos, which are less abundant, compensating for the excess of halos below $~10^{13} \Msun$. The smaller values of $M_1$ compensates for the increase of $M_{\mathrm{min}}$, while the larger value of $\alpha$ is responsible for the heavier right tail of the satellite distributions.
Figure~\ref{fig:GalDistr_MultipolesFit} also shows the host halo mass and velocity distribution but for all the gravity models (GR, F5 and N1) after the fit of the HOD parameters, which are summarised in Table~\ref{tab:RSD_HOD_params}.
We can see that the differences in the velocity distribution of central galaxies in different models are well captured in COLA. 

\begin{figure}[t]
\centering 
\subfloat[][GR]{
\includegraphics[width=.48\textwidth,clip]{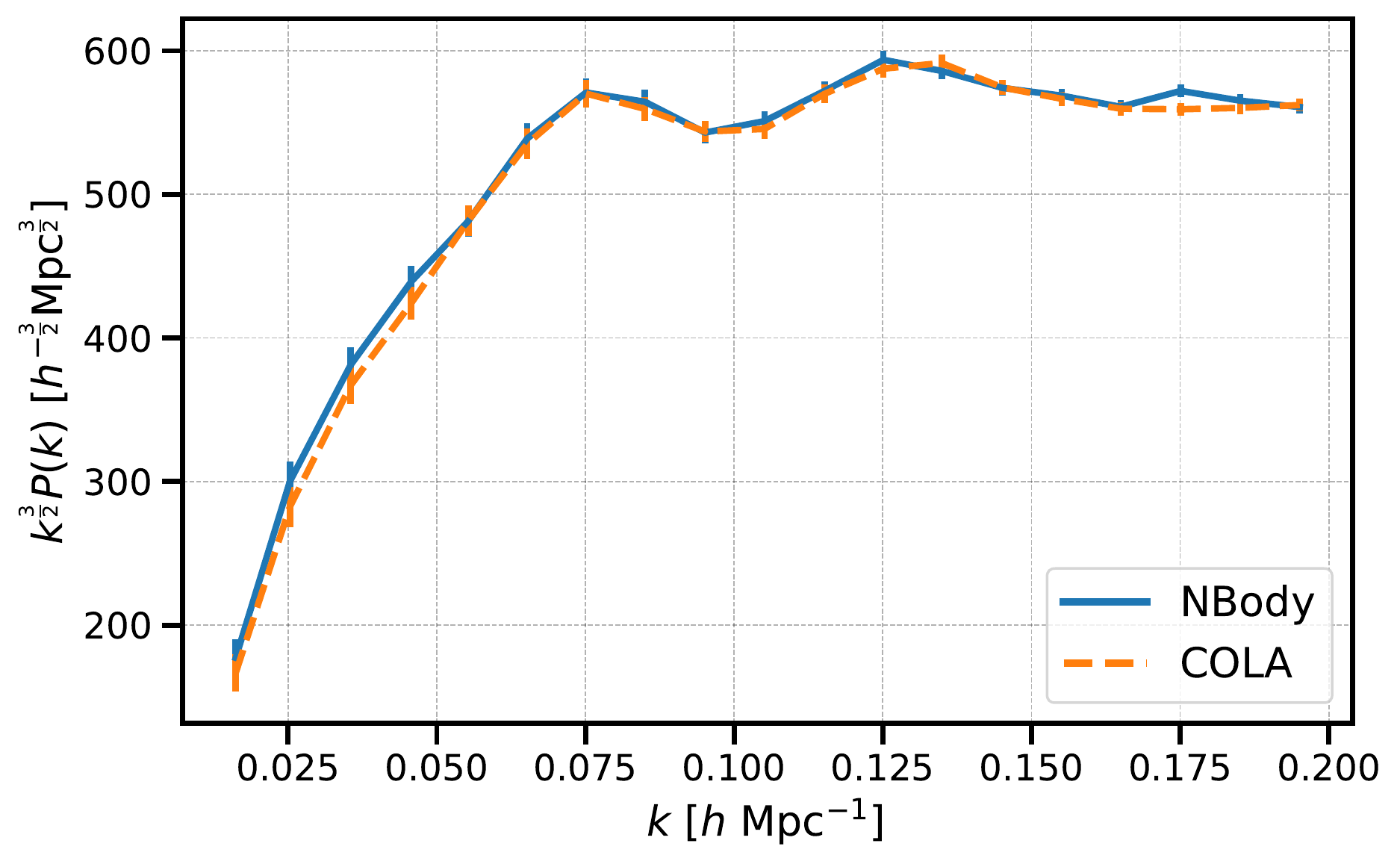}
}
\hfill
\subfloat[][GR ratio]{
\includegraphics[width=.48\textwidth]{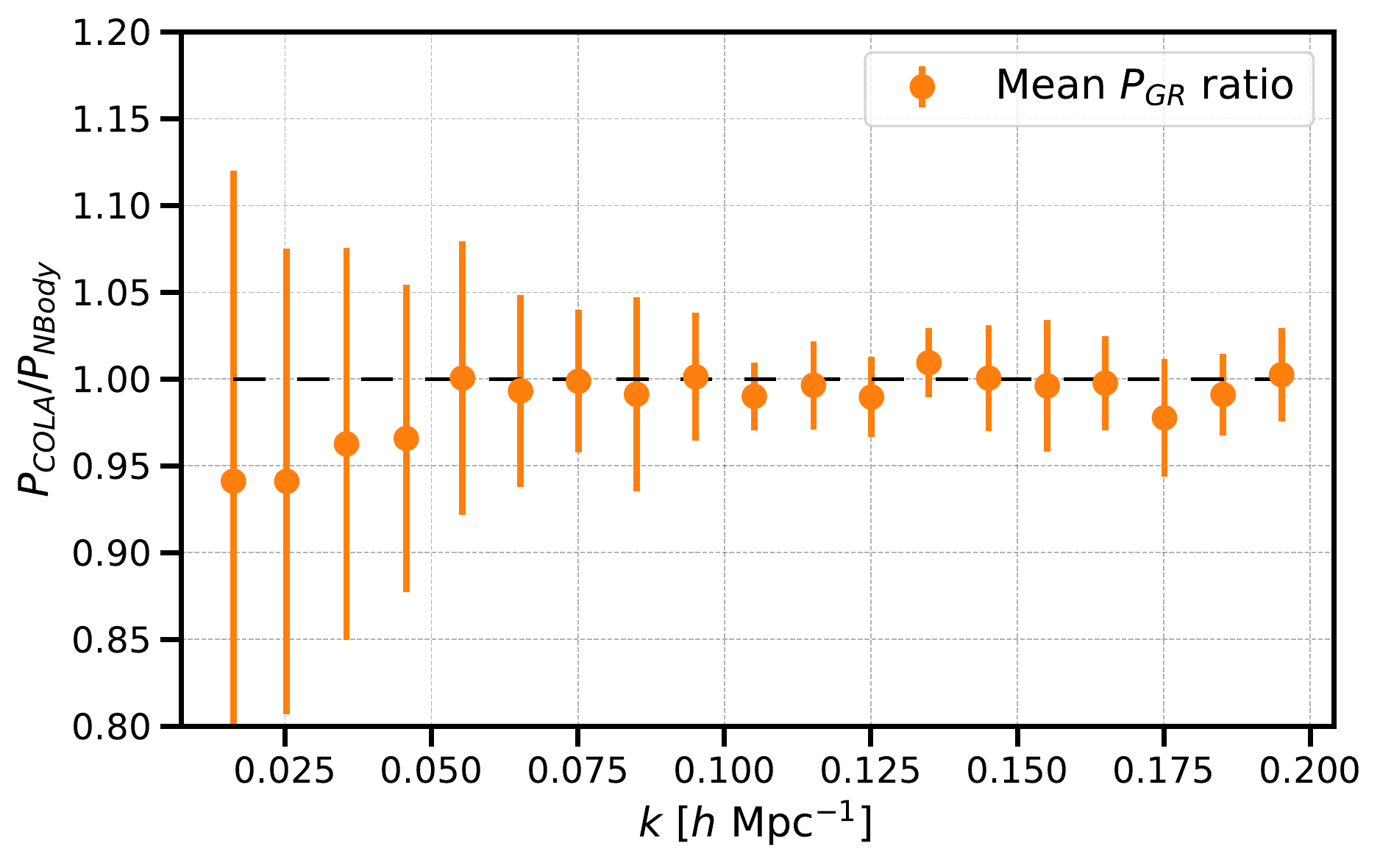}
}
\vfill
\subfloat[][F5 boost factor]{
\includegraphics[width=.48\textwidth]{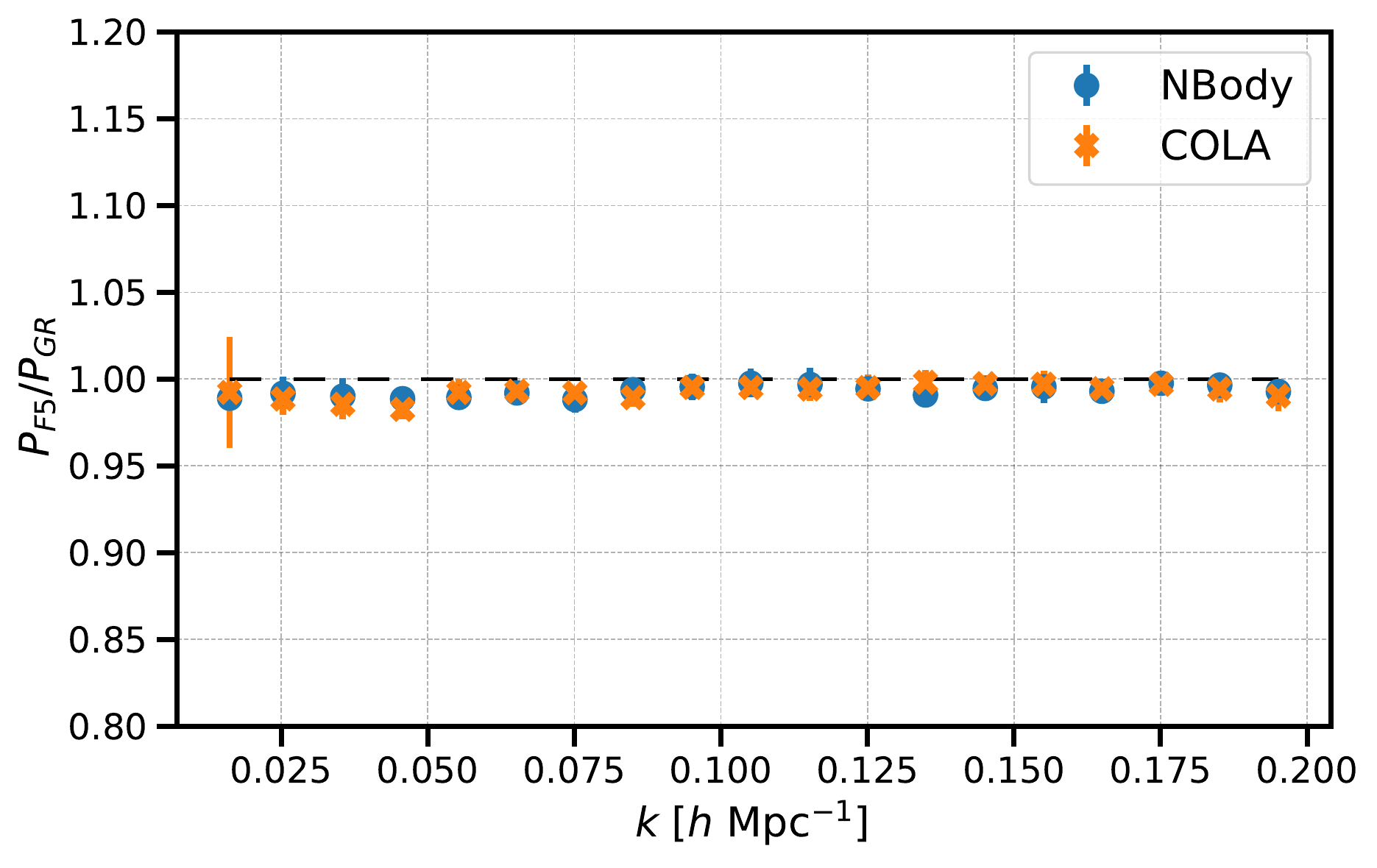}
}
\hfill
\subfloat[][N1 boost factor]{
\includegraphics[width=.48\textwidth]{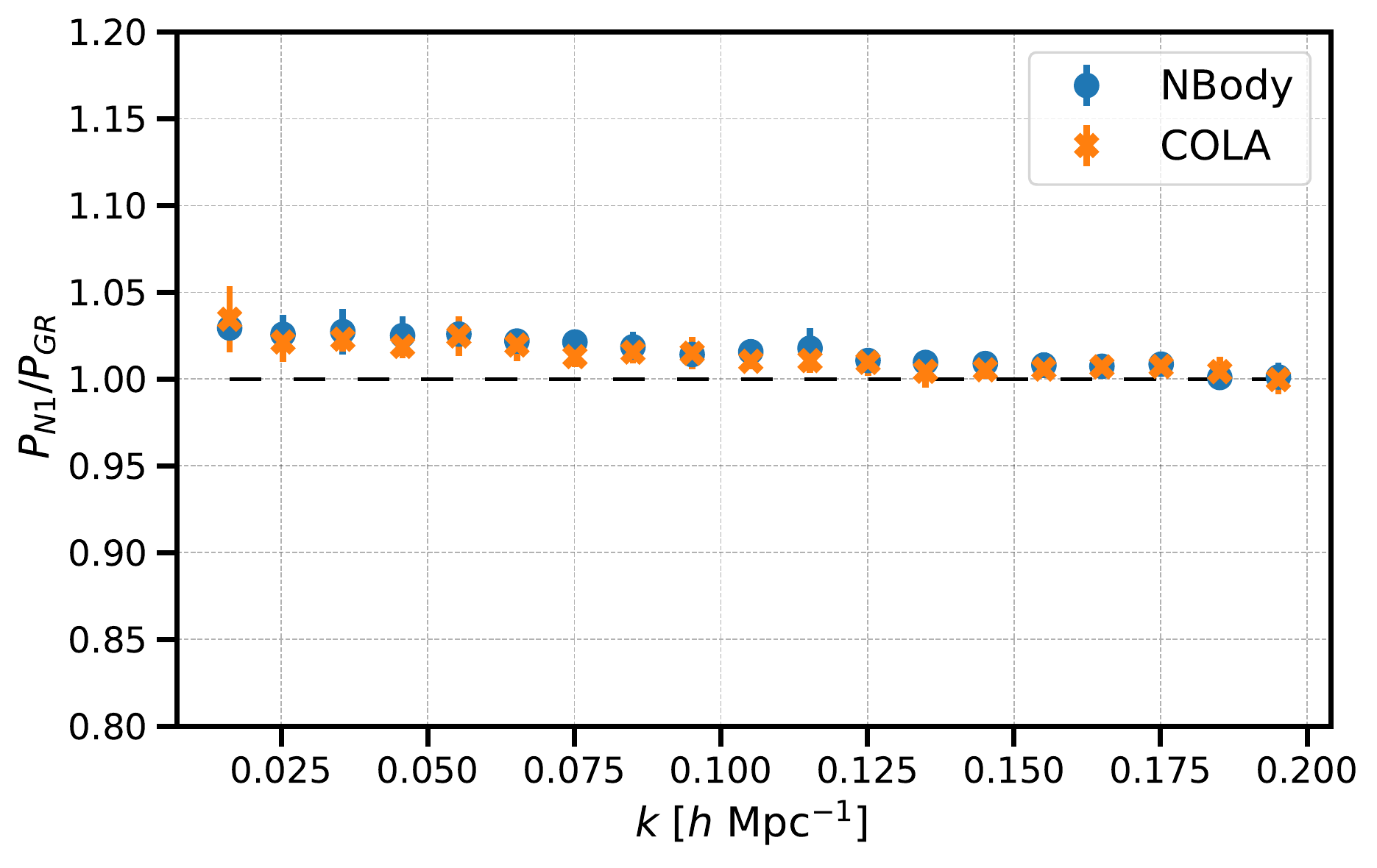}
}
\caption{Comparison of the monopole of the galaxy power spectrum in redshift space between the COLA (in orange) and {\it N}-body (in blue) catalogues. The power spectrum in GR (top panels) and the boost factors in F5 and N1 (bottom panels) show an agreement within the variance between COLA and {\it N}-body.}
\label{fig:RSD_P0}
\end{figure}

\begin{figure}[t]
\centering 
\subfloat[][GR]{
\includegraphics[width=.48\textwidth,clip]{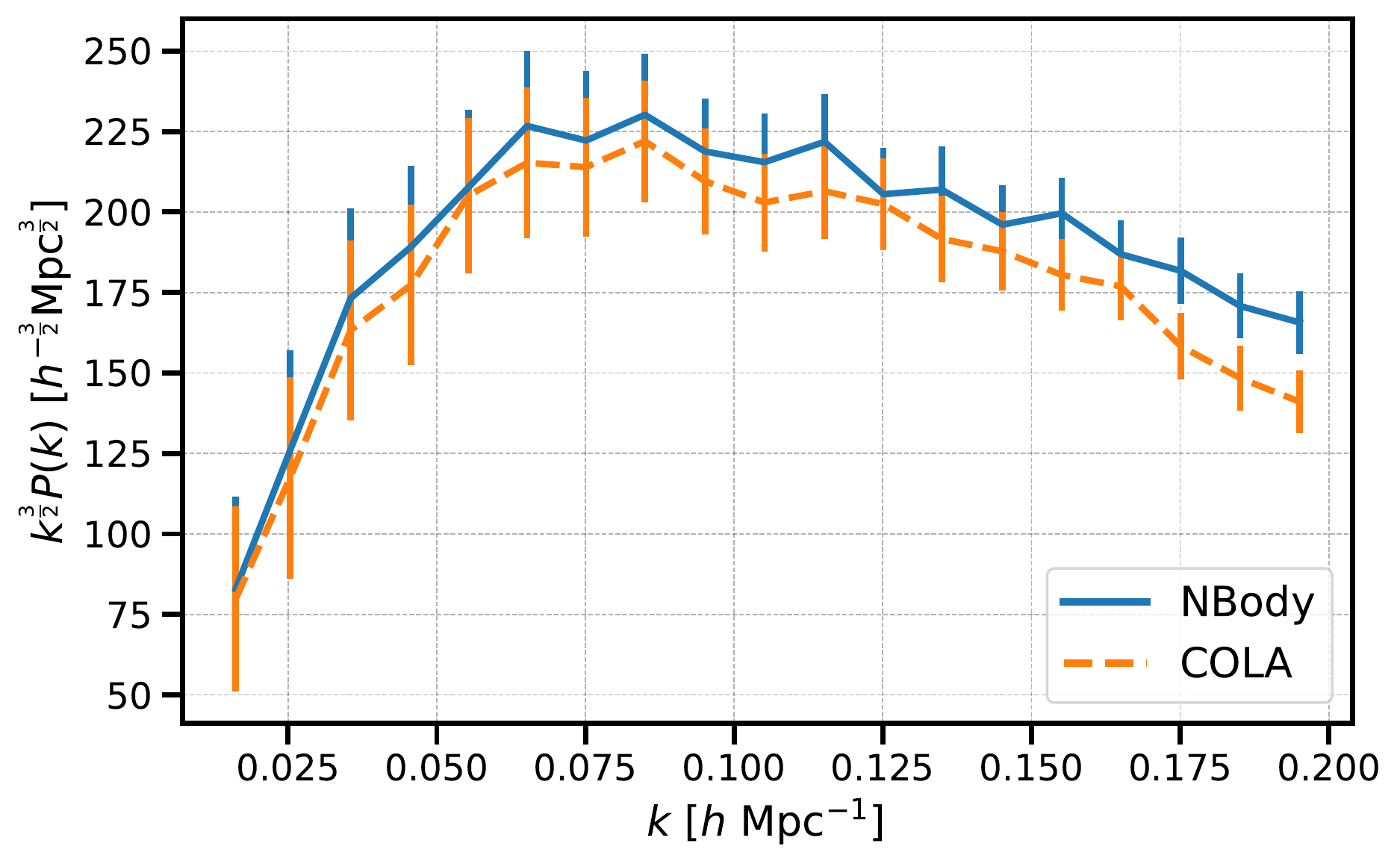}}
\hfill
\subfloat[][GR ratio]{
\includegraphics[width=.48\textwidth]{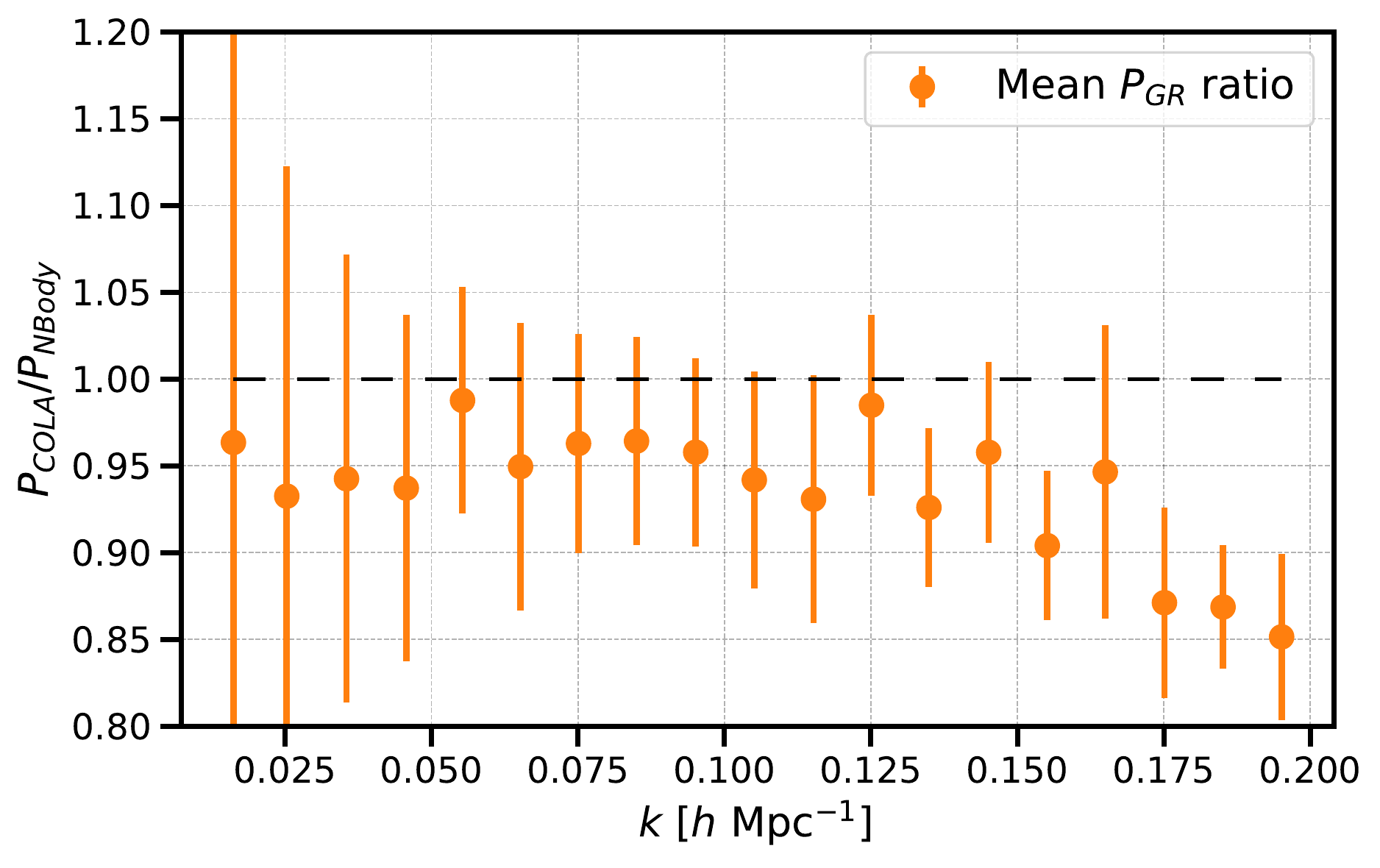}}
\vfill
\subfloat[][F5 boost factor]{
\includegraphics[width=.48\textwidth]{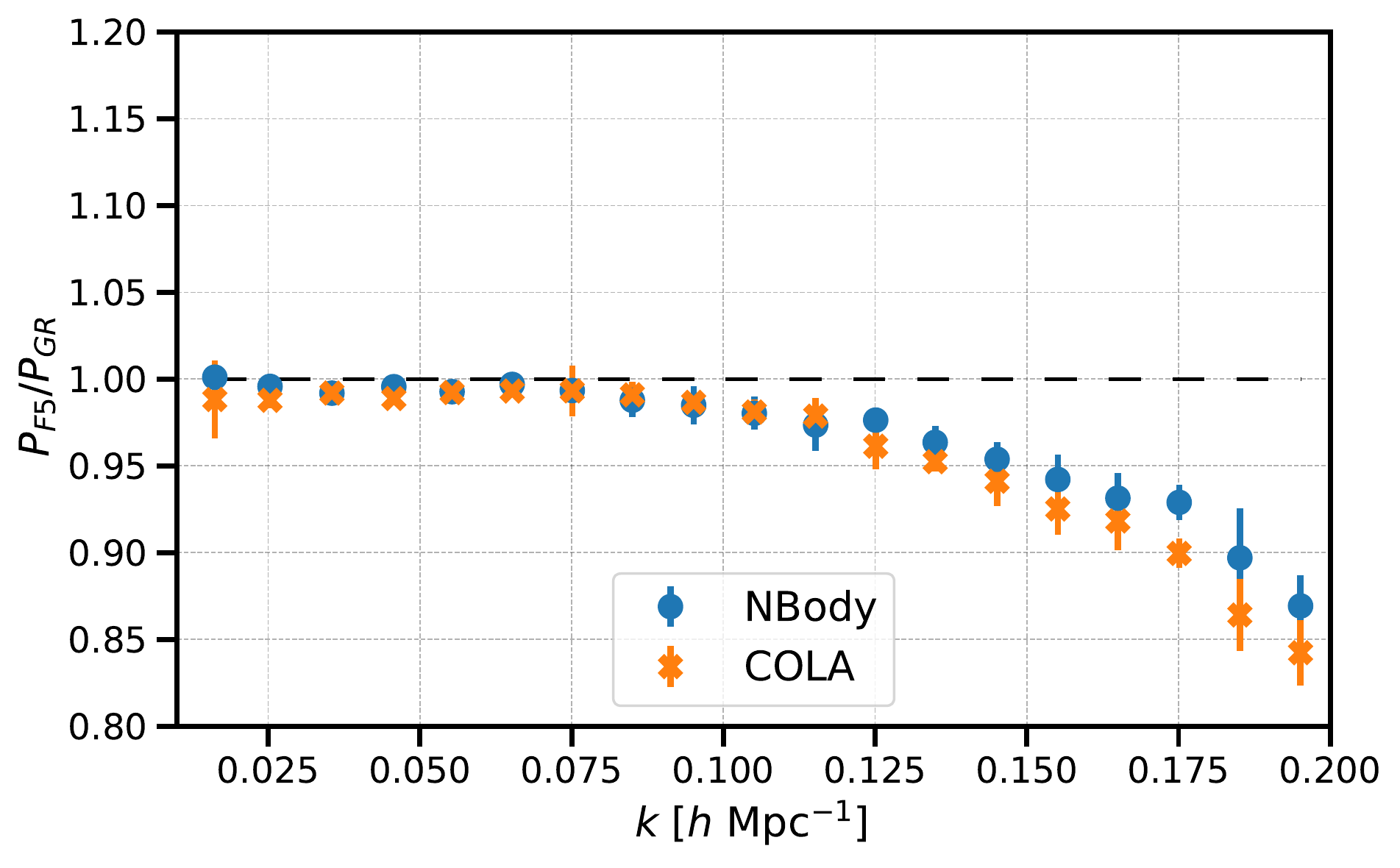}}
\hfill
\subfloat[][N1 boost factor]{
\includegraphics[width=.48\textwidth]{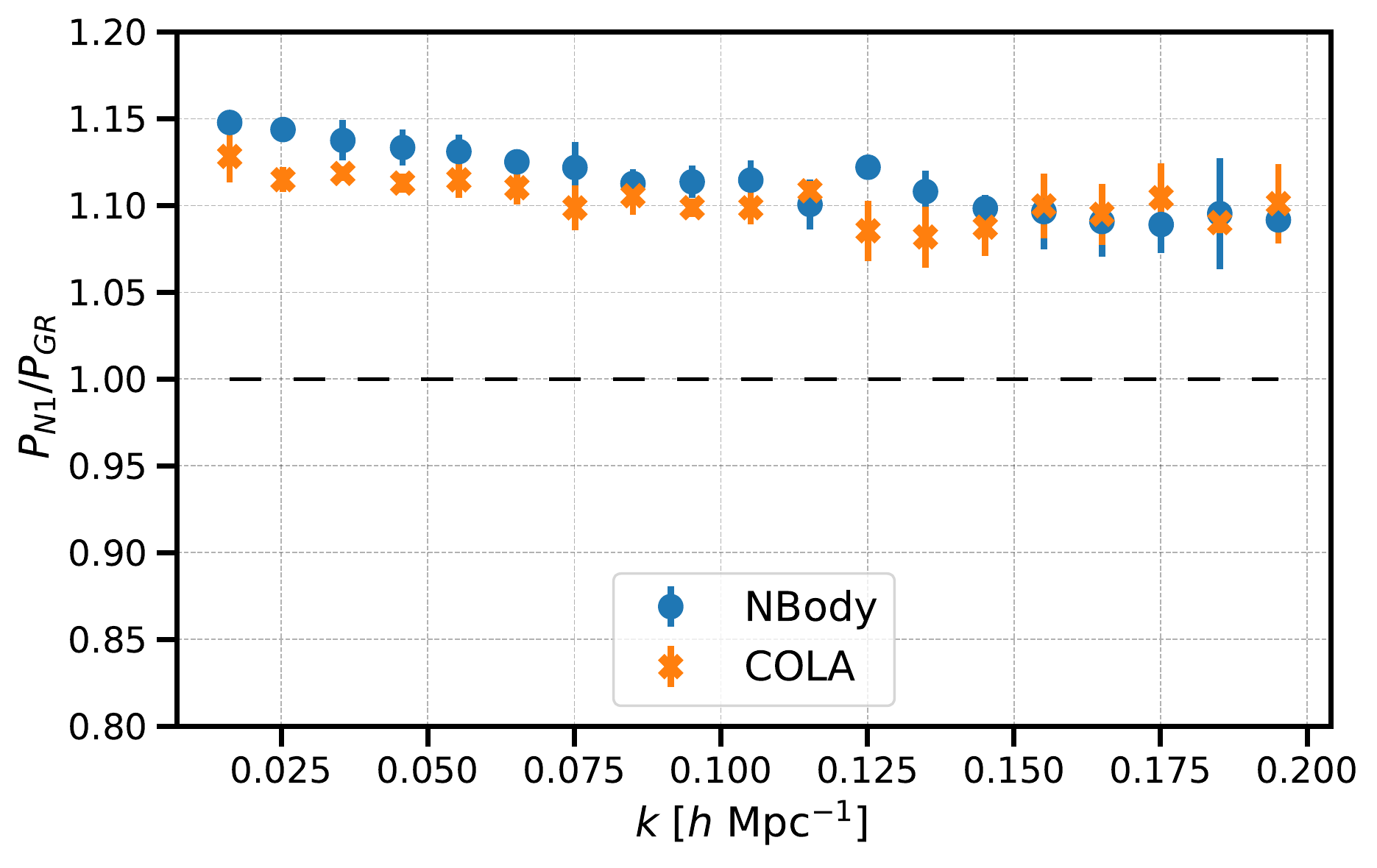}}
\caption{Comparison of the quadrupole of galaxy power spectrum in redshift space between COLA (in orange) and {\it N}-body (in blue) catalogues. The power spectrum in GR (top panels) show a $5\%$ agreement up to $k \sim 0.17 \hompc$. The boost factors in F5 and N1 (bottom panels) show show a better than $5\%$ agreement up to $k \sim 0.2 \hompc$, despite $\sim 10\%$ deviations appear above $k \sim 0.17 \hompc$ in GR.}
\label{fig:RSD_P2}
\end{figure}

We compare the monopole and the quadrupole of the galaxy power spectrum in Figures~\ref{fig:RSD_P0} and \ref{fig:RSD_P2} respectively. The agreement between COLA and {\it N}-body for the monopole is within the variance for all the gravity models up to $k \sim 0.2 \hompc$. The GR quadrupole in COLA agrees with that in {\it N}-body within $5\%$ up to $k \sim 0.17 \hompc$. The agreement for the quadrupole boost factors is within $5\%$ up to $k \sim 0.2 \hompc$ for both F5 and N1. Additionally, the quadrupole boost factor in F5 shows a strong scale dependence, with a small departure from GR on linear scales that increases on non-linear scales. The quadrupole boost factor in N1, instead, shows a large departure from GR already on linear scales.
This different behaviour of the quadrupole boost factors is due to the difference in the linear growth rates of these two models, as well as the different screening mechanisms. This shows that even after the HOD tuning to obtain a similar monopole, the difference between models shows up in the quadrupole. Thus RSD provides a powerful way to distinguish between different MG models \cite{Hernandez-Aguayo:2018oxg}.  

We note that even though we fit multipole moments up to $k_{\rm max}=0.3 \hompc$, the agreement between {\it N}-body and COLA degrades substantially at $k>0.2 \hompc$ for quadrupole. This is because the slight difference in satellite distributions gives a large effect on the higher multipoles at large $k$ due to the Finger-of-God effect \cite{Hikage:2014bza}, which is highly sensitive to velocity dispersion of galaxies on small scales. This is not necessarily an issue for the modelling of cosmological and MG effects in mock galaxies as the velocity dispersion is usually treated as a nuisance parameter in cosmological analyses.

\begin{figure}
\centering 
\subfloat[][F5]{
\includegraphics[width=.48\textwidth]{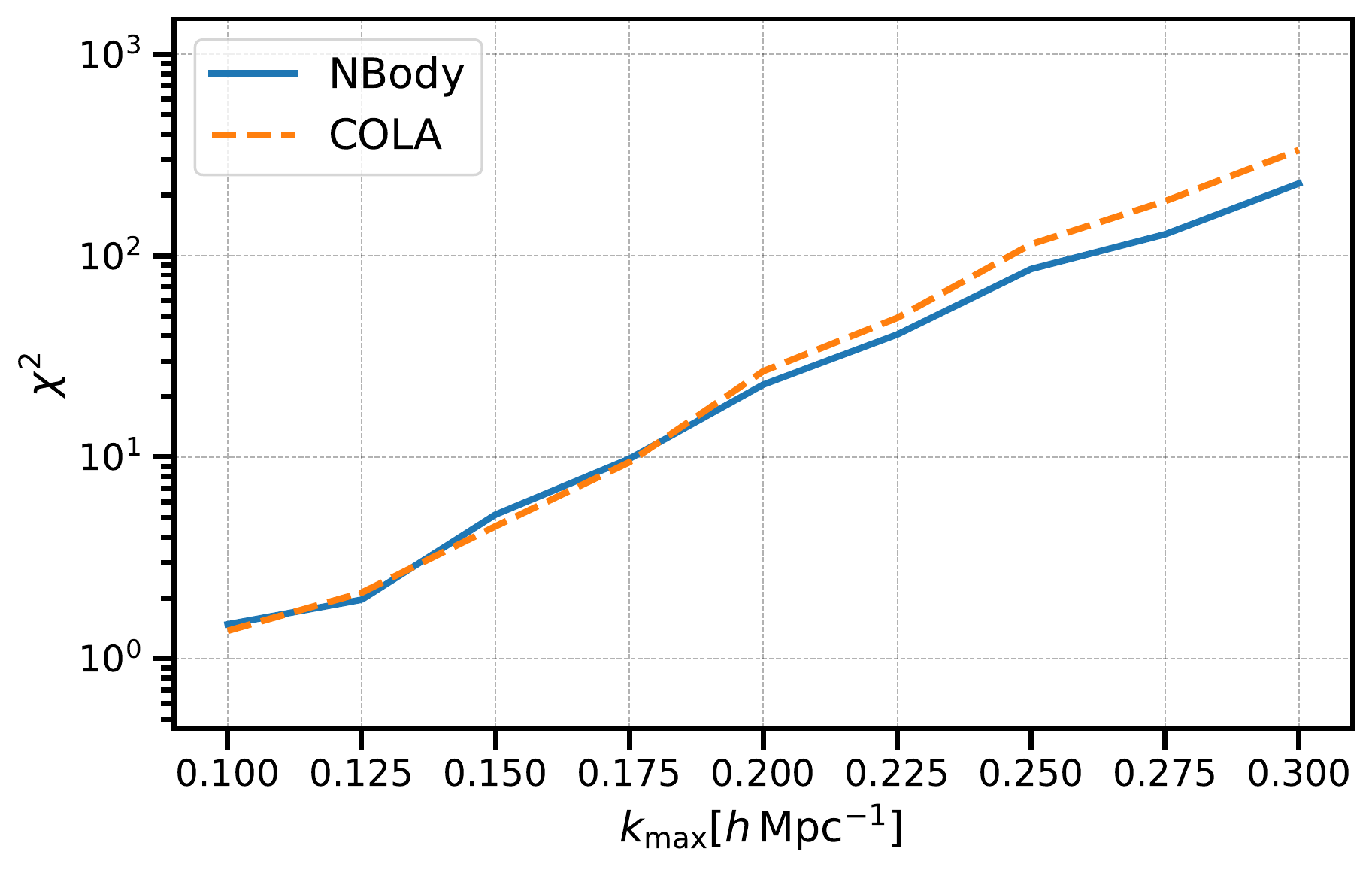}}
\hfill
\subfloat[][N1]{
\includegraphics[width=.48\textwidth]{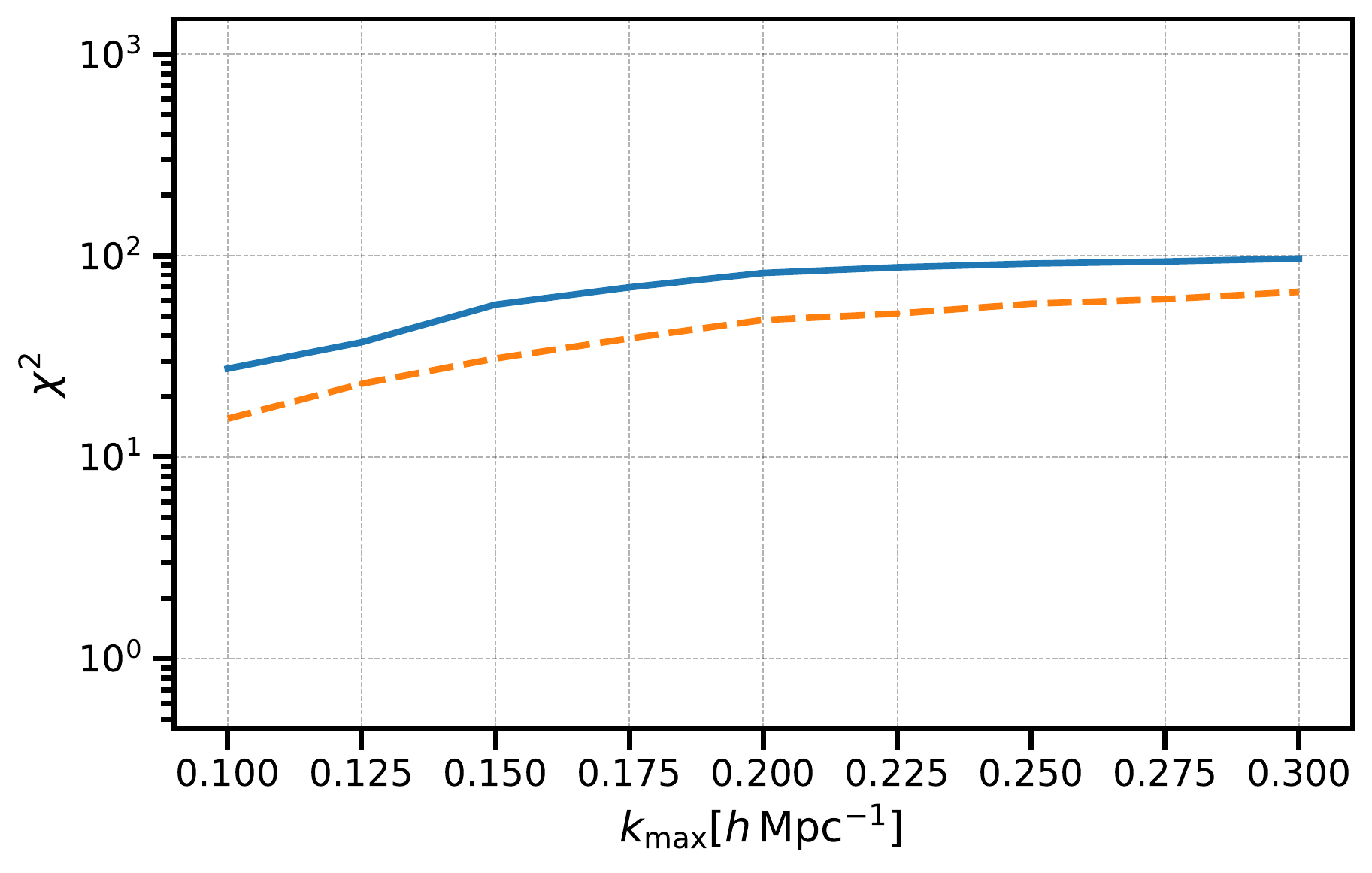}}
\caption{Value of $\chi^2$ as defined in Eq.~\eqref{RSD_ObjFun} in COLA (orange dashed lines) and {\it N}-body (blue solid lines) evaluated using the signal in GR for the corresponding simulation technique as the target function, as a function of the small scales cut-off $k_{\rm max}$. The $\chi^2$ in F5 (left panel) shows a strong scale dependence, with small values on linear scales and larger values on mildly non-linear scales, highlighting the importance of mildly non-linear scales in constraining MG parameters. The $\chi^2$ in N1 (right panel) shows little scale dependence, reflecting the features of Vainshtein screening. The results in COLA and {\it N}-body are in good agreement.}
\label{fig:RSD_chi2}
\end{figure}

To determine the detectability of MG with summary statistics of galaxy catalogues, we compute $\chi^2$ from Eq.~\eqref{RSD_ObjFun} in F5 and N1 for both the COLA and {\it N}-body catalogues while using the clustering signal in GR as a reference.
To investigate the role of the different scales, we evaluate the $\chi^2$ for different small scale cut-off $k_{\rm max}$ in the range between $0.1$ and $0.3 \hompc$.
The results in Figure~\ref{fig:RSD_chi2} highlight the intrinsic difference between F5 and N1, with the former being significantly different from GR only on non-linear scales while the latter showing a strong departure from GR already on linear scales. COLA reproduces the $\chi^2(k_{\rm max})$ behaviour of the {\it N}-body very well, demonstrating that it captures the MG effects accurately. 

\chapter{Large-scale structure probes of modified gravity}\label{chp:statistics}
\begin{quote}
    {\it The content of this chapter is based on the publication \cite{Fiorini:2022srj}. } 
\end{quote}
In chapter~\ref{chp:mocks}, we have extended and validated the HOD formalism to produce galaxy mocks catalogues from COLA simulations in $f(R)$ and nDGP gravity theories, reproducing full {\it N}-body results for the MG signals on the redshift-space multipoles of the galaxy power spectrum. 

Given the large number of parameters that enter the modelling of galaxy clustering (e.g. cosmological parameters, MG parameters, HOD parameters), degeneracies in the high-dimensional theory parameter space can affect the constraining power of the LSS. However, the theory parameters may have different impacts on different summary statistics. An example of this is the degeneracy between massive neutrinos and $\sigma_8$ in the power spectrum which is broken with bispectrum information \cite{Hahn:2019zob,Hahn:2020lou}. 

In this chapter, we exploit the results and simulated data sets discussed in chapter~\ref{chp:mocks} to investigate the impact of MG in statistics beyond the redshift-space multipoles of the power spectrum, and by doing so we validate the COLA method to formulate theoretical predictions on these additional statistics in MG theories. 
As already mentioned, COLA simulations developed in \cite{Winther:2017jof, Wright:2017dkw} use the screening approximation discussed in section~\ref{sec:MG}, which linearises the scalar field equation while including an environmental dependent screening factor. The validity of this approximation needs to be checked for higher-order statistics as they lack contributions from the non-linearity of the scalar field. 
In this chapter, we test the effects of MG theories and assess COLA's accuracy on the power spectrum orthogonal to the line-of-sight in section~\ref{sec:PowerSpectrum}, on the bispectrum in section~\ref{sec:Bispectrum} and on voids in section~\ref{sec:Voids}.
\section{Power spectrum}
\label{sec:PowerSpectrum}
The real space power spectrum
\begin{equation}
    \left\langle\delta\left(\vec{k}\right) \delta\left(\vec{k}'\right)\right\rangle \equiv(2 \pi)^{3} P^{r}\left(k\right) \delta_{D}\left(\vec{k}+\vec{k}'\right) \,.
\end{equation}
is not directly measurable in galaxy surveys because we cannot probe the real space position of galaxies. What we can directly measure is the redshift-space power spectrum $P^{s}(k, \mu)$ where $\mu \equiv \hat{z}\cdot \hat{k}$ and $\hat{z}$ is the line-of-sight in the plane-parallel approximation. The redshift-space power spectrum can be projected on the orthonormal basis of Legendre polynomials
\begin{equation}\label{multipoles}
    P_{\ell} (k) = \frac{2\ell +1}{2} \int_{-1}^{1}P^{s}(k, \mu) \mathcal{L}_{\ell}(\mu) d\mu \, ,
\end{equation}
where $\mathcal{L}_{\ell}$ is the Legendre polynomial of order $\ell$, obtaining the multipole of the power spectrum $P_{\ell}$.

\subsection{Estimator of the real space power spectrum}
\label{sec:Q0}
Before studying statistics beyond the power spectrum, we first study the method to estimate the real space power spectrum from the combination of redshift-space multipoles proposed in \cite{Ivanov:2021fbu}. 

From the definition of the power spectrum multipoles and given the orthogonality of the Legendre polynomials, the redshift-space power spectrum $P^s(k, \mu)$ can be reconstructed using the multipoles $P_{\ell}$
\begin{equation}\label{Ps_LagrangeExp}
    P^s(k, \mu)=\sum_{\ell} P_{\ell}(k) \mathcal{L}_{\ell}(\mu) \, .
\end{equation}
On the other hand, $P^s(k, \mu)$ can also be expanded in the monomial basis as
\begin{equation}
    P^s(k, \mu)=\sum_{n} Q_{n}(k) \mu^n \, .
\end{equation}
An interesting quantity arising from the latter expansion is the first moment $Q_0 = P^s(k, 0)$ which, representing the power spectrum orthogonal to the line-of-sight, is unaffected by RSD and coincides with the real space power spectrum $P^{r}\left(k\right)$. Using Eq.\eqref{Ps_LagrangeExp}, $Q_0$ can be expressed in term of the power spectrum multipoles $P_{\ell}$
\begin{equation}\label{Q0ofPell}
    Q_0 = \sum_{\ell} P_{\ell}(k) \mathcal{L}_{\ell}(0) \, .
\end{equation}
The Kaiser model for RSD (introduced in sub-section~\ref{ssec:RSD_Intro}) predicts that linear velocity inflow towards over-densities produces power transfer from the real space power spectrum to the quadrupole and hexadecapole of the power spectrum in redshift-space, as can be deduced from eq.~\ref{kaiser_delta_gs}.
On top of this, the Fingers-of-God (FoG) effect (also introduced in sub-section~\ref{ssec:RSD_Intro}), due to the random non-linear motion of galaxies on small scales, further transfers power to higher multipoles. When the FoG effect can be neglected, the sum in Eq.~\eqref{Q0ofPell} can be restricted to the multipoles sourced by the Kaiser effect \cite{Scoccimarro:2015bla,Ivanov:2021fbu}
\begin{equation}\label{Q0ofP024}
    Q_0 = \sum_{\ell=0,2,4} P_{\ell}(k) \mathcal{L}(0) = P_{0}-\frac{1}{2} P_{2}+\frac{3}{8} P_{4} \, .
\end{equation}
Otherwise higher multipoles should be considered in the expansion, and since these are more affected by noise than lower multipoles, approaches such as the one of \cite{Ivanov:2021fbu} can be used to mitigate the loss of precision.

Since $Q_0$ is unaffected by RSD, theoretical modelling is easier than for the redshift-space multipoles $P_{\ell}$, and this fact was used to extend the validity range of perturbation theory-based approaches such as EFTofLSS \cite{Ivanov:2021fbu}. We limit our analysis to the $Q_0$ estimate by means of the truncated sum of Eq.~\eqref{Q0ofP024}, which we measure in our halo and galaxy catalogues and compare to the real space power spectrum to gain insight into the effect of MG on this observable and into the validity of COLA simulations to make theoretical predictions for $Q_0$.

\subsection{Measurement and results}
\begin{figure}
\centering
\includegraphics[width=.98\textwidth]{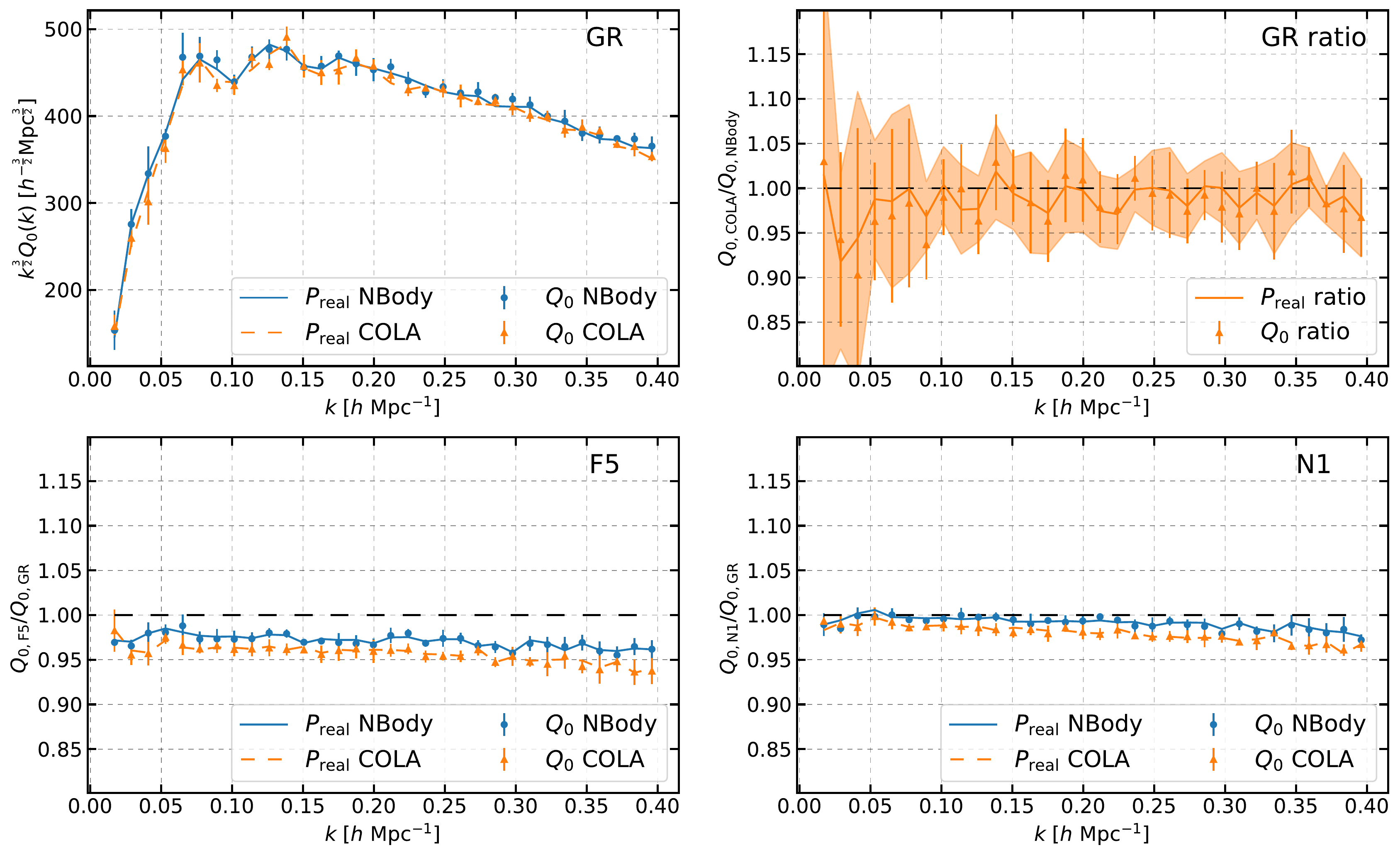}
\caption{\label{fig:Q0_halos} Comparison of the power spectrum orthogonal to the line-of-sight $Q_0$ (dots and error-bars) and the real space power spectrum $P^{r}$ (lines and shaded regions) for halos in COLA (in orange) and {\it N}-body simulations  (in blue). \textit{Top left:} Full signal in GR. \textit{Top right:} Ratio of the GR signal between COLA and {\it N}-body simulations. \textit{Bottom:} Boost factors, the ratio of $Q_0$ in F5 and N1 to that in GR.}
\end{figure}

\begin{figure}
\centering
\includegraphics[width=.98\textwidth]{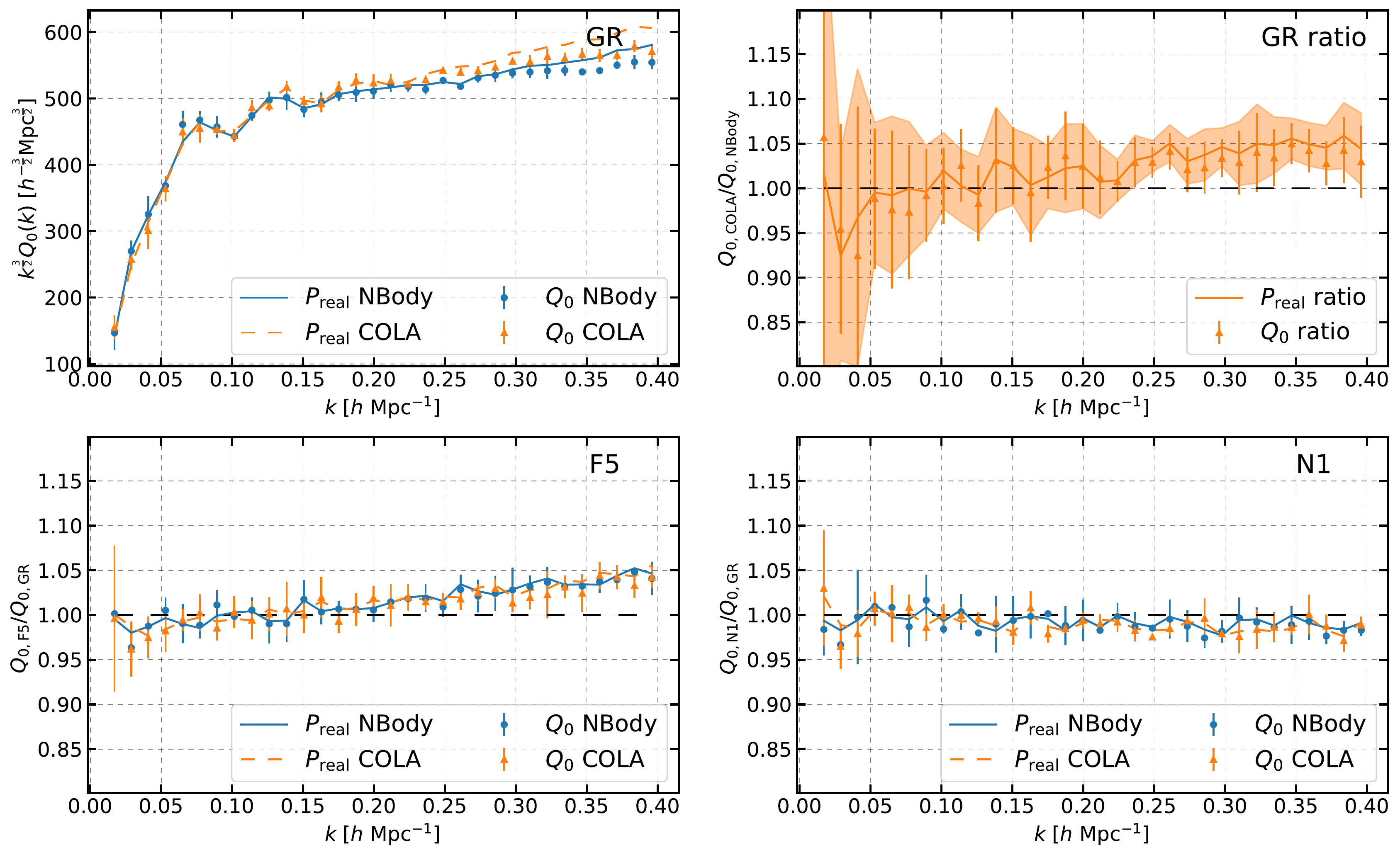}
\caption{\label{fig:Q0_galaxies}Same as figure~\ref{fig:Q0_halos} but for galaxies}
\end{figure}

Using the halo and galaxy catalogues from COLA and N-body simulations discussed in chapter~\ref{chp:mocks}, we compute the real space power spectrum and the multipoles in redshift-space in our halo and galaxy catalogues with the publicly available code \codeword{nbodykit}\footnote{\url{https://nbodykit.readthedocs.io}}. We assign the density to a $512^3$ mesh grid with the Triangular Shaped Clouds (TSC) interpolation \cite{Sefusatti:2015aex} and we correct the resulting power spectra for shot noise and the effects of the interpolation used \cite{Jing:2004fq,Sefusatti:2015aex}. We use bins of width $\Delta k=2 k_f$ between $k_{\rm min}=\frac{3}{2} k_f$ and $k_{\rm max}= 0.4 \hompc$, where $k_f$ is the fundamental frequency of the box. For the multipoles in redshift-space, we add RSD to the real space positions of tracers using in turn lines-of-sight parallel to each of the simulation box axes, and average the results over the three lines-of-sight. In the case of galaxy catalogues, we also average over the 5 HOD realisations for each simulation realisation. We then use the mean and standard deviation over the simulations' realisations to produce the signals and error bars that we plot in figure~\ref{fig:Q0_halos} and figure~\ref{fig:Q0_galaxies}.

In figure~\ref{fig:Q0_halos} we compare $Q_0$ to the real space power spectrum for halos. The top left panel shows that $Q_0$ is in good agreement with $P^{r}$ both in COLA and {\it N}-body halo catalogues. The top right panel shows that COLA and {\it N}-body are consistent within the variance for both $Q_0$ and $P^{r}$. The bottom panels show that the MG boost factors for $Q_0$ and $P^r$ are consistent within the variance and that COLA is able to reproduce the {\it N}-body boost factors, the ratio of the power spectrum between MG and GR, with $\sim2\%$ accuracy at all scales in both F5 and N1. The MG signal in F5 is in general stronger than in N1 where the signal is compatible with GR within the variance.

The comparison between $Q_0$ and $P^{r}$ for the galaxy catalogues is shown in figure~\ref{fig:Q0_galaxies}. The top left panel shows that $Q_0$ is in good agreement with $P^{r}$ up to $k\sim 0.25 \hompc$ in COLA and up to $k\sim 0.3 \hompc$ in {\it N}-body galaxy catalogues, where it starts deviating due to the truncation of the sum in eq.~\eqref{Q0ofP024}. The top right panel shows that COLA and {\it N}-body are consistent within the variance for both $Q_0$ and $P^{r}$ up to $k\sim 0.24 \hompc$ where COLA starts deviating from {\it N}-body. %
The bottom panels show that the MG boost factors for $Q_0$ and $P^r$ are consistent within the variance at all scales as are COLA and {\it N}-body results. We note that, in spite of the limits of the truncation in eq.~\eqref{Q0ofP024} and of the precision of COLA in reproducing the {\it N}-body power spectrum in GR, the MG boost factors are fully consistent at all scales. Overall the MG signal is quite weak in N1 while it reaches the $5\%$ in F5 on small scales.

This suggests that the observable $Q_0$ can be an interesting statistic to include in cosmological analysis to constrain MG theories and that COLA can be used to accurately predict the MG boost factors of $Q_0$. Furthermore, the good agreement of the MG boost factors of $P^r$ and $Q_0$ estimated using only the multipoles $P_0$, $P_2$, $P_4$ suggests that higher order multipoles can be modelled using the GR theory without significantly impacting the MG signal.  
\section{Bispectrum}
\label{sec:Bispectrum}
The bispectrum is the simplest higher-order summary statistics beyond the power spectrum and it holds non-gaussian information of the density field. The study of bispectrum also serves as a test for the screening approximation adopted in COLA simulations.

\subsection{Definitions and measurements}
We define the bispectrum as the three-point function in Fourier space for closed triangle configurations
\begin{equation}
    \left\langle\delta\left(\boldsymbol{k}_{1}\right) \delta\left(\boldsymbol{k}_{2}\right) \delta\left(\boldsymbol{k}_{3}\right)\right\rangle \equiv(2 \pi)^{3} B\left(k_{1}, k_{2}, k_{3}\right) \delta_{D}\left(\boldsymbol{k}_{1}+\boldsymbol{k}_{2}+\boldsymbol{k}_{3}\right) \,.
\end{equation}
and the reduced bispectrum as the ratio
\begin{equation}
    Q(k_{1}, k_{2}, k_{3})=\frac{B(k_{1}, k_{2}, k_{3})}{P(k_{1}) P(k_{2})+P(k_{2}) P(k_{3})+P(k_{3}) P(k_{1})} \, ,
\end{equation}
where $P(k)$ is the power spectrum \cite{Scoccimarro:2000sn}. In the absence of primordial non-gaussianity, the bispectrum at the leading order in perturbation theory shows a $P^2$ scaling which is removed in the reduced bispectrum to highlight the information content beyond the power spectrum \cite{Fry:1984}.

To measure the bispectrum in our simulations we use the bispectrum estimator proposed in \cite{Scoccimarro:2015bla} as implemented in the publicly available library \codeword{FML}\footnote{\href{https://github.com/HAWinther/FML}{https://github.com/HAWinther/FML}}. The density is interpolated on a $360^3$ cells grid with the cloud-in-cell (CIC) mass assignment scheme with interlacing to reduce the aliasing at small scales \cite{Scoccimarro:2015bla,Sefusatti:2015aex}. We estimate the bispectrum in bins of size $\Delta k = 3 k_f$, centring the first bin in $k_{\rm min} = 3 k_f$. The shot noise, which is subtracted from the bispectrum signal, is calculated as 
\begin{equation}
    B^{\mathrm{SN}}\left(k_{1}, k_{2}, k_{3}\right)=\frac{1}{\bar{n}}\left(P\left(k_{1}\right)+P\left(k_{2}\right)+P\left(k_{3}\right)\right)+\frac{1}{\bar{n}^{2}} \, ,
\end{equation}
with $\bar{n}$ being the number density of tracers and $P(k)$ their power spectrum \cite{Matarrese:1997sk}.

To get an insight into the dependence of the bispectrum signals on the triangle shape, we bin all the configurations with $k_1>0.1 \hompc$ in a grid of $x_2 \equiv k_2/k_1$, $x_3 \equiv k_3/k_1$, and average the signals weighted with the number of fundamental triangles falling in each bin
\begin{equation}\label{BispecConf}
    \overline{B}(x_2^i, x_3^j) = \frac{\sum_{k_1= k_{\rm min}}^{k_{\rm max}} \sum_{k_2,k_3}^{\rm bin} B(k_1, k_2, k_3) N^{\mathrm{T}}(k_1, k_2, k_3)}{\sum_{k_1= k_{\rm min}}^{k_{\rm max}} \sum_{k_2,k_3}^{\rm bin} N^{\mathrm{T}}(k_1, k_2, k_3)} \, ,
\end{equation}
where $N^{\mathrm{T}}$ is the number of fundamental closed-triangles
\begin{equation}
    \begin{split}
    N^{\mathrm{T}}(k_1, k_2, k_3)=\prod_{i=1}^{3} \int_{k_{i}}& d^{3} q_{i} \delta_{\mathrm{D}}\left(\Vec{q_1}+\Vec{q_2}+\Vec{q_3}\right)\\ & \cdot \sum_{n_1,n_2,n_3} \delta_{\mathrm{D}}(\Vec{q_1}-n_1 k_f)\delta_{\mathrm{D}}(\Vec{q_2}-n_2 k_f) \delta_{\mathrm{D}}(\Vec{q_3}-n_3 k_f)\, ,
    \end{split}
\end{equation}
and similarly for the reduced bispectrum
\begin{equation}\label{RedBispecConf}
    \overline{Q}(x_2^i, x_3^j) = \frac{\sum_{k_1= k_{\rm min}}^{k_{\rm max}} \sum_{k_2,k_3}^{\rm bin} Q(k_1, k_2, k_3) N^{\mathrm{T}}(k_1, k_2, k_3)}{\sum_{k_1= k_{\rm min}}^{k_{\rm max}} \sum_{k_2,k_3}^{\rm bin} N^{\mathrm{T}}(k_1, k_2, k_3)}.
\end{equation}

\subsection{Results for the bispectrum}
\begin{figure}
\centering
\includegraphics[width=.98\textwidth]{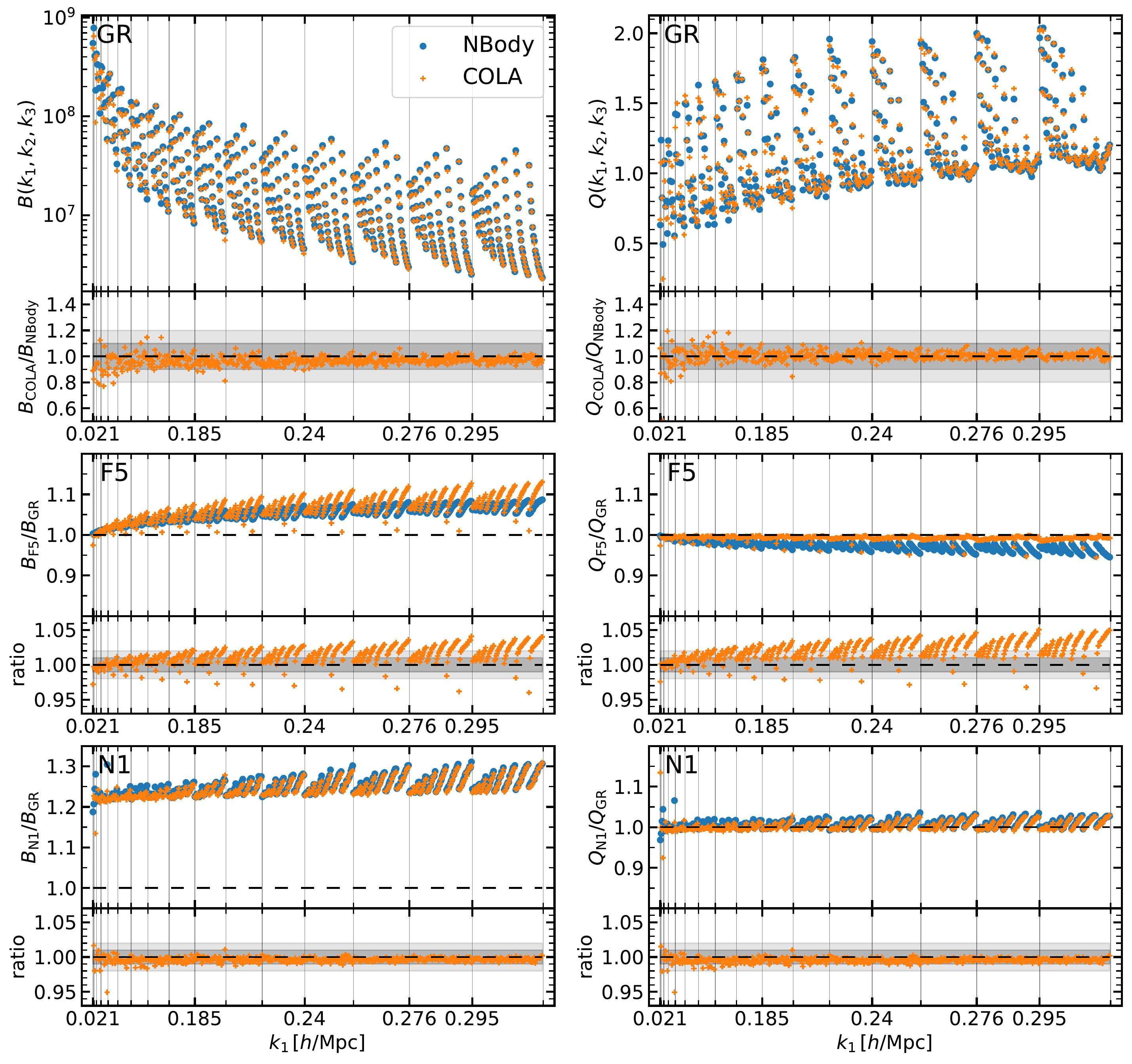}
\caption{\label{fig:DM_bispec} Comparison of bispectrum (left column) and reduced bispectrum (right column) of the matter distribution in COLA (orange crosses) and {\it N}-body (blue dots) simulations. The configurations with $k_1 \ge k_2 \ge k_3$ are displayed in ascending order of the values of the wavenumbers $k_1$, $k_2$, and $k_3$ respectively. The vertical lines denote the value of $k_1$ for the configurations immediately to the right of each line. \textit{Top:} Full signal in GR. \textit{Middle and Bottom:} Boost factors in F5 and N1 respectively. In each panel, the bottom sub-panel shows the ratio between the COLA and {\it N}-body signals displayed in the top sub-panel.}
\end{figure}

\begin{figure}
\centering
\includegraphics[width=.98\textwidth]{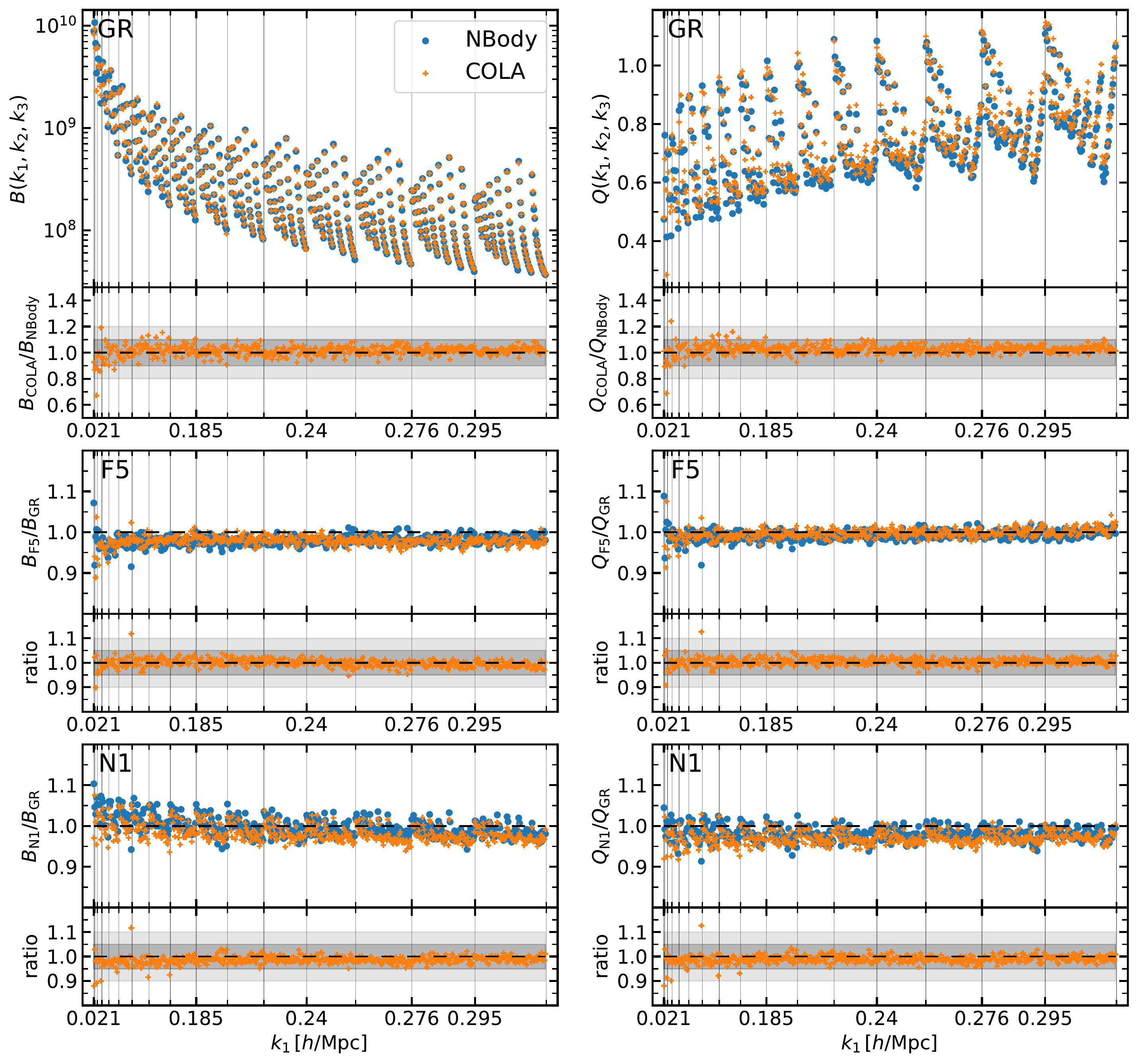}
\caption{\label{fig:Galaxy_bispec_RS} Same as figure~\ref{fig:DM_bispec} but for the monopole of bispectrum and reduced bispectrum of galaxies in redshift-space.}
\end{figure}

Since the bispectrum is a function of three variables, it is non-trivial to represent it in 2D plots, so we follow an approach often used in literature and represent all the configurations of the bispectrum in a single scatter plot by restricting our analysis to the configurations with $k_1 \ge k_2 \ge k_3$ and organising them in ascending order of the values of the wavenumbers $k_1$, $k_2$, and $k_3$, respectively. We show all the configurations of the bispectrum (and the reduced bispectrum) in figure~\ref{fig:DM_bispec} and figure~\ref{fig:Galaxy_bispec_RS} for DM and galaxies in redshift-space respectively. For completeness we show also the plots for halos and galaxies in real space in figure~\ref{fig:Halo_bispec} and figure~\ref{fig:Galaxy_bispec}.
In these figures, the left column shows the full bispectrum signal and the right column the reduced bispectrum. We split the results into three rows, the top one shows the full signal in GR, the middle one shows the boost factor in F5 (i.e. the ratio between F5 and GR) and the bottom one shows the boost factor in N1. To benchmark results with respect to full {\it N}-body, each row contains a sub-panel at the bottom showing the ratio between the COLA and {\it N}-body signals.

\begin{figure}
\centering
    \subfloat[][Full Bispectrum]{
    \includegraphics[width=.48\textwidth,clip]{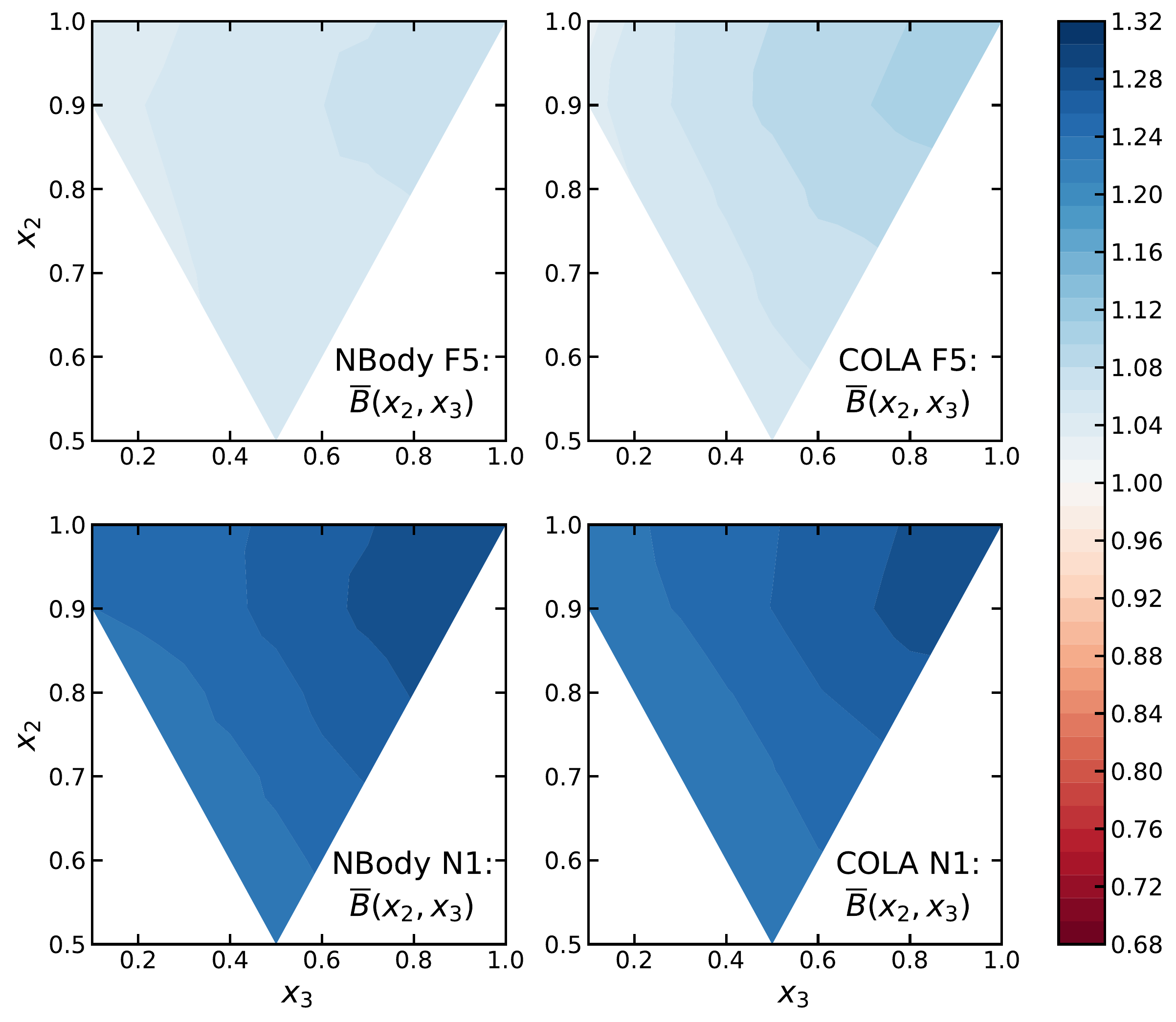}
    }
    \hfill
    \subfloat[][Reduced Bispectrum]{
    \includegraphics[width=.48\textwidth]{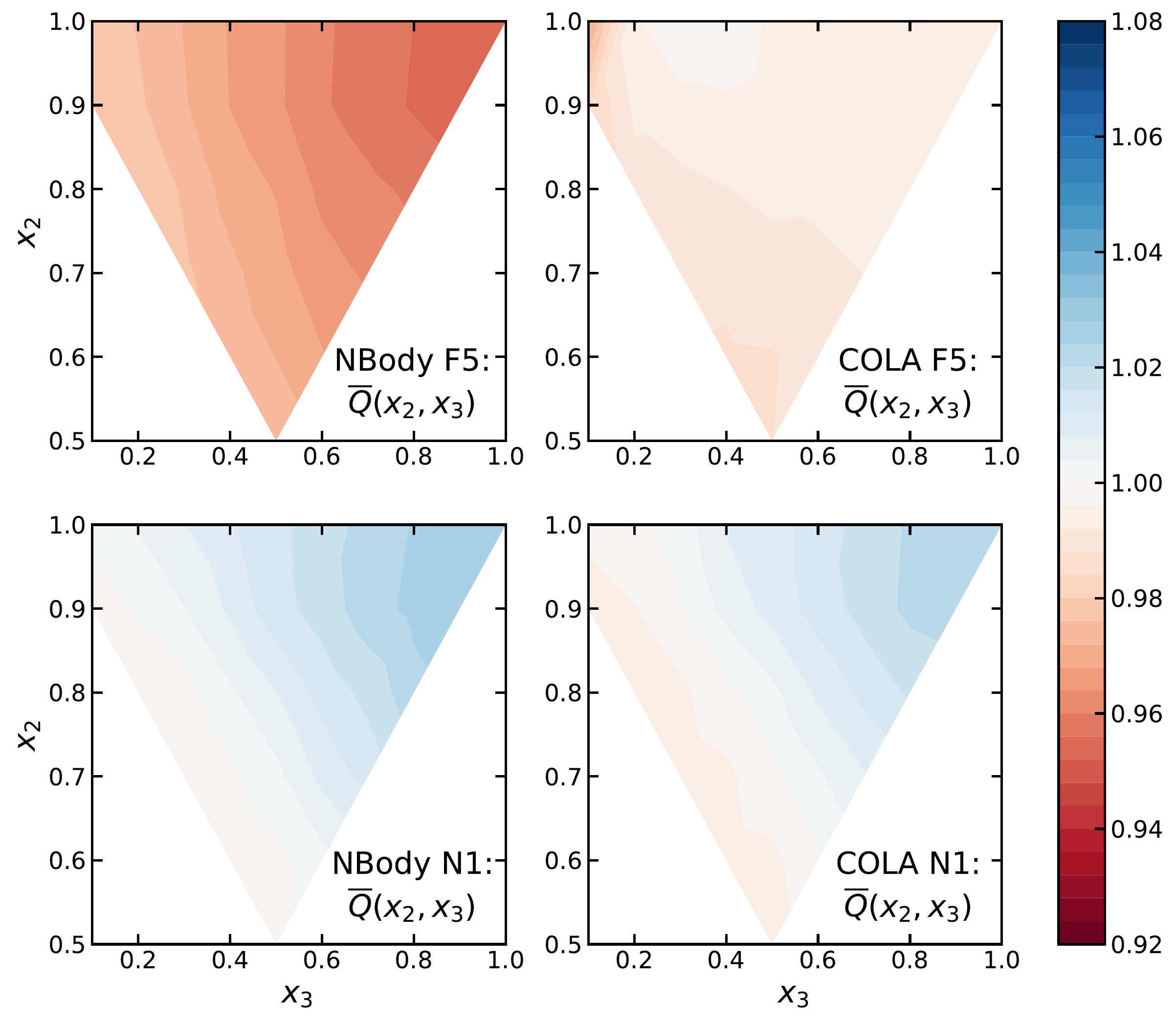}
    }
\caption{\label{fig:DM_bispec_conf}Comparison of the configuration dependence of the F5 (top row) and N1 (bottom row) boost factors of bispectrum (on the left) and reduced-bispectrum (on the right) of DM in {\it N}-body (first and third columns) and COLA simulations (second and fourth columns). The colour bars show the amplitude of the boost factors with blue (red) denoting stronger (weaker) signal in MG with respect to GR. \textit{In each panel:} The top right, top left and bottom corners of the triangle correspond to the equilateral, squeezed and folded configurations respectively. The squeezed configuration is missing from the figure due to the bin's width.}
\end{figure}

\begin{figure}
\centering
    \subfloat[][Full Bispectrum]{
    \includegraphics[width=.48\textwidth,clip]{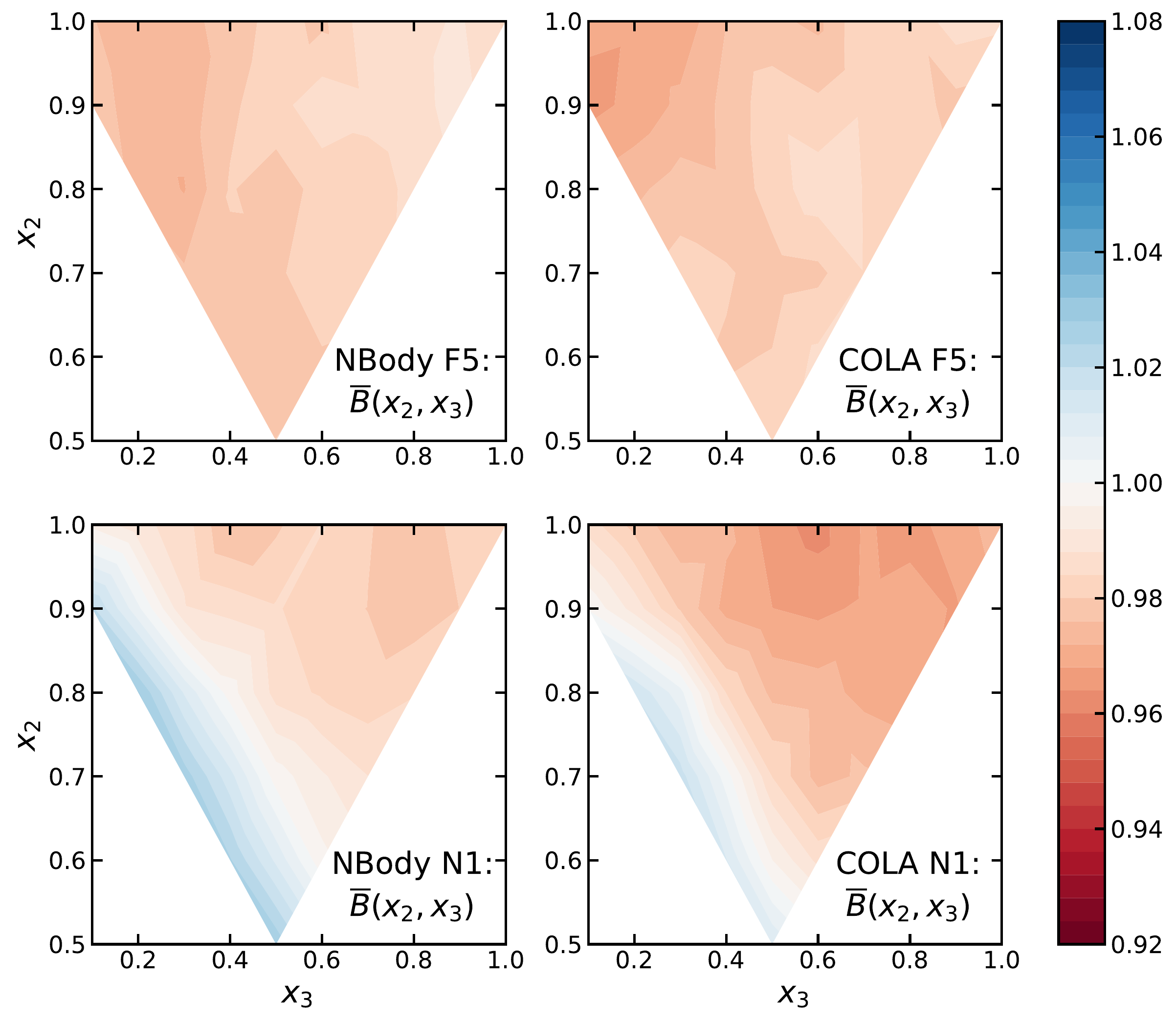}
    }
    \hfill
    \subfloat[][Reduced Bispectrum]{
    \includegraphics[width=.48\textwidth]{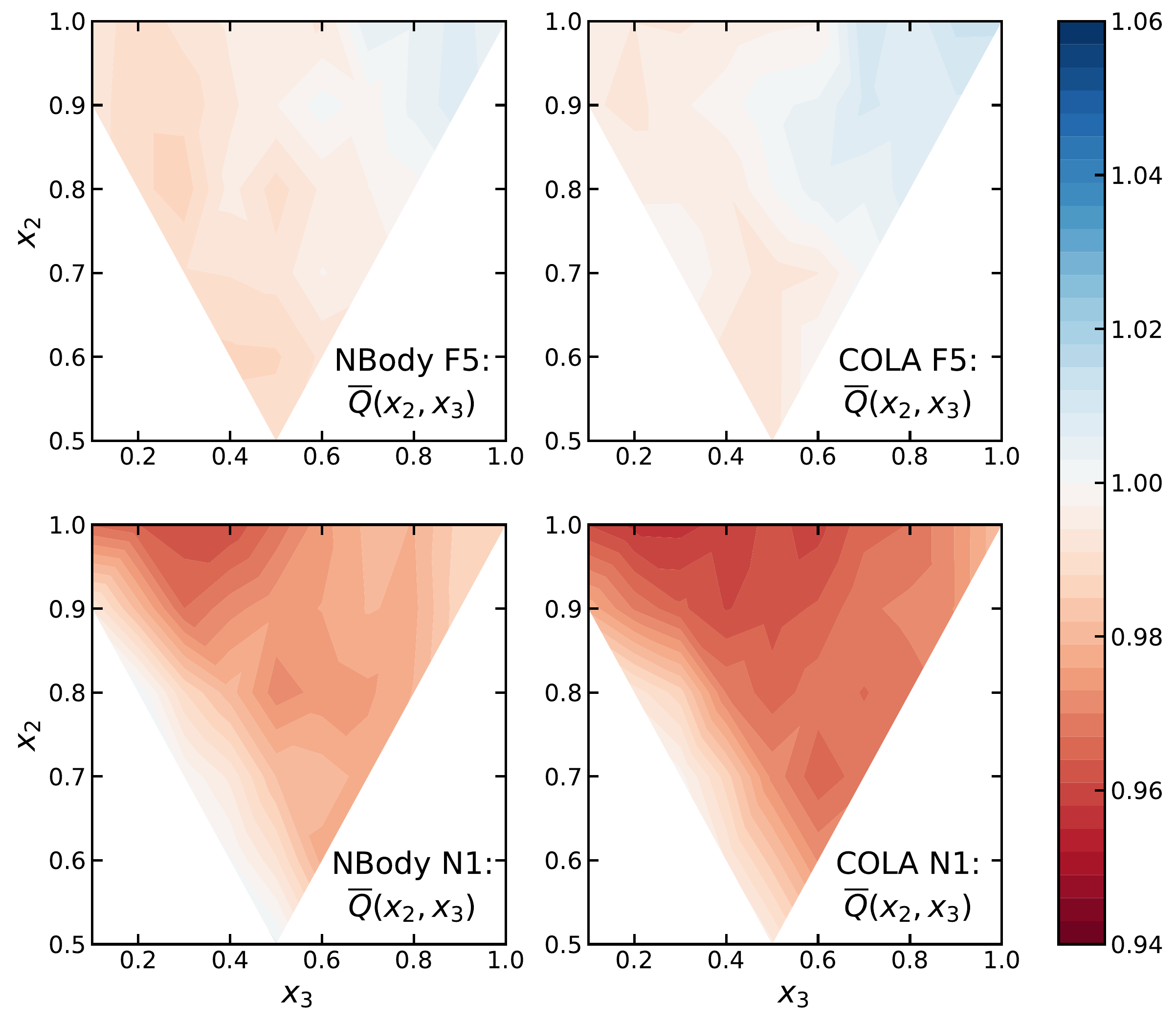}
    }
\caption{\label{fig:Galaxy_bispec_RS_conf}Same as figure~\ref{fig:DM_bispec_conf} but for the monopole of the bispectrum and reduced bispectrum of galaxies in redshift-space.}
\end{figure}

\begin{figure}
\centering
\includegraphics[width=.98\textwidth]{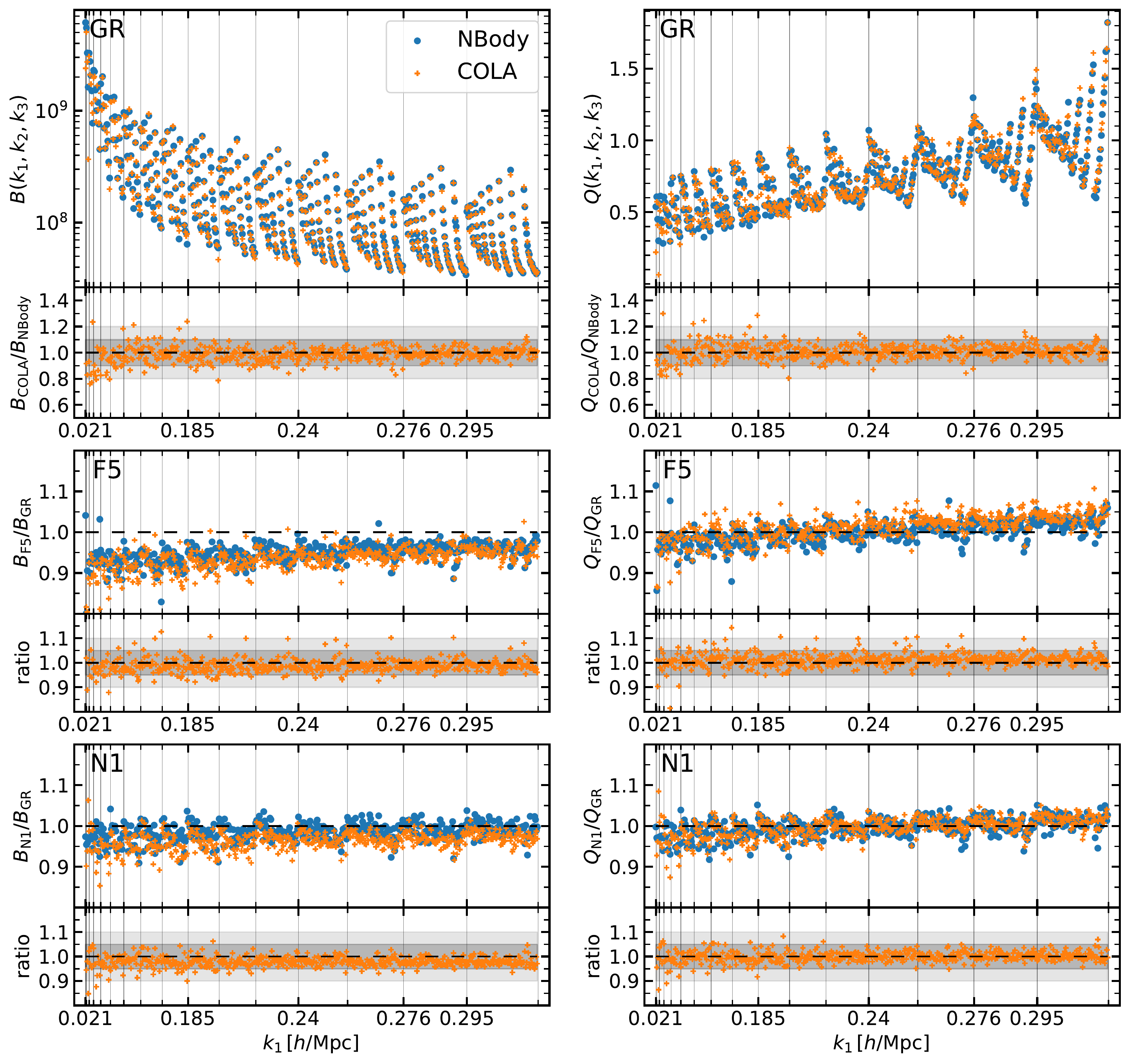}
\caption{\label{fig:Halo_bispec} Comparison of bispectrum (left column) and reduced bispectrum (right column) of halos in COLA (orange crosses) and {\it N}-body (blue dots) simulations. The configurations with $k_1 \ge k_2 \ge k_3$ are displayed in ascending order of the values of the wavenumbers $k_1$, $k_2$, and $k_3$ respectively. The vertical lines denote the value of $k_1$ for the configurations immediately to the right of each line. \textit{Top:} Full signal in GR. \textit{Middle and Bottom:} Boost factors in F5 and N1 respectively. In each panel, the bottom sub-panel shows the ratio between the COLA and {\it N}-body signals displayed in top sub-panel.}
\end{figure}

\begin{figure}
\centering
\includegraphics[width=.98\textwidth]{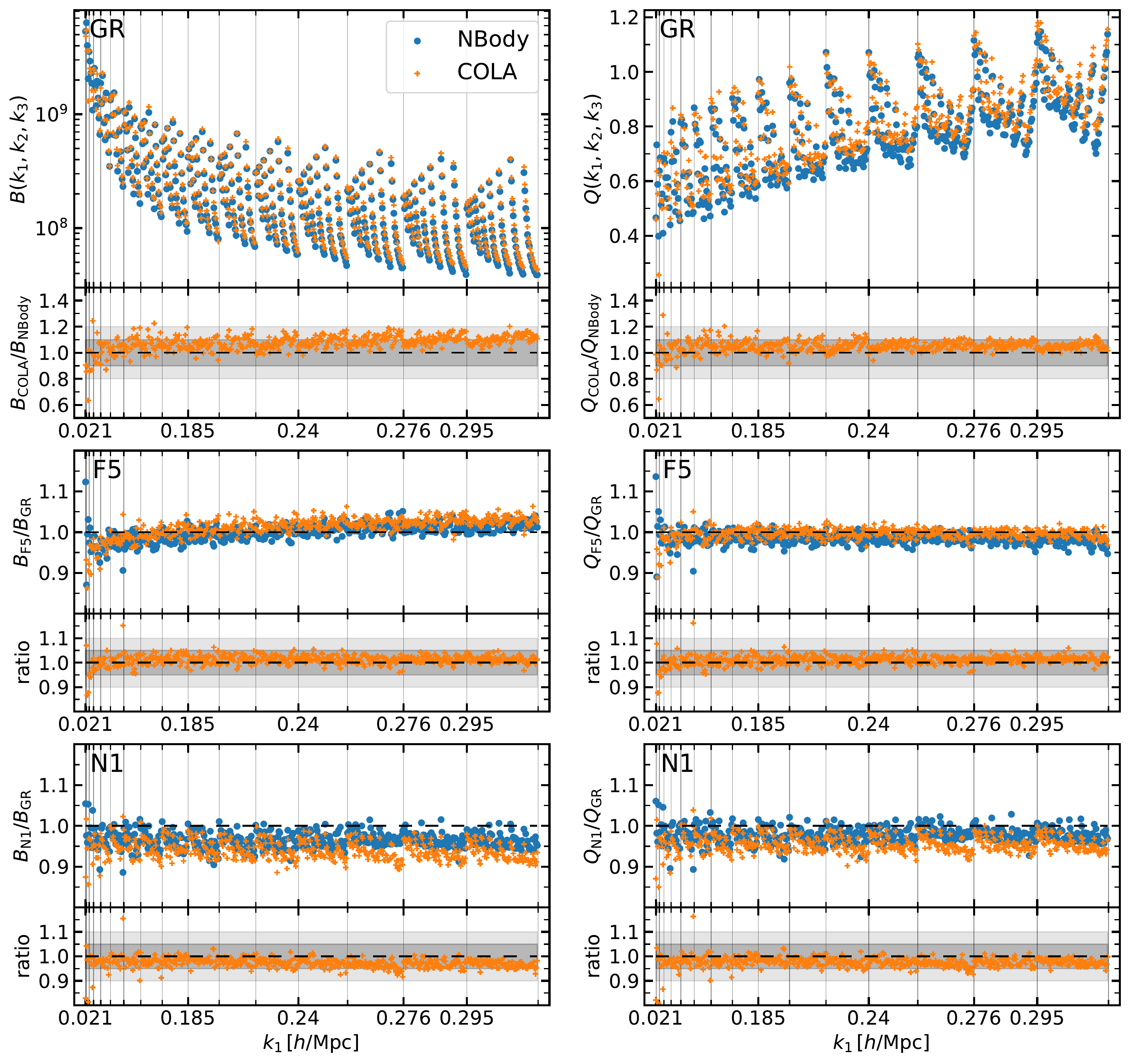}
\caption{\label{fig:Galaxy_bispec} Same as figure~\ref{fig:Halo_bispec} but for galaxies in real space.}
\end{figure}

\begin{figure}
\centering
    \subfloat[][Full Bispectrum]{
    \includegraphics[width=.48\textwidth,clip]{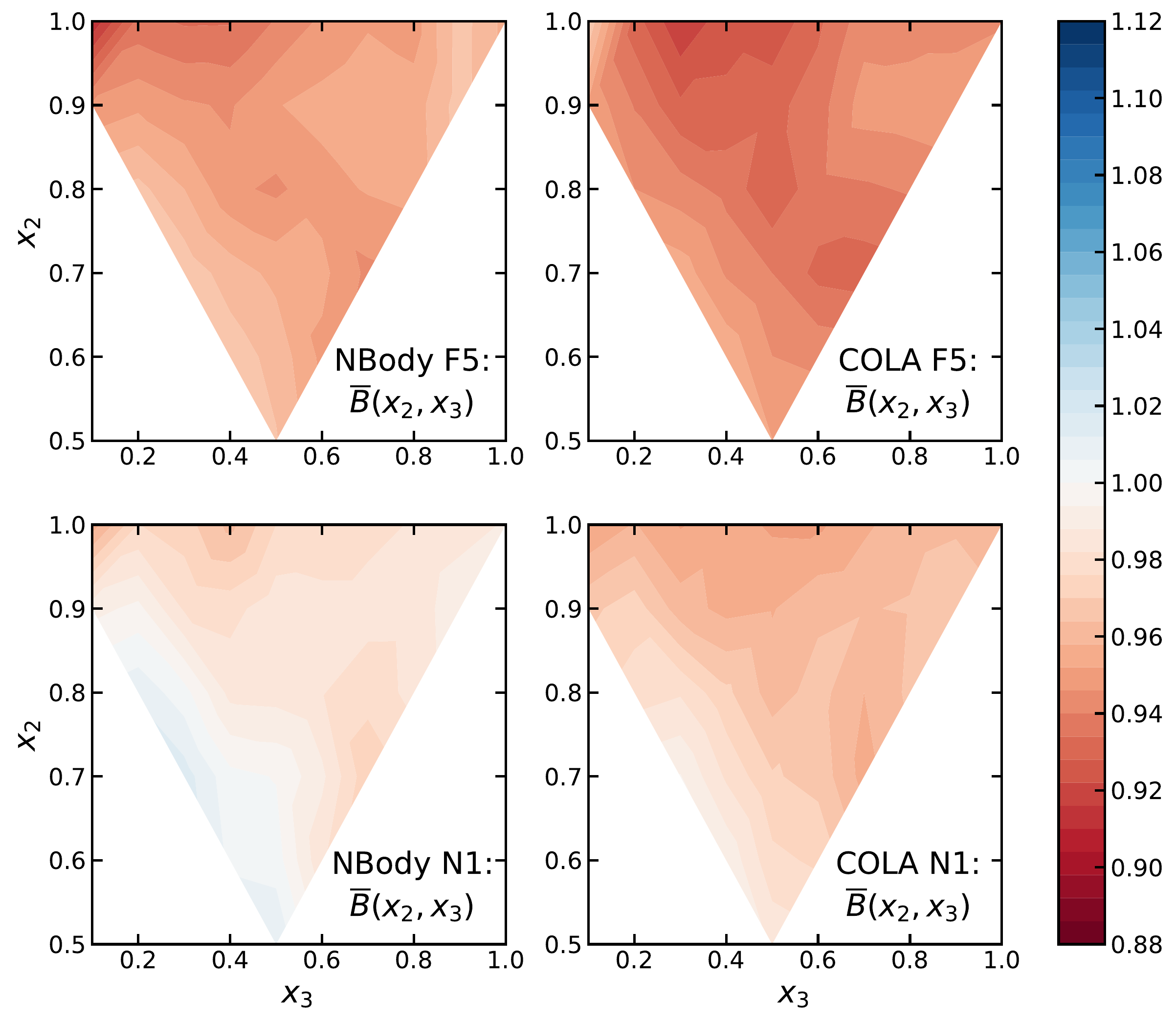}
    }
    \hfill
    \subfloat[][Reduced Bispectrum]{
    \includegraphics[width=.48\textwidth]{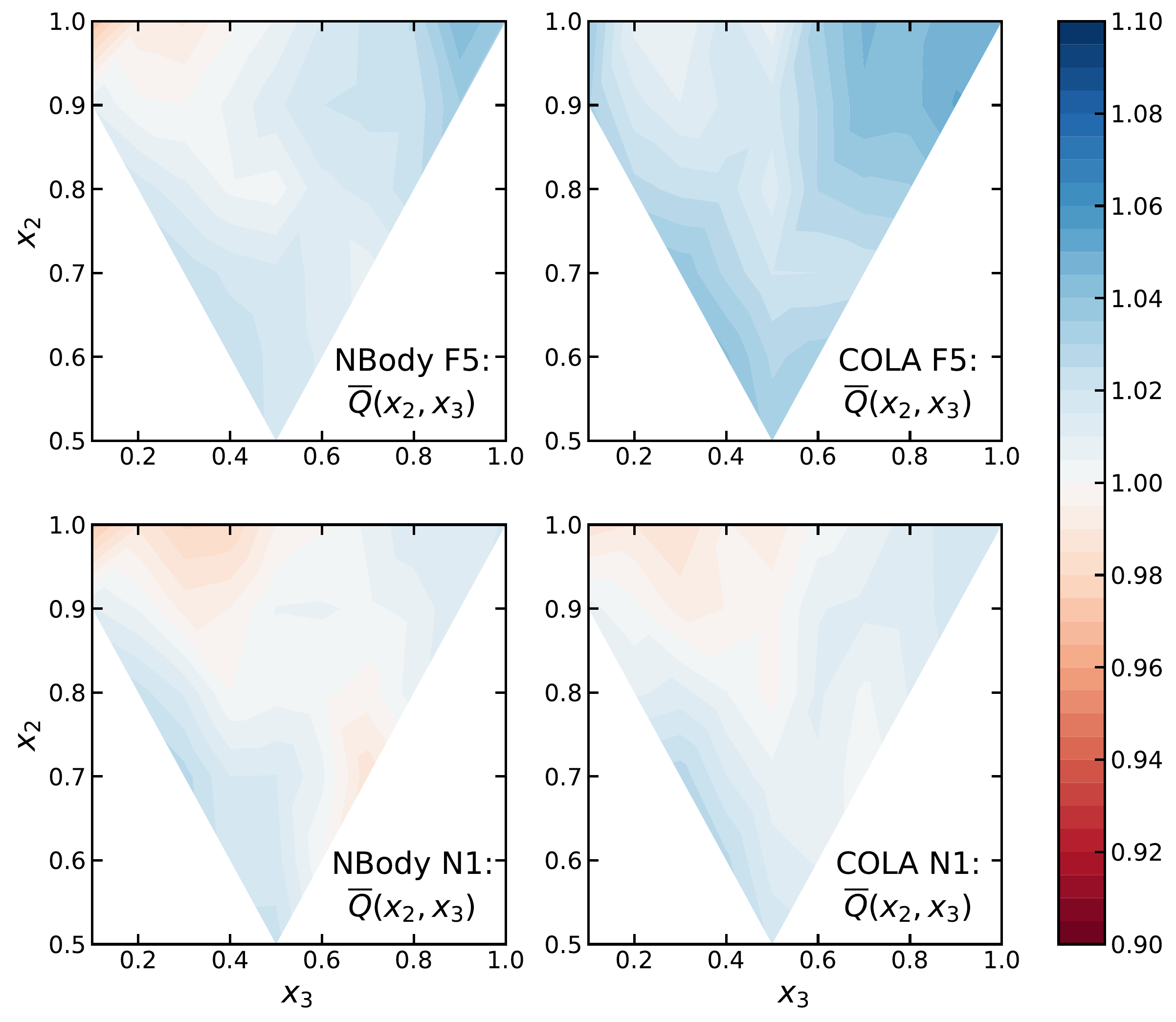}
    }
\caption{\label{fig:Halo_bispec_conf}Comparison of the configuration dependence of the F5 (top row) and N1 (bottom row) boost factors of bispectrum (on the left) and reduced-bispectrum (on the right) of halos in {\it N}-body (first and third columns) and COLA simulations (second and fourth columns). The colour-bars show the amplitude of the boost factors with blue (red) denoting stronger (weaker) signal in MG with respect to GR. \textit{In each panel:} The top right, top left and bottom corners of the triangle correspond to the equilateral, squeezed and folded configurations respectively. The squeezed configuration is missing from the figure due to the bin's width.}
\end{figure}

\begin{figure}
\centering
    \subfloat[][Full Bispectrum]{
    \includegraphics[width=.48\textwidth,clip]{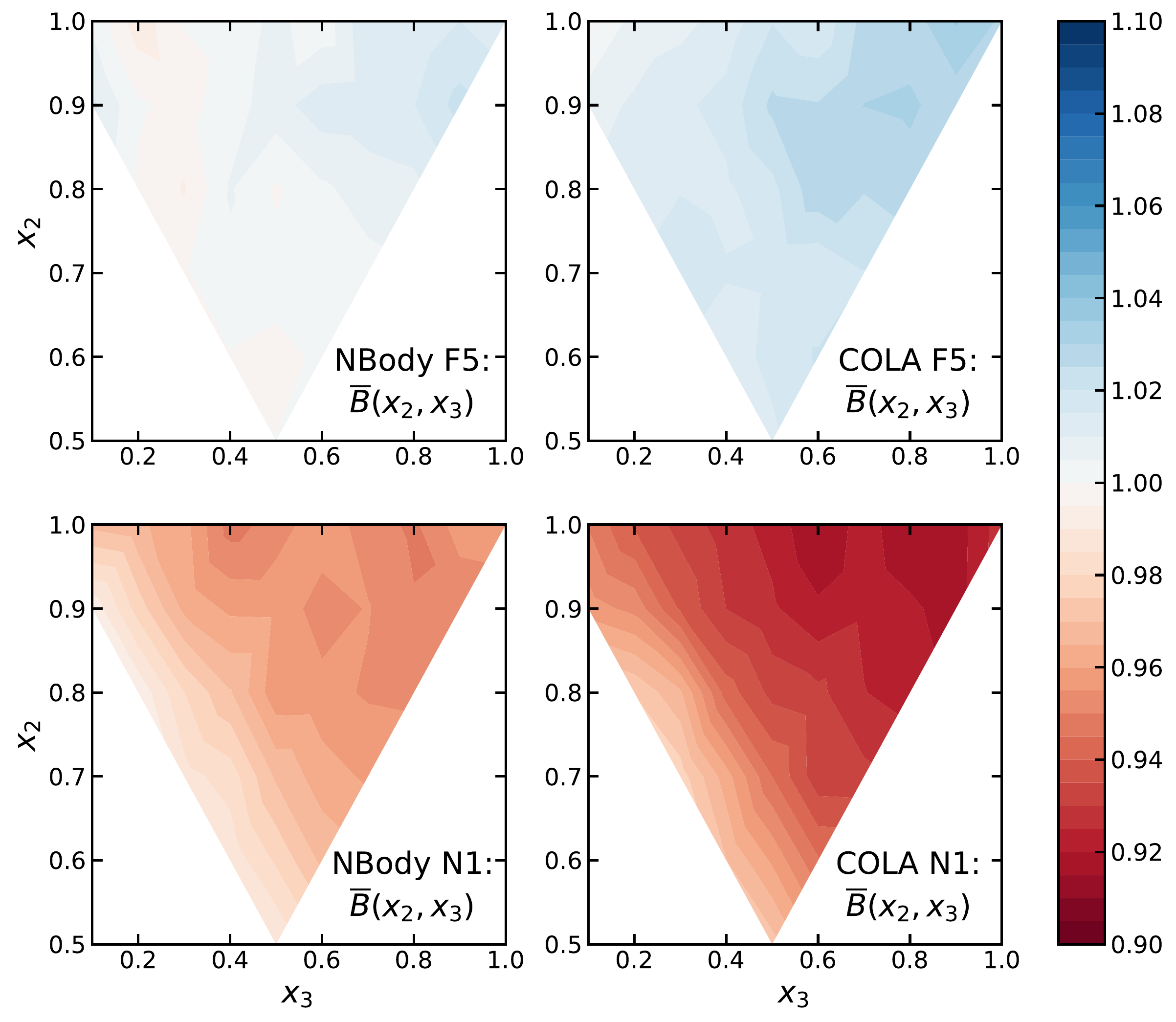}
    }
    \hfill
    \subfloat[][Reduced Bispectrum]{
    \includegraphics[width=.48\textwidth]{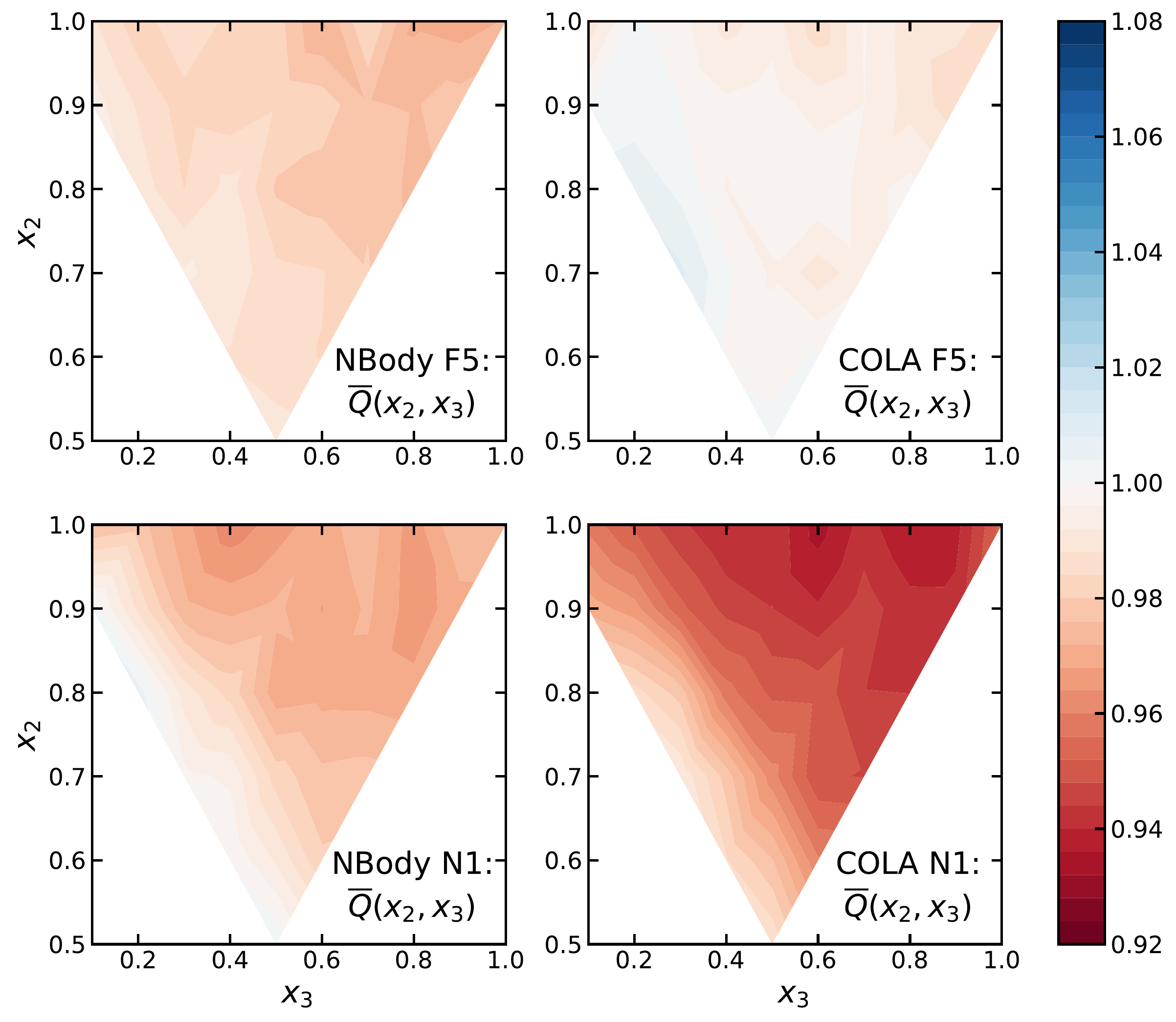}
    }
\caption{\label{fig:Galaxy_bispec_conf}Same as figure~\ref{fig:Halo_bispec_conf} but for galaxies.}
\end{figure}

The MG boost factors of $\bar{B}$ and $\bar{Q}$ (from eq.~\eqref{BispecConf} and eq.~\eqref{RedBispecConf} respectively), 
that is, the ratio of $\bar{B}$ and $\bar{Q}$ between MG and GR, are used to produce figure~\ref{fig:DM_bispec_conf} for DM and figure~\ref{fig:Galaxy_bispec_RS_conf} for galaxies in redshift-space, as well as figure~\ref{fig:Halo_bispec_conf} and figure~\ref{fig:Galaxy_bispec_conf} for halos and galaxies in real space. In each figure, the left panels display the full bispectrum and the right panels the reduced bispectrum. We show the results for  
F5 in the top row and the results for N1 in the bottom row. The colour represents the amplitude of the MG signal, according to the colour bar. 

At the DM level, figure~\ref{fig:DM_bispec} shows that COLA is able to reproduce {\it N}-body results in GR with $\sim 10\%$ accuracy both for the bispectrum and the reduced bispectrum in agreement with the findings in \cite{Colavincenzo_2018}. Focusing on the MG boost factors in {\it N}-body, we notice that the strong MG signal of the bispectrum, reaching $\sim 30\%$ in N1 and $\sim 10\%$ in F5, is suppressed in the reduced bispectrum. This shows that most of the bispectrum signal is due to the enhancement of the power spectrum in MG. The MG boost factors of COLA do not reproduce the information content beyond the power spectrum as highlighted by the reduced bispectrum signal in the case of F5, while it reproduces the {\it N}-body signals with $\sim 2\%$ accuracy in N1. The failure of COLA simulations in F5 gravity can be traced back to the use of the screening approximation, which allows us to recover with good accuracy the power spectrum but produces a DM bispectrum signal consistent with the results of \cite{Gil-Marin:2011iuh} in the case of simulations without the chameleon mechanisms (see their figure 1). 
This is because the screening approximation used in COLA simulations linearises the scalar field equation and thus it lacks the contribution to the bispectrum from the non-linearity of the scalar field. This effect is stronger in F5, where the effect of screening is strong even on quasi-non-linear scales, while it is weak in N1. Concerning the configuration dependence of the DM bispectrum, figure~\ref{fig:DM_bispec_conf} shows that the bispectrum boost factor signals of DM are stronger for equilateral configurations. COLA results reproduce the same configuration dependence of {\it N}-body in N1, but the reduced bispectrum boost factor in F5 clearly shows the failure of COLA simulations in reproducing non-trivial 3-points interactions, due to the screening approximation used in COLA.

Figure~\ref{fig:Galaxy_bispec_RS} shows the bispectrum monopole of the galaxy catalogues in redshift-space. The agreement between COLA and {\it N}-body is within $10\%$ in GR for both the bispectrum and the reduced bispectrum. In the F5 and N1 gravity models, the bispectrum boost factors show deviations from GR at $\sim5\%$ level for some configurations, and the same is true for the reduced bispectrum. COLA reproduces {\it N}-body results for the boost factors with at least $5\%$ accuracy. 
The much weaker MG signal of the redshift-space galaxy bispectrum monopole compared to the DM bispectrum is attributable to the HOD parameters tuning procedure, which tries to reproduce the target multipoles of the galaxy power spectrum, in this case the GR one as discussed in \ref{chp:mocks}. This tuning produces a power spectrum contribution to the bispectrum that is similar for GR and MG theories (as can be seen from the boost-factors of the bispectrum in the left column) and, on the other hand, dilutes the MG signal of the bispectrum coming from higher-order contributions of DM clustering due to the nonlinearity of galaxy bias (as evidenced by the boost-factors of the reduced bispectrum in the right column). This influence of galaxy bias is also visible in figure~\ref{fig:Galaxy_bispec_RS_conf} that shows the configuration dependence of galaxy bispectra boost factors substantially differs from what is found in the DM bispectra boost factors. This explains why the configuration dependence in COLA is qualitatively in agreement with the {\it N}-body result. The contribution from the non-linear galaxy bias dominates over the DM non-linearity hiding the inaccuracy of COLA simulations in reproducing the DM bispectra in F5.



For completeness we show and discuss also the results for halos and galaxies in real space, although they are on the same line of what we have just seen for the bispectrum signal of the galaxies in redshift space.

In the case of halo bispectrum, figure~\ref{fig:Halo_bispec} shows that COLA is able to reproduce the {\it N}-body signals for both bispectrum and reduced bispectrum of halos with $10\%$ accuracy in most of the configurations. The MG signal reaches $\sim 10\%$ in the case of the full bispectrum in F5, while it is limited to $\sim 5\%$ in the other cases. For the boost factors, COLA's accuracy is better than $\sim 5\%$ for most of the configurations.

The configuration dependence of the bispectra boost factors of halos is shown in figure~\ref{fig:Halo_bispec_conf}. Due to the bias of halos, the configuration dependence of DM in figure~\ref{fig:DM_bispec_conf} is not reflected in the configuration dependence of the halo bispectra boost factors. Thanks to this, the configuration dependence of COLA in F5 is qualitatively consistent with the one of {\it N}-body halo catalogues for both full and reduced bispectrum, despite the limitations in reproducing the matter bispectra discussed in section~\ref{sec:Bispectrum}.

The galaxy bispectrum in real-space is shown in figure~\ref{fig:Halo_bispec}. In GR, COLA is able to reproduce the {\it N}-body signals with $20\%$ accuracy for the full bispectrum and $10\%$ accuracy for the reduced bispectrum. The MG signals reach $\sim 10\%$ in the case of the full bispectrum in N1, while it is $\sim 5\%$ or less in the other cases. Despite the lower accuracy in reproducing the full bispectrum in GR, COLA's boost factors are $5\%$ accurate for most of the configurations. Figure~\ref{fig:Halo_bispec_conf} shows the configuration dependence of the bispectra boost factors of galaxies in real space. Again, the bias model dominates the MG signal hiding the clear MG boost factors of DM (figure~\ref{fig:DM_bispec_conf}). The strongest deviations from GR consistently appear towards the equilateral configuration (top right corner), in both full and reduced bispectrum independently of the gravity model. COLA tends to over-predict the MG signal of the bispectra in N1. Nonetheless, the qualitative description of the configuration dependence is consistent between COLA and {\it N}-body. 

Our results show that the MG bispectrum signal is mainly due to MG effects on the power spectrum and less to the higher order terms carrying specific information of the gravity model as shown in \cite{Gil-Marin:2011iuh} for $f(R)$ gravity. Extra care needs to be taken when using COLA simulations with screening approximations to compute the reduced bispectrum of dark matter, in particular in $f(R)$ gravity models as the non-linearity of the scalar field due to screening is neglected. This is relevant for the computation of higher-order statistics in weak lensing, for example. On the other hand, in galaxy clustering, there are large contributions from non-linear galaxy biases to the bispectrum, and this inaccuracy in the reduced bispectrum in COLA is hidden. The bispectrum, when combined with the power spectrum, is useful to break degeneracies in the galaxy biases and COLA simulations can be used to model accurately galaxy bispectrum in MG models, despite the use of screening approximations.

\section{Voids}
\label{sec:Voids}
Voids are low-density regions of space that are useful for cosmological analysis because the weaker gravitational field produces a phase-space distribution that is better described by linear models \cite{Cai:2016jek,Nadathur:2017jos}. Also, screening mechanisms do not operate efficiently in voids, thus voids can be used to test models of gravity
\cite{voidmg1,voidmg2,voidmg3,voidmg4,Cautun:2017tkc,Paillas:2018wxs,Perico:2019obq,Contarini:2020fdu}.
Observationally we cannot directly probe DM, so we have to rely on galaxies to identify voids, keeping in mind that galaxies are biased tracers of the underlying density field and that we only know the redshift-space position of galaxies.

Many void definitions have been proposed in the literature, differing in the properties (e.g. abundances and profiles) of the voids they identify. Some void definitions can prove to be more suited than others in highlighting specific features of the underlying galaxy and DM distributions as discussed in many works (see for example \cite{Colberg:2008qg,Nadathur:2017jos,Cautun:2017tkc,Paillas:2018wxs,Massara:2022lng}).
In particular, the watershed void finding technique (implemented in various codes such as \codeword{ZOBOV} \cite{Neyrinck:2007gy} and \codeword{VIDE} \cite{Sutter:2014haa}) has been shown to identify (real-space) voids whose (stacked) galaxy density profiles in redshift-space can be approximated by linear models sensitive to the linear growth rate of structure \cite{Cai:2016jek,Nadathur:2017jos}. As MG theories may affect the linear growth rate $f$, we want to test if $f$ can also be recovered in MG theories using the RSD model for the void-galaxy Cross-Correlation Function (CCF) in \cite{Nadathur:2017jos}.

In subsection \ref{ssec:RSDinVoids} and \ref{ssec:VelModel} we review the RSD model from \cite{Nadathur:2017jos,Woodfinden:2022bhx}, and in subsection \ref{ssec:VoidMeas} we perform void finding on our simulation suites and measure the void quantities relevant to the RSD model. Finally, in subsection \ref{ssec:VoidResults} we fit the multipoles of the void-galaxy CCF in redshift-space using the RSD model to recover the linear growth rate $f$. Comparing the measurements and results for the fit between COLA and {\it N}-body simulations, we validate COLA simulations for void analysis.

\subsection{Redshift-space distortion model in voids}
\label{ssec:RSDinVoids}

Let $\xi^r(\vec{r})$ be the real-space CCF between voids and galaxies. Assuming a bijective differentiable mapping between real and redshift-space $\vec{s} = \vec{f}(\vec{r})$, the redshift-space CCF $\xi^s(\vec{s})$ is related to the real-space CCF by
\begin{equation}
    \int_V \left[1+\xi^{s}(\mathbf{s})\right] d^{3} s= \int_V \left[1+\xi^{r}(\mathbf{r})\right] d^{3} r .
\end{equation}
Performing the change of variable $\vec{r} \rightarrow \vec{s}$ on the right-hand side we obtain
\begin{equation}
    \int_V \left[1+\xi^{s}(\mathbf{s})\right] d^{3} s= \int_V \left[1+\xi^{r}(\mathbf{r}(\vec{s}))\right] \mathrm{J}_{\vec{r},\vec{s}} \, d^{3} s \, ,
\end{equation}
where ${\rm J}_{\vec{r},\vec{s}}$ is the determinant of the Jacobian matrix $\frac{\partial \vec{r}}{\partial \vec{s}}$. Requiring that this equation holds for every arbitrary volume $V$ implies
\begin{equation}
        \left[1+\xi^{s}(\mathbf{s})\right] =  \left[1+\xi^{r}(\mathbf{r}(\vec{s}))\right] \mathrm{J}_{\vec{r},\vec{s}} \, .
\end{equation}
This relation can be manipulated to express the redshift-space CCF $\xi^s$ in terms of $\xi^r$, once the mapping between real and redshift-space is fixed, determining the two unknowns $\mathbf{r}(\vec{s})$ and $\mathrm{J}_{\vec{r},\vec{s}}$.
This mapping in the distant observer approximation can be described by
\begin{equation} \label{RSD_map}
    \mathbf{s}=\mathbf{r}+\frac{\mathbf{v} \cdot \hat{\mathbf{z}}}{a H} \hat{\mathbf{z}},
\end{equation}
where $\vec{v}$ is the galaxy velocity, $a$ is the scale factor, and $H$ is the Hubble parameter. The void positions are by construction assumed to be invariant under redshift-space remapping, since the CCF in this model always refers to the correlation between real-space voids and galaxies in either real or redshift-space \cite{Nadathur:2017jos}.

Assuming a spherically symmetric model for the galaxy velocity $\vec{v} = v(r)\vec{\hat{r}}$, the Jacobian determinant is given by
\begin{equation} \label{cart_J}
    {\rm J}_{\vec{r},\vec{s}} = \left[ 1+\frac{v(r)}{a H r}+\frac{z^{2} \frac{d}{d r} v\left(r\right)}{a H r^{2}}-\frac{z^{2} v(r)}{a H r^{3}} \right ]^{-1}\, ,
\end{equation} 
with $r = |\vec{r}| =\sqrt{x^2 + y^2 + z^2}$, which can be rewritten as
\begin{equation}
   {\rm J}_{\vec{r},\vec{s}} = \left[1+\frac{\vbar(r)}{r}+\mu_r^{2} \frac{d \vbar(r)}{d r} -\frac{\mu_r^2 \vbar(r)}{r}\right]^{-1}  \, .
\end{equation}
where we defined $\Bar{v} \equiv \frac{v}{aH}$ and $\mu_r \equiv \frac{z}{r}$ for simplicity of notation.

In addition to the spherically symmetric model for the galaxy velocity we consider a random velocity component along the line-of-sight%
\footnote{A more realistic model would be considering a random velocity component with random orientation (and dispersion 3 times larger), but the two models are equivalent for what concerns RSD.}, %
$\vec{v} = v_r(r)\vec{\hat{r}} + v_{\parallel} \vec{\hat{z}}$, where $P(v_{\parallel}) = \mathcal{N}(0,\sigma(r))$ as proposed in \cite{Nadathur:2017jos}.
This makes the mapping between real and redshift-space depends also on $v_{\parallel}$
\begin{equation}\label{RSDmap_disp}
    \begin{aligned}
        \mathbf{s}(\vec{r}, v_{\parallel})&=\mathbf{r}+\frac{\left(v_r(r)\vec{\hat{r}} + v_{\parallel} \vec{\hat{z}}\right) \cdot \hat{\mathbf{z}}}{a H} \, \hat{\mathbf{z}} \\
        &=\mathbf{r}+\frac{\left(v_r(r)\mu_r + v_{\parallel} \right)}{a H} \, \hat{\mathbf{z}}\, ,
    \end{aligned}
\end{equation}
but does not explicitly affect the Jacobian, as $v_{\parallel}$ is an integration variable independent of the real-space radius $r$.
We solve eq.~\eqref{RSDmap_disp} for $\vec{r}$ iteratively to obtain $\vec{r}(\vec{s}, v_{\parallel})$ and integrate over $v_{\parallel}$ to compute the theoretical prediction for the redshift-space CCF
\begin{equation}
    \xi^{s}_{\rm th}(s, \mu_s) =  \int\left(1+\xi^{r}\right) \mathrm{J}_{\vec{r},\vec{s}} e^{-\frac{v_\parallel^2}{2 \sigma_v^2}}\, dv_{\parallel} -1 \, ,
\end{equation}
where $\xi^r$, $\mathrm{J}_{\vec{r},\vec{s}}$ and $\sigma_v$ are expressed in terms of $s$ and $\mu_s$.
Finally we project $\xi^{s}_{\rm th}(s, \mu_s)$ on the Legendre polynomials to compute monopole and quadrupole of the void-galaxy CCF in redshift-space
\begin{equation} \label{theory_xiell}
\begin{gathered}
    \xi^{s}_{0,{\rm th}}(s) = \int_{0}^{1} \xi^{s}_{\rm th}(s, \mu_s) \, d\mu_s \, ,\\ 
    \xi^{s}_{2,{\rm th}}(s) = \frac{5}{2}\int_{0}^{1} \xi^{s}_{\rm th}(s, \mu_s) \left(3 \mu_s^{2}-1\right)\, d\mu_s \,.
\end{gathered}
\end{equation}

In the comparisons of the void-galaxy CCF in redshift-space shown in subsection~\ref{ssec:VoidMeas}, we include these theoretical predictions without any further assumption to test the accuracy of this approach.

\subsection{Velocity model}
\label{ssec:VelModel}
Galaxy velocities correlate with void positions, as galaxies are attracted towards overdensities far from the void centre. In an isotropic universe, despite that voids have in general very irregular shapes, the average matter distribution of stacked voids is isotropic and therefore spherically symmetric. The average velocity of galaxies around voids depends on the matter distribution and it is also spherically symmetric.
This velocity can be estimated in simulations and it has been found to be well described down to scales of $\sim 30$ Mpc$/h$ by the simple linear model
\begin{equation}\label{VelModel}
    v_{r}^{\text {model }}(r)=-\frac{1}{3} f a H r \Delta(r) \, ,
\end{equation}
where $f$ is the linear growth rate and 
$
    \Delta(r) \equiv \frac{3}{r^{3}} \int_{0}^{r} \delta(r') r^{\prime 2} d r'
$
is the integrated matter density profile \cite{Peebles:1994xt,Nadathur:2017jos}. We will show in subsection~\ref{ssec:VoidMeas} a comparison between this velocity model and the radial-velocity profile estimated from our simulations that confirms the same trend of \cite{Nadathur:2017jos} for the accuracy of the velocity model.

\subsection{Velocity-dispersion estimator}\label{sec:AppB}
Since there is no clear treatment of the velocity-dispersion of galaxies around stacked voids in literature, we dissipate here any  ambiguity thanks to a detailed analysis of how to relate the velocity-dispersion of two models for RSD in voids often used in literature that we will refer to as Gaussian streaming model (GSM) \cite{Cai:2016jek} and the change-of-variable model (CVM) \cite{Nadathur:2017jos}. In doing so we obtain two formulae that can be used to estimate the velocity-dispersion from synthetic catalogues in a computationally efficient way.
Let us use the notation of \cite{Woodfinden:2022bhx} where the probability distribution of the velocity in the GSM is 
\begin{equation}
    P_{\rm GSM}\left(v_{\|}, \vec{r}\right) = P_{\rm GSM}\left(v_{\|}, r, \mu \right)
\end{equation}
and the one in the CVM is 
\begin{equation}
P_{\rm CVM}\left(\tilde{v}_{\|}, \vec{r}\right) = P_{\rm CVM}\left(\tilde{v}_{\|}, r\right) \, .
\end{equation}
The velocities of the two probability distributions are connected by 
\begin{equation} \label{strVScv}
    v_{\parallel}(r, \mu)=\tilde{v}_{\parallel}+v_{r}(r) \mu .
\end{equation}
where we changed the side of $v_{\parallel}$ and $\tilde{v}_{\parallel}$ with respect to \cite{Woodfinden:2022bhx} to stress that $\tilde{v}_{\parallel}$ is neither a function of $r$ nor $\mu$.

By hypothesis the probability distribution in the CVM is a Gaussian distribution $\mathcal{N}({\rm mean}, \rm{var})$ with zero mean and $\sigma^2_{\tilde{v}_{\parallel}}$ variance:
\begin{equation} \label{cvProb}
    P_{\rm CVM}\left(\tilde{v}_{\|}, r\right) = \mathcal{N}(0, \sigma_{\tilde{v}_{\parallel}}^2(r)) \, .
\end{equation}
Likewise, we want to express the probability distribution of the GSM as a Gaussian distribution. In the most general case this corresponds to:
\begin{equation}
    P_{\rm GSM}\left(v_{\|}, r, \mu \right) = \mathcal{N}(\left\langle v_{\parallel}(r, \mu) \right\rangle, \left\langle v_{\parallel}^2(r, \mu) \right\rangle - \left\langle v_{\parallel}(r, \mu) \right\rangle^2) \, .
\end{equation}
Now using eq.~\eqref{strVScv} and \eqref{cvProb} we can see that 
\begin{gather}
    \left\langle v_{\parallel}(r, \mu) \right\rangle = v_{r}(r) \mu \\
    \left\langle v_{\parallel}^2(r, \mu) \right\rangle - \left\langle v_{\parallel}(r, \mu) \right\rangle^2 = \sigma_{\tilde{v}_{\parallel}}^2(r)
\end{gather}
so the probability distribution in the streaming model can be expressed as
\begin{equation}
    P_{\rm GSM}\left(v_{\|}, r, \mu \right) = \mathcal{N}(v_{r}(r) \mu, \sigma_{\tilde{v}_{\parallel}}^2(r)) \, .
\end{equation}
The velocity-dispersion can therefore be estimated from synthetic catalogues using only information from one line-of-sight by subtracting the angle averaged contribution of the radial-velocity to the angle averaged velocity-dispersion along the line-of-sight
\begin{equation}
\label{VelDisp1LOS}
\begin{split}
    \sigma_{\tilde{v}_{\parallel}}^2(r) &= \left\langle v_{\parallel}^2(r) \right\rangle - \left\langle v_{\parallel}(r) \right\rangle^2 - \frac{1}{2}\int_{-1}^{1} v_r^2(r) \mu^2 d\mu \\
    &= \left\langle v_{\parallel}^2(r) \right\rangle  - \frac{1}{3} v_r^2(r) \, ,
\end{split}
\end{equation}
where $v(r)$ can either be estimated directly from simulations or from the velocity model of eq~\eqref{VelModel}.

A more realistic description of the velocity of galaxies around (statistically isotropic) stacked voids is by considering a mean radial component and a (gaussianily distributed) random component in a random direction $\hat{n}$
\begin{equation} \label{vgal_dec}
    \vec{v}(\vec{r})=\tilde{v} \, \hat{n} +v_{r}(r) \hat{r} \, ,
\end{equation}
with 
\begin{equation} \label{VrandProb}
    P\left(\tilde{v}, r\right) = \mathcal{N}(0, \sigma_{\tilde{v}}^2(r)) \, ,
\end{equation}
basically releasing the assumption that the random velocity component of the CVM is only along the line-of-sight.

As redshift distortions are sensitive only to the velocity along the line of sight, we project $\vec{v}$ on the line-of-sight $\hat{z}$ ( adopting the distant observer approximation):
\begin{equation}
    \vec{v}(\vec{r}) \cdot \hat{z} = \tilde{v} \, \hat{n} \cdot \hat{z}  +v_{r}(r) \hat{r} \cdot \hat{z}
\end{equation}
which can be written as:
\begin{equation} \label{vz}
    v_{z}(r, \mu) = \tilde{v} \, \hat{n} \cdot \hat{z}  +v_{r}(r) \mu
\end{equation}
As before we express the probability distribution of the velocity parallel to the LOS as a Gaussian distribution 
\begin{equation}
    P \left(v_{z}, r, \mu \right) = \mathcal{N}(\left\langle v_{z}(r, \mu) \right\rangle, \left\langle v_{z}^2(r, \mu) \right\rangle - \left\langle v_{z}(r, \mu) \right\rangle^2) \, .
\end{equation}

Now using eq.~\eqref{VrandProb} and eq.~\eqref{vz} we can see that 
\begin{gather}
    \left\langle v_{z}(r, \mu) \right\rangle = v_{r}(r) \mu \\
    \left\langle v_{z}^2(r, \mu) \right\rangle - \left\langle v_{z}(r, \mu) \right\rangle^2 \equiv \sigma_{\tilde{v}_{z}}^2(r) = \frac{\sigma_{\tilde{v}}^2(r)}{3} \label{VgalVariance}
\end{gather}
since $\langle \hat{n} \cdot \hat{z}\rangle = \frac{1}{3} $.

Now using eq.~\eqref{vgal_dec} and taking the scalar product of each side with itself, we get
\begin{equation}
\begin{aligned}
    \vec{v}(\vec{r}) \cdot \vec{v}(\vec{r}) &= (\tilde{v} \, \hat{n} +v_{r}(r) \hat{r})\cdot (\tilde{v} \, \hat{n} +v_{r}(r) \hat{r}) \\
    &= \tilde{v}^2 + v_{r}(r)^2 + 2 \tilde{v} v_{r}(r) \hat{n} \cdot \hat{r} \, .
\end{aligned}
\end{equation}
Due to the fact that $\hat{n}$ is a random direction, the scalar product $\hat{n} \cdot \hat{r}$ vanishes once taking the ensamble average
\begin{equation}
    \langle \hat{n} \cdot \hat{r} \rangle= 0
\end{equation}
so we can express the dispersion of the random component in terms of the moduli of total and radial velocities
\begin{equation}
    \sigma_{\tilde{v}}^2(r) \equiv \langle \tilde{v}^2\rangle = \left\langle \left| \vec{v}(\vec{r}) \right|^2 \right\rangle - \left\langle v_{r}(r)
\right\rangle^2 \,. 
\end{equation}

Finally we can identify $v_z = v_{\parallel}$ and use eq.~\eqref{VgalVariance} to obtain
\begin{equation}\label{VelDisp3LOS}
    \sigma_{\tilde{v}_{\|}}(r)=\frac{\sqrt{\left\langle\left|\vec{v}(\vec{r})\right|^{2}\right\rangle-v_{\mathrm{r}}^{2}}} {\sqrt{3}} \, ,
\end{equation}
which can be used to estimate the velocity-dispersion along the line-of-sight using the full content of the velocity information, not just the line-of-sight projection of the velocity as done in eq.~\eqref{VelDisp1LOS}.

\subsection{Void finding and measurements}\label{ssec:VoidMeas}
\begin{figure}
\centering
\includegraphics[width=.98\textwidth]{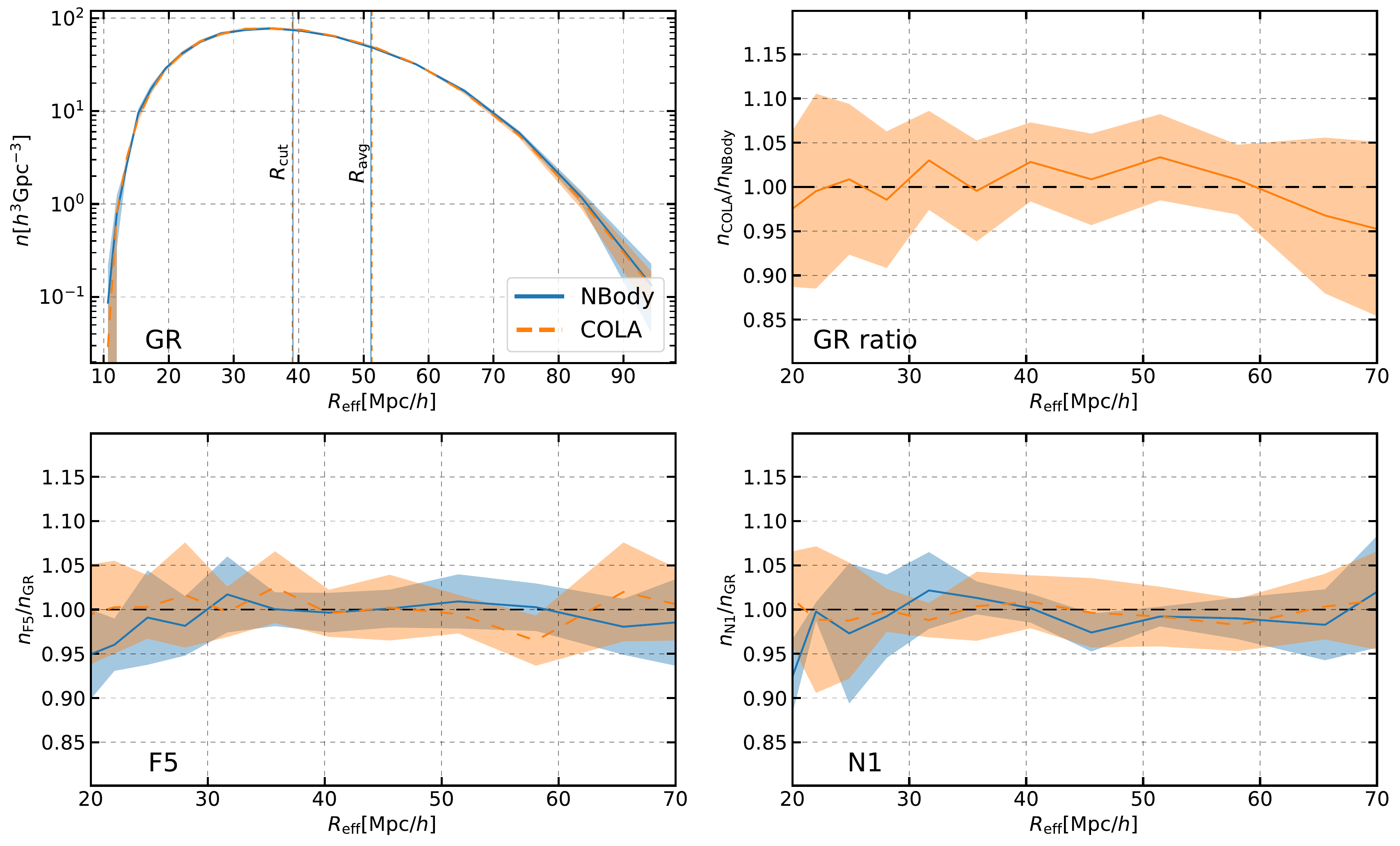}
\caption{\label{fig:VoidSizeFun}Comparison of the void size function in COLA (in orange) and {\it N}-body simulations (in blue). \textit{Top left:} Full signal in GR. The vertical lines denote the minimum and average void radii after applying the median cut. \textit{Top right:} Ratio of the GR signal between COLA and {\it N}-body simulations. \textit{Bottom:} Boost factors in F5 and N1. The panels involving ratios are shown in the range $[20,70] \mpcoh$ as the signal-to-noise ratio deteriorates outside of this range.}
\end{figure}

To perform void-finding for the real-space galaxy mocks, we make use of the publicly available code \codeword{Revolver}\footnote{\href{https://github.com/seshnadathur/Revolver}{https://github.com/seshnadathur/Revolver}} which implements two void-finding techniques: \codeword{ZOBOV} \cite{Neyrinck:2007gy} and \codeword{voxel}. With the aim of performing a similar analysis to \cite{Nadathur:2017jos} we choose \codeword{ZOBOV} void-finding. This algorithm finds voids as the watershed basins of the density field of galaxies in real-space estimated using the Voronoi tessellation \cite{Neyrinck:2007gy}. The void effective radius is determined by the void total volume $V$ via the formula $R_{\rm eff}=(3 V / 4 \pi)^{1 / 3}$. We use this definition of void radius to count the voids in 20 bins logarithmically spaced between 10 and 100 $\mpcoh$ to obtain the void size function displayed in figure~\ref{fig:VoidSizeFun}. This shows that the void-size functions in GR, N1 and F5 are all compatible within the variance, as are the void-size functions of COLA and {\it N}-body simulations. 

We select a sub-sample of voids whose radius is larger than the median radius of the catalogue, which is sufficient to select voids that are uncorrelated and statistically isotropic \cite{Nadathur:2017jos}. %
This results in void catalogues with a minimum void radius $R_{\rm cut} \simeq 39 \mpcoh$ and an average void radius $R_{\rm avg} \simeq 51 \mpcoh$ across all the simulations. We add the vertical lines in the top-left panel of figure~\ref{fig:VoidSizeFun} to highlight these two scales in the void-size function. The void centre is identified as the centre of the largest empty sphere inside the void volume. This is done by finding the circumcentre of the tetrahedron described by the 4 galaxies closest to the under-density peak \cite{Nadathur:2015qua}.
Given the definition of the void centre, it is possible to study several quantities characterising the profile of voids. In particular, we focus on: 
\begin{itemize}
    \item the integrated matter density profile, $\Delta(r) \equiv \frac{3}{r^{3}} \int_{0}^{r} \delta(r') r^{\prime 2} d r'$, where $\delta(r)$ is the matter density-contrast profile of stacked voids;
    \item the galaxy radial-velocity profile, $v_{\rm r}(r) = \left\langle \vec{v}_g \cdot \hat{r} \right\rangle$;
    \item the galaxy velocity-dispersion profile $\sigma_{v_{\parallel}}(r) = \nicefrac{\sqrt{ \left\langle |\vec{v}_g |^2\right\rangle - v_{\rm r}^2 }}{ \sqrt{3}}$;
    \item the CCF of voids and galaxies in real and redshift-space, $\xi^r$ and $\xi^s$ respectively.
\end{itemize} 
These quantities, relevant to the RSD model of the void-galaxy CCF discussed in subsection~\ref{ssec:RSDinVoids}, are estimated using 25 linearly spaced radial bins from the void centre to $120 \mpcoh$. To measure the integrated-density profile and the CCF we use the publicly available code \codeword{pyCUTE}\footnote{\href{https://github.com/seshnadathur/pyCUTE}{https://github.com/seshnadathur/pyCUTE}} based on \codeword{CUTE}\footnote{\href{https://github.com/damonge/CUTE}{https://github.com/damonge/CUTE}} \cite{CUTE}. In the case of real-space (redshift-space) statistics, the signal is first averaged over the 5 HOD realisations (and over the three lines of sight), and then the mean and standard deviations over the 5 boxes are used respectively for the lines and shaded regions in the figures.

\begin{figure}
\centering
\includegraphics[width=.98\textwidth]{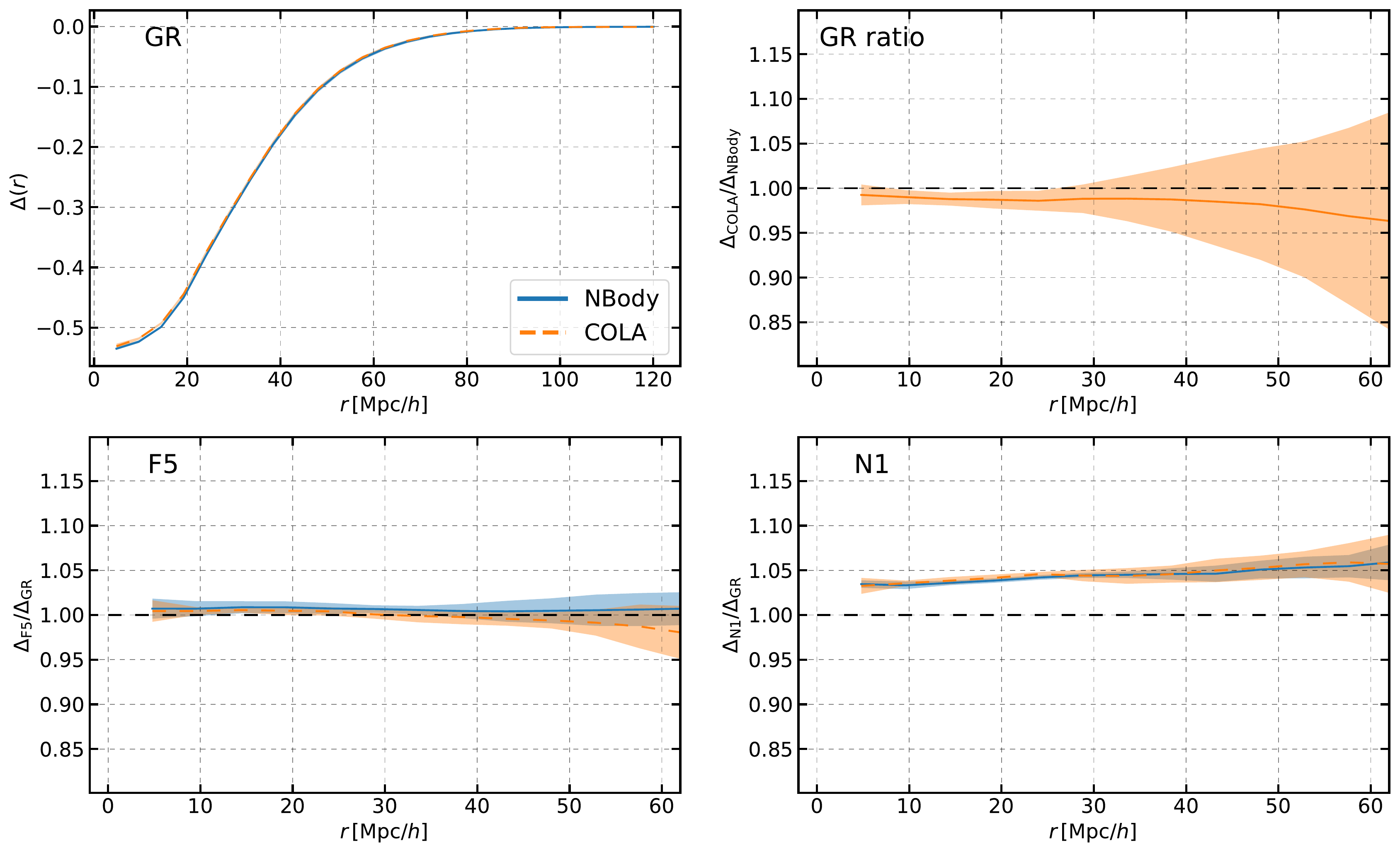}
\caption{\label{fig:DeltaProfile}Comparison of the integrated matter density profile of voids in COLA (in orange) and {\it N}-body simulations (in blue). \textit{Top left:} Full signal in GR. \textit{Top right:} Ratio of the GR signal between COLA and {\it N}-body simulations. \textit{Bottom:} Boost factors in F5 and N1. The panels involving ratios are shown up to $r\simeq 60 \mpcoh$ as the signal-to-noise ratio deteriorates for larger separations.}
\end{figure}

A comparison of the integrated matter density profile measured in our simulations is shown in figure~\ref{fig:DeltaProfile}. The signal is characterised by a good signal-to-noise ratio in the void interior (due to the amplitude of the signal and the abundance of DM tracers) which deteriorates above $\sim60 \mpcoh$ due to the signal approaching the large-scales limit of 0. COLA and {\it N}-body results are very consistent (within the variance) in all gravity models. The MG signal in the integrated-density profile is weak in the case of F5 (compatible with GR) while it shows a $\sim5\%$ enhancement in the case of N1 gravity. 

\begin{figure}
\centering
\includegraphics[width=.98\textwidth]{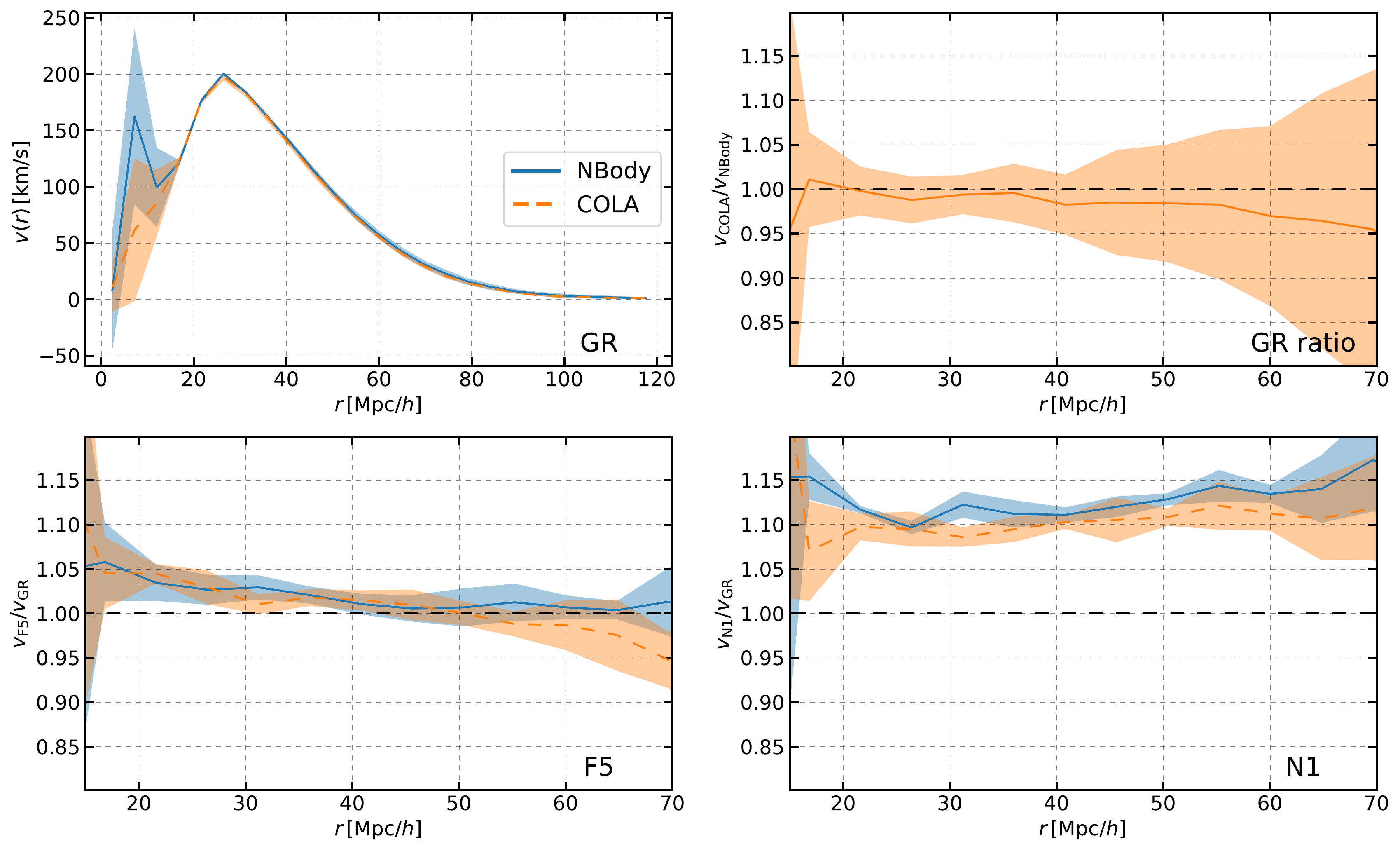}
\caption{\label{fig:VelProfile} Comparison of the galaxy radial-velocity profile around voids in COLA (in orange) and {\it N}-body simulations (in blue). \textit{Top left:} Full signal in GR. \textit{Top right:} Ratio of the GR signal between COLA and {\it N}-body simulations. \textit{Bottom:} Boost factors in F5 and N1. The panels involving ratios are displayed in the range $[20,70] \mpcoh$ as the signal-to-noise ratio deteriorates outside of the range.}
\end{figure}

The average radial velocity of galaxies as a function of the separation between the galaxies and the void centre is represented in figure~\ref{fig:VelProfile}. The bins below $15 \mpcoh$ are heavily affected by shot noise due to the low galaxy counts. %
On large scales, the signal approaching 0 is responsible for the large scatter in the ratios plots, which are therefore shown in the range $[15,70] \mpcoh$.  In GR, COLA is in agreement with {\it N}-body within the variance at all scales. A small scale-dependent MG signal is present in F5, with velocities larger than in GR for small separations and approaching GR results on large scales. The boost factor in N1 shows a $\sim10\%$ enhancement of the radial-velocity profile with respect to GR. The different behaviours of the velocity profiles in N1 and F5 are consistent with the different nature of their fifth force, long-ranged in N1, and short-ranged in F5.

\begin{figure}
\centering
\includegraphics[width=.98\textwidth]{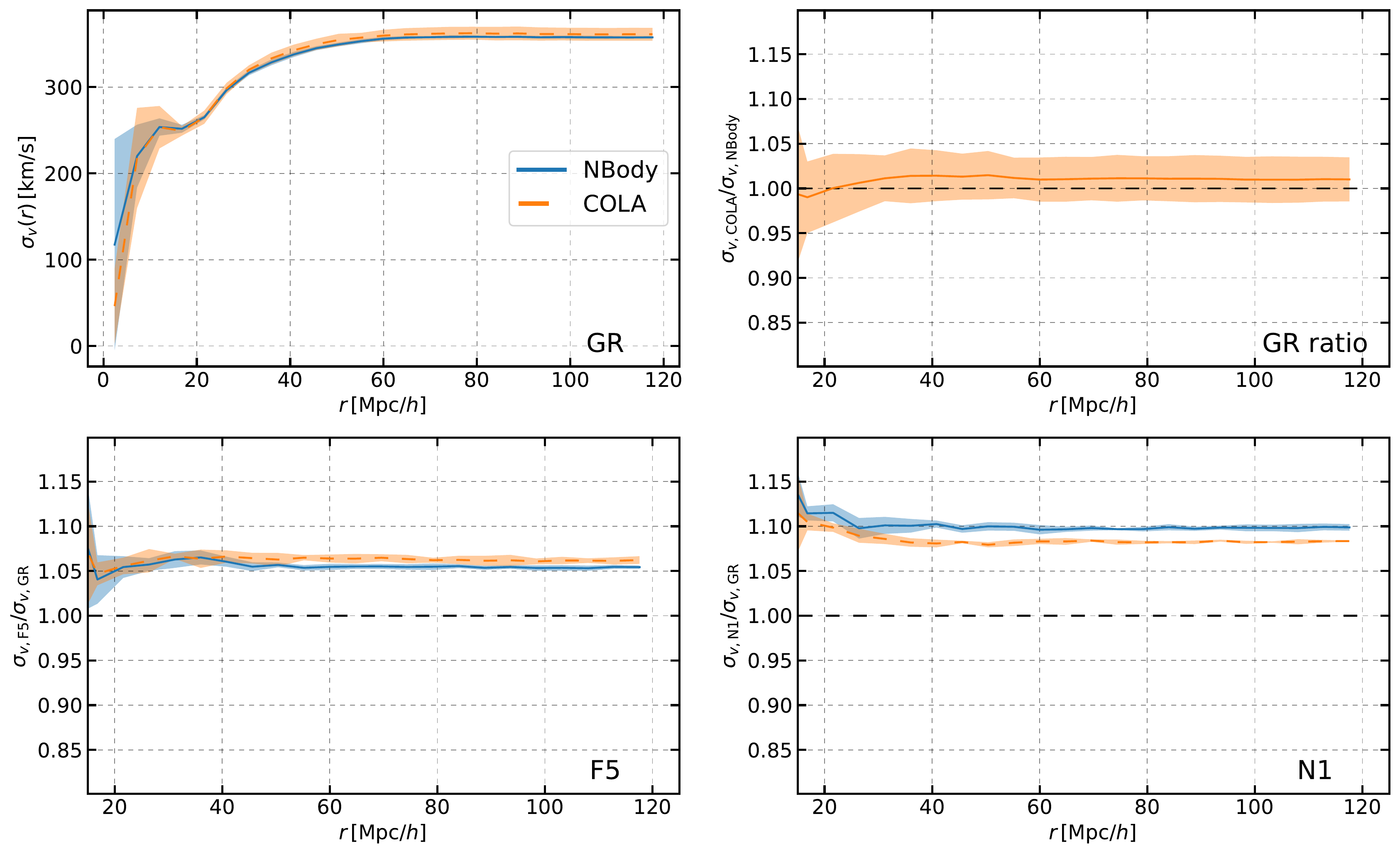}
\caption{\label{fig:VelDispProfile} Comparison of the galaxy velocity-dispersion profile around voids in COLA (in orange) and {\it N}-body simulations (in blue). \textit{Top left:} Full signal in GR. \textit{Top right:} Ratio of the GR signal between COLA and {\it N}-body simulations. \textit{Bottom:} Boost factors in F5 and N1. The panels involving ratios are displayed down to $15 \mpcoh$ as the signal-to-noise ratio deteriorates for smaller separations.}
\end{figure}

Figure~\ref{fig:VelDispProfile} displays the dispersion of the random galaxy velocity component along the line of sight. The large scatter in the first bins is again due to the low number of galaxies in the proximity of the void centre. The velocity-dispersion shows a mild dependence on separations at small separations and plateaus at $\sim 370 \, {\rm km/s}$. COLA results are in agreement with the {\it N}-body results in GR within the variance at all scales. The boost factors in F5 and N1 show a clear enhancement with respect to GR, of $\sim 5\%$ and $\sim 10\%$ respectively. The accuracy of COLA in reproducing the boost factors is $1\%$ in F5 and $2\%$ in N1.

\begin{figure}
\centering
\includegraphics[width=.98\textwidth]{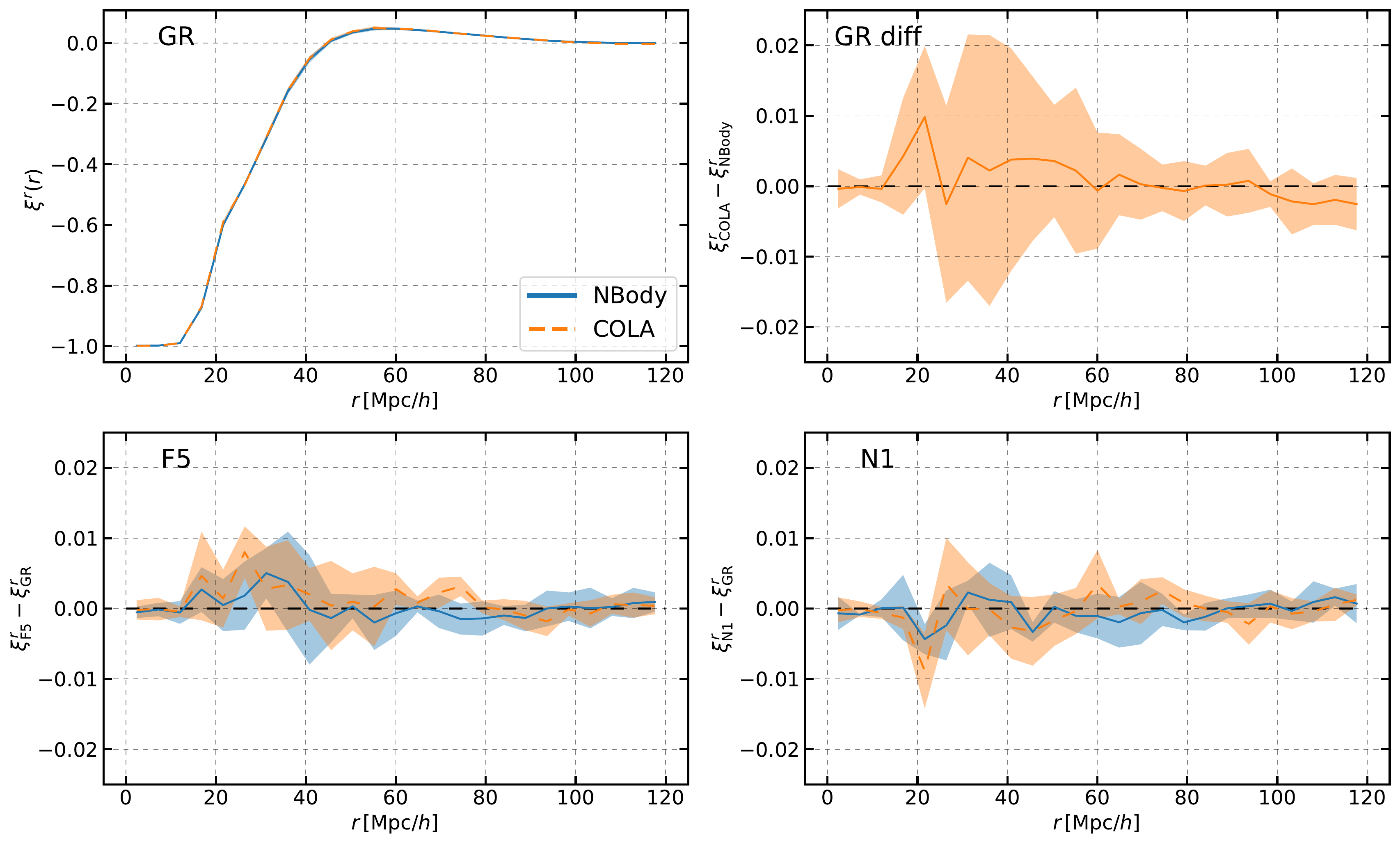}
\caption{\label{fig:CCF_real} Comparison of the galaxy-void real-space CCF in COLA (in orange) and {\it N}-body simulations (in blue). \textit{Top left:} Full signal in GR. \textit{Top right:} Difference of the GR signal between COLA and {\it N}-body simulations. \textit{Bottom:} Difference between the CCF in MG and GR for F5 and N1 gravity. The comparison uses differences instead of usual ratios since the signal crosses 0.}
\end{figure}

The real-space void-galaxy CCF in figure~\ref{fig:CCF_real} shows that COLA and {\it N}-body results are in agreement within the variance in all gravity models. Note that the top right panel shows the difference between COLA and {\it N}-body in GR instead of the usual ratio because the signal goes to zero on large scales. Similarly, the bottom panels show the differences between MG and GR results instead of the boost factors for the same reason. In both F5 and N1, there is no clear MG signal.

\begin{figure}
\centering
\includegraphics[width=.98\textwidth]{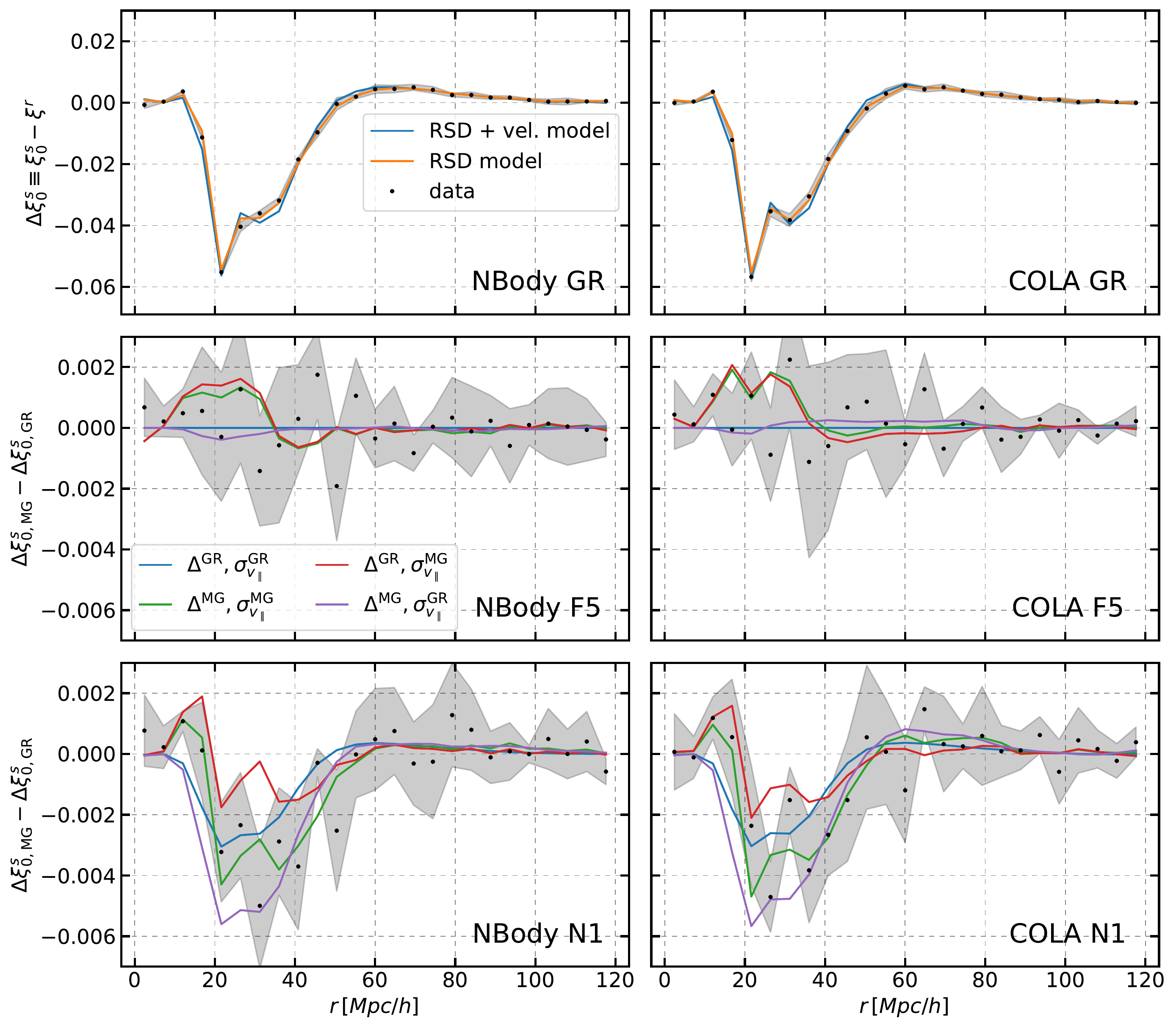}
\caption{\label{fig:xi0_COLAvsNBody_theory} Comparison of the monopole of the CCF of voids and galaxies in redshift-space between {\it N}-body (left column) and COLA (right column) simulations. \textit{Top panels:} The difference between the redshift-space monopole of the CCF and the real-space CCF, $\Delta\xi_{0}^s \equiv\xi_{0}^s-\xi^r$, measured in simulations (black dots and shaded region) is compared with the theory predictions (lines) formulated using the radial-velocity profile from simulations ("RSD model", orange) and using the velocity model of eq.~\eqref{VelModel} ("RSD + vel. model", blue). \textit{Middle and bottom panels:} The difference between $\Delta\xi_{0}^s$ in MG and GR (black dots and shaded regions) is compared to the theory predictions (lines) formulated using different combinations of inputs: integrated-density and velocity-dispersion from GR simulations (blue), integrated-density from MG and velocity-dispersion from GR simulations (violet), integrated-density from GR and velocity-dispersion from MG simulations (red), both integrated-density and velocity-dispersion from MG simulations (green). In all cases, we use the real-space CCF from GR simulations in the theory predictions, even in the MG panels to reduce the noise on the MG signals. The velocity model of eq.~\eqref{VelModel} is evaluated with the reference growth rate from linear theory (see table~\ref{tab:fBestFit}) in GR and N1 panels. In the case of F5, we use the reference growth rate from the GR linear theory.}
\end{figure}

\begin{figure}
\centering
\includegraphics[width=.98\textwidth]{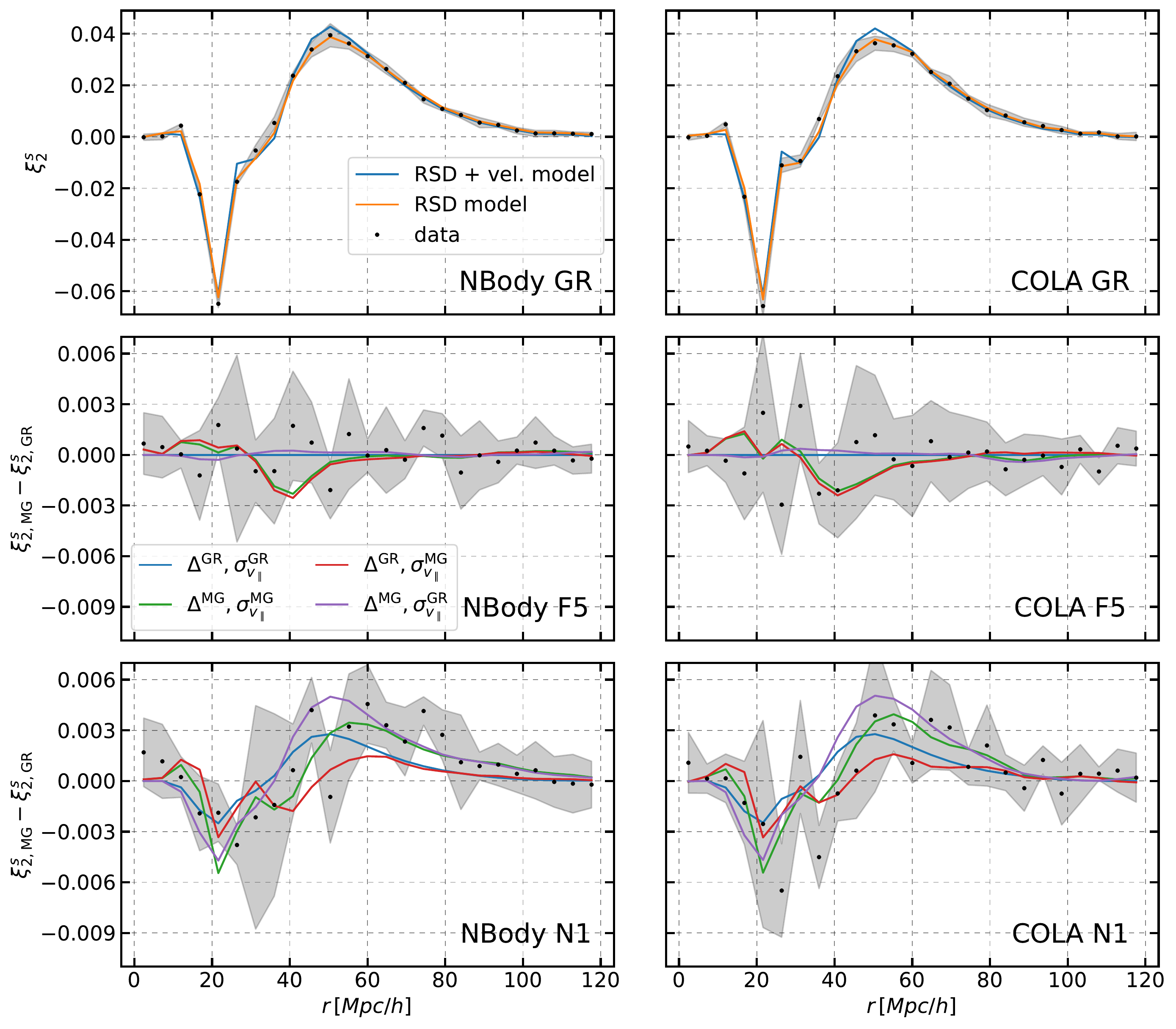}
\caption{\label{fig:xi2_COLAvsNBody_theory}Same as figure~\ref{fig:xi0_COLAvsNBody_theory} but for the quadrupole.}
\end{figure}

\begin{figure}
\centering
\includegraphics[width=.60\textwidth]{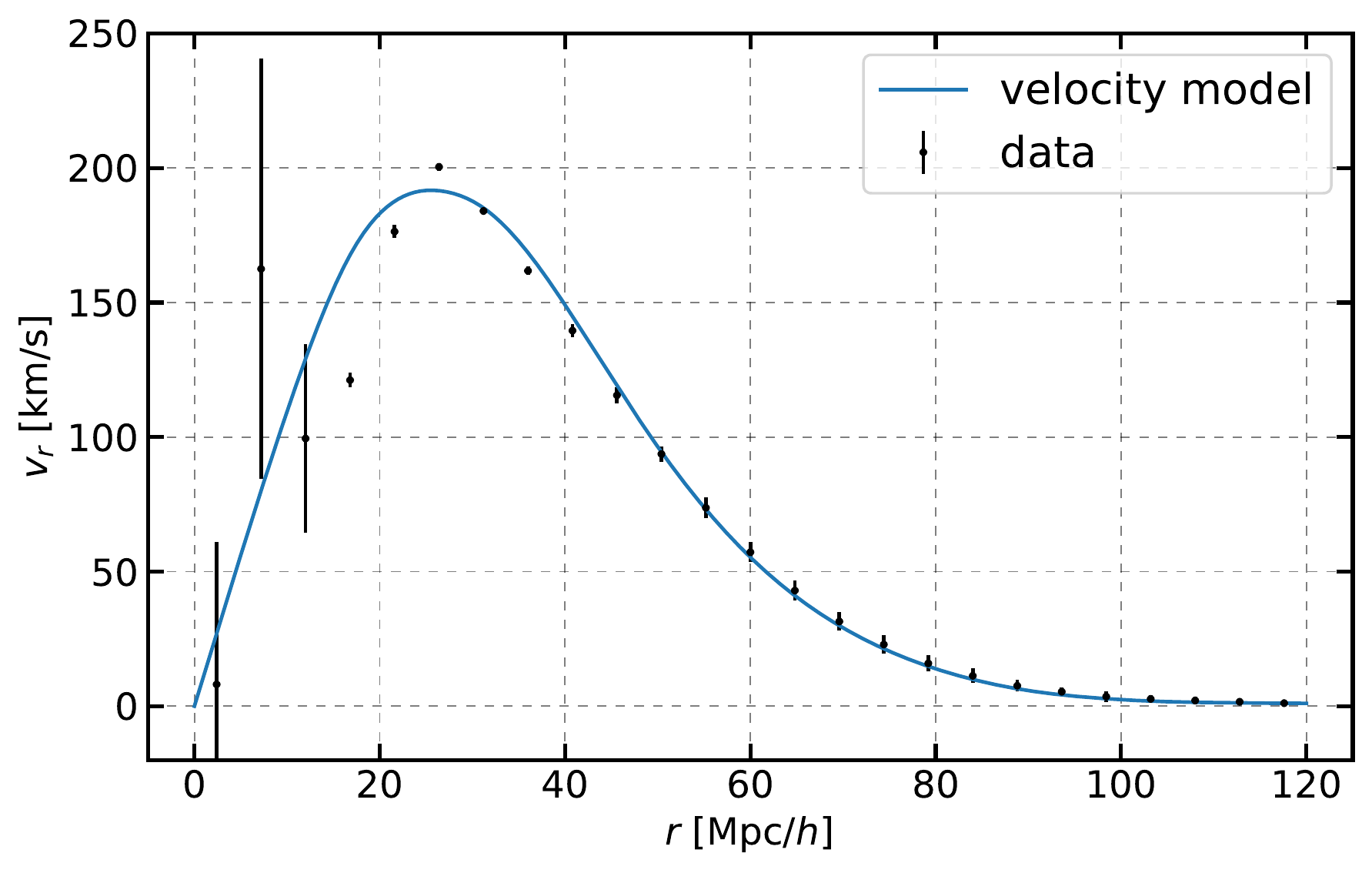}
\caption{\label{fig:DeltaVsVelProf} Comparison of the galaxy radial-velocity profile in GR {\it N}-body simulations with the linear model in eq.~\eqref{VelModel} using the fiducial growth rate $f(z = 0.5057) = 0.736$, as a function of the distance from the void centre.}
\end{figure}

In figure~\ref{fig:xi0_COLAvsNBody_theory} and figure~\ref{fig:xi2_COLAvsNBody_theory} we compare the monopole and the quadrupole of the void-galaxy CCF in redshift-space. In the top panels, the monopoles of the CCF ("data", black dots) are consistent within the variance (shaded regions) with the ones predicted using eq.~\eqref{theory_xiell} and the radial-velocity measured in the simulations ("RSD model", orange lines). This proves the accuracy of the RSD model discussed in section~\ref{ssec:RSDinVoids}. On the other hand, using the radial-velocity from eq.~\eqref{VelModel}%
\footnote{The value of the linear growth rate $f$ is taken from the linear theory.} 
in eq.~\eqref{theory_xiell} ("RSD + vel. model", blue lines) produces theoretical predictions that are still a good description of the data but with some discrepancies arising in the void interior that are traceable to the inaccuracies of the velocity model in reproducing the radial-velocity profile of galaxies in voids. 
The inaccuracies of the velocity model are visible in figure~\ref{fig:DeltaVsVelProf} where, comparing the velocity model with the radial-velocity profile estimated from {\it N}-body simulations in GR, we confirm the results of \cite{Nadathur:2017jos} that the model is a good description of the galaxy radial-velocity profile above $\sim 30 \mpcoh$ with some discrepancies arising below that distance from the void centre.

The bottom two rows of figure~\ref{fig:xi0_COLAvsNBody_theory} and figure~\ref{fig:xi2_COLAvsNBody_theory} show the impact of MG on monopole and quadrupole of the void-galaxy CCF in redshift space. In the case of F5, the signal is compatible with GR within the variance. Interestingly, N1 shows a clear enhancement of the distortions with respect to GR. This signal is well reproduced by using eq.~\eqref{theory_xiell} with the radial-velocity model computed in terms of $f^{\rm N1}$ from linear theory and $\Delta^{\rm N1}(r)$ and $\sigma_{v_\parallel}^{\rm N1}(r)$ estimated from N1 simulations (green lines). However, even ignoring the effects of MG on $\Delta(r)$ and $\sigma_{v_\parallel}(r)$ (i.e. using $\Delta^{\rm GR}(r)$ and $\sigma_{v_\parallel}^{\rm GR}(r)$ from GR simulations) but using the growth factor $f^{\rm N1}$ from N1 linear theory (blue lines) produces theoretical predictions that catch most of the signal (although representing a worse fit to the data). This is somehow surprising as we have seen from figure~\ref{fig:DeltaProfile} and figure~\ref{fig:VelDispProfile} that N1 gravity has impacts on both the integrated-density profile and the velocity-dispersion profile. To gain insight into this aspect, we add to the comparison the theoretical predictions formulated using the growth factor $f^{\rm N1}$ from N1 linear theory and, in one case $\Delta^{\rm GR}(r)$ and $\sigma_{v_\parallel}^{\rm N1}(r)$ (red lines), while in the other $\Delta^{\rm N1}(r)$ and $\sigma_{v_\parallel}^{\rm GR}(r)$ (violet lines). Comparing the two cases we can see that they have somehow competing effects on the redshift-space CCF and this explains why the main effect of N1 gravity is due to the enhancement of the linear growth factor.

For the monopole and quadrupole of the void-galaxy CCF in redshift-space (figure~\ref{fig:xi0_COLAvsNBody_theory} and figure~\ref{fig:xi2_COLAvsNBody_theory} respectively), results in COLA (right column) are consistent with the ones from {\it N}-body (left column) in all gravity models.

\subsection{Results}
\label{ssec:VoidResults}
Using the RSD model discussed in subsection~\ref{ssec:RSDinVoids} with the velocity model of subsection~\ref{ssec:VelModel}, we leave the linear growth rate $f$ in eq.~\eqref{VelModel} as a free parameter and we test if we can recover the growth rate predictions from linear theory. 
We apply the delete-1 jackknife re-sampling technique \cite{Wu_Jackknife,Norberg:2008tg, 2011MNRAS.418.2435N} to $n_{\rm sv}=125$ sub-volumes in which we split each realisation and average over the $n_{\rm r}=5$ realisations to estimate the covariance matrix for the multipoles of the CCF in redshift-space
\begin{equation}
    \matr{Cov}_{\ell, \ell', i,j} = \frac{1}{n_{\rm r}}\sum_{m=1}^{n_{\rm r}} \left[\frac{n_{\rm sv}-1}{n_{\rm sv}} \sum_{k=1}^{n_{\rm sv}} \left(\xi^s_{k}-\overline{\xi^s}\right)_{\ell,i}\left(\xi_{k}^{s}-\overline{\xi^s}\right)_{\ell',j} \right]_m \, ,
\end{equation}
where $\overline{\xi^s} = \frac{1}{n_{\rm sv}}\sum_{k=1}^{n_{\rm sv}} \xi^s_{k}$ is the average over the jackknife realisations of the CCF  $\xi^s_{k}$ estimated using 5 HOD realisations and 3 lines of sight. The indices $\ell, \ell'$ run over the multipoles $\{0,2\}$, while $i,j$ run over the radial bins. Given the periodic boundary conditions of the simulation box, we use the natural 2-point estimator
\begin{equation}\label{Natural2pEst}
    \widehat{\xi}_{\mathrm{N}}=\frac{D D}{R R}-1 \, .
\end{equation}

\begin{figure}
\centering
\includegraphics[width=.66\textwidth]{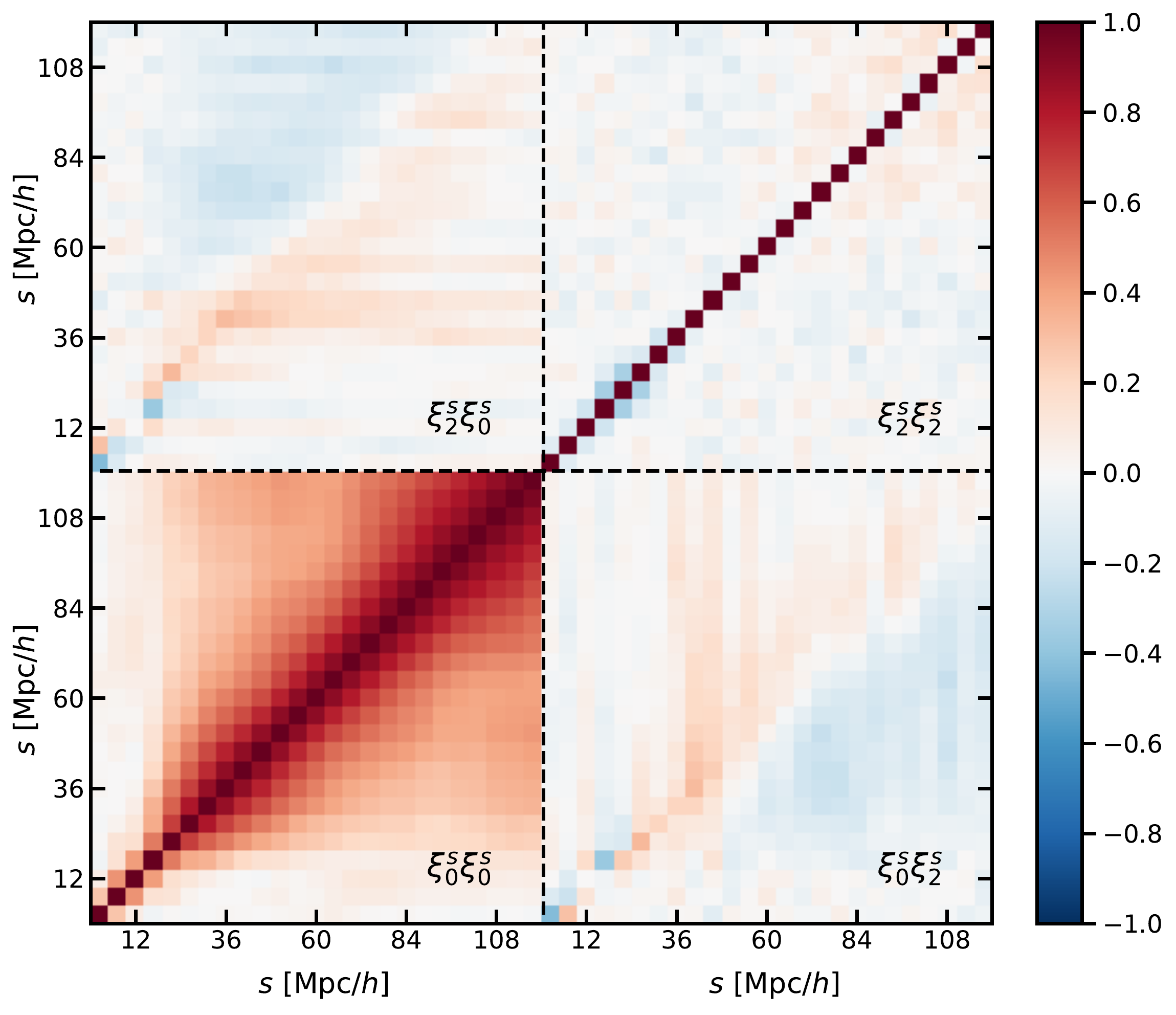}
\caption{\label{fig:CorrMat} Correlation of the covariance matrix estimated from jackknife re-sampling in {\it N}-body simulations in GR. The black dashed lines are used to distinguish between the monopole (bottom/left) and quadrupole (top/right) terms. The correlation matrices for the other gravity models as well as the ones for {\it N}-body simulations appear visually equivalent.}
\end{figure}

The correlation matrix obtained from the covariance matrix (figure~\ref{fig:CorrMat}) shows strong off-diagonal correlations between different bins in the monopole while the quadrupole is almost diagonal. From a visual comparison of our correlation matrix with the one in figure 6 of \cite{Nadathur:2017jos} we notice that in our case the off-diagonal correlations in the quadrupole are smaller. This is due to the fact that we average the signal over three lines of sight while \cite{Nadathur:2017jos} was using only 1 line of sight. Using only 1 line of sight likely produces a more realistic error estimate for a real survey but, since our goal in this work is to test the consistency of the RSD model in beyond-LCDM models and to validate approximate simulations, we choose to leverage on the average over three lines of sight (as well as the average over 5 HOD realisations) to increase the signal-to-noise ratio.

The recent work \cite{Mohammad:2021aqc}, has proposed the use of the Landy-Szalay 2-point estimator \cite{1993ApJ...412...64L}
\begin{equation}
    \widehat{\xi}_{\mathrm{LS}}=\frac{D D-2 D R+R R}{R R} \, ,
\end{equation}
to deal with the mask applied by construction by the jackknife re-sampling, as well as a correction to account for proper weighting of the "cross-pairs" (pairs with only one point falling in the sub-volume masked by the jackknife), which has been shown to produce more realistic covariance matrix estimates for the galaxy correlation function. In light of these results, we compute the jackknife covariance using the publicly available code \codeword{pycorr}\footnote{\href{https://github.com/cosmodesi/pycorr}{https://github.com/cosmodesi/pycorr}} and notice that:
\begin{itemize}
    \item Using the Landy-Szalay estimator produces smaller off-diagonal correlations in the covariance matrix for the monopole and larger errors on scales smaller than $\sim 20 \mpcoh$. The amplitude of the errors on these scales seems to depend on the number density of the random catalogues used (having tested for $20\times$ and $50\times$ the average number density of our galaxy and void catalogues) and shrinks for larger number density. This is compatible with shot-noise effects.
    \item Using the cross-pairs correction shrinks the errors on large scales in the monopole but does not significantly affect the correlations of the covariance matrix. 
\end{itemize}
Since this technique has not been tested yet against mocks in the specific case of the void-galaxy CCF, we make the conservative choice of adopting the same technique originally used to validate the RSD model in voids \cite{Nadathur:2017jos}, i.e. using the natural 2-point estimator in eq.~\eqref{Natural2pEst} and removing all the pairs involving an object falling in the masked sub-volume.

With the covariance obtained with these settings, we perform the fit for the linear growth rate $f$ by minimising the objective function
\begin{equation}\label{chi2}
    \chi^2= \sum_{\ell, \ell'}\sum_{i,j} (\xi^s_{\ell, i} - \xi^{\rm th}_{\ell, i}) \matr{Cov}_{\ell, \ell', i,j}^{-1} (\xi^s_{\ell', j} - \xi^{\rm th}_{\ell', j}) \, ,
\end{equation}
where $\xi^s$ is obtained by averaging over the 3 lines of sight, the 5 HOD realisations and the 5 realisations for each gravity model in COLA and {\it N}-body simulations. The theoretical predictions are based on $\xi^r$, $\Delta$ and $\sigma_{v_{\parallel}}$ estimated from the average of the signal over all the realisations in each gravity model in COLA and {\it N}-body simulations. 
We estimate the error on the best fit growth rate by finding the solutions to
\begin{equation}\label{sigmaf}
    \chi^2(f \pm \sigma_f) = \chi^2(f_{\rm fit})+1 \, .
\end{equation}
The results of the fit are shown in table~\ref{tab:fBestFit}. In GR, both COLA and {\it N}-body recover the reference linear growth rate within $\sim1$ standard deviation. 
\begin{table}
    \centering
        \begin{tabular}{cccccccc}
        \toprule
        {} & {} & \multicolumn{2}{c}{GR} & \multicolumn{2}{c}{F5} & \multicolumn{2}{c}{N1} \\
        {} & {} & COLA & {\it N}-body & COLA & {\it N}-body & COLA & {\it N}-body \\
        \midrule
        {} & $f_{\rm ref}$ & \multicolumn{2}{c}{0.736} & \multicolumn{2}{c}{--} & \multicolumn{2}{c}{0.777} \\
        \midrule
        {GR theory} & $f_{\rm fit}$ &  0.721 &  0.726 &  0.717 &  0.726 &  0.772 &  0.784 \\
        {} & $\sigma_f$ &  0.012 &  0.012 &  0.013 &  0.013 &  0.013 &  0.013 \\
        {} & $\Delta f /\sigma_f$ & -1.2 & -0.8 & -- & -- & -0.4 & 0.5 \\
        {} & $\chi^2/{\rm d.o.f.}$ & 0.96 & 1.08 & 0.88 & 0.92 & 1.53 & 1.69 \\
        \midrule
        {MG theory} & $f_{\rm fit}$ & -- & -- &  0.735 &  0.734 &  0.763 &  0.774 \\
        {} & $\sigma_f$ & -- & -- &  0.013 &  0.013 &  0.013 &  0.013 \\
        {} & $\Delta f /\sigma_f$ & -- & -- & -- & -- & -1.1 & -0.2 \\
        {} & $\chi^2/{\rm d.o.f.}$ & -- & -- & 0.67 & 0.77 & 0.78 & 0.95 \\
        \bottomrule
        \end{tabular}
    \caption{Results of the optimisation of eq.~\eqref{chi2} and eq.~\eqref{sigmaf} using theory predictions from GR simulations (top) and using theory predictions from MG simulations (bottom). COLA and {\it N}-body results are compared side by side in each gravity model (different columns). The additional row at the top of the table shows the linear theory predictions for the linear growth rate  at redshift $z = 0.5057$ in GR and N1. The linear theory growth rate in F5 is scale-dependent so we omit it from the table.}
    \label{tab:fBestFit}
\end{table}
In the case of N1 and F5 theories we investigate the impact of MG on the RSD model by performing the fit with different theoretical inputs:
\begin{itemize}
    \item using $(\xi^r$, $\Delta$, $\sigma_{v_{\parallel}})^{\rm GR}$ estimated in GR simulations  ("GR theory") ;
    \item using $(\xi^r$, $\Delta$, $\sigma_{v_{\parallel}})^{\rm MG}$ estimated in MG simulations  ("MG theory").
\end{itemize}
In both cases, we recover the reference growth rates within $\sim 1$ standard deviation in N1. In F5, we do not have a reference growth rate to compare our results with since the linear theory solution is scale-dependent, but the best fit values are consistent with the reference value in GR. Interestingly, comparing best-fit values obtained using "MG theory" and "GR theory", we notice a shift in the growth rate of $\sim 1 \sigma_f$ in both COLA and {\it N}-body simulations, towards larger values in the case of F5 and towards smaller values in the case of N1 using the GR theory. We also find an increase of the reduced $\chi^2$ in the GR theory. If this shift is in fact due to the effects of MG on the components entering the theoretical model, we might need to take into account these effects to get an accurate determination of the linear growth rate with $\sigma_f \lesssim 0.01$ precision with the void-galaxy CCF. At that level of precision, however, the inaccuracies of the velocity model of subsection~\ref{ssec:VelModel} (see also figure~\ref{fig:DeltaVsVelProf}) could become the limiting factor \cite{Massara:2022lng}. In such a scenario, simulation-based emulation of the void radial-velocity profile may be a way to avoid systematic errors of theoretical origin, and the COLA method could certainly be considered a viable option given its computational efficient nature and its accuracy in reproducing full {\it N}-body results as proven by this work. With less precision on the growth rate estimate, e.g. $\sigma_f \simeq 0.02$, voids analysis can still help constrain nDGP theories by providing independent information on the amplitude of the linear growth rate. In the particular case of N1, MG effects on the integrated-density profile and on the velocity-dispersion can be ignored in the light of the cancellation discussed in the previous subsection, but in general, this cancellation depends on the specific gravity model (and screening mechanism) and it needs to be carefully checked whether the same conclusion applies to other theories of gravity.

We performed the fit also using the covariance matrices corrected for the cross-pairs and/or computing the jackknife CCF with the Landy-Szalay estimator and confirmed that we obtained the same results.

\chapter{Matter power spectrum emulation}\label{chp:powerspectrum}
\begin{quote}
    {\it The content of sub-section~\ref{ssec:CosmoEmu} and section~\ref{sec:Response} in this chapter are based on the author's contribution to the publication \cite{Brando:2022gvg}. The power spectra from Arepo simulations used in this chapter were provided by César Hernández-Aguayo.} 
\end{quote}
In order to get fast predictions of the matter power spectrum, emulation techniques have gained much attraction in cosmology, and they are now seen as viable alternatives for extracting parameter constraints using the data from upcoming surveys.  

Emulators are numerical interpolations that are trained using accurate {\it N}-body simulation outputs based on machine learning algorithms to quickly predict the matter power spectrum from linear to non-linear scales in the vast cosmological parameter space. In other words, emulators have free fitting parameters that are tuned to reproduce a finite number of theoretical predictions from {\it N}-body simulations for an equal number of points in the theory parameter space and, when after fixing the fitting parameters, the emulators produce quick approximate predictions for the emulated quantity in a dense region of the theory parameter space. Among the emulators already available in the literature, we will focus on two that have been considered highly effective and validated, the Euclid Emulator 2\footnote{\href{https://github.com/miknab/EuclidEmulator2}{https://github.com/miknab/EuclidEmulator2}}~\cite{Euclid:2020rfv} (EE2) and Bacco\footnote{\href{https://baccoemu.readthedocs.io/en/latest/}{https://baccoemu.readthedocs.io/en/latest/}}~\cite{Angulo:2020vky} in this chapter, and use them to check the accuracy of predicting non-linear matter power spectra. Both have an accuracy of about $1\%$ on small scales for $\Lambda$CDM cosmologies in predicting the non-linear power spectrum. 

The process of training these emulators heavily relies on the use of computationally expensive and time-consuming full {\it N}-body simulations. To overcome these limitations, there are several well-established methods that allow us to quickly generate approximate non-linear realisations of the matter density field, such as the COLA (COmoving Lagrangian Acceleration) approach~\cite{Tassev:2013pn,Howlett:2015hfa}, PINOCCHIO~\cite{Monaco:2001jg} and
GLAM~\cite{Klypin:2017iwu}, to name a few. While PINOCCHIO uses the 2LPT solution to estimate the density field (with limited accuracy on non-linear scales) and GLAM uses an optimised implementation of the PM technique (requiring a large number of time-steps to be accurate on large scales), COLA is an hybrid version of the 2LPT and PM techniques which allows the user to trade computational complexity with the accuracy on non-linear scales without sacrificing the accuracy on large linear scales. Specifically, in this work, we will consider the first of these examples, the COLA approach. This method has been well-validated and is known to give a good agreement on quasi-non-linear scales in $\Lambda$CDM and beyond-$\Lambda$CDM cosmologies when comparing its prediction for the matter power spectrum to the ones from full {\it N}-body simulations \cite{Izard:2015dja,Valogiannis:2016ane, Winther:2017jof,Wright:2017dkw}. Additionally, a new avenue using the COLA method was presented in~\cite{Kaushal:2021hqv}. In this paper, it was shown that the mapping from displacements in COLA simulations and those in full {\it N}-body simulations can be trained on simulations with a fixed value of the cosmological parameters, and this model can be used to correct the output of COLA simulations with different values of cosmological parameters including different masses of massive neutrinos with very high accuracy: the power spectrum and the cross-correlation coefficients are within $1\%$ down to $k=1 \hompc$. 

This indicates that the inaccuracy in COLA's predictions is fairly cosmological parameter independent, and this can be corrected by a cosmological parameter independent model. After studying the convergence of COLA simulations for the matter power spectrum in section~\ref{sec:Convergence}, we show in section~\ref{sec:Response} that COLA is capable of describing the non-linear response of the matter power spectrum to the change of cosmological parameters down to $k=1 \hompc$ with high accuracy. This is because the inaccuracy of COLA is largely cancelled by taking the ratio of power spectra in two different cosmologies. The accurate and fast computation of this response function is of great importance when building emulators, as it allows us to obtain quickly the expected non-linear prediction when parameters are varied with respect to a fixed pre-defined reference cosmology. Given the prediction of non-linear power spectra in a few sparsely sampled reference cosmologies by full {\it N}-body simulations, we can provide the prediction of the matter spectrum in a wide parameter space. This can be used to further extend the reach of the already accurate \lcdm \ emulators to beyond-\lcdm \ cosmologies, for example, where running full {\it N}-body simulations is very expensive. Finally, in section~\ref{Sec:Emulator}, we give the explicit example of how this can be done in nDGP gravity by designing a COLA simulation suite that we use to train and test an MG boost factor emulator for the matter power spectrum which achieves sub-percent accuracy up to $k=1 \hompc$ in our tests. 
\section{Convergence tests}
\label{sec:Convergence}

Cosmological simulations are subject to errors due to the finite resolution characterising their numerical implementation. In the case of {\it N}-body simulations, time-resolution, force-resolution and mass-resolution are important quantities which control the simulation accuracy \cite{Klypin:2017iwu,DeRose:2018xdj,Euclid:2020rfv}. To validate simulations results it is necessary to compare the output of simulations against well-established results like analytic predictions or previous simulations already validated (e.g., in chapter~\ref{chp:mocks} and chapter~\ref{chp:statistics} we validated COLA simulations by comparing them with ELEPHANT simulations). Furthermore, the self-consistency of a simulation technique can be established by performing relative-convergence tests, where simulations with different resolutions are compared to understand the reliability of specific simulation settings. 

The COLA method exploits the 2LPT and the Particle-Mesh (PM) method to evolve the DM distribution. The use of 2LPT allows us to reduce the time resolution of simulations without losing accuracy on large scales \cite{Izard:2015dja}. Conversely, increasing the time resolution makes the use of 2LPT less important and with sufficient time resolution, the simulations evolve like PM-only simulations. In this sense, we will refer to COLA simulations with high time resolution as PM simulations in the following sections. 
This equivalence is important since the accuracy of PM simulations (and therefore of COLA simulations in $\Lambda$CDM) is only limited by resolution and they converge to full {\it N}-body simulations results with increasing resolution \cite{Hernandez-Aguayo:2021kuh}. 
However, when comparing COLA with PM-only algorithms, it is not fair to compare their computational cost using the same number of time-steps as done in \cite{Klypin:2017iwu}. In fact, COLA does not necessarily need as many time-steps as PM-only simulations to achieve comparable accuracy. This is evident in the low time-resolution case, where the 2LPT guarantees that COLA provides accurate results on large scales while PM-only simulations already show significant deviations \cite{Izard:2015dja}. Even in the high time-resolution case, COLA can achieve the same accuracy as PM-only codes with fewer time-steps. Indeed, the latter needs a finer time-stepping at high redshift to deal with the rapidly changing growth factor \cite{Klypin:2017iwu} while the 2LPT guarantees good accuracy at high redshift in COLA even with larger time-steps \cite{Tassev:2013pn,Izard:2015dja}. For instance, the PM code \codeword{GLAM} needs 147 time-steps to have $\Delta a / a \approx 0.014$ at $z=0$ \cite{Klypin:2017iwu} while COLA needs only $\approx70$ time-steps for the same low-redshift time-resolution.

The focus of this section is to validate COLA simulations for the predictions of the matter power spectrum. To do so, we first introduce the cosmological emulators in sub-section~\ref{ssec:CosmoEmu}, then we compare COLA results with full {\it N}-body simulations and cosmological emulators in sub-section~\ref{ssec:COLAvsCosmoEmuandArepo}, and finally, we study the relative-convergence of COLA simulations for the matter power spectrum in sub-section~\ref{ssec:Convergence}.

\subsection{Emulators}
\label{ssec:CosmoEmu}

In sight of Stage-IV LSS surveys, efforts have been made toward producing fast and accurate theoretical predictions of summary statistics by means of emulators. Emulation methods interpolate the results of cosmological simulations in a broad range of models and cosmological parameters using machine learning techniques~\cite{agarwal2012,habib2007}. Among the various emulators produced so far, the Bacco emulator \cite{Angulo:2020vky} and the Euclid Emulator 2 (EE2) \cite{Euclid:2020rfv} are setting the standards in terms of the accuracy and parameters space coverage.

The Bacco emulator takes advantage of Principal Components Analysis (PCA) to reduce the dimensionality of the interpolation problem and applies Gaussian Process Regression to emulate each of the dimensions selected in the PCA \cite{Angulo:2020vky}. It has been trained on a set of non-linear boost factors produced from $16000$ power spectra spanning the parameter space schematised in Table~\ref{table:bacco_params}.
\begin{table}[t]
    \begin{center} 
        \begin{tabular}{lcc} 
            \toprule
            Parameter & Min. & Max. \\
            \midrule
            $h$  & $0.6$  & $0.8$ \\
            $\Omega_{\rm b}$  & $0.04$  & $0.06$ \\
            $\Omega_{\rm cdm + b}$  & $0.23$  & $0.40$  \\
            $n_{\rm s}$  & $0.92$  & $1.01$ \\
            $\sigma_{8}$  & $0.73$ & $0.9$  \\
            $\sum_{\nu} m_{\nu}$  & $0.0$ eV & $0.4$ eV \\
            $w_{0}$  & $-1.15$  & $-0.85$  \\
            $w_{a}$  & $-0.3$  & $0.3$ \\
            \bottomrule						
        \end{tabular}
    \end{center}
    \caption{Bacco training set cosmological parameter space.}
    \label{table:bacco_params} 
\end{table}
These power spectra have been obtained using the cosmology-rescaling algorithm~\cite{angulo2009} on a small suite of only $6$ \lcdm \ simulations obtained with Gadget3 (see section~\ref{sec:Nbody} for more on this code). The cosmology-rescaling algorithm applies a space-time rescaling of simulations which is optimised for a specific range of scales to map a simulation in the original cosmology to simulations in target cosmologies \cite{angulo2009}. The non-linear modes in the original cosmology are isolated by subtracting the 2LPT displacements from the particles' positions, rescaled, and added back to the 2LPT displacements in order to approximate the non-linear modes in the target cosmology \cite{Contreras:2020kbv}. This algorithm enables much faster production of the training set at the expense of a modest loss of accuracy. It has been proven to be $1\%$ accurate in \lcdm \, and $3\%$ accurate for dynamical dark energy ($w_{0}-w_{a})$ and massive neutrinos implementations. The emulator intrinsic accuracy is $\sim 2 \%$ up to scales of $5$ $h$/Mpc, so the overall accuracy is $\sim 2 \%$ in $\Lambda$CDM and $\sim 3 \%$ in the DE and massive neutrino cases \cite{Angulo:2020vky}. 

Similarly to the Bacco emulator, EE2 performs dimensionality reduction using PCA, but, then relies on a Polynomial Chaos Expansion to emulate the resulting components \cite{Euclid:2020rfv}. The power spectra are measured at $100$ time-steps between $z=10$ and $z=0$ in a suite of $108$ paired-and-fixed simulations~\cite{pair_fixed} performed with PKDGRAV3 (see section~\ref{sec:Nbody} for more on this code). The cosmological parameters space spanned by EE2 is illustrated in Table~\ref{table:EE2_params}. The non-linear boost factors for the training and test sets are computed by taking the ratio of the simulations' power spectra with the linear power spectra from \CLASS~\cite{class1}. The EE2 provides $\sim 1 \%$ accuracy up to $k=10 \ h/{\rm Mpc}$ in the ellipsoid centred on the reference cosmology and extending to the edges of the interpolation range \cite{Euclid:2020rfv}.

\begin{table}[t]
    \begin{center} 
        \begin{tabular}{lccc} 
            \toprule
            Parameter & Min. & Max. & Center \\
            \midrule
            $h$  & $0.61$  & $0.73$  & $0.67$ \\
            $\Omega_{\rm b}$  & $0.04$  & $0.06$  & $0.05$ \\
            $\Omega_{\rm m}$  & $0.24$  & $0.40$  & $0.32$ \\
            $n_{\rm s}$  & $0.92$  & $1.0$  & $0.96$ \\
            $A_{\rm s}$  & $1.7\times 10^{-9}$  & $2.5\times 10^{-9}$  & $2.1\times 10^{-9}$ \\
            $\sum_{\nu} m_{\nu}$ & $0.0$ eV & $0.15$ eV & $0.075$ eV \\
            $w_{0}$  & $-1.3$  & $-0.7$  & $-1.0$ \\
            $w_{a}$  & $-0.7$  & $0.7$  & $0.0$ \\
            \bottomrule						
        \end{tabular}
    \end{center}
    \caption{EE2 parameters.}
    \label{table:EE2_params} 
\end{table}

\begin{figure}
\centering 
\includegraphics[width=.85\textwidth]{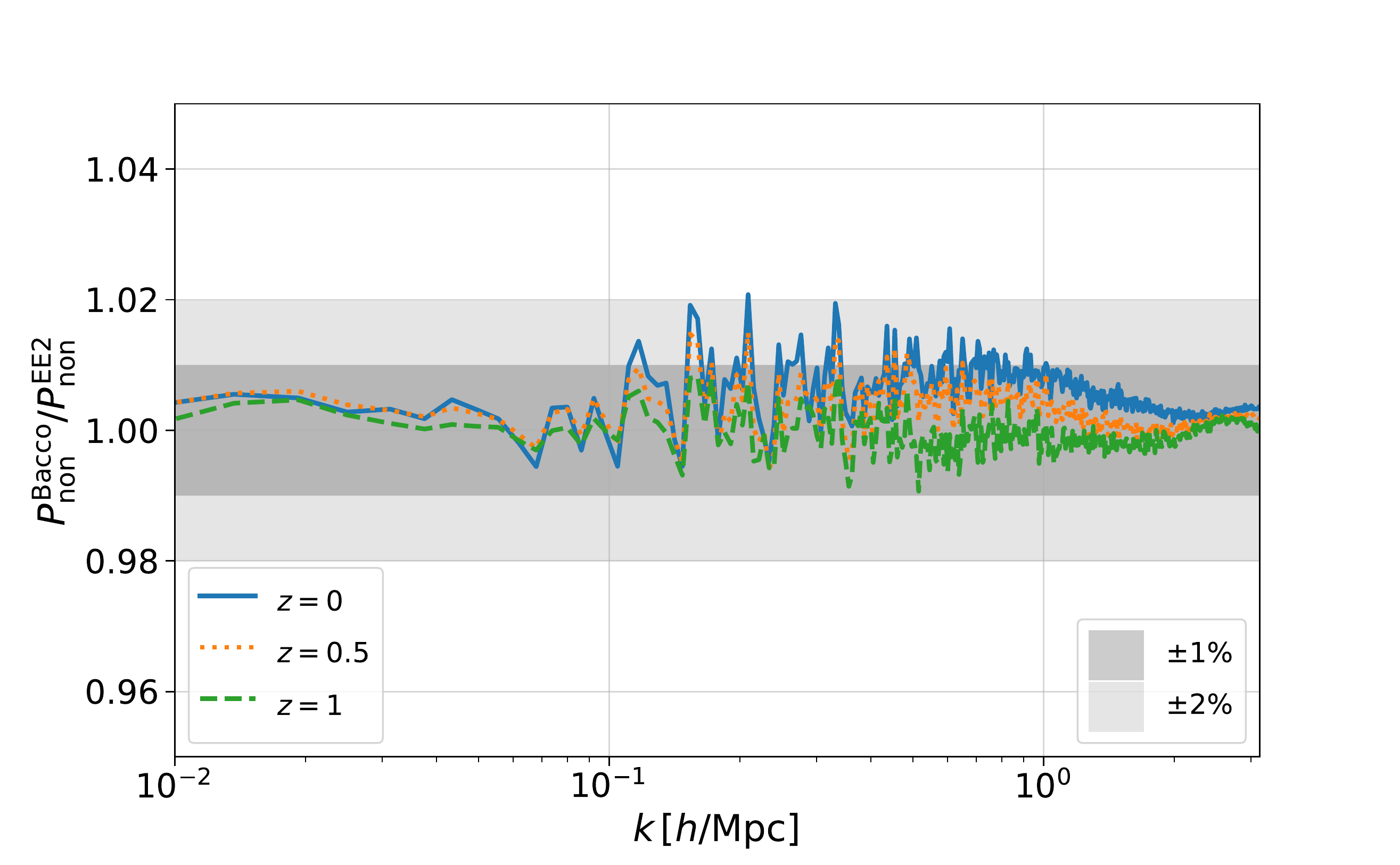}
\caption{\label{Fig:Pk_BaccoVsEE2} Ratio of Bacco and EE2 power spectra for the cosmology defined in eq~\eqref{ArepoCosmo}.}
\end{figure}

In figure~\ref{Fig:Pk_BaccoVsEE2} we show the ratio of the matter power spectra estimated with Bacco and EE2 at redshifts $0$, $0.5$ and $1$ for the cosmology (see eq~\ref{ArepoCosmo}) of the {\it N}-body simulation suite in subsection~\ref{ssec:COLAvsCosmoEmuandArepo}. The agreement is within $2\%$ at all the scales considered. At $z=1$, the agreement is well within $1\%$. This is consistent with the accuracy claimed by the two emulators and gives additional reliability to the emulators' results, in particular at redshift $z=1$, that we use in the next sub-section to prove the accuracy of COLA simulations for predictions of the matter power spectrum.  

\subsection{Comparison with emulators and {\it N}-body simulations}
\label{ssec:COLAvsCosmoEmuandArepo}
As reference results to compare COLA with, we make use of the power spectra obtained from a suite of simulations in GR and nDGP gravity\footnote{The nDGP results (for $H_0 r_c$ values of 0.5, 1, 2, 5) are used in section~\ref{sec:Boost}.} ran with \codeword{MG-AREPO} which uses the tree-particle-mesh technique for gravitational interactions and a moving mesh to solve the hydrodynamic equations (see section~\ref{sec:Nbody} for more on this code). The simulations were started at redshift $z=127$ using the cosmological parameters
\begin{equation}
\begin{array}{ccc}
\Omega_{m,0}=0.3089 \, , & \Omega_{\Lambda,0}=0.6911 \, , & \Omega_{b,0}=0.0486 \, , \\
n_{s}=0.9667 \, , & \sigma_{8}=0.8159 \, , & h=0.6774 \, ,
\end{array}
\label{ArepoCosmo}
\end{equation}
and evolved up to redshift $z=0$ using $1024^3$ particles in a $(1{\rm Gpc}/h)^3$ box.

To assess the accuracy of COLA simulations we produce a suite of simulations in a $L = 512$Mpc$/h$ box, with the same cosmology of the Arepo simulations and with different mass resolutions ($\NP=512, 1024, 2048$), time resolutions ($\NS=23\times[1, 2, 3]= 23, 46, 69$) and with force resolutions depending on the mass resolutions ($\NM=[1,2,3,4]\times\NP$ for $\NP=512,1024$ and $\NM=[1,1.5,2]\times\NP$ for $\NP=2048$). We start the simulations at redshift $z=19$ and stop them at redshift $z=1$. The choice of 23 time-steps as the low time-resolution case is motivated by the time-step size of typical COLA configurations ($\Delta a \approx 0.02$). In the high time-resolution case, where the late-time evolution is PM-like, 69 time-steps produce a time-step size of $\Delta a \approx 0.06$. This is the same step-size of the reference simulations used in the convergence tests of \codeword{GLAM} at $z<3$ \cite{Klypin:2017iwu}.

\begin{figure}[t]
\centering 
\includegraphics[width=.85\textwidth]{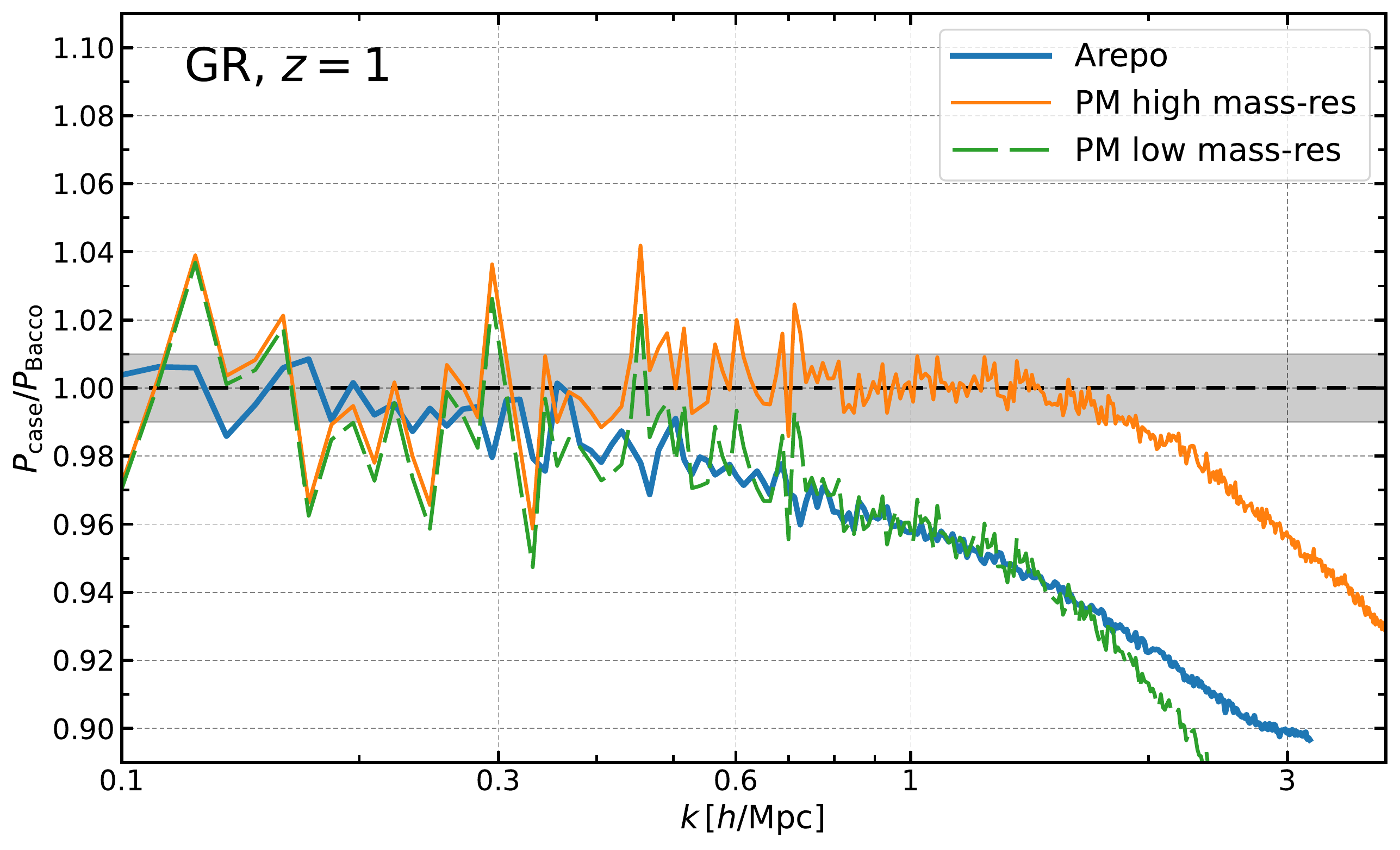}
\caption{\label{Fig:COLA_Arepo_over_Bacco} Comparison of PM simulations matter power spectra with Bacco and Arepo power spectra. The high mass-resolution PM run agrees with Bacco up to $k\sim2h/$Mpc within $1\%$. Arepo does not agree with Bacco but is in better agreement with the low mass-resolution PM run.}
\end{figure}

 In the next sub-section, we will use the highest resolution PM run ($\NP=2048$, $\NM=4096$, $\NS=23\times 3$) as the reference to assess the accuracy of all the other runs. Before that, we validate the highest resolution run by comparing its matter power spectrum with the ones from the Bacco emulator and from the Arepo simulations. In Figure~\ref{Fig:COLA_Arepo_over_Bacco} %
we plot the ratio of the matter power spectra of PM and Arepo simulations with the Bacco power spectrum for the cosmology of eq.~\eqref{ArepoCosmo}, in GR at redshift 1. The first thing that comes to the eye, is that the Arepo power spectrum (green line) departs from Bacco already at $k\sim 0.6h$/Mpc, and at $k\sim 1h/$Mpc the discrepancy is already at $4\%$. We attribute this to the lower mass resolutions of Arepo simulations \cite{Euclid:2020rfv}, and to motivate this claim we add to the comparison the PM simulation for $\NP=512$, $\NM=2048$, $\NS=23\times 3$ (blue line) which has a similar mass resolution of Arepo simulations ($M_{\rm part} \approx8\cdot 10^{10} \Msun$). This is in good agreement with Arepo up to  $k\sim 2h/$Mpc. The blue line instead shows that the highest resolution PM run is in sub-percent agreement with Bacco up to $k\sim 2 \hompc$. 

\subsection{Relative convergence}
\label{ssec:Convergence}
\begin{figure}[t]
\centering 
\includegraphics[width=.99\textwidth]{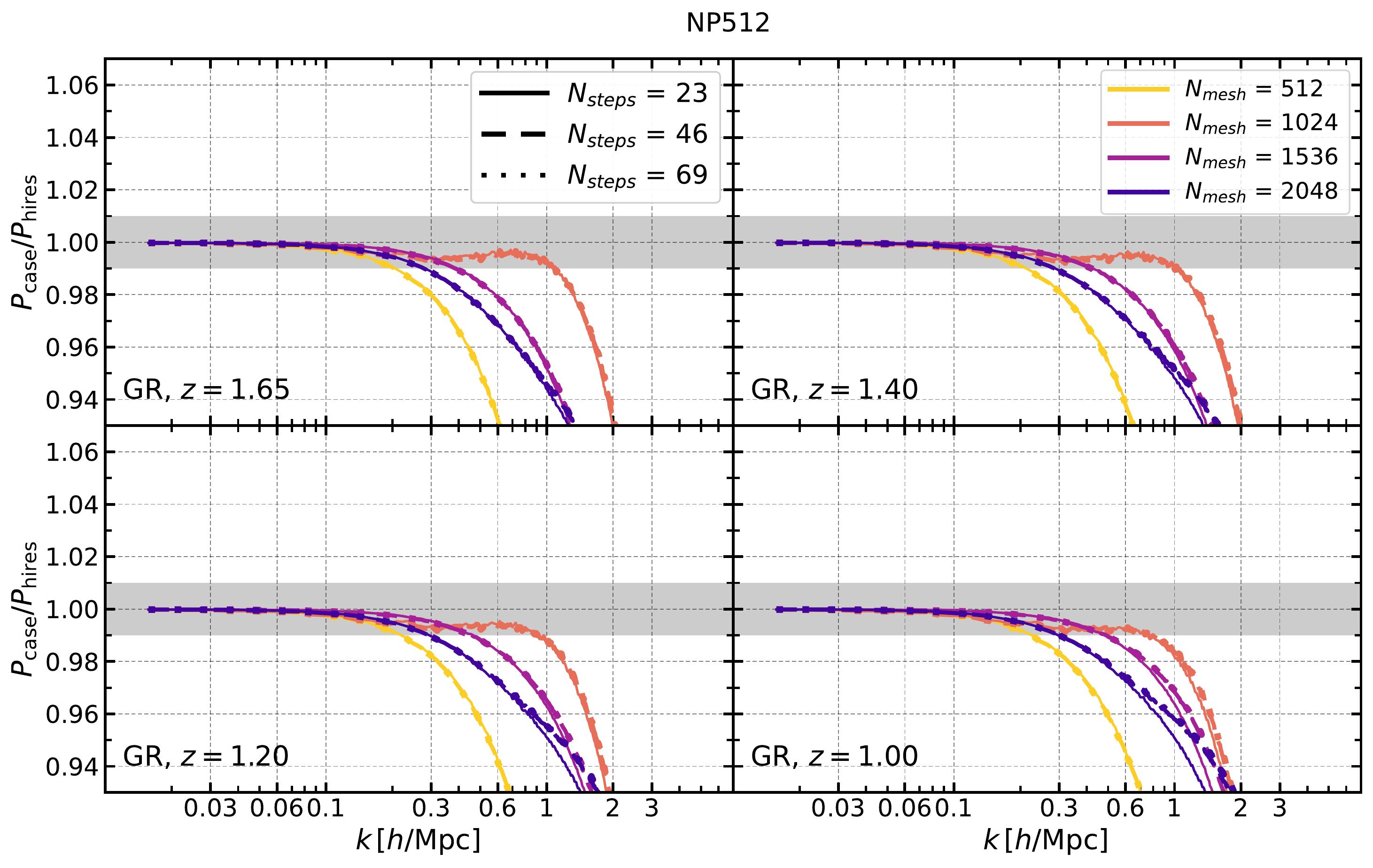}
\caption{\label{Fig:GR_NP512} Ratio of the matter power spectrum of simulations with 512 particles for the force and time resolutions shown in the legend with that of the high-resolution run.}
\end{figure}

\begin{figure}[t]
\centering 
\includegraphics[width=.99\textwidth]{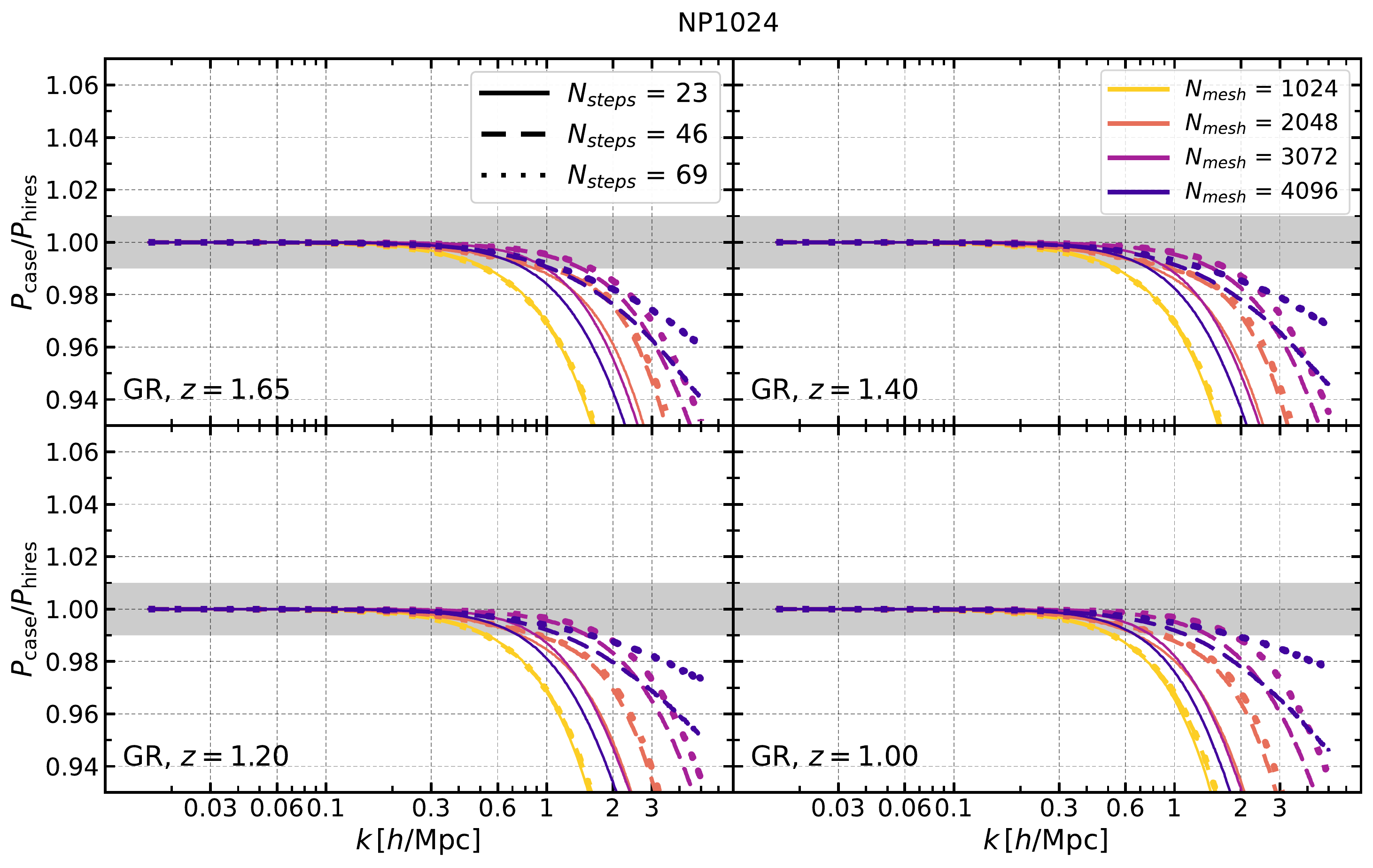}
\caption{\label{Fig:GR_NP1024} Ratio of the matter power spectrum of simulations with 1024 particles for the force and time resolutions shown in the legend with that of the high-resolution run.}
\end{figure}

\begin{figure}[t]
\centering 
\includegraphics[width=.99\textwidth]{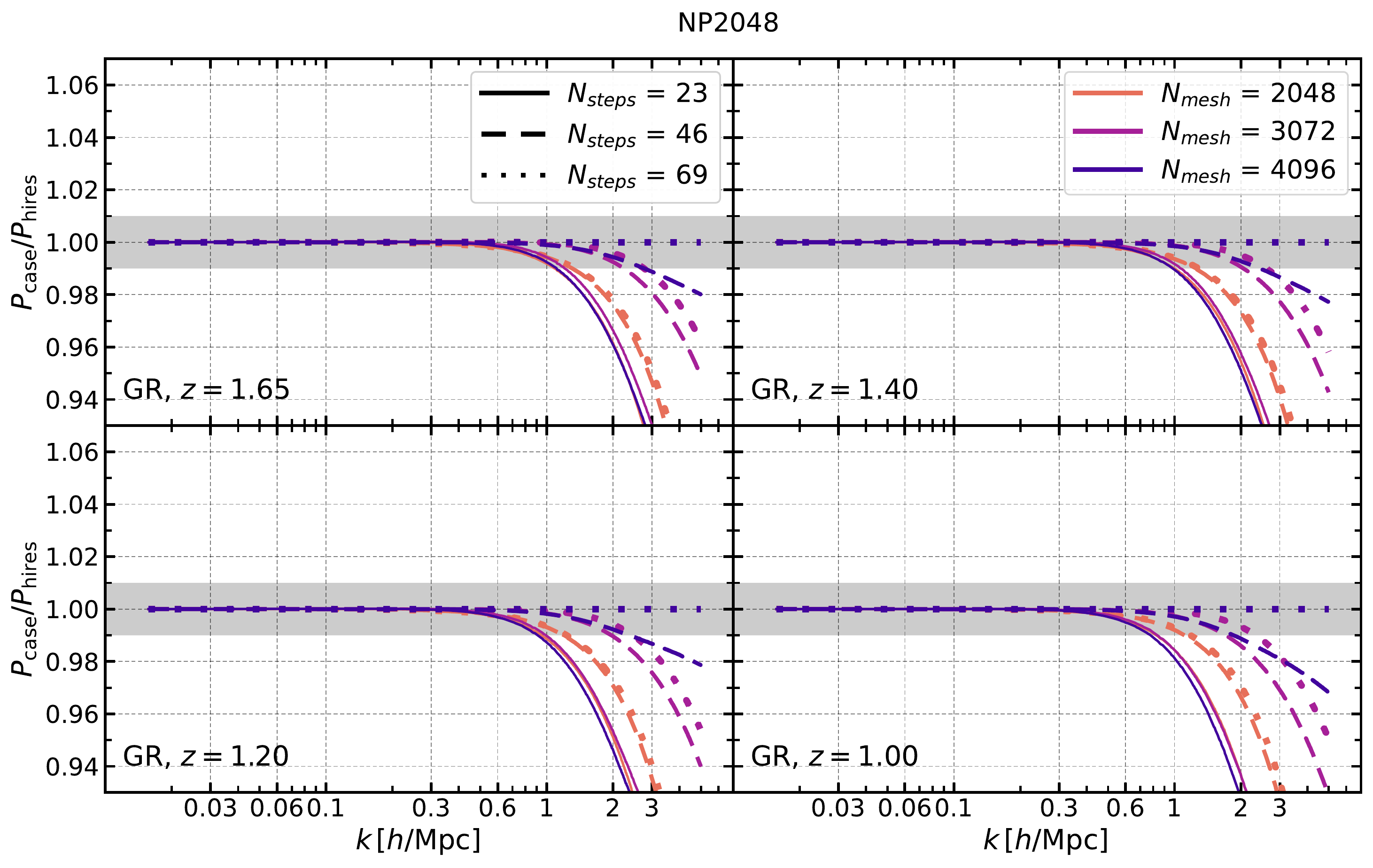}
\caption{\label{Fig:GR_NP2048} Ratio of the matter power spectrum of simulations with 2048 particles for the force and time resolutions shown in the legend with that of the high-resolution run.}
\end{figure}

With all the simulations of the convergence suite, we compare the accuracy of all the different settings using the highest-resolution run as the reference.
To show all the results, at the 4 redshift values relevant for this project, we show in different figures different mass resolutions, figure~\ref{Fig:GR_NP512} for $\NP=512$, figure~\ref{Fig:GR_NP1024} for $\NP=1024$, figure~\ref{Fig:GR_NP2048} for $\NP=2048$, and split each figure in 4 panels, one for each redshift, $z=1.65,1.4,1.2,1$%
\footnote{ This choice of redshift values is motivated by the expected redshift localisation of H$\alpha$ emitters in the Euclid spectroscopic survey \cite{Euclid:2019clj}.} %
from top left to bottom right. In each panel, we show the ratio of the power spectra in the COLA run with that of the high-resolution run for the various force and time resolutions described in the legends.

For the $\NP=512$ case, we notice that the results are almost insensitive to the time resolution and do not converge towards the high-resolution result. This is attributed to the fact that the mass resolution has a strong effect on the scales considered in the comparison, especially for higher redshifts.
For the $\NP=1024$ and $2048$ cases, the convergence towards the high-resolution PM result happens only if the force and time resolutions are increased accordingly. This condition is particularly evident in the simulations with $\NP = 1024$, where increasing the number of time-steps from 23 to 69 keeping the fixed force resolution of $\NM = 1024$ does not improve the convergence toward the high-resolution run. Similarly, increasing the force resolution above $\NM = 2048$ but using only 23 time-steps does not improve the agreement with the highest-resolution run. The agreement between the power spectra of lower-resolution simulations with the highest-resolution one indicates that the latter is converged at $~1\%$ level up to $k\sim2\hompc$. This confirms the result of the comparison with the Bacco emulator.

\section{Response function}
\label{sec:Response}
In section~\ref{sec:Convergence} we have shown that the power spectrum of COLA simulations with very low resolution (e.g., $\NP=512$, $\NM=512$, $\NS=23\times 1$) can depart from high-resolutions results already at $k\sim 0.3 \hompc$ (see figure~\ref{Fig:GR_NP512}). Despite this inaccuracy,   in~\cite{Kaushal:2021hqv} it was shown that the mapping from displacements in low-resolution COLA simulations to those in full {\it N}-body simulations can be trained on simulations with a fixed value of the cosmological parameters, and this model can be used to correct the output of COLA simulations with different values of cosmological parameters including different masses of massive neutrinos with very high accuracy: the power spectrum and the cross-correlation coefficients are within $1\%$ down to $k=1 \ h/$Mpc. This indicates that the inaccuracy in COLA's predictions is fairly cosmological parameter independent, and this can be corrected by a cosmological parameter independent model. 

In this section, we show that COLA is capable of describing the non-linear response of the matter power spectrum to the change of cosmological parameters down to $k=1 \ h/$Mpc with high accuracy, as long as the change of the matter power spectrum is not too large. This is because the inaccuracy of COLA is largely cancelled by taking the ratio of power spectra in two different cosmologies. The accurate and fast computation of this response function is of great importance when building emulators, as it allows us to obtain quickly the expected non-linear prediction when parameters are varied with respect to a fixed pre-defined reference cosmology. Given the prediction of non-linear power spectra in a few sparsely sampled reference cosmologies by full {\it N}-body simulations, we can provide the prediction of the matter spectrum in a wide parameter space. This can be used to further extend the reach of the already accurate \lcdm \ emulators to beyond-\lcdm \ cosmologies, for example, where running full {\it N}-body simulations is very expensive. 

COLA's accuracy depends on a number of settings, such as the number of time steps and the number of grids for the PM part, and it is always a trade-off between accuracy and speed. To study the response function we use COLA simulations with $1024^3$ particles in a $(1024 \mpcoh)^3$ box. The simulations are started from the same IC at $z=19$ and evolved until redshift $z=0$ using 50 time-steps with (approximately) constant size in the scale factor, $\Delta a \approx 0.02$. We emphasize that all the comparisons and results using our COLA simulations shown in this work can be further improved by changing these specifications at the cost of speed. We have run high-resolution PM simulations in different cosmologies and confirmed the robustness of our results for the response function.

Therefore, we will compare the ratio between the linear and non-linear matter power spectrum in different cosmologies with respect to a pre-defined reference cosmology, which in our case will be $\Lambda$CDM with the cosmological parameters shown in Table~\ref{table:ref_paarms}.
\begin{table}
    \begin{center} 
        \begin{tabular}{lc} 
            \toprule
            Parameter & Value \\
            \midrule
            $h$  & $0.67$ \\
            $\Omega_{\rm b}$ & $0.049$ \\
            $\Omega_{\rm m}$ & $0.319$ \\
            $n_{\rm s}$ & $0.96$ \\
            $A_{\rm s}$ & $2.1 \times 10^{-9}$ \\
            \bottomrule						
        \end{tabular}
    \end{center}
    \caption{Reference parameters.}
    \label{table:ref_paarms} 
\end{table}
We fix the dark energy equation of state to that of a cosmological constant, i.e., $w_{0}=-1$ and $w_{a}=0$, and restrict our analysis to massless neutrinos, $\sum_{\nu} m_{\nu} = 0 \, \mathrm{eV}$. For an extension of this analysis to the case of massive neutrinos see \cite{Brando:2022gvg}.
We define the linear and non-linear response functions as
\begin{align}\label{eq:R_functions}
    \Rlin(k,z) = \frac{\Plin^{\rm case}(k,z)}{\Plinref(k,z)}, \ \ \ \Rnon(k,z) = \frac{\Pnon^{\rm case}(k,z)}{\Pnonref(k,z)} 
\end{align}
respectively, where the superscript ``case" refers to a given case cosmology being investigated, and the superscript ``ref" always refers to predictions of GR with parameters from Table~\ref{table:ref_paarms}. We will also define the non-linear boost\footnote{This is the quantity used to produce the training set of emulators like Bacco and EE2.} as the function that maps the linear matter power spectrum to the non-linear one
\begin{equation}
    \Pnon^{\rm case/ref}(k,z) = B^{\rm case/ref}(k,z)\times \Plin^{\rm case/ref}(k,z),
\end{equation}
and then we can get the non-linear boost in a different cosmology from the reference boost and the ratio of the response functions:
\begin{equation}\label{eq:B_function}
    B^{\rm case}(k,z) = \Bref(k,z)\times \frac{\Rnon^{\rm case}(k,z)}{\Rlin^{\rm case}(k,z)}.
\end{equation}

In the following sub-sections, we will check the validity of using COLA to compute $\Rnon^{\rm case}(k,z) / \Rlin^{\rm case}(k,z)$ against emulators in \lcdm \ and then compute them in nDGP gravity in section~\ref{sec:Boost}. 



\subsection{Variation of cosmological parameters}\label{sec:LCDM_var_par}

In this section, we compare COLA simulations with EE2 in terms of the response function defined by Equation (\ref{eq:R_functions}), by varying cosmological parameters one at a time. Throughout this section, we fixed the Hubble constant and baryon energy density to their reference values, $h=0.67$ and $\Omega_{\rm b}=0.049$.
The reference cosmology is defined by Table~\ref{table:ref_paarms}. 

\begin{figure}
\centering
\includegraphics[width=1\textwidth]{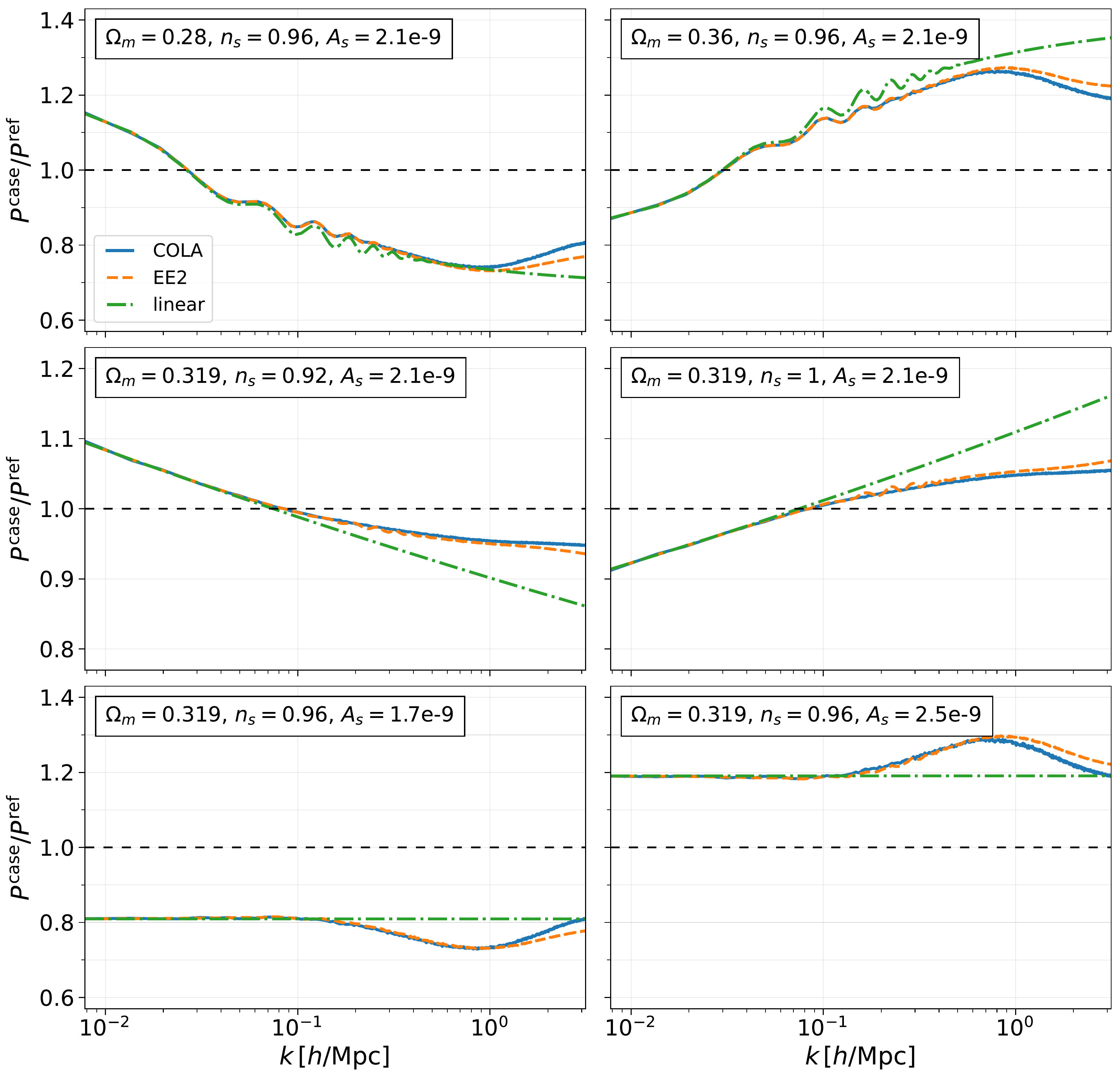}
\caption{Non-linear and linear response functions for each variation of the cosmological parameters with respect to the reference cosmology. Blue solid lines are computed from COLA simulations, orange dashed lines are obtained from EE2, and dash-dotted green lines are the linear predictions computed using CLASS. 
}
\label{fig:cosmo_response_COLA_EE2_lin}
\end{figure}

After making these choices we are left with only three cosmological parameters, $\Omega_{\rm m}$, $n_{\rm s}$ and $A_{\rm s}$ 
, which when varied independently alongside the fixed choices of parameters, impact differently the matter power spectrum.
That is, increasing (reducing) the value of the amplitude of the primordial scalar perturbations, $A_{\rm s}$, leads to a re-scaling of the matter power spectrum amplitude up (down), while the variation of the spectral index, $n_{\rm s}$, enhances or suppresses power at small scales. Augmenting the total amount of matter in the Universe, while keeping the baryon densities fixed, leads to increasing the value of the dark matter density. This affects the matter power spectrum by first changing the scale of equality between matter and radiation era, $k_{\rm eq}$, and by tilting the spectrum at small scales, i.e., if we have a bigger $\Omega_{\rm cdm}$ we will have steeper gravitational potentials, leading to more matter clustering at small scales, while a smaller value for $\Omega_{\rm cdm}$ gives you the opposite. To get a better perspective of these features, Figure~\ref{fig:cosmo_response_COLA_EE2_lin} shows the linear response and non-linear response from COLA simulations and EE2 at $z=0$. 
At large scales (small $k$ values), all curves agree with each other, while at higher $k$ values we see non-linear corrections in the solid blue and dashed orange curves.

In our COLA simulations, cosmological parameters were varied in the range shown in table~\ref{table:varied_params}. However, in this section, we present the results and comparisons only for the cases to which we will refer as ``large'', that is, the minimum and maximum values shown in the same table. 

\begin{table}
    \begin{center} 
        \begin{tabular}{lcc} 
            \toprule
            Parameter & Min. & Max. \\
            \midrule
            $\Omega_{\rm m}$  & $0.28$  & $0.36$ \\
            $n_{\rm s}$  & $0.92$  & $1.0$ \\
            $A_{\rm s}$  & $1.7\times 10^{-9}$  & $2.5\times 10^{-9}$ \\
            \bottomrule
        \end{tabular}
    \end{center}
    \caption{``Large'' variation of parameters.}
    \label{table:varied_params} 
\end{table}

\begin{figure}[t]
\centering
\includegraphics[width=1\textwidth]{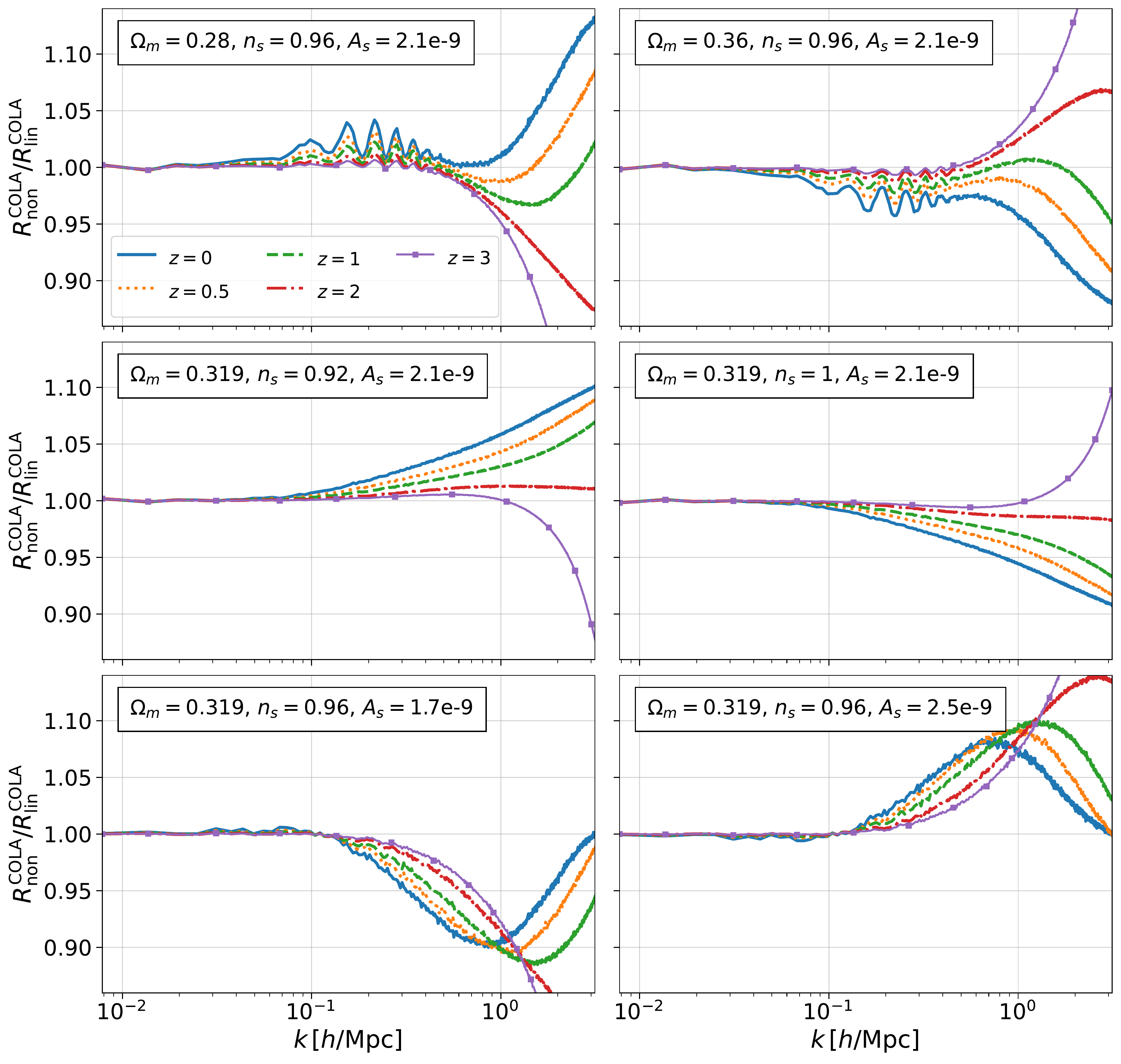}
\caption{Ratio between the non-linear and linear response functions computed from COLA simulations. We show this quantity for five different redshifts: $z=0$ (solid blue), $z=0.5$ (dotted orange), $z=1$ (dashed green), $z=2$ (dash-dotted red) and $z=3$ (solid-squared purple). We use the same convention in all other figures when we show the results at these five redshifts.}
\label{fig:mnu0_COLA_Rnon_Rlin}
\end{figure}

The difference between linear and non-linear predictions for $P^{\rm case}/P^{\rm ref}$ in  Figure~\ref{fig:cosmo_response_COLA_EE2_lin} is characterised by $R_{\rm non}/R_{\rm lin}$, which needs to be computed by simulations. 
Figure~\ref{fig:mnu0_COLA_Rnon_Rlin} shows the predictions for this function by COLA at different redshifts. In our COLA implementation, the linear prediction from COLA and the one from \hiclass \ have a $0.1\%$ agreement with each other. For the change of $\Omega_{\rm m}$, we see oscillations on quasi-non-linear scales, which describe the smoothing of BAO oscillations in $P^{\rm case}/P^{\rm ref}$ by non-linearity. For the change of $n_{\rm s}$ and $A_{\rm s}$, the non-linearity gives a scale-dependent enhancement or suppression at large $k$.  

\begin{figure}[t] 
\centering
\includegraphics[width=1\textwidth]{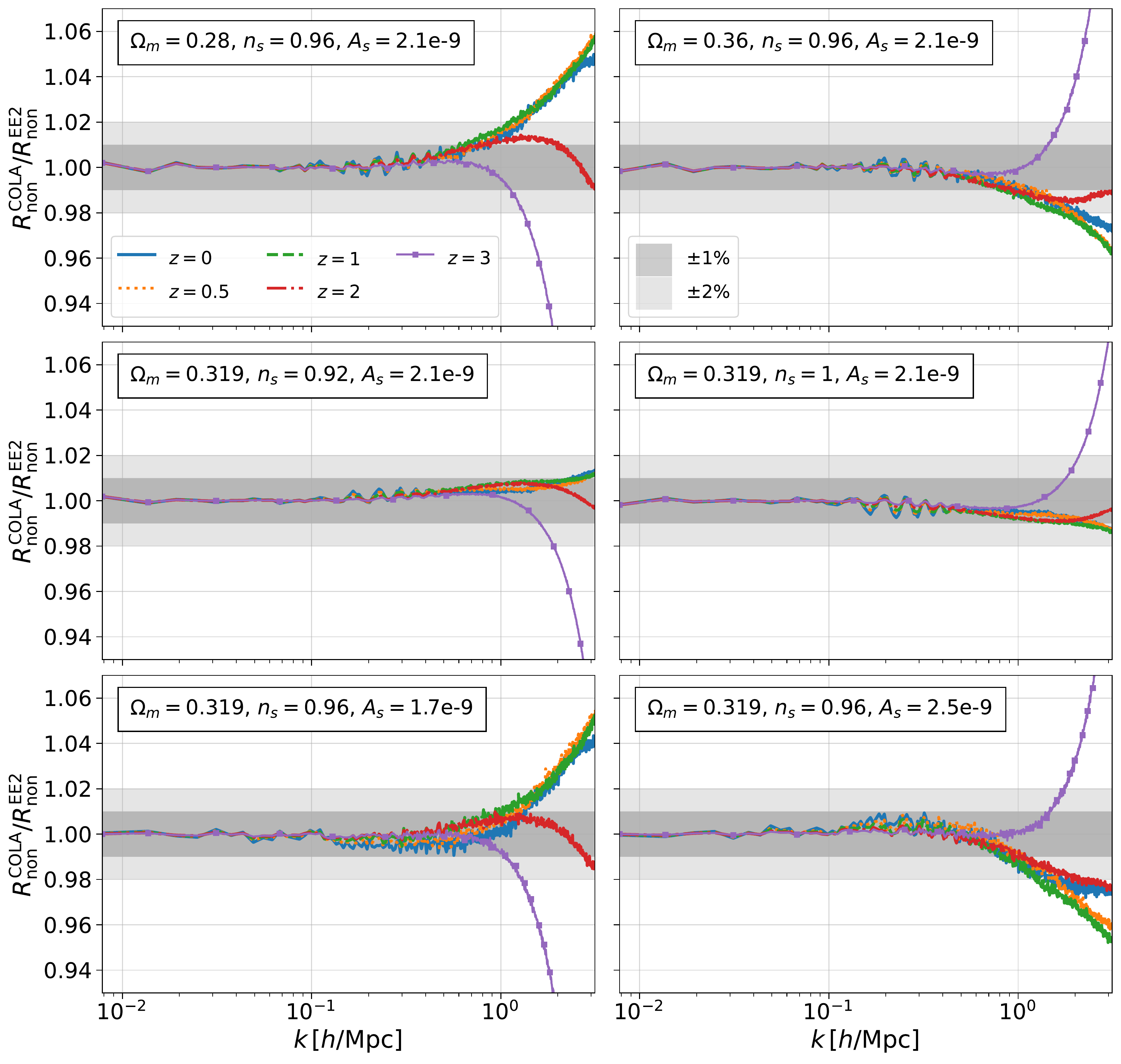} 
\caption{Ratio between the non-linear response function computed using COLA and the EE2.} 
\label{fig:mnu0_COLA_EE2_ratio_Rs}
\end{figure}

\begin{figure}[t] 
\centering
\includegraphics[width=1.0\textwidth]{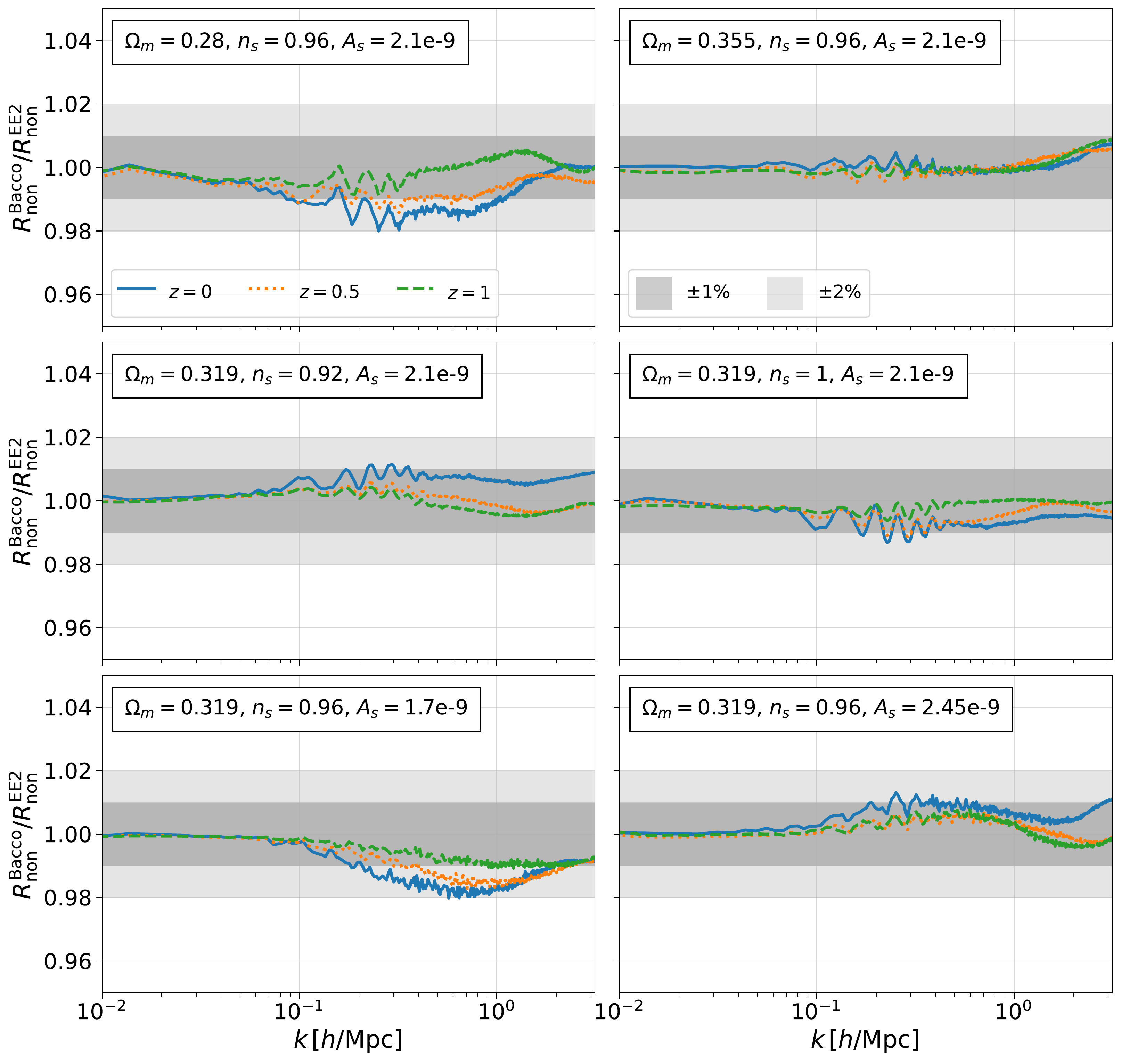} 
\caption{Ratio between the non-linear response function computed using Bacco and the EE2.}
\label{fig:Rnon_BaccovsEE2_LargeVar}
\end{figure}

To investigate how well COLA fares with EE2 in predicting $R_{\rm non}/R_{\rm lin}$ , in Figure~\ref{fig:mnu0_COLA_EE2_ratio_Rs} we plot the ratio of the non-linear response from COLA and that from EE2. We can see that we get $2\%$ agreements up to $k \sim 1 \ h$/Mpc when varying $\Omega_{\rm m}$. When we vary $n_{\rm s}$ we get $1\%$ agreements, and for $A_{\rm s}$ we obtain $2\%$ agreements at higher redshifts, while at $z \leq 1$, they become $1\%$ up to $k \sim 1 \ h$/Mpc.
In figure~\ref{fig:Rnon_BaccovsEE2_LargeVar}, we show a similar comparison for the response function, but between EE2 and Bacco. Note that Bacco does not cover the largest $\Omega_{\rm m}$ and $A_{\rm s}$ used in this analysis. At $k < 1 \ h$/Mpc, the agreement between EE2 and Bacco is comparable to that between EE2 and COLA although the agreement is much better at $k > 1 \ h$/Mpc as expected. 

In the above studies, we consider the cases where the matter power spectrum changes up to $30 \%$ compared with the reference cosmology as shown in figure~\ref{fig:cosmo_response_COLA_EE2_lin}. 
Future surveys have the ability to constrain the power spectrum at $1\%$ level. In the following subsection, we show the results for small variations of the cosmological parameters, 
where there are $1 \%$ variations in the matter power spectrum. 



\subsection{Small variations}\label{sec:SmallVar}
\begin{figure}[t] 
\centering
\includegraphics[width=1.0\textwidth]{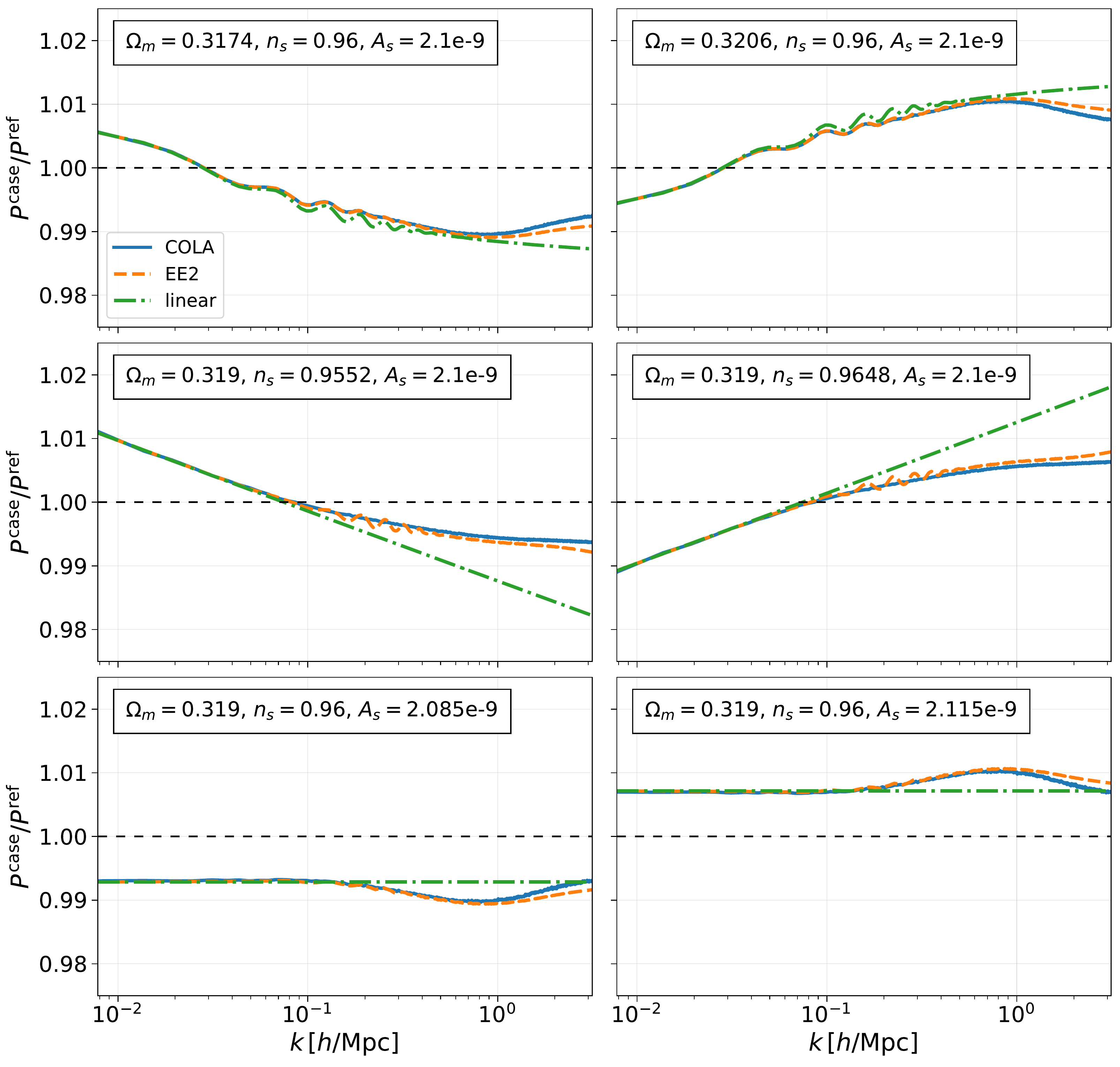} 
\caption{Non-linear and linear response functions computed using COLA, EE2 and \hiclass, for small variations of the cosmological parameters.}
\label{fig:cosmo_response_COLA_EE2_lin_small_var}
\end{figure}

We have seen that COLA can be sub-percent accurate in predicting the response function for a wide range of cosmologies where the response function reaches $\sim 30\%$ deviations from unity. This suggests (but does not guarantee) that COLA can achieve higher accuracy for cosmologies where the response function shows smaller deviations from the identity.  
Here we confirm this statement by showing the results for small variations of cosmological parameters between COLA and EE2. Our choice of small variations represents increasing and decreasing $0.5\%$ of the reference value of $\Omega_{\rm m}$, $A_{\rm s}$ and $n_{\rm s}$. This small change in the parameters has smaller effects on the linear and non-linear response functions, as opposed to the large variation cases considered in Figure~\ref{fig:cosmo_response_COLA_EE2_lin}. Their impact on the matter power spectrum is shown in Figure~\ref{fig:cosmo_response_COLA_EE2_lin_small_var}. It is important to check that COLA can reproduce the response function with much better accuracy in the case. 

\begin{figure}[t] 
\centering
\includegraphics[width=1.0\textwidth]{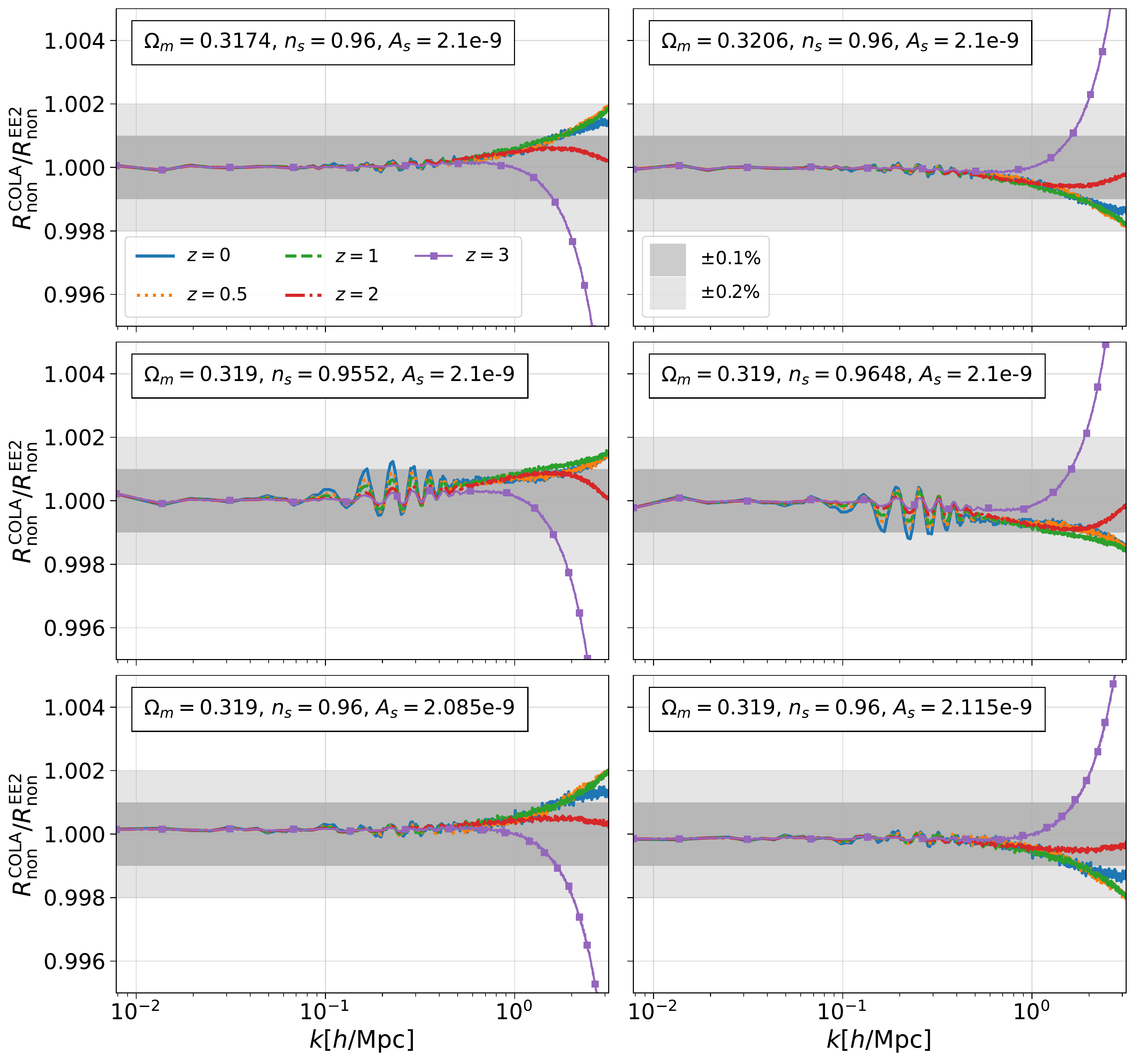} 
\caption{The same ratio as in Figure~\ref{fig:mnu0_COLA_EE2_ratio_Rs}, but for small variations of the cosmological parameters. 
}
\label{fig:Rnon_COLAvsEE2_SmallVar}
\end{figure}



\begin{figure}[t] 
\centering
\includegraphics[width=1.0\textwidth]{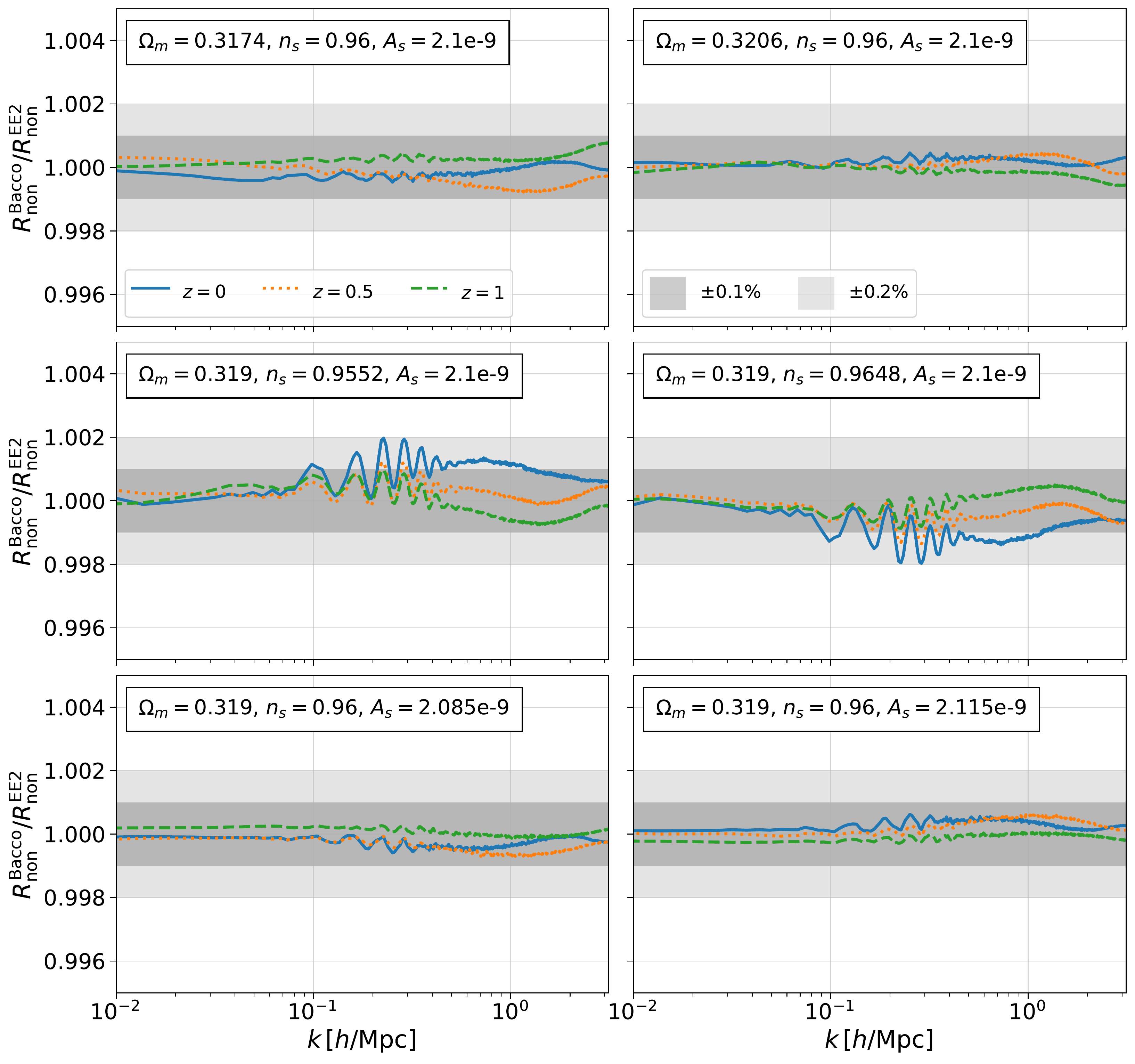} 
\caption{The same ratio as in Figure~\ref{fig:Rnon_BaccovsEE2_LargeVar}, but for small variations of the cosmological parameters.}
\label{fig:Rnon_BaccovsEE2_SmallVar}
\end{figure}

We show the comparisons of the response functions for small variations of cosmological parameters of COLA and EE2 in figure~\ref{fig:Rnon_COLAvsEE2_SmallVar}, and of Bacco and EE2 in figure~\ref{fig:Rnon_BaccovsEE2_SmallVar}. 
In figure~\ref{fig:Rnon_COLAvsEE2_SmallVar}, we can see that COLA is in $\sim 0.1 \%$ agreement with EE2 up to $k = 1\hompc$ for each variation of cosmological parameters and at all redshift values under consideration. 
Figure~\ref{fig:Rnon_BaccovsEE2_SmallVar} shows that the agreement between EE2 and Bacco is comparable to that between EE2 and COLA up to $k < 1 \ \hompc$ even for small-variation of cosmological parameters.

\subsection{Discussion}
We have shown that the agreement of COLA and EE2 in predicting the response function is within $2\%$ up to $k=1\hompc$ for large variations of cosmological parameters and within $0.1\%$ up to $k=1\hompc$ for small variations, validating COLA predictions in these range of scales.
In order to go beyond $k=1 \ h/$Mpc, we need to improve the PM part of the simulations. There are several methods proposed to improve the accuracy of COLA and PM simulations \cite{Kaushal:2021hqv,Dai:2018vvv,Lanzieri:2022zvv}. In addition, baryonic effects become significant beyond $k=1 \ h/$Mpc~\cite{Chisari:2019tus} and these effects need to be added to dark matter only simulations using the methods proposed by \cite{Schneider:2015wta,Dai:2018vvv} for example, where parametric models for the baryonic feedback are introduced to map the density field of DM-only {\it N}-body simulations to that of hydro-dynamical simulations. It is still an open question whether we gain any information on cosmology and modified gravity by marginalising over parameters describing these baryonic effects beyond $k=1 \ h/$Mpc. Even for emulators using more sophisticated {\it N}-body codes to go beyond $k=1 \ h/$Mpc such as the FORGE emulator in $f(R)$ gravity \cite{Arnold:2021xtm}, our study indicates that it is easier to emulate the response function to avoid the effect of resolution issues.
\section{Modified gravity boost factor}
\label{sec:Boost}

Another interesting application of COLA simulations is to predict the matter power spectrum in MG \cite{Winther:2019mus}. As non-linear matter power spectrum emulators are already available, it is possible to obtain the matter power spectrum in MG by multiplying the matter power spectrum in GR with the MG boost factor
\begin{equation}
    B^{\rm MG}(k,z) = \frac{\Pnon^{\rm MG}(k,z)}{\Pnon^{\rm GR}(k,z)} \, . 
\end{equation}
In general, the MG boost factor depends on MG parameters as well as on cosmological parameters. In the following, we focus on the MG boost factor of the nDGP gravity model and use COLA simulations to train an emulator for the matter power spectrum in nDGP theory. To do so, we start by studying the simulation's requirements to accurately predict the nDGP boost factor with COLA and investigate the sensitivity of the boost factor to the theory parameters in subsection~\ref{Sec:AccSens}. Then we present the simulations suite and the data processing used to create the emulator's data sets in subsection~\ref{ssec:SimsAndData} and finally we discuss the emulator's training and performance in subsection~\ref{Sec:Emulator}.

\subsection{Accuracy and sensitivity}
\label{Sec:AccSens}
The computational cost of cosmological simulations depends on a number of parameters. In particular, it is very sensitive to volume and resolution. With this in mind, we want to find the optimal simulation settings that allow us to have the target accuracy while probing large enough scales but without requiring excessive computational resources.

To define the simulation requirements for the nDGP emulator, we start by assessing the accuracy of COLA in predicting the boost factor of nDGP theories against Arepo simulations. To this extent, we run high-resolution PM simulations\footnote{We emphasize that PM here refers to the high time-resolution used by these COLA simulations at late time, which makes the 2LPT contribution important only at early times as discussed in section~\ref{sec:Convergence}. The screening mechanism is implemented thanks to the approximation normally used in COLA simulations in MG and discussed in section~\ref{sec:MG}.} with the same cosmology of Arepo simulations for several values of the nDGP parameter. We simulate the evolution of $2048^3$ particles in a box of side $L = 1024 \mpcoh$ using $6048^3$ mesh-grids and 69 time-steps from redshift $z_{\rm ini}=127$ to $z_{\rm fin}=1$. The mass resolutions that we use in PM simulations is $M_{\rm part} \sim 1 \cdot 10^{10} \Msun$, roughly 8 times higher than the one of Arepo simulations. The force resolution is $\LF \equiv L/\NM \approx 0.17 \mpcoh$ and the time resolution is $\Delta a \approx 0.006$. We run the simulations in GR and in nDGP from the same IC. For nDGP simulations we use the 4 values of the $H_0 r_c$ parameter ($H_0 r_c= $0.5, 1, 2, 5) used in the Arepo simulations suite.

\begin{figure}
\centering
\includegraphics[width=.8\textwidth]{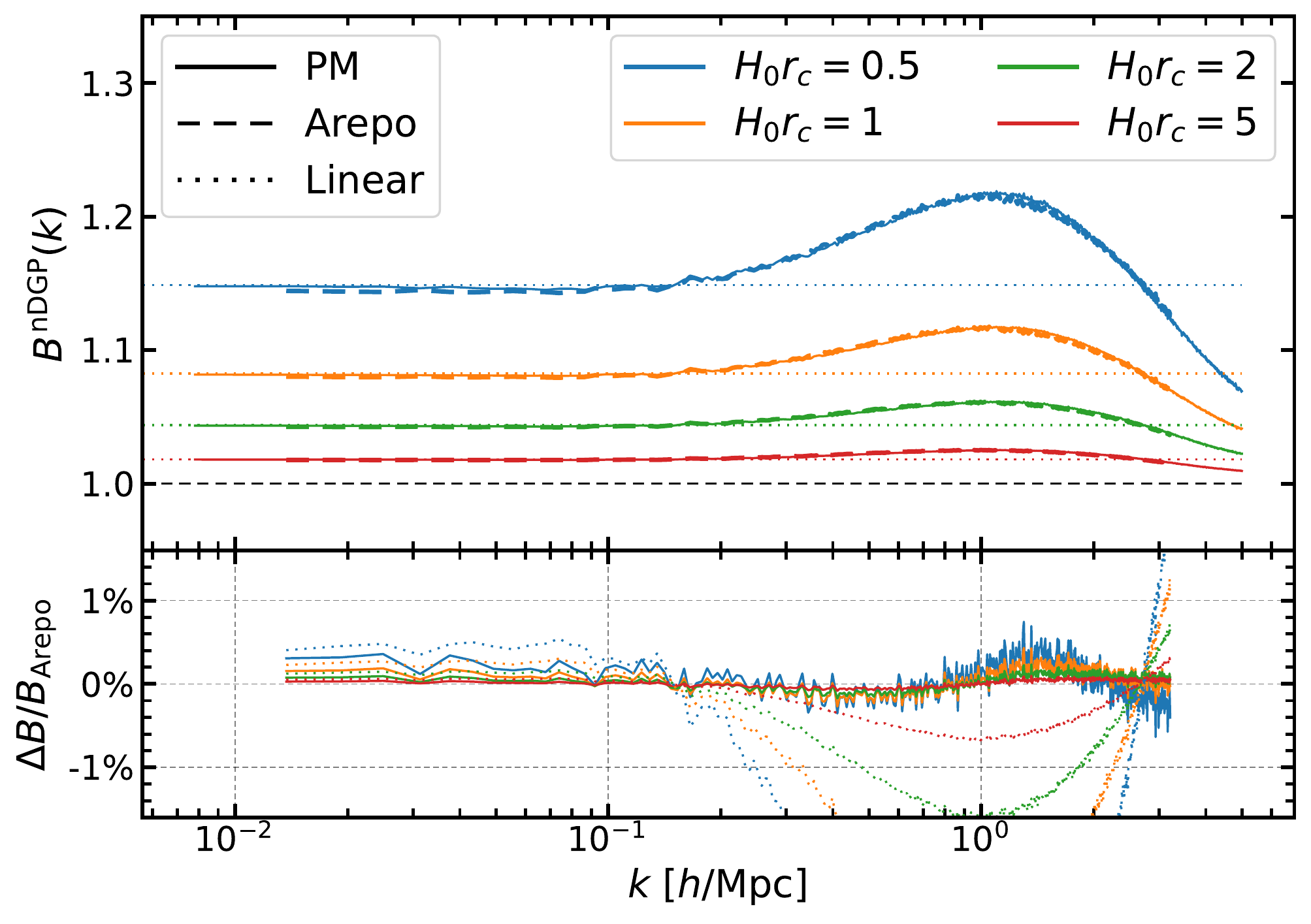}
\caption{\label{fig:BnDGP_COLAvsArepo} Ratio of the nDGP boost factors at redshift $z=1$ between PM simulations and Arepo simulations in 4 nDGP gravity models as indicated in the legend.}
\end{figure}

Figure~\ref{fig:BnDGP_COLAvsArepo} shows a comparison between the nDGP boost factors of PM and Arepo simulations for the different values of the $H_0 r_c$ parameter. We also include in the comparison the nDGP boost factor estimated from linear theory. In the top panel, we show the nDGP boost factors, while in the bottom we show the relative difference between the PM and linear boost factors from Arepo boost factors. On large scales, PM results seem to be more consistent with linear theory than the ones from Arepo simulations. Overall, we find $\lesssim 0.5 \%$ agreement in all gravity models up to $k \sim 3 \hompc$ between the high-resolution simulations and Arepo simulations. This means that the screening approximation used in COLA simulations in nDGP theory (see section~\ref{dgp} and in particular eq.~\eqref{Geff_DGP}), once tuned on full {\it N}-body simulations for a single value of $H_0 r_c$, is sub-percent accurate in reproducing the power spectrum boost factor of nDGP theories for a wide range of the parameter $H_0 r_c$ and for different cosmologies. 

\begin{figure}
\centering
\includegraphics[width=.98\textwidth]{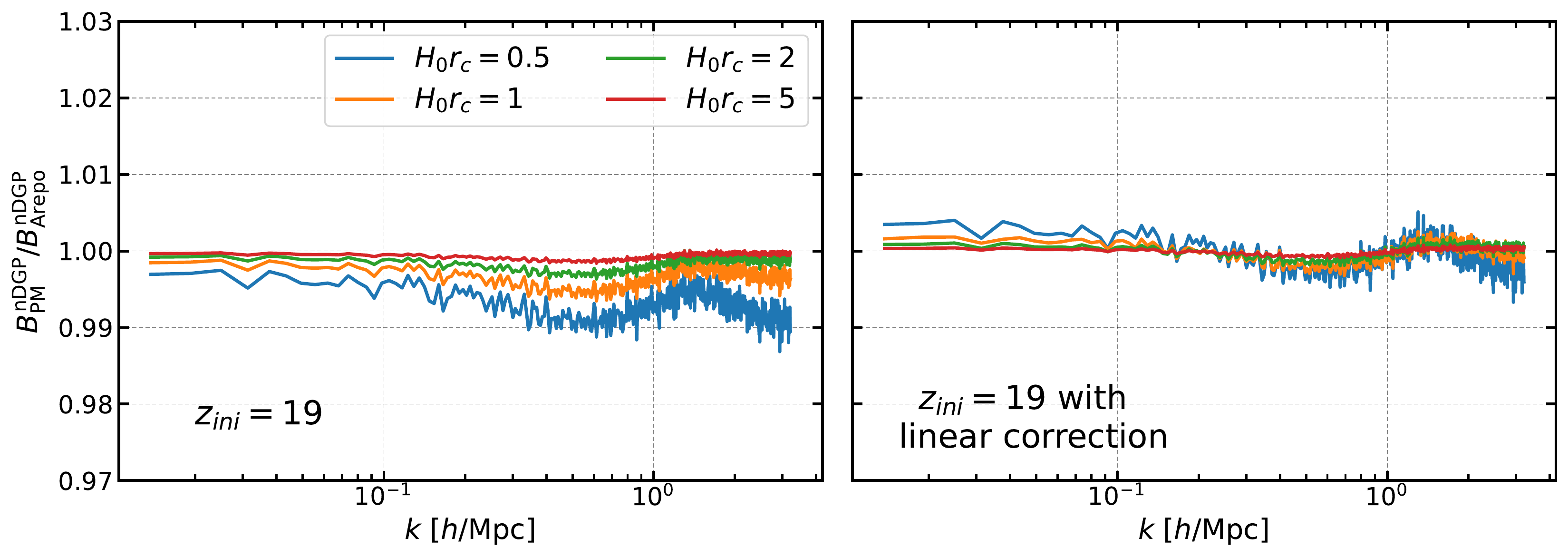}
\caption{\label{fig:BnDGP_COLAz19LinCorr} Ratio of the nDGP boost factors at redshift $z=1$ between PM simulations started at redshift 19 and Arepo simulations in 4 nDGP gravity models as indicated in the legend. The left panel shows the raw results from simulations, the right panel shows the results after applying the linear correction in eq~\eqref{GrowthLinCorr}.}
\end{figure}

The COLA simulations that we run in nDGP use IC produced with the back-scaling approach. The IC are set in such a way that the IC power spectrum is the same for nDGP and GR simulations. This is done by integrating the growth factors from matter domination at redshift $z=499$ and by normalising them at the redshift of IC $z_{\rm ini}$
\begin{equation}
    D^{z_{\rm ini}} (z_{\rm ini}) = 1 \, .
\end{equation}
Due to this choice of normalisation, COLA simulations ignore the fact that nDGP can be different from GR also at higher redshift. In particular, COLA simulations are usually started at redshift $z_{\rm ini}=19$, while nDGP effects can be important up to redshift $z \sim 100$. We can correct this by multiplying the final power spectrum by
\begin{equation}\label{GrowthLinCorr}
    \mathcal{A}_{\rm lin} (z) = \left(\frac{D_{\rm nDGP}^{z_{\rm ini}=127}(z) \cdot D_{\rm GR}^{z_{\rm ini}=19}(z) }{D_{\rm GR}^{z_{\rm ini}=127}(z) \cdot D_{\rm nDGP}^{z_{\rm ini}=19}(z)} \right)^2 \, ,
\end{equation}
which has the effect of re-scaling the amplitude of the power spectrum as if the simulations were started at redshift $z=127$. This approach neglects the effects of non-linearity before $z_{\rm ini}=19$, but these are sub-dominant due to the weakness of the nDGP fifth force at early times. The importance of this correction is evident from figure~\ref{fig:BnDGP_COLAz19LinCorr}, where we show that the agreement with Arepo of the nDGP boost factors from PM simulations started at $z_{ini}=19$ is improved when applying the linear correction. In fact, the raw results of $z_{ini}=19$ PM simulations show $1\%$ deviations from Arepo (left panel), while they are in better than $0.5\%$ agreement with Arepo once corrected using eq~\eqref{GrowthLinCorr} (right panel).

\begin{figure}[t]
\centering
\includegraphics[width=.98\textwidth]{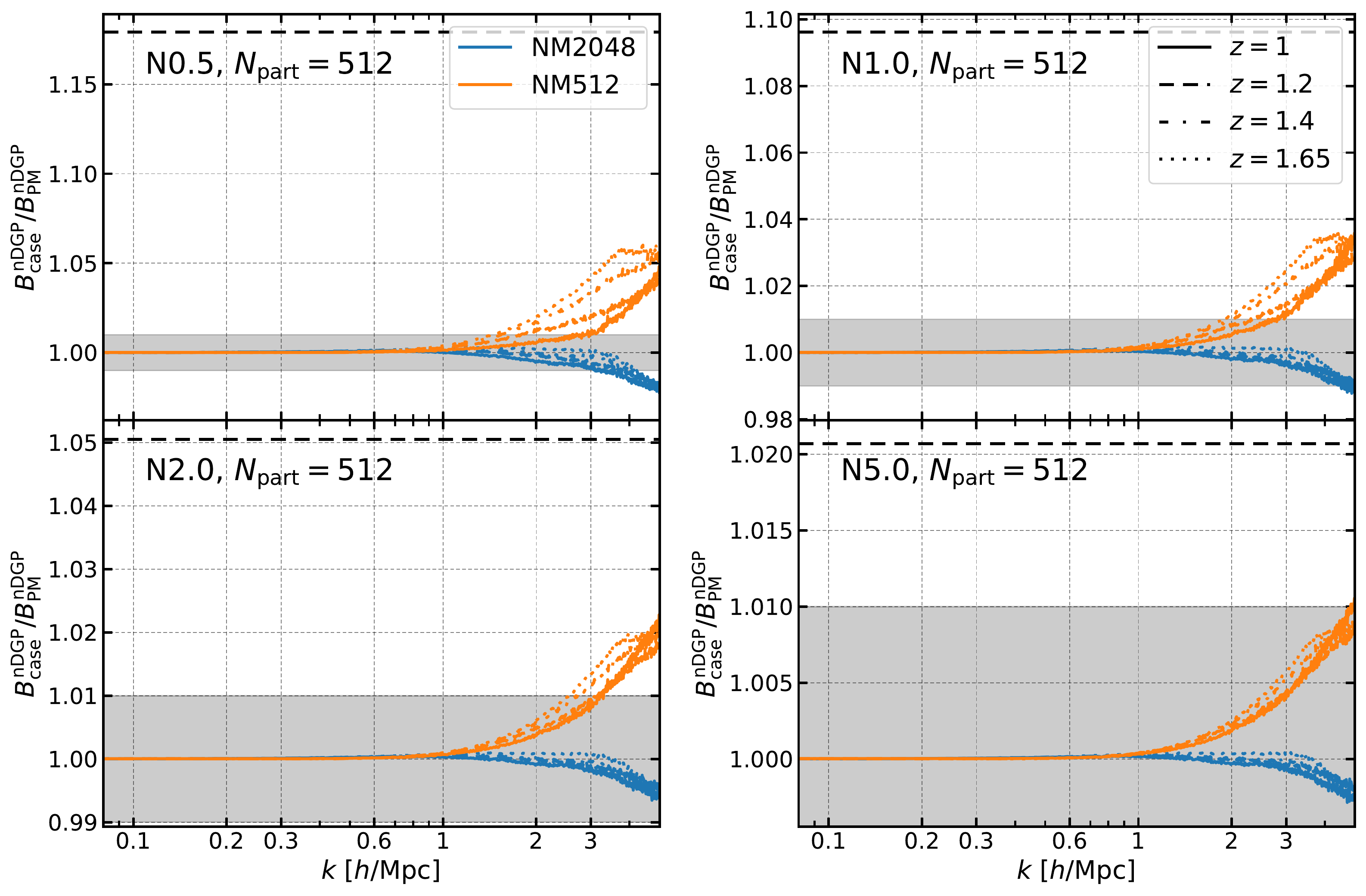}
\caption{\label{fig:BnDGP_LowVsHires} Ratio of the nDGP boost factors in low mass-resolution COLA simulations with that in high-resolution PM simulations for two different force resolutions, $\NM = 512$ and $\NM = 2048$. In different panels, we show results for different nDGP gravity models as indicated in the top-left corner of each panel. The black dashed lines show the typical amplitude of the boost factor in the different gravity models for visual reference.}
\end{figure}

Having tested the accuracy of COLA simulations in the PM-limit in reproducing the nDGP boost factor, we want to test the impact of force and mass resolution on the nDGP boost factor in COLA simulations to find the optimal setup to span the theory parameter space running a large number of simulations. To do so, we run more COLA simulations with lower mass-resolution ($M_{\rm part} \sim 8 \cdot 10^{10} \Msun$, similarly to Arepo simulations), with time-resolution $\Delta a \approx 0.02$ and with two force resolutions: using $\NM=512$ (low force-resolution) and using $\NM=2048$ (high force-resolution). We compare the results for the nDGP boost factor of low mass-resolutions COLA simulations with high-resolutions PM simulations in figure~\ref{fig:BnDGP_LowVsHires} where we can see that the results are converged at sub-percent level up to $k\sim 1 \hompc$ and the force resolution has a significant impact on the boost factor only on very small scales, $k> 1 \hompc$. 

Before choosing a strategy to sample the parameter space for producing the emulator's training set, we explore the impact that each parameter has on the boost factor. To do so we first investigate the range of values that the parameter $H_0 r_c$ should span: for stronger deviations from GR we choose $\sim 50\%$ deviations as the upper limit, while for the other limit we want to "smoothly" connect to GR so we look for models where the deviation is  $< 1\%$ \footnote{Which is our target accuracy.}. Based on previous results in the literature \cite{Ruan:2021wup}, we select $H_0 r_c = 0.2$ as the strong deviation case and $H_0 r_c = 20$ as the weak deviation case.

\begin{figure}
\centering
\includegraphics[width=.98\textwidth]{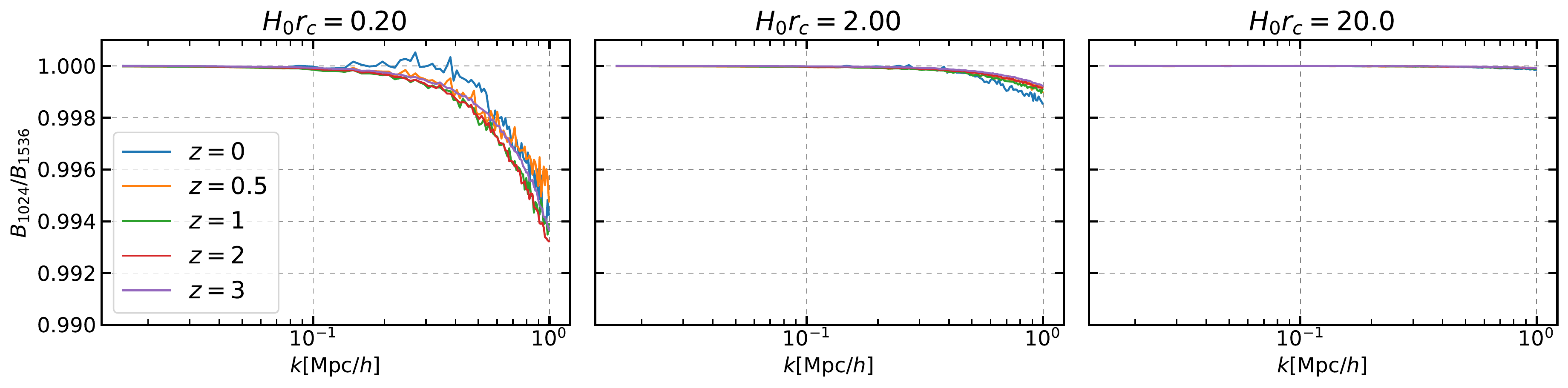}
\caption{\label{fig:nDGP_BoostConv_WideRange} Ratio of the nDGP boost factors obtained with $\NM = 1024$ COLA simulations with respect to the same boost factors but obtained with $\NM = 1536$ COLA simulations. The legend describes the redshift value of each line. }
\end{figure}

In the previous section, we have seen that the response function estimated with COLA agrees with the emulators in a wide range of cosmologies when using $\LF = 0.5 \mpcoh$. We use the same force resolution here for the boost factor since we want to predict it for a similar range of cosmologies. As an additional test to check the accuracy of the boost factor for the wider range of gravity parameters we produce $\LF = 1/3 \mpcoh$ simulations and use them to make a comparison with the boost factors of $\LF = 0.5 \mpcoh$ simulations.  
Assuming a default box side of $512 \mpcoh$ with $512^3$ particles, we actually compare the boost factors obtained using $\NM= 1024$ with the ones obtained using $\NM= 1536$. We show the ratio of the two boost factors for 5 redshift values (legend) and 3 values of the parameter $H_0 r_c$ (plot titles) in figure~\ref{fig:nDGP_BoostConv_WideRange}. In the most extreme case, the difference in the two boost factors is less than $1\%$ so we opt for the lower resolution settings in our simulations. 

\begin{figure}[t]
\centering 
\includegraphics[width=.98\textwidth]{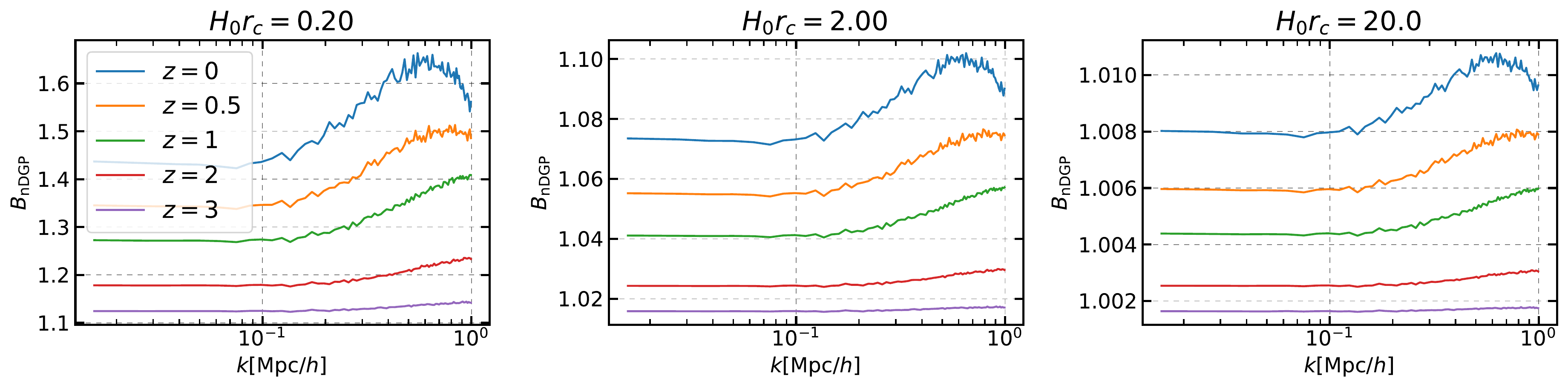}
\caption{\label{fig:Boost} nDGP boost factor for three values of the parameter $H_0 r_c$. The legend describes the redshift value of each line. }
\end{figure}

\begin{figure}[t]
\centering
\includegraphics[width=.8\textwidth]{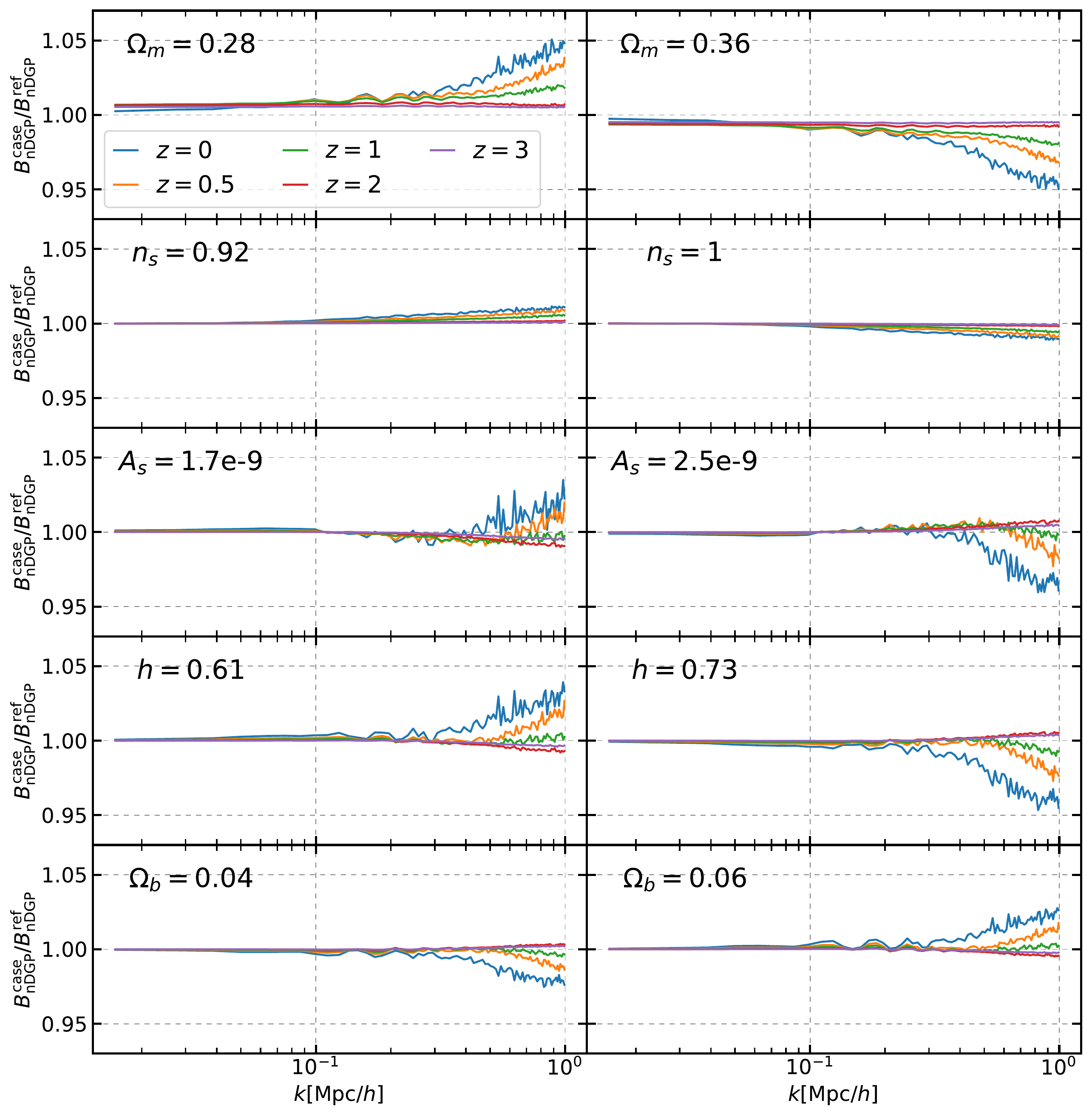}
\caption{\label{fig:CosmoEffect1} Change on the nDGP boost factor for $H_0 r_c = 0.2$ due the variation of each cosmological parameter (specified in each panel) with respect to the reference cosmology. The redshift values of each line are as described in the legend.}
\end{figure}

\begin{figure}[t]
\centering
\includegraphics[width=.8\textwidth]{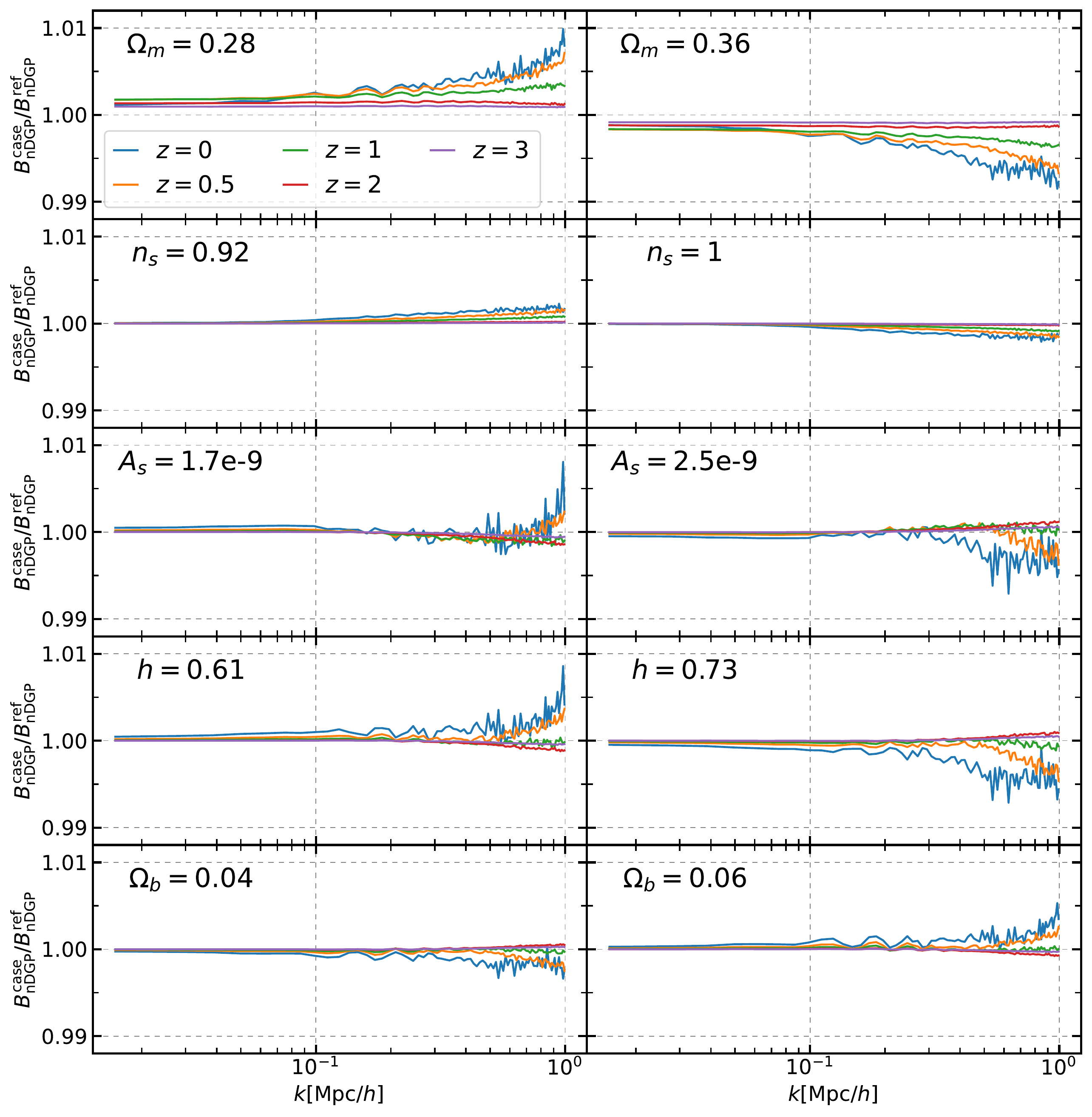}
\caption{\label{fig:CosmoEffect2} Same as figure~\ref{fig:CosmoEffect1} but for $H_0 r_c=2$}
\end{figure}

\begin{figure}[t]
\centering
\includegraphics[width=.8\textwidth]{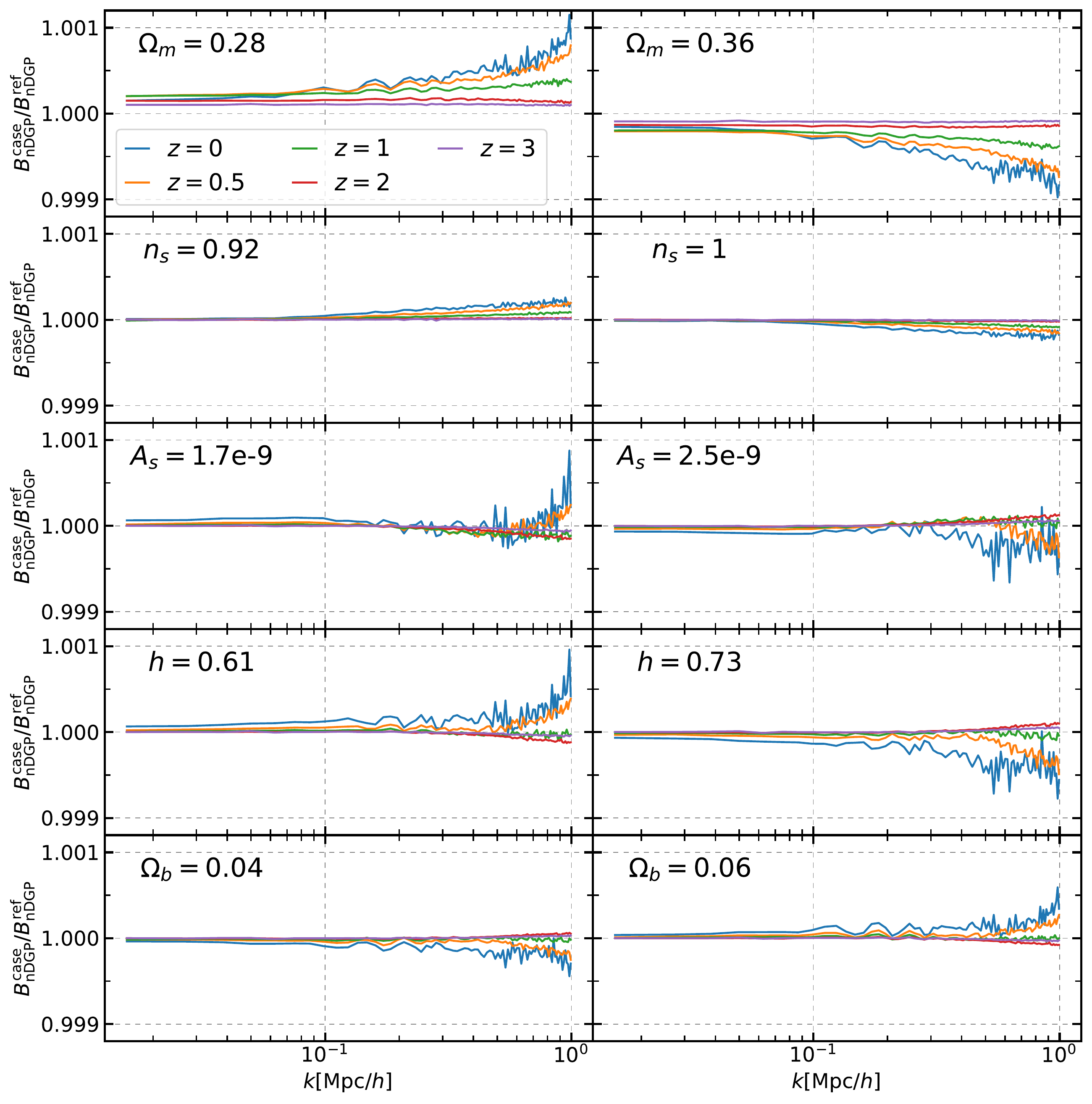}
\caption{\label{fig:CosmoEffect3} Same as figure~\ref{fig:CosmoEffect1} but for $H_0 r_c=20$}
\end{figure}

The boost factor of nDGP is strongly dependent on the theory parameter $H_0 r_c$ and on the redshift as shown in figure~\ref{fig:Boost}, where we plot the boost factors for three values of $H_0 r_c$ (titles) and for 5 redshift values (legend). However, the cosmological dependence of the nDGP boost factor is quite weak. 
In support of this statement, we show the effect of changing each cosmological parameter on the nDGP boost factor for $H_0 r_c =0.2$ in figure~\ref{fig:CosmoEffect1}, $H_0 r_c =2$ in figure~\ref{fig:CosmoEffect2} and $H_0 r_c =20$ in figure~\ref{fig:CosmoEffect3}. 
These figures not only show that the response is limited to $5\%$ in the most extreme case (variations of $\Omega_m$ for $H_0 r_c=0.2$ at redshift $z=0$), but also that its scale dependence is similar across the different cosmological parameters and values of $H_0 r_c$. The amplitude of the effect is smaller for larger $H_0 r_c$ values.

\subsection{Simulations and data sets}
\label{ssec:SimsAndData}
We produce two independent simulation suites, one for the training set and the other for the test set. 
Both training and test set simulations use $512^3$ particles to track the evolution of the matter density field in a $(512\mpcoh)^3$ box. The forces are calculated using $1024^3$ mesh grids. The simulations are started at $z_{\rm ini}=19$ and evolved up to $z_{\rm fin}=0$ using 50 time-steps linearly spaced in the scale factor. 
We sample the cosmological parameter space using the Latin Hypercube Sampling (LHS) technique \cite{LatinHypercubeSampling}, a space-filling sampling method widely used in computer experiments to produce near-random sets of points which represent the real variability of multi-dimensional sample spaces. The range of cosmological parameters spanned by the training and test sets is described in table~\ref{tab:EmuCosmoRange}. We use 20 cosmologies for the training set and 10 for the test set, which are listed in table~\ref{tab:TrainingCosmos} and table~\ref{tab:TestCosmos} respectively. For the modified gravity parameter $H_0 r_c$ we opt for a logarithmic sampling with 21 points in the range $[0.2,20]$ to produce the training set, and with 10 log-random points in the same range for the test set. The redshift sampling is determined by the time-stepping of COLA simulations, which is linear in the scale factor with 41 points between $z_{\rm max}=3$ and $z_{\rm min}=0$ for the settings used in this case. For each cosmology in the training set (test set) we run a vanilla GR simulation and nDGP simulations in all the gravity models of the training set (test set) using the same IC. 

\begin{table}[t]
    \begin{center} 
        \begin{tabular}{lcc} 
            \toprule
            Parameter & Min. & Max. \\
            \midrule
            $h$  & $0.61$  & $0.73$ \\
            $\Omega_{\rm b}$  & $0.04$  & $0.06$ \\
            $\Omega_{\rm m}$  & $0.28$  & $0.36$ \\
            $n_{\rm s}$  & $0.92$  & $1.0$ \\
            $A_{\rm s}$  & $1.7\times 10^{-9}$  & $2.5\times 10^{-9}$ \\
            \bottomrule						
        \end{tabular}
    \end{center}
    \caption{Range of cosmological parameters spanned by the nDGP emulator.}
    \label{tab:EmuCosmoRange} 
\end{table}

Each simulation takes roughly 28 minutes on a single node of the Sciama High-Performance Compute (HPC) cluster using 32 tasks\footnote{The specific partition used to run the simulations mounts \href{https://www.intel.com/content/www/us/en/products/sku/91766/intel-xeon-processor-e52683-v4-40m-cache-2-10-ghz/specifications.html}{Intel Xeon E5-2683v4} processors.}. This corresponds to $\approx 15$ CPU hours. In total, we ran $22\times20=440$ simulations for the training set and $11\times10=110$ simulations for the test set. This means that the total computational cost for the simulations is $\approx 8 \cdot 10^3$ CPU hours. In particular, we have used a partition counting 12 nodes with 32 cores each for a total of 384 cores (a typical value for an HPC cluster) obtaining all the power spectra in less than one wall-clock day.

\begin{table}[t]
    \footnotesize
    \begin{tabular}{ccccc}
    \toprule
    $\Omega_{\rm m}$ &      $\Omega_{\rm b}$ &      $h$ &     $n_s$ &            $A_s$ [$\cdot 10^{-9}$]\\
    \midrule
    0.282 &  0.0515 &  0.667 &  0.950 &  2.12 \\
    0.286 &  0.0445 &  0.655 &  0.990 &  2.44 \\
    0.294 &  0.0585 &  0.697 &  0.938 &  2.00 \\
    0.298 &  0.0415 &  0.619 &  0.978 &  2.04 \\
    0.290 &  0.0545 &  0.625 &  0.986 &  1.80 \\
    0.302 &  0.0485 &  0.709 &  0.922 &  1.72 \\
    0.306 &  0.0435 &  0.649 &  0.942 &  2.36 \\
    0.314 &  0.0595 &  0.721 &  0.958 &  2.28 \\
    0.318 &  0.0575 &  0.661 &  0.962 &  2.32 \\
    0.310 &  0.0405 &  0.673 &  0.966 &  1.88 \\
    \bottomrule
    \end{tabular}
    \hfill
    \begin{tabular}{ccccc}
    \toprule
    $\Omega_{\rm m}$ &      $\Omega_{\rm b}$ &      $h$ &     $n_s$ &            $A_s$ [$\cdot 10^{-9}$]\\
    \midrule
    0.322 &  0.0465 &  0.727 &  0.982 &  1.96 \\
    0.326 &  0.0455 &  0.637 &  0.994 &  2.08 \\
    0.334 &  0.0535 &  0.631 &  0.974 &  2.16 \\
    0.338 &  0.0555 &  0.703 &  0.954 &  1.92 \\
    0.330 &  0.0525 &  0.613 &  0.934 &  2.24 \\
    0.342 &  0.0495 &  0.685 &  0.998 &  2.20 \\
    0.346 &  0.0565 &  0.691 &  0.970 &  2.48 \\
    0.354 &  0.0505 &  0.643 &  0.930 &  1.76 \\
    0.358 &  0.0425 &  0.679 &  0.926 &  2.40 \\
    0.350 &  0.0475 &  0.715 &  0.946 &  1.84 \\
    \bottomrule
    \end{tabular}
    \caption{List of cosmologies used to create the training set for the nDGP emulator.}
    \label{tab:TrainingCosmos}
\end{table}

\begin{table}[t]
    \footnotesize
    \begin{tabular}{ccccc}
    \toprule
    $\Omega_{\rm m}$ &      $\Omega_{\rm b}$ &      $h$ &     $n_s$ &            $A_s$ [$\cdot 10^{-9}$]\\
    \midrule
    0.284 &  0.0410 &  0.700 &  0.956 &  1.82 \\
    0.292 &  0.0490 &  0.628 &  0.940 &  1.74 \\
    0.308 &  0.0510 &  0.652 &  0.964 &  2.22 \\
    0.316 &  0.0430 &  0.688 &  0.996 &  2.30 \\
    0.324 &  0.0470 &  0.712 &  0.932 &  2.06 \\
    \bottomrule
    \end{tabular}
    \hfill
    \begin{tabular}{ccccc}
    \toprule
    $\Omega_{\rm m}$ &      $\Omega_{\rm b}$ &      $h$ &     $n_s$ &            $A_s$ [$\cdot 10^{-9}$]\\
    \midrule
    0.332 &  0.0530 &  0.724 &  0.980 &  2.38 \\
    0.348 &  0.0450 &  0.616 &  0.972 &  1.98 \\
    0.340 &  0.0550 &  0.664 &  0.948 &  1.90 \\
    0.356 &  0.0570 &  0.640 &  0.988 &  2.46 \\
    0.300 &  0.0590 &  0.676 &  0.924 &  2.14 \\
    \bottomrule
    \end{tabular}
    \caption{List of cosmologies used to create the test set for the nDGP emulator.}
    \label{tab:TestCosmos}
\end{table}


The power spectra are computed by interpolating the DM distribution on a $1024^3$ mesh grid with the cloud-in-cell mass-assignment scheme. The power spectra are corrected for the window function used in the interpolation and for shot-noise. We use bins of width $\Delta k= k_f$ between $k_{\rm min}=\frac{1}{2} k_f$ and $k_{\rm Nyquist}= \pi \hompc$, where $k_f$ is the fundamental frequency of the box. As COLA simulations with these specifications have $\sim1\%$ reliable determination of the cosmological response function only up to $k \sim 1\hompc$ we cut the power spectra at $k_{\rm max}= 1 \hompc$. Similarly, as we have seen that mass-resolution is a limiting factor at high-redshift, we make the conservative choice of restricting our data sets to the power spectra at $z_{\rm max} = 3$. We take the ratio of the power spectra in nDGP theories with the respective power spectrum in GR for each redshift and cosmology to obtain the nDGP boost factors. As the nDGP and GR simulations are run from the same IC, the sample variance is largely cancelled out when computing the nDGP boost factors. Since our simulations are started at redshift $z_{\rm ini}=19$ we multiply the nDGP boost factors with the linear theory correction in eq.~\eqref{GrowthLinCorr}.

\begin{figure}[t]
\centering 
\includegraphics[width=.48\textwidth]{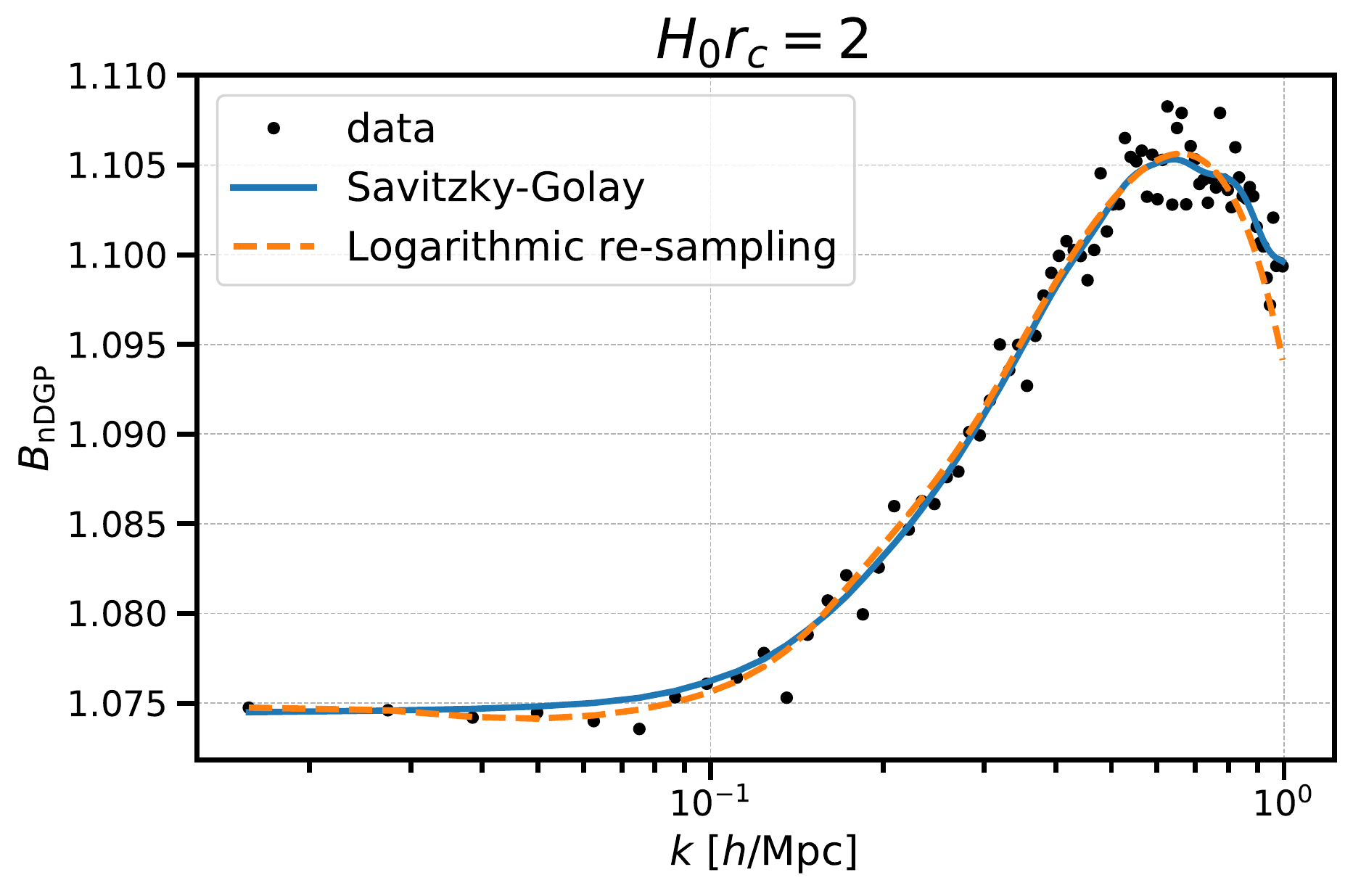}
\includegraphics[width=.48\textwidth]{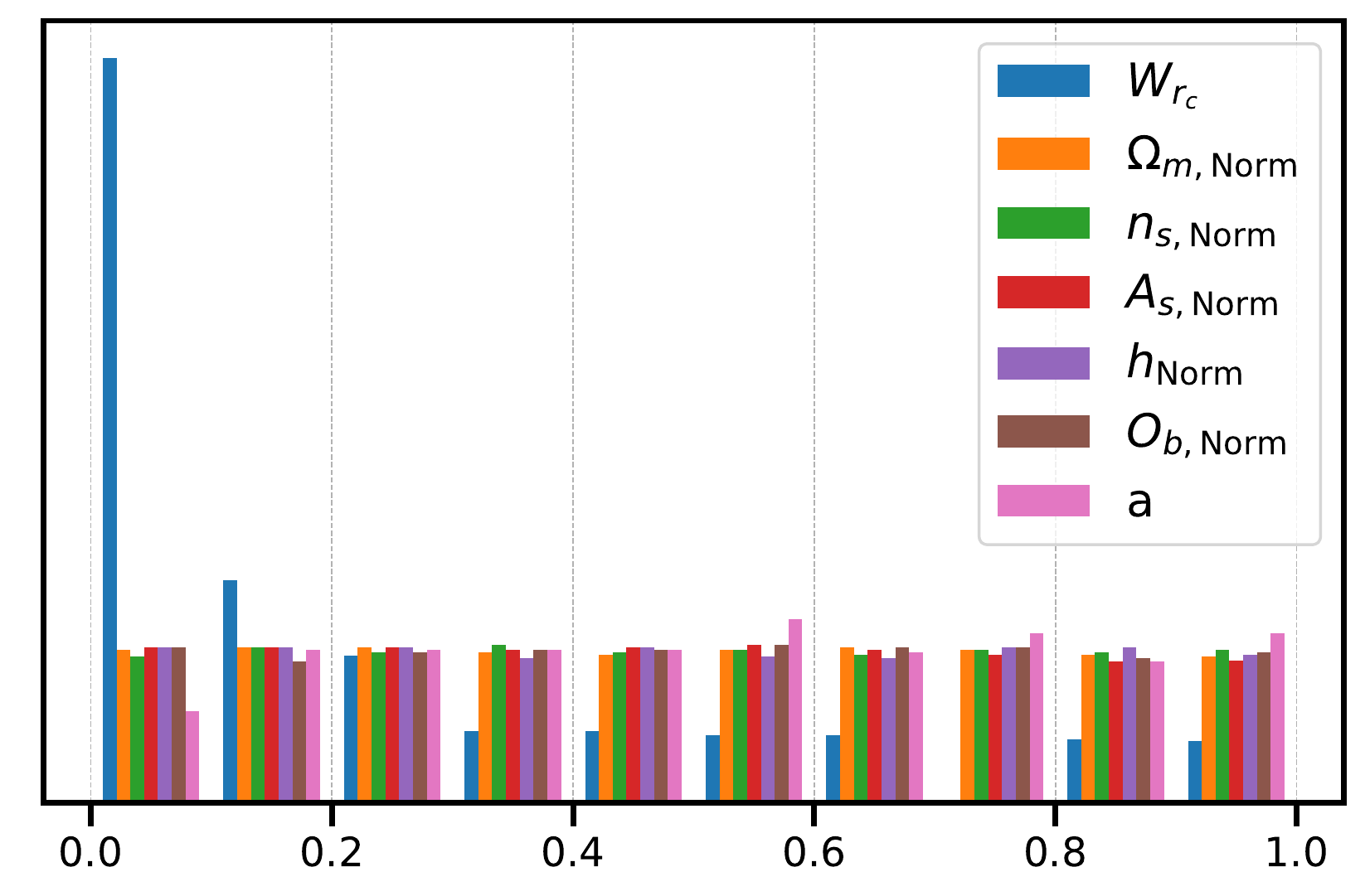}
\caption{\label{fig:SmoothAndNorm} \textit{On the left.} Comparison of the two smoothing techniques for the boost factor of nDGP with $H_0 r_c=2$. \textit{On the right.} Distribution of the normalised parameters in 10 bins between 0 and 1.}
\end{figure}

Due to the numerical methods used to estimate the power spectrum of the {\it N}-body particles, the nDGP boost factors that we obtain are sampled with a linear binning in $k$-space and are affected by finite resolution noise at small scales. This noise is non-physical and may worsen the interpolation accuracy, therefore we decide to smooth the boost factors. We compare two smoothing techniques: one based on a logarithmic re-sampling of the data, the other based on the Savitzky-Golay filter \cite{SavitzkyGolay}. In the left plot of Fig~\ref{fig:SmoothAndNorm} is shown a comparison of the two smoothing techniques for the central case of $H_0 r_c = 2$ and $z=0$. The two techniques give consistent results so we decide to perform the rest of the analysis with the Savitzky-Golay filter. 

Multi-Layer Perceptrons (MLP) work better when they deal with normalised input values \cite{NNbook}, hence we re-scale our parameters accordingly. We map the cosmological parameters in such a way that the minimum and maximum values they assume in parameter space are 0 and 1. The redshift is naturally mapped in the range $[0.25,1]$ by computing the scale factor. We finally convert the MG parameter $H_0 r_c$ into $W_{r_c} \equiv \frac{0.02}{H_0 r_c}$ which spans $[0,1]$ for our $H_0 r_c$ range. In the right plot of Fig~\ref{fig:SmoothAndNorm}, the distribution of the normalised parameters is summarised in 10 linear bins between 0 and 1.

We convert the boost factor with the formula
\begin{equation}
     B_{\rm Norm}(k) = -\frac{\log_{10}(B(k) - 0.999)}{3} \, ,
\end{equation}
to give more importance to small deviations from GR than to larger deviations.
The simplicity and self-similarity of the nDGP boost factor curve suggest that the 81 $k$-modes that we are using to represent the boost factor are probably unnecessary. After subtracting the mean value of the boost factor from each element of the dataset, we perform a Principal Component Analysis (PCA) \cite{Jolliffe2002} to study if the variance of the training set is significantly larger in some $k$-modes-space directions than in others. We find that the component with the greatest variance accounts for more than $99.9 \%$ of the variance, while by using the first and second components it is possible to describe $\sim 99.99 \%$ of the variance. This means that we can reduce the dimensionality of the problem from 81 output values to 2 output values losing only $\sim 0.01 \%$ of the information. 
\subsection{Emulator}
\label{Sec:Emulator}
\begin{figure}
\centering 
\includegraphics[width=.8\textwidth]{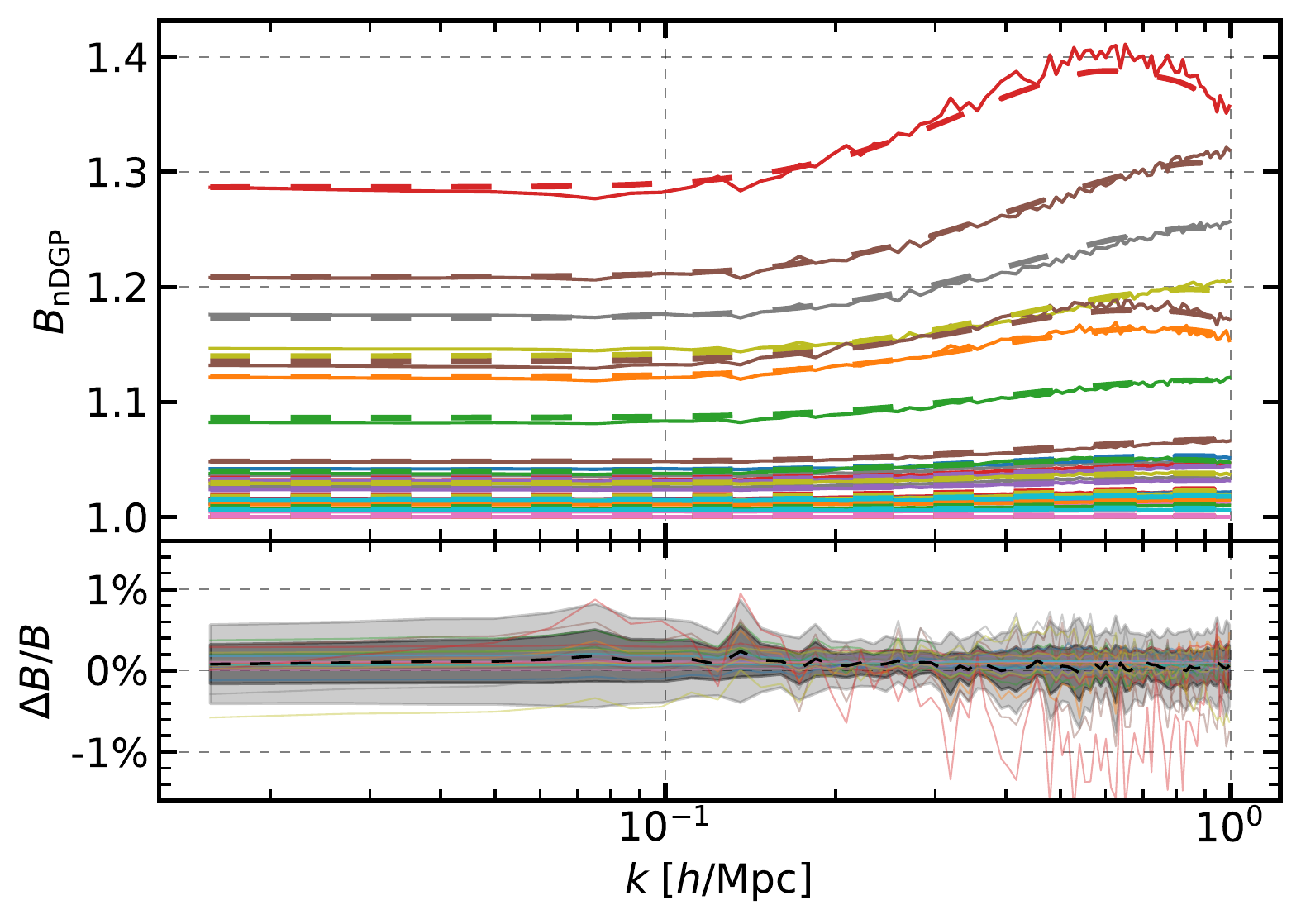}
\caption{\label{fig:EmulatorPerformance} Comparison between boost factors of the test set (solid lines) and emulated boost factors (dashed lines) for 30 examples randomly chosen in the test set. The residuals between emulated and true boost factors (coloured solid lines) are shown in the bottom panel together with the mean (dashed black line) and variance ($1 \sigma$ and $2 \sigma$ shaded regions) of the emulation error on the test-set.}
\end{figure}

We train an MLP with one hidden layer of 100 nodes activated by hyperbolic tangent functions with the limited-memory Broyden–Fletcher–Goldfarb–Shanno \cite{Fletcher1988PracticalMO} optimiser\footnote{This iterative optimisation technique exploits a numerical estimate of the Hessian based only on gradient evaluations to determine the descent direction, which makes the technique computationally more efficient than the Newton's method.} on the training set. This converges in less than 5000 iterations, producing the nDGP emulator that we benchmark on the test set. In the top panel of figure~\ref{fig:EmulatorPerformance} we show some boost factors examples taken from the test-set (solid lines) and compared with the emulator's predictions (dashed lines). It is worth noticing that the emulated boost factors are free of high-frequency noise thanks to the smoothing discussed in sub-section~\ref{ssec:SimsAndData}. In the bottom panel of figure~\ref{fig:EmulatorPerformance} we show the mean (black dashed line) and variance ($1 \sigma$ and $2 \sigma$ shaded regions) of the emulation error on the test-set together with the residuals for the 30 examples shown in the top panel. The $2 \sigma$ contours are well within the $1\%$ threshold at all scales. 

\begin{figure}
\centering 
\includegraphics[width=.8\textwidth]{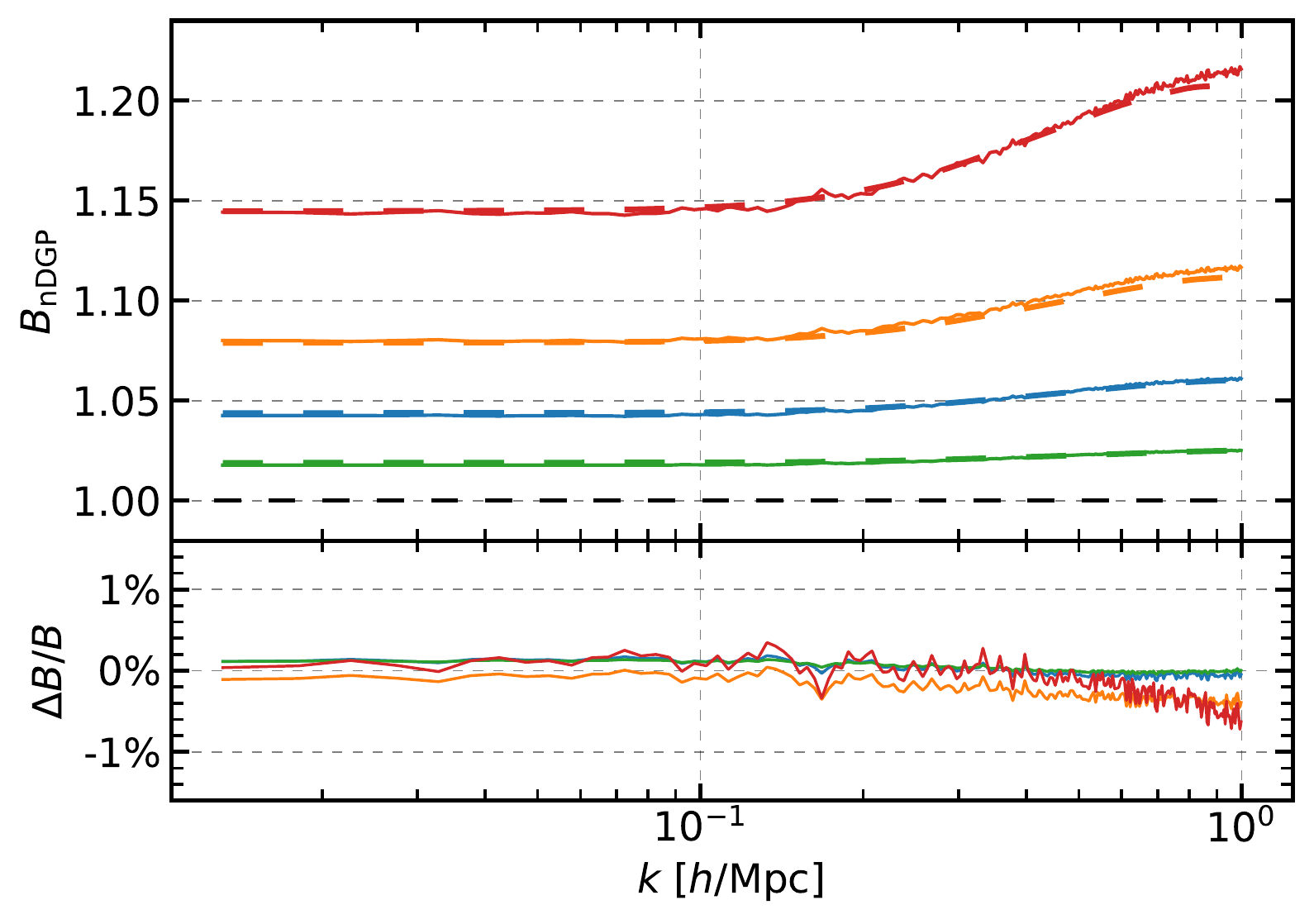}
\caption{\label{fig:EmulatorVsArepo} Comparison between emulator predictions (dashed lines) and Arepo results (solid lines) of the nDGP boost factor at redshift $z=1$ for 4 values of the parameter $H_0 r_c$: 0.5 (in red), 1 (in orange), 2 (in blue) and 5 (in green). The bottom panel shows the residuals.}
\end{figure}

As an additional test of the emulation accuracy, we compare the emulator's predictions with the boost factors computed from Arepo simulations in nDGP gravity using 4 values of $H_0 r_c$: 0.5, 1, 2, 5. The results are shown in figure~\ref{fig:EmulatorVsArepo}. The emulator predicts the nDGP boost factor of Arepo with sub-percent accuracy in all cases and at all the scales considered. This is made possible by the excellent agreement between the nDGP boost factors of COLA and Arepo simulations but also by the linear-theory correction of eq~\eqref{GrowthLinCorr} applied to the nDGP boost factors of COLA simulations started at redshift $z_{ini}=19$.

The biggest computational cost behind the nDGP emulator comes from running the COLA simulations necessary to produce the power spectra in GR and nDGP theories ($\sim 10^4$ CPU hours). The data processing applied to the power spectra in order to produce the training set has a minor impact on the computational cost ($\sim10$ CPU minutes). Also, the training of the emulator is relatively fast ($\sim1$ CPU hour). Once the model is trained, predictions are extremely fast. On an average single-core CPU, the model can produce $\sim 5000$ predictions in a second. The prediction speed of emulators is key to enabling fast cosmological inference with Monte Carlo techniques which require $10^5-10^6$ evaluations \cite{Donald-McCann:2021nxc,Donald-McCann:2022pac}. 

As we have shown here, DM power spectrum emulators can be extended to MG models, and in particular to nDGP gravity, with modest computational cost thanks to COLA simulations, enabling percent accurate and extremely fast predictions of the matter power spectrum up to $k=1 \mpcoh$.

\chapter{Summary and conclusions} 
\label{chp:conclusion}

\begin{quote}
    {\it The content of this chapter is based on the publications \cite{Fiorini:2021dzs, Fiorini:2022srj, Brando:2022gvg}. } 
\end{quote}

In this thesis, we discussed our strategy to bridge the gap between theory and observation, in order to constrain gravity with the upcoming stage IV galaxy surveys. 

In chapter~\ref{chp:introduction}, after introducing the $\Lambda$CDM model and the challenges it faces we focused on a possible solution to the late time accelerated expansion, modifications of gravity, and studied the common features of the yet unconstrained theories, the screening mechanisms. In a Newtonian context, we studied how structure formation is driven by gravity in the linear regime and discussed how the large-scale structure that we observe today has formed, focusing on the collapse of dark matter halos and the models connecting halos and galaxies, explaining why they are biased tracers of the underlying matter density field. We then analysed the problems and opportunities connected with the fact that we directly observe the redshift-space position of galaxies which is affected by their peculiar velocities. Finally, we briefly reviewed the techniques and the codes that allow us to study structure formation in the non-linear regime, the \nbody{} techniques. An approximate version of these is the main focus of our approach, the COLA simulations method that we used in place of \nbody{} simulations to produce theoretical predictions of structure formation in the non-linear regime.


In chapter~\ref{chp:mocks}, we extended, applied and validated the HOD formalism to COLA simulations in MG, enabling the fast generation of mock galaxy catalogues that are required to test gravity on cosmological scales in Stage IV surveys. 


Using a full {\it N}-body simulation suite as a benchmark, we produced an equivalent simulation suite with \textcode{mg-picola}, focusing on the following gravity models: GR, braneworld nDGP model (N1) and $n=1$ Hu-Sawicki $f(R)$ gravity (F5). Looking at the power spectra of the density and velocity divergence fields, we confirmed the good accuracy of COLA in reproducing the density field (agreement within the variance up to $k \sim 1 \hompc$) and have given the first measure of its accuracy in the velocity field (within $4 \%$ up to $k \sim 0.4 \hompc$).

To apply the HOD formalism to COLA simulations, we first need to employ a halo-finding algorithm to find dark matter halos. We found that the \textcode{rockstar} halo-finder, which uses the six-dimensional phase-space information to find halos, did not perform well with our COLA simulations with the default setting.
On the other hand, by finding halos with a simple FoF finder and converting the halo mass with the results of \cite{Lukic:2008ds} and \cite{Dutton:2014xda}, we have produced halo catalogues in COLA and {\it N}-body which are in good agreement, as shown by the hmf (better than $10\%$ agreement) and the halo power spectrum (agreement within the variance).

Since COLA simulations do not resolve the internal halo structure, we have assumed the NFW profile and the concentration-mass relation of \cite{Dutton:2014xda} in GR. To take into account the effect of the fifth force on the internal structure of halos, we have measured the boost factors of the concentration and velocity dispersion mass relations in MG with \textcode{rockstar} halo catalogues in {\it N}-body. We have found that no significant difference is present in the N1 boost factors, which allows us to use the standard NFW profile in N1. In F5, instead, there is a clear transition from screened to un-screened halos both in the concentration and the velocity dispersion boost factors. This is in line with the findings of \cite{Mitchell:2018qrg, Mitchell:2019qke}. Using simple fitting formulae that we calibrated on these results, we have been able to assign realistic concentration and velocity dispersion to halos in F5 depending on their mass.

With the HOD model proposed in \cite{Zheng:2007zg}, we have fitted the HOD parameters of the different models to match the galaxy clustering of {\it N}-body catalogues in GR. The fit performed with a simplex search algorithm is mostly driven by the central galaxies, whose final distributions in velocity are different in each model but similar between COLA and {\it N}-body. Despite that the objective function for the fit is designed to minimise the difference of the monopole and the quadrupole of the galaxy power spectrum with the target signals, we have shown that MGs leave a characteristic imprint in RSD statistics, especially in the quadrupole. To estimate the constraining power of galaxy statistics, we have computed the $\chi^2$ of the difference between MG and GR multipoles. 
We confirmed that non-linear scales are important in constraining F5, while N1 can be constrained on large scales due to the scale-independent modification of the growth rate. We also found a good agreement between COLA and {\it N}-body galaxy statistics. 

In chapter~\ref{chp:statistics}, using the simulation output of chapter~\ref{chp:mocks}, we have investigated the effects of MG on the estimator for the real space power spectrum $Q_0$, bispectrum and voids while validating COLA simulations for these probes.

In section \ref{sec:Q0} we compared the power spectrum orthogonal to the line-of-sight estimated by means of the truncated sum of eq.~\eqref{Q0ofP024}, $Q_0$, with the real space power spectrum $P^r$ in our halo and galaxy catalogues. 
For halos, we showed that $Q_0$ is a good estimator for the real space power spectrum in the full range of scales considered and for all gravity models.
For galaxies, we showed that the agreement between $Q_0$ and $P^r$ deteriorated after $k\sim 0.25 \hompc$, meaning that higher multipoles need to be considered to recover $P^r$ at smaller scales. However, we found that the MG boost factors of $Q_0$ and $P^r$ were in excellent agreement at all scales, indicating that most of the MG information is contained in the first three even multipole moments. We determined that the accuracy of COLA is within the variance up to $k \sim 0.24 \hompc$ in GR and at all scales in the MG boost factors.

We examined the bispectrum of DM and galaxies in redshift space in section~\ref{sec:Bispectrum}, to understand how much of the MG bispectrum information filters through the HOD tuning. We compared results for bispectrum and reduced bispectrum to discriminate the part of the bispectrum signal due to the power spectrum from the one due to non-linearity. At the DM level, we showed that the significant MG signals in the bispectrum are due to the enhanced MG power spectrum as proven by the much weaker signal in the reduced bispectrum. We found that COLA achieved $\sim 10\%$ accuracy in reproducing the {\it N}-body results in GR and $\sim 2\%$ in the N1 boost factor. In F5, we discovered that COLA simulations did not accurately reproduce the {\it N}-body bispectrum boost factors. We concluded that they lacked the contribution coming from non-linearity in the screening mechanism, which is neglected in the screening approximation used in COLA by linearising the scalar field equation. We highlighted this also by showing that COLA was inaccurate in the configuration dependence in F5, while it captured the {\it N}-body configuration dependence in N1. 
For galaxies in redshift-space, we have confirmed that HOD tuning can hide the strong MG signal in the DM bispectrum coming from the power spectrum (only $\sim 5 \%$ deviations from GR after the HOD tuning). We found that COLA's accuracy is better than $10 \%$ in GR and $\sim 5\%$ in the MG boost factors. By looking at the configuration dependence of the boost factors for the bispectrum of galaxies in redshift space, we highlighted that this was dissimilar to the one observed in DM and concluded that the MG signal was dominated by non-linearity coming from galaxy bias. Due to this, the inaccuracy of COLA simulations in F5 for DM is largely hidden and COLA reproduced the configuration dependence of {\it N}-body simulations well.

In section~\ref{sec:Voids} we analysed the voids obtained from COLA and {\it N}-body galaxy catalogues with the \codeword{ZOBOV} void-finder. In order to test the applicability of the linear model of \cite{Nadathur:2017jos} to the case of MG, we carefully tracked the MG effects in each key component involved in the modelling. We showed that significant MG signals were present in the integrated-density, radial-velocity and velocity-dispersion profiles in N1, while in F5 only the velocity-dispersion profile showed a substantial enhancement compared to GR. We have not found any significant MG effect on the real space void-galaxy CCF, but by examining the monopole and quadrupole of the void-galaxy CCF in redshift space, we discovered that the RSD in N1 were significantly stronger than in GR, while in F5 they were not substantially different from the GR case. To understand the effects of MG in the RSD modelling, we compared the predictions formulated using different combinations of theory components estimated from GR and MG simulations. We concluded that the enhancements of the integrated-density profile and velocity-dispersion profile observed in N1 partially cancelled out and the linear growth rate played the leading role in enhancing the RSD in the void-galaxy CCF. Using the linear growth rate as a free parameter of the theory, we fitted the monopole and the quadrupole of the void-galaxy CCF in redshift space. We found that the best fit values were recovered within $\sim 1 \sigma$ of the reference value computed from the linear theory in GR and N1, even when using the void profiles from GR simulations in the theoretical predictions.
Concerning the accuracy of COLA in reproducing {\it N}-body results, we determined that COLA and {\it N}-body were consistent within the variance for all the void summary statistics except for the velocity-dispersion boost factor in N1 where COLA's accuracy is at $\sim 2\%$ level. We have found that the best fit values of the linear growth rate in COLA were consistent within $\sim 1\sigma$ with {\it N}-body and with theory predictions.
We note that we used voids identified in real space galaxy mocks so that the real space void positions were known. When we use real observational data, they need to be recovered by applying a density-field reconstruction method \cite{Nadathur:2018pjn}. The reconstruction requires an assumption on the gravitational theory and there will be an additional MG effect in this process.

In chapter~\ref{chp:powerspectrum} we have studied how COLA simulations can be used to extend matter power spectrum emulators to MG theories. To this end, we performed convergence tests, explored the ability of COLA in reproducing the response to changes in cosmological parameters and presented an extension of matter power spectrum emulators to nDGP theories.

We studied the convergence of COLA simulations for predictions of the matter power spectrum in section~\ref{sec:Convergence}. We argued that COLA converges to PM-only simulations increasing the time resolution, and referred to results obtained in this limit as PM results. In passing, we stressed how a relevant comparison in literature \cite{Klypin:2017iwu} ignored the intrinsic difference between COLA and PM-only simulations, making the comparison unfair. By comparing high force-resolution PM simulations with Bacco emulator (trained with high mass-resolution simulations) and low mass-resolution N-body simulations (Arepo) for predictions of the matter power spectrum at redshift $z=1$, we highlighted the importance of mass resolution, responsible for $~4\%$ differences at $k=1\hompc$, while showing that PM simulations reproduce the reference results in both the low mass-resolution and high-mass-resolution cases with $1\%$ accuracy up to $k\sim 2 \hompc$. After that, we produced a suite of simulations in $\Lambda$CDM with the same cosmology but with different numbers of particles, force-grids and time-steps, and using high-resolution PM simulations as a reference, we tested the relative convergence of COLA simulations. We concluded that the power spectra converge towards the reference only when mass-resolution, force-resolution and time-resolution are increased accordingly.

In order to test a new way of extending cosmological emulators with COLA simulations, in section~\ref{sec:Response}, we have made use of the non-linear response to the change of each of the following cosmological parameters: the total matter energy-density, spectral index and amplitude of the primordial fluctuations. We analysed how the ratio between the non-linear and linear response functions, computed using COLA, is affected by varying the cosmological parameters to large deviation values. To investigate the agreement between COLA, Bacco and EE2 we showed the ratio between the non-linear response function computed using each method, and we concluded that COLA and EE2 predictions agree at the $2\%$ level up until $k=1 \ \hompc$, while Bacco and EE2 predictions agree at the $2\%$ level at all scales considered. For small variations of the cosmological parameters (0.5$\%$ variations), the agreement is even better, being of $\sim 0.1\%$ with a similar agreement present also between Bacco and EE2. We concluded that emulating the response function can mitigate the resolution issues that affect the full power spectrum and suggested that future emulation projects adopt this approach. 

To showcase the potentiality of COLA for extending cosmological emulators to MG theories, we developed an nDGP boost-factor emulator of the matter power spectrum in section~\ref{sec:Boost}. By comparing the nDGP boost-factor for the matter power spectrum between high-resolution PM simulations and Arepo simulations, we showed that their agreement is better than $0.5\%$ for all the models and scales considered when the PM simulations are started at the same redshift ($z_{\rm ini} = 127$) of the Arepo simulations. To get accurate boost factors from COLA simulations started at the usual redshift ($z_{\rm ini} = 19$), we proposed and validated the linear theory correction in eq~\eqref{GrowthLinCorr} for the nDGP boost-factors to include MG effects on the linear growth factor at $z > z_{\rm ini}$ in the back-scaling approach. We then investigated the minimum simulation requirements by using high-resolution PM simulations as a reference. Thanks to relative-convergence comparisons we concluded that the nDGP boost factors of low-resolution COLA simulations were already converged in our target parameter space. 
We also studied the response of the nDGP boost factors to the change of cosmological parameters finding weak cosmological dependence of the nDGP boost factor and confirmed that the cosmological dependence becomes weaker for smaller deviations from $\Lambda$CDM (larger values of $H_0 r_c$). In sub-section~\ref{ssec:SimsAndData}, we detailed our simulation strategy to compute the nDGP boost factors and examined the computational cost consisting of a total of $\approx 8 \cdot 10^{3}$ CPU hours. 
After smoothing the signals and remapping the parameters in a unitary interval, we performed a PCA and selected only 2 components while preserving $99.99 \%$ of the information. We then used the resulting data set to train a neural network and bench-marked its accuracy on an independent test set ($<1\%$ at all scales considered). As an additional test, we compared the emulator predictions with the Arepo results and found sub-percent agreement. We concluded that COLA simulations can be used to accurately extend matter power spectrum emulators to MG theories up to (at least) $k=1\hompc$.

The accuracy of COLA galaxy catalogues may be further improved by tuning the COLA halo catalogues to reproduce an analytic halo mass function. 
This can be achieved by introducing a mass-dependent conversion factor when converting $M_{\rm FoF}$ to $M_{\rm200c}$. However, this requires an accurate measurement of the mass function from {\it N}-body simulations in MG, which undermines the predictability of COLA, one of the major advantages that COLA has over other approximate methods.

The use of multi-probes analysis will be key in breaking the degeneracies of the theory parameter space, enabling tight constraints of gravity on cosmological scales with next-generation experiments. The findings of chapter~\ref{chp:statistics} are important to understanding the potential of $Q_0$, bispectrum and void-galaxy CCF in reducing the freedom of the galaxy bias model and in increasing the sensitivity to the gravity model. Furthermore, the validation of COLA simulations for these probes gives us a computationally cheaper alternative to full {\it N}-body simulations to formulate reliable theoretical predictions, bearing in mind the limitations of the approximations used in COLA simulations. This opens up the possibility to train beyond-$\Lambda$CDM cosmological emulators for these additional observables in sight of Stage IV surveys.

To maximise the scientific return of stage IV surveys it is crucial to access the deep non-linear regime. As we proved in chapter~\ref{chp:powerspectrum}, the parameters of COLA simulations can be tweaked to reach the desired precision (within the limits of the computing power available) for predictions of the matter power spectrum in GR. The accuracy of COLA in MG theories, instead, can be affected by the inaccuracies of the screening approximations employed. These depend on the specific screening mechanism and are based on the spherical approximation as discussed in section~\ref{ScreeningMechanisms}. It would be interesting in future to study more accurate screening approximations in order to overcome this limit.

Finally, the recent extension of COLA to Horndeski theories~\cite{Wright:2022krq} gives us access to a wider class of gravity models that are yet unconstrained on non-linear scales. In future, we plan to extend our pipeline and validate COLA simulations for the Horndeski models, as well as train cosmological emulators in a wider region of the modified gravity parameter space.


\begin{spacing}{0.9}


\bibliographystyle{JHEP}
\cleardoublepage
\bibliography{References} 

\providecommand{\href}[2]{#2}\begingroup\raggedright\begin{thebibliography}{100}

\bibitem{Angulo:2020vky}
R.E.~Angulo, M.~Zennaro, S.~Contreras, G.~Aric\`o, M.~Pellejero-Iba\~nez and
  J.~St\"ucker, \emph{{The BACCO simulation project: exploiting the full power
  of large-scale structure for cosmology}},
  \href{https://doi.org/10.1093/mnras/stab2018}{\emph{Mon. Not. Roy. Astron.
  Soc.} {\bfseries 507} (2021) 5869}
  [\href{https://arxiv.org/abs/2004.06245}{{\ttfamily 2004.06245}}].

\bibitem{Manera:2012sc}
M.~Manera et~al., \emph{{The clustering of galaxies in the SDSS-III Baryon
  Oscillation Spectroscopic Survey: a large sample of mock galaxy catalogues}},
  \href{https://doi.org/10.1093/mnras/sts084}{\emph{Mon. Not. Roy. Astron.
  Soc.} {\bfseries 428} (2012) 1036}
  [\href{https://arxiv.org/abs/1203.6609}{{\ttfamily 1203.6609}}].

\bibitem{Fiorini:2021dzs}
B.~Fiorini, K.~Koyama, A.~Izard, H.A.~Winther, B.S.~Wright and B.~Li,
  \emph{{Fast generation of mock galaxy catalogues in modified gravity models
  with COLA}}, \href{https://doi.org/10.1088/1475-7516/2021/09/021}{\emph{JCAP}
  {\bfseries 09} (2021) 021}
  [\href{https://arxiv.org/abs/2106.05197}{{\ttfamily 2106.05197}}].

\bibitem{Planck:2018vyg}
{\scshape Planck} collaboration, \emph{{Planck 2018 results. VI. Cosmological
  parameters}},
  \href{https://doi.org/10.1051/0004-6361/201833910}{\emph{Astron. Astrophys.}
  {\bfseries 641} (2020) A6}
  [\href{https://arxiv.org/abs/1807.06209}{{\ttfamily 1807.06209}}].

\bibitem{Beutler:2011hx}
F.~Beutler, C.~Blake, M.~Colless, D.H.~Jones, L.~Staveley-Smith, L.~Campbell
  et~al., \emph{{The 6dF Galaxy Survey: Baryon Acoustic Oscillations and the
  Local Hubble Constant}},
  \href{https://doi.org/10.1111/j.1365-2966.2011.19250.x}{\emph{Mon. Not. Roy.
  Astron. Soc.} {\bfseries 416} (2011) 3017}
  [\href{https://arxiv.org/abs/1106.3366}{{\ttfamily 1106.3366}}].

\bibitem{Ross:2014qpa}
A.J.~Ross, L.~Samushia, C.~Howlett, W.J.~Percival, A.~Burden and M.~Manera,
  \emph{{The clustering of the SDSS DR7 main Galaxy sample \textendash{} I. A 4
  per cent distance measure at $z = 0.15$}},
  \href{https://doi.org/10.1093/mnras/stv154}{\emph{Mon. Not. Roy. Astron.
  Soc.} {\bfseries 449} (2015) 835}
  [\href{https://arxiv.org/abs/1409.3242}{{\ttfamily 1409.3242}}].

\bibitem{BOSS:2016wmc}
{\scshape BOSS} collaboration, \emph{{The clustering of galaxies in the
  completed SDSS-III Baryon Oscillation Spectroscopic Survey: cosmological
  analysis of the DR12 galaxy sample}},
  \href{https://doi.org/10.1093/mnras/stx721}{\emph{Mon. Not. Roy. Astron.
  Soc.} {\bfseries 470} (2017) 2617}
  [\href{https://arxiv.org/abs/1607.03155}{{\ttfamily 1607.03155}}].

\bibitem{SupernovaSearchTeam:1998fmf}
{\scshape Supernova Search Team} collaboration, \emph{{Observational evidence
  from supernovae for an accelerating universe and a cosmological constant}},
  \href{https://doi.org/10.1086/300499}{\emph{Astron. J.} {\bfseries 116}
  (1998) 1009} [\href{https://arxiv.org/abs/astro-ph/9805201}{{\ttfamily
  astro-ph/9805201}}].

\bibitem{SupernovaCosmologyProject:1998vns}
{\scshape Supernova Cosmology Project} collaboration, \emph{{Measurements of
  $\Omega$ and $\Lambda$ from 42 high redshift supernovae}},
  \href{https://doi.org/10.1086/307221}{\emph{Astrophys. J.} {\bfseries 517}
  (1999) 565} [\href{https://arxiv.org/abs/astro-ph/9812133}{{\ttfamily
  astro-ph/9812133}}].

\bibitem{Riess:2020fzl}
A.G.~Riess, S.~Casertano, W.~Yuan, J.B.~Bowers, L.~Macri, J.C.~Zinn et~al.,
  \emph{{Cosmic Distances Calibrated to 1\% Precision with Gaia EDR3 Parallaxes
  and Hubble Space Telescope Photometry of 75 Milky Way Cepheids Confirm
  Tension with $\Lambda$CDM}},
  \href{https://doi.org/10.3847/2041-8213/abdbaf}{\emph{Astrophys. J. Lett.}
  {\bfseries 908} (2021) L6}
  [\href{https://arxiv.org/abs/2012.08534}{{\ttfamily 2012.08534}}].

\bibitem{Troster:2019ean}
T.~Tr\"oster et~al., \emph{{Cosmology from large-scale structure: Constraining
  $\Lambda$CDM with BOSS}},
  \href{https://doi.org/10.1051/0004-6361/201936772}{\emph{Astron. Astrophys.}
  {\bfseries 633} (2020) L10}
  [\href{https://arxiv.org/abs/1909.11006}{{\ttfamily 1909.11006}}].

\bibitem{DES:2017qwj}
{\scshape DES} collaboration, \emph{{Dark Energy Survey Year 1 results:
  Cosmological constraints from cosmic shear}},
  \href{https://doi.org/10.1103/PhysRevD.98.043528}{\emph{Phys. Rev. D}
  {\bfseries 98} (2018) 043528}
  [\href{https://arxiv.org/abs/1708.01538}{{\ttfamily 1708.01538}}].

\bibitem{Hildebrandt:2018yau}
H.~Hildebrandt et~al., \emph{{KiDS+VIKING-450: Cosmic shear tomography with
  optical and infrared data}},
  \href{https://doi.org/10.1051/0004-6361/201834878}{\emph{Astron. Astrophys.}
  {\bfseries 633} (2020) A69}
  [\href{https://arxiv.org/abs/1812.06076}{{\ttfamily 1812.06076}}].

\bibitem{HSC:2018mrq}
{\scshape HSC} collaboration, \emph{{Cosmology from cosmic shear power spectra
  with Subaru Hyper Suprime-Cam first-year data}},
  \href{https://doi.org/10.1093/pasj/psz010}{\emph{Publ. Astron. Soc. Jap.}
  {\bfseries 71} (2019) 43} [\href{https://arxiv.org/abs/1809.09148}{{\ttfamily
  1809.09148}}].

\bibitem{Raveri:2019mxg}
M.~Raveri, \emph{{Reconstructing Gravity on Cosmological Scales}},
  \href{https://doi.org/10.1103/PhysRevD.101.083524}{\emph{Phys. Rev. D}
  {\bfseries 101} (2020) 083524}
  [\href{https://arxiv.org/abs/1902.01366}{{\ttfamily 1902.01366}}].

\bibitem{Lovelock:1971yv}
D.~Lovelock, \emph{{The Einstein tensor and its generalizations}},
  \href{https://doi.org/10.1063/1.1665613}{\emph{J. Math. Phys.} {\bfseries 12}
  (1971) 498}.

\bibitem{Lovelock:1972vz}
D.~Lovelock, \emph{{The four-dimensionality of space and the einstein tensor}},
  \href{https://doi.org/10.1063/1.1666069}{\emph{J. Math. Phys.} {\bfseries 13}
  (1972) 874}.

\bibitem{Li:2020uaz}
B.~Li and K.~Koyama, \emph{{Modified Gravity}}, WSP (2020),
  \href{https://doi.org/10.1142/11090}{10.1142/11090}.

\bibitem{Will:2014kxa}
C.M.~Will, \emph{{The Confrontation between General Relativity and
  Experiment}}, \href{https://doi.org/10.12942/lrr-2014-4}{\emph{Living Rev.
  Rel.} {\bfseries 17} (2014) 4}
  [\href{https://arxiv.org/abs/1403.7377}{{\ttfamily 1403.7377}}].

\bibitem{Joyce:2014kja}
A.~Joyce, B.~Jain, J.~Khoury and M.~Trodden, \emph{{Beyond the Cosmological
  Standard Model}},
  \href{https://doi.org/10.1016/j.physrep.2014.12.002}{\emph{Phys. Rept.}
  {\bfseries 568} (2015) 1} [\href{https://arxiv.org/abs/1407.0059}{{\ttfamily
  1407.0059}}].

\bibitem{Koyama:2015vza}
K.~Koyama, \emph{{Cosmological Tests of Modified Gravity}},
  \href{https://doi.org/10.1088/0034-4885/79/4/046902}{\emph{Rept. Prog. Phys.}
  {\bfseries 79} (2016) 046902}
  [\href{https://arxiv.org/abs/1504.04623}{{\ttfamily 1504.04623}}].

\bibitem{Koyama:2018som}
K.~Koyama, \emph{{Gravity beyond general relativity}},
  \href{https://doi.org/10.1142/S0218271818480012}{\emph{Int. J. Mod. Phys. D}
  {\bfseries 27} (2018) 1848001}.

\bibitem{Khoury:2003aq}
J.~Khoury and A.~Weltman, \emph{{Chameleon fields: Awaiting surprises for tests
  of gravity in space}},
  \href{https://doi.org/10.1103/PhysRevLett.93.171104}{\emph{Phys. Rev. Lett.}
  {\bfseries 93} (2004) 171104}
  [\href{https://arxiv.org/abs/astro-ph/0309300}{{\ttfamily
  astro-ph/0309300}}].

\bibitem{Khoury:2003rn}
J.~Khoury and A.~Weltman, \emph{{Chameleon cosmology}},
  \href{https://doi.org/10.1103/PhysRevD.69.044026}{\emph{Phys. Rev. D}
  {\bfseries 69} (2004) 044026}
  [\href{https://arxiv.org/abs/astro-ph/0309411}{{\ttfamily
  astro-ph/0309411}}].

\bibitem{Brax:2010gi}
P.~Brax, C.~van~de Bruck, A.-C.~Davis and D.~Shaw, \emph{{The Dilaton and
  Modified Gravity}},
  \href{https://doi.org/10.1103/PhysRevD.82.063519}{\emph{Phys. Rev. D}
  {\bfseries 82} (2010) 063519}
  [\href{https://arxiv.org/abs/1005.3735}{{\ttfamily 1005.3735}}].

\bibitem{Hinterbichler:2010es}
K.~Hinterbichler and J.~Khoury, \emph{{Symmetron Fields: Screening Long-Range
  Forces Through Local Symmetry Restoration}},
  \href{https://doi.org/10.1103/PhysRevLett.104.231301}{\emph{Phys. Rev. Lett.}
  {\bfseries 104} (2010) 231301}
  [\href{https://arxiv.org/abs/1001.4525}{{\ttfamily 1001.4525}}].

\bibitem{Babichev:2009ee}
E.~Babichev, C.~Deffayet and R.~Ziour, \emph{{k-Mouflage gravity}},
  \href{https://doi.org/10.1142/S0218271809016107}{\emph{Int. J. Mod. Phys. D}
  {\bfseries 18} (2009) 2147}
  [\href{https://arxiv.org/abs/0905.2943}{{\ttfamily 0905.2943}}].

\bibitem{Vainshtein:1972sx}
A.I.~Vainshtein, \emph{{To the problem of nonvanishing gravitation mass}},
  \href{https://doi.org/10.1016/0370-2693(72)90147-5}{\emph{Phys. Lett. B}
  {\bfseries 39} (1972) 393}.

\bibitem{Noller:2013wca}
J.~Noller, F.~von Braun-Bates and P.G.~Ferreira, \emph{{Relativistic scalar
  fields and the quasistatic approximation in theories of modified gravity}},
  \href{https://doi.org/10.1103/PhysRevD.89.023521}{\emph{Phys. Rev. D}
  {\bfseries 89} (2014) 023521}
  [\href{https://arxiv.org/abs/1310.3266}{{\ttfamily 1310.3266}}].

\bibitem{Winther:2014cia}
H.A.~Winther and P.G.~Ferreira, \emph{{Fast route to nonlinear clustering
  statistics in modified gravity theories}},
  \href{https://doi.org/10.1103/PhysRevD.91.123507}{\emph{Phys. Rev. D}
  {\bfseries 91} (2015) 123507}
  [\href{https://arxiv.org/abs/1403.6492}{{\ttfamily 1403.6492}}].

\bibitem{Burrage:2017qrf}
C.~Burrage and J.~Sakstein, \emph{{Tests of Chameleon Gravity}},
  \href{https://doi.org/10.1007/s41114-018-0011-x}{\emph{Living Rev. Rel.}
  {\bfseries 21} (2018) 1} [\href{https://arxiv.org/abs/1709.09071}{{\ttfamily
  1709.09071}}].

\bibitem{Desmond:2020gzn}
H.~Desmond and P.G.~Ferreira, \emph{{Galaxy morphology rules out
  astrophysically relevant Hu-Sawicki $f(R)$ gravity}},
  \href{https://doi.org/10.1103/PhysRevD.102.104060}{\emph{Phys. Rev. D}
  {\bfseries 102} (2020) 104060}
  [\href{https://arxiv.org/abs/2009.08743}{{\ttfamily 2009.08743}}].

\bibitem{Bartlett:2020tjd}
D.J.~Bartlett, H.~Desmond and P.G.~Ferreira, \emph{{Constraints on galileons
  from the positions of supermassive black holes}},
  \href{https://doi.org/10.1103/PhysRevD.103.023523}{\emph{Phys. Rev. D}
  {\bfseries 103} (2021) 023523}
  [\href{https://arxiv.org/abs/2010.05811}{{\ttfamily 2010.05811}}].

\bibitem{Barreira:2016ovx}
A.~Barreira, A.G.~S\'anchez and F.~Schmidt, \emph{{Validating estimates of the
  growth rate of structure with modified gravity simulations}},
  \href{https://doi.org/10.1103/PhysRevD.94.084022}{\emph{Phys. Rev. D}
  {\bfseries 94} (2016) 084022}
  [\href{https://arxiv.org/abs/1605.03965}{{\ttfamily 1605.03965}}].

\bibitem{Starobinsky:1980te}
A.A.~Starobinsky, \emph{{A New Type of Isotropic Cosmological Models Without
  Singularity}},
  \href{https://doi.org/10.1016/0370-2693(80)90670-X}{\emph{Phys. Lett. B}
  {\bfseries 91} (1980) 99}.

\bibitem{Hu:2007nk}
W.~Hu and I.~Sawicki, \emph{{Models of f(R) Cosmic Acceleration that Evade
  Solar-System Tests}},
  \href{https://doi.org/10.1103/PhysRevD.76.064004}{\emph{Phys. Rev. D}
  {\bfseries 76} (2007) 064004}
  [\href{https://arxiv.org/abs/0705.1158}{{\ttfamily 0705.1158}}].

\bibitem{Pogosian:2007sw}
L.~Pogosian and A.~Silvestri, \emph{{The pattern of growth in viable f(R)
  cosmologies}}, \href{https://doi.org/10.1103/PhysRevD.77.023503}{\emph{Phys.
  Rev. D} {\bfseries 77} (2008) 023503}
  [\href{https://arxiv.org/abs/0709.0296}{{\ttfamily 0709.0296}}].

\bibitem{winther15}
H.A.~{Winther} and P.G.~{Ferreira}, \emph{{Fast route to nonlinear clustering
  statistics in modified gravity theories}},
  \href{https://doi.org/10.1103/PhysRevD.91.123507}{\emph{\prd} {\bfseries 91}
  (2015) 123507} [\href{https://arxiv.org/abs/1403.6492}{{\ttfamily
  1403.6492}}].

\bibitem{Dvali:2000hr}
G.R.~Dvali, G.~Gabadadze and M.~Porrati, \emph{{4-D gravity on a brane in 5-D
  Minkowski space}},
  \href{https://doi.org/10.1016/S0370-2693(00)00669-9}{\emph{Phys. Lett. B}
  {\bfseries 485} (2000) 208}
  [\href{https://arxiv.org/abs/hep-th/0005016}{{\ttfamily hep-th/0005016}}].

\bibitem{Schmidt:2009sg}
F.~Schmidt, \emph{{Self-Consistent Cosmological Simulations of DGP Braneworld
  Gravity}}, \href{https://doi.org/10.1103/PhysRevD.80.043001}{\emph{Phys. Rev.
  D} {\bfseries 80} (2009) 043001}
  [\href{https://arxiv.org/abs/0905.0858}{{\ttfamily 0905.0858}}].

\bibitem{Bag:2018jle}
S.~Bag, S.S.~Mishra and V.~Sahni, \emph{{Emulating a
  \ensuremath{\Lambda}CDM-like expansion on the phantom brane}},
  \href{https://doi.org/10.1103/PhysRevD.97.123537}{\emph{Phys. Rev. D}
  {\bfseries 97} (2018) 123537}
  [\href{https://arxiv.org/abs/1807.00684}{{\ttfamily 1807.00684}}].

\bibitem{Koyama:2005kd}
K.~Koyama and R.~Maartens, \emph{{Structure formation in the dgp cosmological
  model}}, \href{https://doi.org/10.1088/1475-7516/2006/01/016}{\emph{JCAP}
  {\bfseries 01} (2006) 016}
  [\href{https://arxiv.org/abs/astro-ph/0511634}{{\ttfamily
  astro-ph/0511634}}].

\bibitem{Schmidt:2009yj}
F.~Schmidt, W.~Hu and M.~Lima, \emph{{Spherical Collapse and the Halo Model in
  Braneworld Gravity}},
  \href{https://doi.org/10.1103/PhysRevD.81.063005}{\emph{Phys. Rev. D}
  {\bfseries 81} (2010) 063005}
  [\href{https://arxiv.org/abs/0911.5178}{{\ttfamily 0911.5178}}].

\bibitem{Bertschinger:1998tv}
E.~Bertschinger, \emph{{Simulations of structure formation in the universe}},
  \href{https://doi.org/10.1146/annurev.astro.36.1.599}{\emph{Ann. Rev. Astron.
  Astrophys.} {\bfseries 36} (1998) 599}.

\bibitem{Bernardeau:2001qr}
F.~Bernardeau, S.~Colombi, E.~Gaztanaga and R.~Scoccimarro, \emph{{Large scale
  structure of the universe and cosmological perturbation theory}},
  \href{https://doi.org/10.1016/S0370-1573(02)00135-7}{\emph{Phys. Rept.}
  {\bfseries 367} (2002) 1}
  [\href{https://arxiv.org/abs/astro-ph/0112551}{{\ttfamily
  astro-ph/0112551}}].

\bibitem{Dodelson:2003ft}
S.~Dodelson, \emph{{Modern Cosmology}}, Academic Press, Amsterdam (2003).

\bibitem{Wechsler:2018pic}
R.H.~Wechsler and J.L.~Tinker, \emph{{The Connection between Galaxies and their
  Dark Matter Halos}},
  \href{https://doi.org/10.1146/annurev-astro-081817-051756}{\emph{Ann. Rev.
  Astron. Astrophys.} {\bfseries 56} (2018) 435}
  [\href{https://arxiv.org/abs/1804.03097}{{\ttfamily 1804.03097}}].

\bibitem{Zeldovich:1969sb}
Y.B.~Zeldovich, \emph{{Gravitational instability: An Approximate theory for
  large density perturbations}}, {\emph{Astron. Astrophys.} {\bfseries 5}
  (1970) 84}.

\bibitem{Valogiannis:2016ane}
G.~Valogiannis and R.~Bean, \emph{{Efficient simulations of large scale
  structure in modified gravity cosmologies with comoving Lagrangian
  acceleration}}, \href{https://doi.org/10.1103/PhysRevD.95.103515}{\emph{Phys.
  Rev. D} {\bfseries 95} (2017) 103515}
  [\href{https://arxiv.org/abs/1612.06469}{{\ttfamily 1612.06469}}].

\bibitem{Winther:2017jof}
H.A.~Winther, K.~Koyama, M.~Manera, B.S.~Wright and G.-B.~Zhao, \emph{{COLA
  with scale-dependent growth: applications to screened modified gravity
  models}}, \href{https://doi.org/10.1088/1475-7516/2017/08/006}{\emph{JCAP}
  {\bfseries 08} (2017) 006}
  [\href{https://arxiv.org/abs/1703.00879}{{\ttfamily 1703.00879}}].

\bibitem{Aviles:2017aor}
A.~Aviles and J.L.~Cervantes-Cota, \emph{{Lagrangian perturbation theory for
  modified gravity}},
  \href{https://doi.org/10.1103/PhysRevD.96.123526}{\emph{Phys. Rev. D}
  {\bfseries 96} (2017) 123526}
  [\href{https://arxiv.org/abs/1705.10719}{{\ttfamily 1705.10719}}].

\bibitem{Press:1973iz}
W.H.~Press and P.~Schechter, \emph{{Formation of galaxies and clusters of
  galaxies by selfsimilar gravitational condensation}},
  \href{https://doi.org/10.1086/152650}{\emph{Astrophys. J.} {\bfseries 187}
  (1974) 425}.

\bibitem{Bond:1990iw}
J.R.~Bond, S.~Cole, G.~Efstathiou and N.~Kaiser, \emph{{Excursion set mass
  functions for hierarchical Gaussian fluctuations}},
  \href{https://doi.org/10.1086/170520}{\emph{Astrophys. J.} {\bfseries 379}
  (1991) 440}.

\bibitem{Navarro:1995iw}
J.F.~Navarro, C.S.~Frenk and S.D.M.~White, \emph{{The Structure of cold dark
  matter halos}}, \href{https://doi.org/10.1086/177173}{\emph{Astrophys. J.}
  {\bfseries 462} (1996) 563}
  [\href{https://arxiv.org/abs/astro-ph/9508025}{{\ttfamily
  astro-ph/9508025}}].

\bibitem{NFW_VelDisp}
S.~More, F.C.v.d.~Bosch and M.~Cacciato, \emph{{Satellite Kinematics I: A New
  Method to Constrain the Halo Mass-Luminosity Relation of Central Galaxies}},
  \href{https://doi.org/10.1111/j.1365-2966.2008.14114.x}{\emph{Mon. Not. Roy.
  Astron. Soc.} {\bfseries 392} (2009) 917}
  [\href{https://arxiv.org/abs/0807.4529}{{\ttfamily 0807.4529}}].

\bibitem{Jenkins:2000bv}
A.~Jenkins, C.S.~Frenk, S.D.M.~White, J.M.~Colberg, S.~Cole, A.E.~Evrard
  et~al., \emph{{The Mass function of dark matter halos}},
  \href{https://doi.org/10.1046/j.1365-8711.2001.04029.x}{\emph{Mon. Not. Roy.
  Astron. Soc.} {\bfseries 321} (2001) 372}
  [\href{https://arxiv.org/abs/astro-ph/0005260}{{\ttfamily
  astro-ph/0005260}}].

\bibitem{Sheth:2001dp}
R.K.~Sheth and G.~Tormen, \emph{{An Excursion Set Model of Hierarchical
  Clustering : Ellipsoidal Collapse and the Moving Barrier}},
  \href{https://doi.org/10.1046/j.1365-8711.2002.04950.x}{\emph{Mon. Not. Roy.
  Astron. Soc.} {\bfseries 329} (2002) 61}
  [\href{https://arxiv.org/abs/astro-ph/0105113}{{\ttfamily
  astro-ph/0105113}}].

\bibitem{Warren:2005ey}
M.S.~Warren, K.~Abazajian, D.E.~Holz and L.~Teodoro, \emph{{Precision
  determination of the mass function of dark matter halos}},
  \href{https://doi.org/10.1086/504962}{\emph{Astrophys. J.} {\bfseries 646}
  (2006) 881} [\href{https://arxiv.org/abs/astro-ph/0506395}{{\ttfamily
  astro-ph/0506395}}].

\bibitem{Heitmann:2015xma}
K.~Heitmann et~al., \emph{{The Mira\textendash{}Titan Universe: Precision
  Predictions for Dark Energy Surveys}},
  \href{https://doi.org/10.3847/0004-637X/820/2/108}{\emph{Astrophys. J.}
  {\bfseries 820} (2016) 108}
  [\href{https://arxiv.org/abs/1508.02654}{{\ttfamily 1508.02654}}].

\bibitem{McClintock:2018uyf}
T.~McClintock, E.~Rozo, M.R.~Becker, J.~DeRose, Y.-Y.~Mao, S.~McLaughlin
  et~al., \emph{{The Aemulus Project II: Emulating the Halo Mass Function}},
  \href{https://doi.org/10.3847/1538-4357/aaf568}{\emph{Astrophys. J.}
  {\bfseries 872} (2019) 53}
  [\href{https://arxiv.org/abs/1804.05866}{{\ttfamily 1804.05866}}].

\bibitem{Bocquet:2020tes}
S.~Bocquet, K.~Heitmann, S.~Habib, E.~Lawrence, T.~Uram, N.~Frontiere et~al.,
  \emph{{The Mira-Titan Universe. III. Emulation of the Halo Mass Function}},
  \href{https://doi.org/10.3847/1538-4357/abac5c}{\emph{Astrophys. J.}
  {\bfseries 901} (2020) 5} [\href{https://arxiv.org/abs/2003.12116}{{\ttfamily
  2003.12116}}].

\bibitem{Tinker:2008ff}
J.L.~Tinker, A.V.~Kravtsov, A.~Klypin, K.~Abazajian, M.S.~Warren, G.~Yepes
  et~al., \emph{{Toward a halo mass function for precision cosmology: The
  Limits of universality}},
  \href{https://doi.org/10.1086/591439}{\emph{Astrophys. J.} {\bfseries 688}
  (2008) 709} [\href{https://arxiv.org/abs/0803.2706}{{\ttfamily 0803.2706}}].

\bibitem{Klypin:2013rsa}
A.~Klypin, F.~Prada, G.~Yepes, S.~Hess and S.~Gottlober, \emph{{Halo Abundance
  Matching: accuracy and conditions for numerical convergence}},
  \href{https://arxiv.org/abs/1310.3740}{{\ttfamily 1310.3740}}.

\bibitem{Kravtsov:2003sg}
A.V.~Kravtsov, A.A.~Berlind, R.H.~Wechsler, A.A.~Klypin, S.~Gottloeber,
  B.~Allgood et~al., \emph{{The Dark side of the halo occupation
  distribution}}, \href{https://doi.org/10.1086/420959}{\emph{Astrophys. J.}
  {\bfseries 609} (2004) 35}
  [\href{https://arxiv.org/abs/astro-ph/0308519}{{\ttfamily
  astro-ph/0308519}}].

\bibitem{Vale:2004yt}
A.~Vale and J.P.~Ostriker, \emph{{Linking halo mass to galaxy luminosity}},
  \href{https://doi.org/10.1111/j.1365-2966.2004.08059.x}{\emph{Mon. Not. Roy.
  Astron. Soc.} {\bfseries 353} (2004) 189}
  [\href{https://arxiv.org/abs/astro-ph/0402500}{{\ttfamily
  astro-ph/0402500}}].

\bibitem{Nagai:2004ac}
D.~Nagai and A.V.~Kravtsov, \emph{{The radial distribution of galaxies in lcdm
  clusters}}, \href{https://doi.org/10.1086/426016}{\emph{Astrophys. J.}
  {\bfseries 618} (2005) 557}
  [\href{https://arxiv.org/abs/astro-ph/0408273}{{\ttfamily
  astro-ph/0408273}}].

\bibitem{Conroy:2005aq}
C.~Conroy, R.H.~Wechsler and A.V.~Kravtsov, \emph{{Modeling
  luminosity-dependent galaxy clustering through cosmic time}},
  \href{https://doi.org/10.1086/503602}{\emph{Astrophys. J.} {\bfseries 647}
  (2006) 201} [\href{https://arxiv.org/abs/astro-ph/0512234}{{\ttfamily
  astro-ph/0512234}}].

\bibitem{Wechsler:1997fz}
R.H.~Wechsler, M.A.K.~Gross, J.R.~Primack, G.R.~Blumenthal and A.~Dekel,
  \emph{{Implications of spikes in the redshift distribution of z
  \textasciitilde{} 3 galaxies}},
  \href{https://doi.org/10.1086/306229}{\emph{Astrophys. J.} {\bfseries 506}
  (1998) 19} [\href{https://arxiv.org/abs/astro-ph/9712141}{{\ttfamily
  astro-ph/9712141}}].

\bibitem{Klypin:1997fb}
A.A.~Klypin, S.~Gottlober and A.V.~Kravtsov, \emph{{Galaxies in N body
  simulations: Overcoming the overmerging problem}},
  \href{https://doi.org/10.1086/307122}{\emph{Astrophys. J.} {\bfseries 516}
  (1999) 530} [\href{https://arxiv.org/abs/astro-ph/9708191}{{\ttfamily
  astro-ph/9708191}}].

\bibitem{Graham:2005xx}
A.W.~Graham, D.~Merritt, B.~Moore, J.~Diemand and B.~Terzic, \emph{{Empirical
  models for Dark Matter Halos. I. Nonparametric Construction of Density
  Profiles and Comparison with Parametric Models}},
  \href{https://doi.org/10.1086/508988}{\emph{Astron. J.} {\bfseries 132}
  (2006) 2685} [\href{https://arxiv.org/abs/astro-ph/0509417}{{\ttfamily
  astro-ph/0509417}}].

\bibitem{Kaiser:1987qv}
N.~Kaiser, \emph{{Clustering in real space and in redshift space}}, {\emph{Mon.
  Not. Roy. Astron. Soc.} {\bfseries 227} (1987) 1}.

\bibitem{Jackson:1971sky}
J.C.~Jackson, \emph{{Fingers of God: A critique of Rees' theory of primoridal
  gravitational radiation}},
  \href{https://doi.org/10.1093/mnras/156.1.1P}{\emph{Mon. Not. Roy. Astron.
  Soc.} {\bfseries 156} (1972) 1P}
  [\href{https://arxiv.org/abs/0810.3908}{{\ttfamily 0810.3908}}].

\bibitem{Hahn:2016kiy}
C.~Hahn, R.~Scoccimarro, M.R.~Blanton, J.L.~Tinker and
  S.A.~Rodr\'\i{}guez-Torres, \emph{{The effect of fibre collisions on the
  galaxy power spectrum multipoles}},
  \href{https://doi.org/10.1093/mnras/stx185}{\emph{Mon. Not. Roy. Astron.
  Soc.} {\bfseries 467} (2017) 1940}
  [\href{https://arxiv.org/abs/1609.01714}{{\ttfamily 1609.01714}}].

\bibitem{2012MNRAS.424..564R}
{\scshape BOSS} collaboration, \emph{{The clustering of galaxies in the
  SDSS-III Baryon Oscillation Spectroscopic Survey: Analysis of potential
  systematics}},
  \href{https://doi.org/10.1111/j.1365-2966.2012.21235.x}{\emph{Mon. Not. Roy.
  Astron. Soc.} {\bfseries 424} (2012) 564}
  [\href{https://arxiv.org/abs/1203.6499}{{\ttfamily 1203.6499}}].

\bibitem{2DFGRS:2001zay}
{\scshape 2DFGRS} collaboration, \emph{{The 2dF Galaxy Redshift Survey: Spectra
  and redshifts}},
  \href{https://doi.org/10.1046/j.1365-8711.2001.04902.x}{\emph{Mon. Not. Roy.
  Astron. Soc.} {\bfseries 328} (2001) 1039}
  [\href{https://arxiv.org/abs/astro-ph/0106498}{{\ttfamily
  astro-ph/0106498}}].

\bibitem{2dFGRS:2005yhx}
{\scshape 2dFGRS} collaboration, \emph{{The 2dF Galaxy Redshift Survey:
  Power-spectrum analysis of the final dataset and cosmological implications}},
  \href{https://doi.org/10.1111/j.1365-2966.2005.09318.x}{\emph{Mon. Not. Roy.
  Astron. Soc.} {\bfseries 362} (2005) 505}
  [\href{https://arxiv.org/abs/astro-ph/0501174}{{\ttfamily
  astro-ph/0501174}}].

\bibitem{2dFGRS:2001csf}
{\scshape 2dFGRS} collaboration, \emph{{The 2dF Galaxy Redshift Survey: The
  Power spectrum and the matter content of the Universe}},
  \href{https://doi.org/10.1046/j.1365-8711.2001.04827.x}{\emph{Mon. Not. Roy.
  Astron. Soc.} {\bfseries 327} (2001) 1297}
  [\href{https://arxiv.org/abs/astro-ph/0105252}{{\ttfamily
  astro-ph/0105252}}].

\bibitem{Jones:2004zy}
D.H.~Jones et~al., \emph{{The 6dF Galaxy Survey: Samples, observational
  techniques and the first data release}},
  \href{https://doi.org/10.1111/j.1365-2966.2004.08353.x}{\emph{Mon. Not. Roy.
  Astron. Soc.} {\bfseries 355} (2004) 747}
  [\href{https://arxiv.org/abs/astro-ph/0403501}{{\ttfamily
  astro-ph/0403501}}].

\bibitem{Drinkwater:2009sd}
M.J.~Drinkwater et~al., \emph{{The WiggleZ Dark Energy Survey: Survey Design
  and First Data Release}},
  \href{https://doi.org/10.1111/j.1365-2966.2009.15754.x}{\emph{Mon. Not. Roy.
  Astron. Soc.} {\bfseries 401} (2010) 1429}
  [\href{https://arxiv.org/abs/0911.4246}{{\ttfamily 0911.4246}}].

\bibitem{Blake:2012pj}
C.~Blake et~al., \emph{{The WiggleZ Dark Energy Survey: Joint measurements of
  the expansion and growth history at z \ensuremath{<} 1}},
  \href{https://doi.org/10.1111/j.1365-2966.2012.21473.x}{\emph{Mon. Not. Roy.
  Astron. Soc.} {\bfseries 425} (2012) 405}
  [\href{https://arxiv.org/abs/1204.3674}{{\ttfamily 1204.3674}}].

\bibitem{SDSS:2000hjo}
{\scshape SDSS} collaboration, \emph{{The Sloan Digital Sky Survey: Technical
  Summary}}, \href{https://doi.org/10.1086/301513}{\emph{Astron. J.} {\bfseries
  120} (2000) 1579} [\href{https://arxiv.org/abs/astro-ph/0006396}{{\ttfamily
  astro-ph/0006396}}].

\bibitem{eBOSS:2020yzd}
{\scshape eBOSS} collaboration, \emph{{Completed SDSS-IV extended Baryon
  Oscillation Spectroscopic Survey: Cosmological implications from two decades
  of spectroscopic surveys at the Apache Point Observatory}},
  \href{https://doi.org/10.1103/PhysRevD.103.083533}{\emph{Phys. Rev. D}
  {\bfseries 103} (2021) 083533}
  [\href{https://arxiv.org/abs/2007.08991}{{\ttfamily 2007.08991}}].

\bibitem{DESI:2016fyo}
{\scshape DESI} collaboration, \emph{{The DESI Experiment Part I:
  Science,Targeting, and Survey Design}},
  \href{https://arxiv.org/abs/1611.00036}{{\ttfamily 1611.00036}}.

\bibitem{Euclid:2021icp}
{\scshape Euclid} collaboration, \emph{{Euclid preparation - I. The Euclid Wide
  Survey}}, \href{https://doi.org/10.1051/0004-6361/202141938}{\emph{Astron.
  Astrophys.} {\bfseries 662} (2022) A112}
  [\href{https://arxiv.org/abs/2108.01201}{{\ttfamily 2108.01201}}].

\bibitem{Pozzetti:2016cch}
L.~Pozzetti, C.M.~Hirata, J.E.~Geach, A.~Cimatti, C.~Baugh, O.~Cucciati et~al.,
  \emph{{Modelling the number density of H\ensuremath{\alpha} emitters for
  future spectroscopic near-IR space missions}},
  \href{https://doi.org/10.1051/0004-6361/201527081}{\emph{Astron. Astrophys.}
  {\bfseries 590} (2016) A3}
  [\href{https://arxiv.org/abs/1603.01453}{{\ttfamily 1603.01453}}].

\bibitem{Euclid:2019clj}
{\scshape Euclid} collaboration, \emph{{Euclid preparation: VII. Forecast
  validation for Euclid cosmological probes}},
  \href{https://doi.org/10.1051/0004-6361/202038071}{\emph{Astron. Astrophys.}
  {\bfseries 642} (2020) A191}
  [\href{https://arxiv.org/abs/1910.09273}{{\ttfamily 1910.09273}}].

\bibitem{Akeson:2019biv}
R.~Akeson et~al., \emph{{The Wide Field Infrared Survey Telescope: 100 Hubbles
  for the 2020s}},  \href{https://arxiv.org/abs/1902.05569}{{\ttfamily
  1902.05569}}.

\bibitem{Wang:2021oec}
Y.~Wang et~al., \emph{{The High Latitude Spectroscopic Survey on the Nancy
  Grace Roman Space Telescope}},
  \href{https://doi.org/10.3847/1538-4357/ac4973}{\emph{Astrophys. J.}
  {\bfseries 928} (2022) 1} [\href{https://arxiv.org/abs/2110.01829}{{\ttfamily
  2110.01829}}].

\bibitem{Fidler:2015npa}
C.~Fidler, C.~Rampf, T.~Tram, R.~Crittenden, K.~Koyama and D.~Wands,
  \emph{{General relativistic corrections to $N$-body simulations and the
  Zel'dovich approximation}},
  \href{https://doi.org/10.1103/PhysRevD.92.123517}{\emph{Phys. Rev. D}
  {\bfseries 92} (2015) 123517}
  [\href{https://arxiv.org/abs/1505.04756}{{\ttfamily 1505.04756}}].

\bibitem{Tram:2018znz}
T.~Tram, J.~Brandbyge, J.~Dakin and S.~Hannestad, \emph{{Fully relativistic
  treatment of light neutrinos in $N$-body simulations}},
  \href{https://doi.org/10.1088/1475-7516/2019/03/022}{\emph{JCAP} {\bfseries
  03} (2019) 022} [\href{https://arxiv.org/abs/1811.00904}{{\ttfamily
  1811.00904}}].

\bibitem{Dakin:2019vnj}
J.~Dakin, S.~Hannestad, T.~Tram, M.~Knabenhans and J.~Stadel, \emph{{Dark
  energy perturbations in $N$-body simulations}},
  \href{https://doi.org/10.1088/1475-7516/2019/08/013}{\emph{JCAP} {\bfseries
  08} (2019) 013} [\href{https://arxiv.org/abs/1904.05210}{{\ttfamily
  1904.05210}}].

\bibitem{Brando:2020ouk}
G.~Brando, K.~Koyama and D.~Wands, \emph{{Relativistic Corrections to the
  Growth of Structure in Modified Gravity}},
  \href{https://doi.org/10.1088/1475-7516/2021/01/013}{\emph{JCAP} {\bfseries
  01} (2021) 013} [\href{https://arxiv.org/abs/2006.11019}{{\ttfamily
  2006.11019}}].

\bibitem{Brando:2021jga}
G.~Brando, K.~Koyama, D.~Wands, M.~Zumalac\'arregui, I.~Sawicki and E.~Bellini,
  \emph{{Fully relativistic predictions in Horndeski gravity from standard
  Newtonian N-body simulations}},
  \href{https://doi.org/10.1088/1475-7516/2021/09/024}{\emph{JCAP} {\bfseries
  09} (2021) 024} [\href{https://arxiv.org/abs/2105.04491}{{\ttfamily
  2105.04491}}].

\bibitem{Schneider:2015yka}
A.~Schneider, R.~Teyssier, D.~Potter, J.~Stadel, J.~Onions, D.S.~Reed et~al.,
  \emph{{Matter power spectrum and the challenge of percent accuracy}},
  \href{https://doi.org/10.1088/1475-7516/2016/04/047}{\emph{JCAP} {\bfseries
  04} (2016) 047} [\href{https://arxiv.org/abs/1503.05920}{{\ttfamily
  1503.05920}}].

\bibitem{Ewald1921}
P.P.~Ewald, \emph{Die berechnung optischer und elektrostatischer
  gitterpotentiale},
  \href{https://doi.org/https://doi.org/10.1002/andp.19213690304}{\emph{Annalen
  der Physik} {\bfseries 369} (1921) 253}
  [\href{https://arxiv.org/abs/https://onlinelibrary.wiley.com/doi/pdf/10.1002/andp.19213690304}{{\ttfamily
  https://onlinelibrary.wiley.com/doi/pdf/10.1002/andp.19213690304}}].

\bibitem{Sefusatti:2015aex}
E.~Sefusatti, M.~Crocce, R.~Scoccimarro and H.~Couchman, \emph{{Accurate
  Estimators of Correlation Functions in Fourier Space}},
  \href{https://doi.org/10.1093/mnras/stw1229}{\emph{Mon. Not. Roy. Astron.
  Soc.} {\bfseries 460} (2016) 3624}
  [\href{https://arxiv.org/abs/1512.07295}{{\ttfamily 1512.07295}}].

\bibitem{Klypin:2017iwu}
A.~Klypin and F.~Prada, \emph{{Dark matter statistics for large galaxy
  catalogues: power spectra and covariance matrices}},
  \href{https://doi.org/10.1093/mnras/sty1340}{\emph{Mon. Not. Roy. Astron.
  Soc.} {\bfseries 478} (2018) 4602}
  [\href{https://arxiv.org/abs/1701.05690}{{\ttfamily 1701.05690}}].

\bibitem{1986Natur.324..446B}
J.~{Barnes} and P.~{Hut}, \emph{{A hierarchical O(N log N) force-calculation
  algorithm}}, \href{https://doi.org/10.1038/324446a0}{\emph{Nature} {\bfseries
  324} (1986) 446}.

\bibitem{pkdgrav}
D.~Potter, J.~Stadel and R.~Teyssier, \emph{{PKDGRAV3: Beyond Trillion Particle
  Cosmological Simulations for the Next Era of Galaxy Surveys}},
  \href{https://arxiv.org/abs/1609.08621}{{\ttfamily 1609.08621}}.

\bibitem{Teyssier:2001cp}
R.~Teyssier, \emph{{Cosmological hydrodynamics with adaptive mesh refinement: a
  new high resolution code called ramses}},
  \href{https://doi.org/10.1051/0004-6361:20011817}{\emph{Astron. Astrophys.}
  {\bfseries 385} (2002) 337}
  [\href{https://arxiv.org/abs/astro-ph/0111367}{{\ttfamily
  astro-ph/0111367}}].

\bibitem{gadget2}
V.~Springel, \emph{{The Cosmological simulation code GADGET-2}},
  \href{https://doi.org/10.1111/j.1365-2966.2005.09655.x}{\emph{Mon. Not. Roy.
  Astron. Soc.} {\bfseries 364} (2005) 1105}
  [\href{https://arxiv.org/abs/astro-ph/0505010}{{\ttfamily
  astro-ph/0505010}}].

\bibitem{angulo2012}
R.E.~{Angulo}, V.~{Springel}, S.D.M.~{White}, A.~{Jenkins}, C.M.~{Baugh} and
  C.S.~{Frenk}, \emph{{Scaling relations for galaxy clusters in the
  Millennium-XXL simulation}},
  \href{https://doi.org/10.1111/j.1365-2966.2012.21830.x}{\emph{\mnras}
  {\bfseries 426} (2012) 2046}
  [\href{https://arxiv.org/abs/1203.3216}{{\ttfamily 1203.3216}}].

\bibitem{Arepo}
V.~Springel, \emph{{E pur si muove: Galiliean-invariant cosmological
  hydrodynamical simulations on a moving mesh}},
  \href{https://doi.org/10.1111/j.1365-2966.2009.15715.x}{\emph{Mon. Not. Roy.
  Astron. Soc.} {\bfseries 401} (2010) 791}
  [\href{https://arxiv.org/abs/0901.4107}{{\ttfamily 0901.4107}}].

\bibitem{Llinares:2018maz}
C.~Llinares, \emph{{Simulation techniques for modified gravity}},
  \href{https://doi.org/10.1142/S0218271818480036}{\emph{Int. J. Mod. Phys. D}
  {\bfseries 27} (2018) 1848003}
  [\href{https://arxiv.org/abs/2103.10890}{{\ttfamily 2103.10890}}].

\bibitem{Winther:2015wla}
H.A.~Winther et~al., \emph{{Modified Gravity N-body Code Comparison Project}},
  \href{https://doi.org/10.1093/mnras/stv2253}{\emph{Mon. Not. Roy. Astron.
  Soc.} {\bfseries 454} (2015) 4208}
  [\href{https://arxiv.org/abs/1506.06384}{{\ttfamily 1506.06384}}].

\bibitem{Li:2011vk}
B.~Li, G.-B.~Zhao, R.~Teyssier and K.~Koyama, \emph{{ECOSMOG: An Efficient Code
  for Simulating Modified Gravity}},
  \href{https://doi.org/10.1088/1475-7516/2012/01/051}{\emph{JCAP} {\bfseries
  01} (2012) 051} [\href{https://arxiv.org/abs/1110.1379}{{\ttfamily
  1110.1379}}].

\bibitem{Li:2013nua}
B.~Li, G.-B.~Zhao and K.~Koyama, \emph{{Exploring Vainshtein mechanism on
  adaptively refined meshes}},
  \href{https://doi.org/10.1088/1475-7516/2013/05/023}{\emph{JCAP} {\bfseries
  05} (2013) 023} [\href{https://arxiv.org/abs/1303.0008}{{\ttfamily
  1303.0008}}].

\bibitem{Puchwein:2013lza}
E.~Puchwein, M.~Baldi and V.~Springel, \emph{{Modified Gravity-GADGET: A new
  code for cosmological hydrodynamical simulations of modified gravity
  models}}, \href{https://doi.org/10.1093/mnras/stt1575}{\emph{Mon. Not. Roy.
  Astron. Soc.} {\bfseries 436} (2013) 348}
  [\href{https://arxiv.org/abs/1305.2418}{{\ttfamily 1305.2418}}].

\bibitem{Oyaizu:2008sr}
H.~Oyaizu, \emph{{Non-linear evolution of f(R) cosmologies I: methodology}},
  \href{https://doi.org/10.1103/PhysRevD.78.123523}{\emph{Phys. Rev. D}
  {\bfseries 78} (2008) 123523}
  [\href{https://arxiv.org/abs/0807.2449}{{\ttfamily 0807.2449}}].

\bibitem{Llinares:2013jza}
C.~Llinares, D.F.~Mota and H.A.~Winther, \emph{{ISIS: a new N-body cosmological
  code with scalar fields based on RAMSES. Code presentation and application to
  the shapes of clusters}},
  \href{https://doi.org/10.1051/0004-6361/201322412}{\emph{Astron. Astrophys.}
  {\bfseries 562} (2014) A78}
  [\href{https://arxiv.org/abs/1307.6748}{{\ttfamily 1307.6748}}].

\bibitem{Llinares:2013qbh}
C.~Llinares and D.~Mota, \emph{{Releasing scalar fields: cosmological
  simulations of scalar-tensor theories for gravity beyond the static
  approximation}},
  \href{https://doi.org/10.1103/PhysRevLett.110.161101}{\emph{Phys. Rev. Lett.}
  {\bfseries 110} (2013) 161101}
  [\href{https://arxiv.org/abs/1302.1774}{{\ttfamily 1302.1774}}].

\bibitem{Arepo_fR}
C.~Arnold, M.~Leo and B.~Li, \emph{{Realistic simulations of galaxy formation
  in $f(R)$ modified gravity}},
  \href{https://doi.org/10.1038/s41550-019-0823-y}{\emph{Nature Astron.}
  {\bfseries 3} (2019) 945} [\href{https://arxiv.org/abs/1907.02977}{{\ttfamily
  1907.02977}}].

\bibitem{Arepo_nDGP}
C.~Hern\'andez-Aguayo, C.~Arnold, B.~Li and C.M.~Baugh, \emph{{Galaxy formation
  in the brane world I: overview and first results}},
  \href{https://doi.org/10.1093/mnras/stab694}{\emph{Mon. Not. Roy. Astron.
  Soc.} {\bfseries 503} (2021) 3867}
  [\href{https://arxiv.org/abs/2006.15467}{{\ttfamily 2006.15467}}].

\bibitem{Hernandez-Aguayo:2021kuh}
C.~Hern\'andez-Aguayo, C.-Z.~Ruan, B.~Li, C.~Arnold, C.M.~Baugh, A.~Klypin
  et~al., \emph{{Fast full N-body simulations of generic modified gravity:
  derivative coupling models}},
  \href{https://doi.org/10.1088/1475-7516/2022/01/048}{\emph{JCAP} {\bfseries
  01} (2022) 048} [\href{https://arxiv.org/abs/2110.00566}{{\ttfamily
  2110.00566}}].

\bibitem{Ruan:2021wup}
C.-Z.~Ruan, C.~Hern\'andez-Aguayo, B.~Li, C.~Arnold, C.M.~Baugh, A.~Klypin
  et~al., \emph{{Fast full N-body simulations of generic modified gravity:
  conformal coupling models}},
  \href{https://doi.org/10.1088/1475-7516/2022/05/018}{\emph{JCAP} {\bfseries
  05} (2022) 018} [\href{https://arxiv.org/abs/2110.00328}{{\ttfamily
  2110.00328}}].

\bibitem{Alam:2020jdv}
S.~Alam et~al., \emph{{Towards testing the theory of gravity with DESI: summary
  statistics, model predictions and future simulation requirements}},
  \href{https://doi.org/10.1088/1475-7516/2021/11/050}{\emph{JCAP} {\bfseries
  11} (2021) 050} [\href{https://arxiv.org/abs/2011.05771}{{\ttfamily
  2011.05771}}].

\bibitem{Monaco16}
P.~{Monaco}, \emph{{Approximated methods for the generation of dark matter halo
  catalogs in the age of precision cosmology}}, {\emph{ArXiv e-prints} (2016) }
  [\href{https://arxiv.org/abs/1605.07752}{{\ttfamily 1605.07752}}].

\bibitem{Chuang:2014toa}
C.-H.~Chuang et~al., \emph{{nIFTy Cosmology: Galaxy/halo mock catalogue
  comparison project on clustering statistics}},
  \href{https://doi.org/10.1093/mnras/stv1289}{\emph{Mon. Not. Roy. Astron.
  Soc.} {\bfseries 452} (2015) 686}
  [\href{https://arxiv.org/abs/1412.7729}{{\ttfamily 1412.7729}}].

\bibitem{Tassev:2013pn}
S.~Tassev, M.~Zaldarriaga and D.~Eisenstein, \emph{{Solving Large Scale
  Structure in Ten Easy Steps with COLA}},
  \href{https://doi.org/10.1088/1475-7516/2013/06/036}{\emph{JCAP} {\bfseries
  06} (2013) 036} [\href{https://arxiv.org/abs/1301.0322}{{\ttfamily
  1301.0322}}].

\bibitem{Koda:2015mca}
J.~Koda, C.~Blake, F.~Beutler, E.~Kazin and F.~Marin, \emph{{Fast and accurate
  mock catalogue generation for low-mass galaxies}},
  \href{https://doi.org/10.1093/mnras/stw763}{\emph{Mon. Not. Roy. Astron.
  Soc.} {\bfseries 459} (2016) 2118}
  [\href{https://arxiv.org/abs/1507.05329}{{\ttfamily 1507.05329}}].

\bibitem{Izard:2015dja}
A.~Izard, M.~Crocce and P.~Fosalba, \emph{{ICE-COLA: Towards fast and accurate
  synthetic galaxy catalogues optimizing a quasi $N$-body method}},
  \href{https://doi.org/10.1093/mnras/stw797}{\emph{Mon. Not. Roy. Astron.
  Soc.} {\bfseries 459} (2016) 2327}
  [\href{https://arxiv.org/abs/1509.04685}{{\ttfamily 1509.04685}}].

\bibitem{Howlett:2015hfa}
C.~Howlett, M.~Manera and W.J.~Percival, \emph{{L-PICOLA: A parallel code for
  fast dark matter simulation}},
  \href{https://doi.org/10.1016/j.ascom.2015.07.003}{\emph{Astron. Comput.}
  {\bfseries 12} (2015) 109}
  [\href{https://arxiv.org/abs/1506.03737}{{\ttfamily 1506.03737}}].

\bibitem{Wright:2017dkw}
B.S.~Wright, H.A.~Winther and K.~Koyama, \emph{{COLA with massive neutrinos}},
  \href{https://doi.org/10.1088/1475-7516/2017/10/054}{\emph{JCAP} {\bfseries
  10} (2017) 054} [\href{https://arxiv.org/abs/1705.08165}{{\ttfamily
  1705.08165}}].

\bibitem{Moretti:2019bob}
C.~Moretti, S.~Mozzon, P.~Monaco, E.~Munari and M.~Baldi, \emph{{Fast numerical
  method to generate halo catalogues in modified gravity (part I): second-order
  Lagrangian perturbation theory}},
  \href{https://doi.org/10.1093/mnras/staa312}{\emph{Mon. Not. Roy. Astron.
  Soc.} {\bfseries 493} (2020) 1153}
  [\href{https://arxiv.org/abs/1909.06282}{{\ttfamily 1909.06282}}].

\bibitem{Monaco:2001jg}
P.~Monaco, T.~Theuns and G.~Taffoni, \emph{{Pinocchio: pinpointing
  orbit-crossing collapsed hierarchical objects in a linear density field}},
  \href{https://doi.org/10.1046/j.1365-8711.2002.05162.x}{\emph{Mon. Not. Roy.
  Astron. Soc.} {\bfseries 331} (2002) 587}
  [\href{https://arxiv.org/abs/astro-ph/0109323}{{\ttfamily
  astro-ph/0109323}}].

\bibitem{Knebe:2011rx}
A.~Knebe et~al., \emph{{Haloes gone MAD: The Halo-Finder Comparison Project}},
  \href{https://doi.org/10.1111/j.1365-2966.2011.18858.x}{\emph{Mon. Not. Roy.
  Astron. Soc.} {\bfseries 415} (2011) 2293}
  [\href{https://arxiv.org/abs/1104.0949}{{\ttfamily 1104.0949}}].

\bibitem{Zheng:2007zg}
Z.~Zheng, A.L.~Coil and I.~Zehavi, \emph{{Galaxy Evolution from Halo Occupation
  Distribution Modeling of DEEP2 and SDSS Galaxy Clustering}},
  \href{https://doi.org/10.1086/521074}{\emph{Astrophys. J.} {\bfseries 667}
  (2007) 760} [\href{https://arxiv.org/abs/astro-ph/0703457}{{\ttfamily
  astro-ph/0703457}}].

\bibitem{Mitchell:2018qrg}
M.A.~Mitchell, J.-h.~He, C.~Arnold and B.~Li, \emph{{A general framework to
  test gravity using galaxy clusters \textendash{} I. Modelling the dynamical
  mass of haloes in $f(R)$ gravity}},
  \href{https://doi.org/10.1093/mnras/sty636}{\emph{Mon. Not. Roy. Astron.
  Soc.} {\bfseries 477} (2018) 1133}
  [\href{https://arxiv.org/abs/1802.02165}{{\ttfamily 1802.02165}}].

\bibitem{Mitchell:2019qke}
M.A.~Mitchell, C.~Arnold, J.-h.~He and B.~Li, \emph{{A general framework to
  test gravity using galaxy clusters II: A universal model for the halo
  concentration in $f(R)$ gravity}},
  \href{https://doi.org/10.1093/mnras/stz1389}{\emph{Mon. Not. Roy. Astron.
  Soc.} {\bfseries 487} (2019) 1410}
  [\href{https://arxiv.org/abs/1901.06392}{{\ttfamily 1901.06392}}].

\bibitem{Lippich_2018}
M.~Lippich, A.G.~Sánchez, M.~Colavincenzo, E.~Sefusatti, P.~Monaco, L.~Blot
  et~al., \emph{Comparing approximate methods for mock catalogues and
  covariance matrices – i. correlation function},
  \href{https://doi.org/10.1093/mnras/sty2757}{\emph{Monthly Notices of the
  Royal Astronomical Society} {\bfseries 482} (2018) 1786–1806}.

\bibitem{Blot_2019}
L.~Blot, M.~Crocce, E.~Sefusatti, M.~Lippich, A.G.~Sánchez, M.~Colavincenzo
  et~al., \emph{Comparing approximate methods for mock catalogues and
  covariance matrices ii: power spectrum multipoles},
  \href{https://doi.org/10.1093/mnras/stz507}{\emph{Monthly Notices of the
  Royal Astronomical Society} {\bfseries 485} (2019) 2806–2824}.

\bibitem{Colavincenzo_2018}
M.~Colavincenzo, E.~Sefusatti, P.~Monaco, L.~Blot, M.~Crocce, M.~Lippich
  et~al., \emph{Comparing approximate methods for mock catalogues and
  covariance matrices – iii: bispectrum},
  \href{https://doi.org/10.1093/mnras/sty2964}{\emph{Monthly Notices of the
  Royal Astronomical Society} {\bfseries 482} (2018) 4883–4905}.

\bibitem{Cautun:2017tkc}
M.~Cautun, E.~Paillas, Y.-C.~Cai, S.~Bose, J.~Armijo, B.~Li et~al., \emph{{The
  Santiago\textendash{}Harvard\textendash{}Edinburgh\textendash{}Durham void
  comparison \textendash{} I. SHEDding light on chameleon gravity tests}},
  \href{https://doi.org/10.1093/mnras/sty463}{\emph{Mon. Not. Roy. Astron.
  Soc.} {\bfseries 476} (2018) 3195}
  [\href{https://arxiv.org/abs/1710.01730}{{\ttfamily 1710.01730}}].

\bibitem{Hernandez-Aguayo:2018yrp}
C.~Hern\'andez-Aguayo, C.M.~Baugh and B.~Li, \emph{{Marked clustering
  statistics in $f(R)$ gravity cosmologies}},
  \href{https://doi.org/10.1093/mnras/sty1822}{\emph{Mon. Not. Roy. Astron.
  Soc.} {\bfseries 479} (2018) 4824}
  [\href{https://arxiv.org/abs/1801.08880}{{\ttfamily 1801.08880}}].

\bibitem{Hernandez-Aguayo:2018oxg}
C.~Hern\'andez-Aguayo, J.~Hou, B.~Li, C.M.~Baugh and A.G.~S\'anchez,
  \emph{{Large-scale redshift space distortions in modified gravity theories}},
  \href{https://doi.org/10.1093/mnras/stz516}{\emph{Mon. Not. Roy. Astron.
  Soc.} {\bfseries 485} (2019) 2194}
  [\href{https://arxiv.org/abs/1811.09197}{{\ttfamily 1811.09197}}].

\bibitem{Devi:2019swk}
N.C.~Devi, A.~Rodr\'\i{}guez-Puebla, O.~Valenzuela, V.~Avila-Reese,
  C.~Hern\'andez-Aguayo and B.~Li, \emph{{The galaxy\textendash{}halo
  connection in modified gravity cosmologies: environment dependence of galaxy
  luminosity function}},
  \href{https://doi.org/10.1093/mnras/stz1664}{\emph{Mon. Not. Roy. Astron.
  Soc.} {\bfseries 488} (2019) 782}
  [\href{https://arxiv.org/abs/1901.02121}{{\ttfamily 1901.02121}}].

\bibitem{Cautun:2011gf}
M.C.~Cautun and R.~van~de Weygaert, \emph{{The DTFE public software: The
  Delaunay Tessellation Field Estimator code}},
  \href{https://arxiv.org/abs/1105.0370}{{\ttfamily 1105.0370}}.

\bibitem{Behroozi13}
P.S.~{Behroozi}, R.H.~{Wechsler} and H.-Y.~{Wu}, \emph{{The ROCKSTAR
  Phase-space Temporal Halo Finder and the Velocity Offsets of Cluster Cores}},
  \href{https://doi.org/10.1088/0004-637X/762/2/109}{\emph{\apj} {\bfseries
  762} (2013) 109} [\href{https://arxiv.org/abs/1110.4372}{{\ttfamily
  1110.4372}}].

\bibitem{Davis85}
M.~{Davis}, G.~{Efstathiou}, C.S.~{Frenk} and S.D.M.~{White}, \emph{{The
  evolution of large-scale structure in a universe dominated by cold dark
  matter}}, \href{https://doi.org/10.1086/163168}{\emph{\apj} {\bfseries 292}
  (1985) 371}.

\bibitem{Howlett:2014opa}
C.~Howlett, A.~Ross, L.~Samushia, W.~Percival and M.~Manera, \emph{{The
  clustering of the SDSS main galaxy sample \textendash{} II. Mock galaxy
  catalogues and a measurement of the growth of structure from redshift space
  distortions at $z = 0.15$}},
  \href{https://doi.org/10.1093/mnras/stu2693}{\emph{Mon. Not. Roy. Astron.
  Soc.} {\bfseries 449} (2015) 848}
  [\href{https://arxiv.org/abs/1409.3238}{{\ttfamily 1409.3238}}].

\bibitem{Lukic:2008ds}
Z.~Lukic, D.~Reed, S.~Habib and K.~Heitmann, \emph{{The Structure of Halos:
  Implications for Group and Cluster Cosmology}},
  \href{https://doi.org/10.1088/0004-637X/692/1/217}{\emph{Astrophys. J.}
  {\bfseries 692} (2009) 217}
  [\href{https://arxiv.org/abs/0803.3624}{{\ttfamily 0803.3624}}].

\bibitem{Dutton:2014xda}
A.A.~Dutton and A.V.~Macci\`o, \emph{{Cold dark matter haloes in the Planck
  era: evolution of structural parameters for Einasto and NFW profiles}},
  \href{https://doi.org/10.1093/mnras/stu742}{\emph{Mon. Not. Roy. Astron.
  Soc.} {\bfseries 441} (2014) 3359}
  [\href{https://arxiv.org/abs/1402.7073}{{\ttfamily 1402.7073}}].

\bibitem{Hearin:2016uxs}
A.P.~Hearin et~al., \emph{{Forward Modeling of Large-Scale Structure: An
  open-source approach with Halotools}},
  \href{https://doi.org/10.5281/zenodo.835895}{\emph{Astron. J.} {\bfseries
  154} (2017) 190} [\href{https://arxiv.org/abs/1606.04106}{{\ttfamily
  1606.04106}}].

\bibitem{More:2008yy}
S.~More, F.C.v.d.~Bosch and M.~Cacciato, \emph{{Satellite Kinematics I: A New
  Method to Constrain the Halo Mass-Luminosity Relation of Central Galaxies}},
  \href{https://doi.org/10.1111/j.1365-2966.2008.14114.x}{\emph{Mon. Not. Roy.
  Astron. Soc.} {\bfseries 392} (2009) 917}
  [\href{https://arxiv.org/abs/0807.4529}{{\ttfamily 0807.4529}}].

\bibitem{Hikage:2014bza}
C.~Hikage, \emph{{Constraining Halo Occupation Distribution and Cosmic Growth
  Rate using Multipole Power Spectrum}},
  \href{https://doi.org/10.1093/mnrasl/slu038}{\emph{Mon. Not. Roy. Astron.
  Soc.} {\bfseries 441} (2014) L21}
  [\href{https://arxiv.org/abs/1401.1246}{{\ttfamily 1401.1246}}].

\bibitem{Taruya:2010mx}
A.~Taruya, T.~Nishimichi and S.~Saito, \emph{{Baryon Acoustic Oscillations in
  2D: Modeling Redshift-space Power Spectrum from Perturbation Theory}},
  \href{https://doi.org/10.1103/PhysRevD.82.063522}{\emph{Phys. Rev. D}
  {\bfseries 82} (2010) 063522}
  [\href{https://arxiv.org/abs/1006.0699}{{\ttfamily 1006.0699}}].

\bibitem{Nelder:1965zz}
J.A.~Nelder and R.~Mead, \emph{{A Simplex Method for Function Minimization}},
  \href{https://doi.org/10.1093/comjnl/7.4.308}{\emph{Comput. J.} {\bfseries 7}
  (1965) 308}.

\bibitem{Fiorini:2022srj}
B.~Fiorini, K.~Koyama and A.~Izard, \emph{{Studying large-scale structure
  probes of modified gravity with COLA}},
  \href{https://arxiv.org/abs/2208.01345}{{\ttfamily 2208.01345}}.

\bibitem{Hahn:2019zob}
C.~Hahn, F.~Villaescusa-Navarro, E.~Castorina and R.~Scoccimarro,
  \emph{{Constraining $M_\nu$ with the bispectrum. Part I. Breaking parameter
  degeneracies}},
  \href{https://doi.org/10.1088/1475-7516/2020/03/040}{\emph{JCAP} {\bfseries
  03} (2020) 040} [\href{https://arxiv.org/abs/1909.11107}{{\ttfamily
  1909.11107}}].

\bibitem{Hahn:2020lou}
C.~Hahn and F.~Villaescusa-Navarro, \emph{{Constraining $M_\nu$ with the
  bispectrum. Part II. The information content of the galaxy bispectrum
  monopole}}, \href{https://doi.org/10.1088/1475-7516/2021/04/029}{\emph{JCAP}
  {\bfseries 04} (2021) 029}
  [\href{https://arxiv.org/abs/2012.02200}{{\ttfamily 2012.02200}}].

\bibitem{Ivanov:2021fbu}
M.M.~Ivanov, O.H.E.~Philcox, M.~Simonovi\'c, M.~Zaldarriaga, T.~Nischimichi and
  M.~Takada, \emph{{Cosmological constraints without nonlinear redshift-space
  distortions}}, \href{https://doi.org/10.1103/PhysRevD.105.043531}{\emph{Phys.
  Rev. D} {\bfseries 105} (2022) 043531}
  [\href{https://arxiv.org/abs/2110.00006}{{\ttfamily 2110.00006}}].

\bibitem{Scoccimarro:2015bla}
R.~Scoccimarro, \emph{{Fast Estimators for Redshift-Space Clustering}},
  \href{https://doi.org/10.1103/PhysRevD.92.083532}{\emph{Phys. Rev. D}
  {\bfseries 92} (2015) 083532}
  [\href{https://arxiv.org/abs/1506.02729}{{\ttfamily 1506.02729}}].

\bibitem{Jing:2004fq}
Y.P.~Jing, \emph{{Correcting for the alias effect when measuring the power
  spectrum using FFT}}, \href{https://doi.org/10.1086/427087}{\emph{Astrophys.
  J.} {\bfseries 620} (2005) 559}
  [\href{https://arxiv.org/abs/astro-ph/0409240}{{\ttfamily
  astro-ph/0409240}}].

\bibitem{Scoccimarro:2000sn}
R.~Scoccimarro, \emph{{The bispectrum: from theory to observations}},
  \href{https://doi.org/10.1086/317248}{\emph{Astrophys. J.} {\bfseries 544}
  (2000) 597} [\href{https://arxiv.org/abs/astro-ph/0004086}{{\ttfamily
  astro-ph/0004086}}].

\bibitem{Fry:1984}
J.N.~{Fry}, \emph{{The Galaxy correlation hierarchy in perturbation theory}},
  \href{https://doi.org/10.1086/161913}{\emph{\apj} {\bfseries 279} (1984)
  499}.

\bibitem{Matarrese:1997sk}
S.~Matarrese, L.~Verde and A.F.~Heavens, \emph{{Large scale bias in the
  universe: Bispectrum method}},
  \href{https://doi.org/10.1093/mnras/290.4.651}{\emph{Mon. Not. Roy. Astron.
  Soc.} {\bfseries 290} (1997) 651}
  [\href{https://arxiv.org/abs/astro-ph/9706059}{{\ttfamily
  astro-ph/9706059}}].

\bibitem{Gil-Marin:2011iuh}
H.~Gil-Marin, F.~Schmidt, W.~Hu, R.~Jimenez and L.~Verde, \emph{{The Bispectrum
  of f(R) Cosmologies}},
  \href{https://doi.org/10.1088/1475-7516/2011/11/019}{\emph{JCAP} {\bfseries
  11} (2011) 019} [\href{https://arxiv.org/abs/1109.2115}{{\ttfamily
  1109.2115}}].

\bibitem{Cai:2016jek}
Y.-C.~Cai, A.~Taylor, J.A.~Peacock and N.~Padilla, \emph{{Redshift-space
  distortions around voids}},
  \href{https://doi.org/10.1093/mnras/stw1809}{\emph{Mon. Not. Roy. Astron.
  Soc.} {\bfseries 462} (2016) 2465}
  [\href{https://arxiv.org/abs/1603.05184}{{\ttfamily 1603.05184}}].

\bibitem{Nadathur:2017jos}
S.~Nadathur and W.J.~Percival, \emph{{An accurate linear model for redshift
  space distortions in the void-galaxy correlation function}},
  \href{https://doi.org/10.1093/mnras/sty3372}{\emph{Mon. Not. Roy. Astron.
  Soc.} {\bfseries 483} (2019) 3472}
  [\href{https://arxiv.org/abs/1712.07575}{{\ttfamily 1712.07575}}].

\bibitem{voidmg1}
P.~Zivick, P.M.~Sutter, B.D.~Wandelt, B.~Li and T.Y.~Lam, \emph{{Using cosmic
  voids to distinguish f(R) gravity in future galaxy surveys}},
  \href{https://doi.org/10.1093/mnras/stv1209}{\emph{Mon. Not. Roy. Astron.
  Soc.} {\bfseries 451} (2015) 4215}
  [\href{https://arxiv.org/abs/1411.5694}{{\ttfamily 1411.5694}}].

\bibitem{voidmg2}
Y.-C.~Cai, N.~Padilla and B.~Li, \emph{{Testing Gravity using Cosmic Voids}},
  \href{https://doi.org/10.1093/mnras/stv777}{\emph{Mon. Not. Roy. Astron.
  Soc.} {\bfseries 451} (2015) 1036}
  [\href{https://arxiv.org/abs/1410.1510}{{\ttfamily 1410.1510}}].

\bibitem{voidmg3}
B.~Falck, K.~Koyama, G.-B.~Zhao and M.~Cautun, \emph{{Using Voids to Unscreen
  Modified Gravity}}, \href{https://doi.org/10.1093/mnras/stx3288}{\emph{Mon.
  Not. Roy. Astron. Soc.} {\bfseries 475} (2018) 3262}
  [\href{https://arxiv.org/abs/1704.08942}{{\ttfamily 1704.08942}}].

\bibitem{voidmg4}
P.~Cataldi, S.~Pedrosa, N.~Padilla, S.~Landau, C.~Arnold and B.~Li,
  \emph{{Fingerprints of modified gravity on galaxies in voids}},
  \href{https://arxiv.org/abs/2207.12917}{{\ttfamily 2207.12917}}.

\bibitem{Paillas:2018wxs}
E.~Paillas, M.~Cautun, B.~Li, Y.-C.~Cai, N.~Padilla, J.~Armijo et~al.,
  \emph{{The
  Santiago\textendash{}Harvard\textendash{}Edinburgh\textendash{}Durham void
  comparison II: unveiling the Vainshtein screening using weak lensing}},
  \href{https://doi.org/10.1093/mnras/stz022}{\emph{Mon. Not. Roy. Astron.
  Soc.} {\bfseries 484} (2019) 1149}
  [\href{https://arxiv.org/abs/1810.02864}{{\ttfamily 1810.02864}}].

\bibitem{Perico:2019obq}
E.L.D.~Perico, R.~Voivodic, M.~Lima and D.F.~Mota, \emph{{Cosmic voids in
  modified gravity scenarios}},
  \href{https://doi.org/10.1051/0004-6361/201935949}{\emph{Astron. Astrophys.}
  {\bfseries 632} (2019) A52}
  [\href{https://arxiv.org/abs/1905.12450}{{\ttfamily 1905.12450}}].

\bibitem{Contarini:2020fdu}
S.~Contarini, F.~Marulli, L.~Moscardini, A.~Veropalumbo, C.~Giocoli and
  M.~Baldi, \emph{{Cosmic voids in modified gravity models with massive
  neutrinos}}, \href{https://doi.org/10.1093/mnras/stab1112}{\emph{Mon. Not.
  Roy. Astron. Soc.} {\bfseries 504} (2021) 5021}
  [\href{https://arxiv.org/abs/2009.03309}{{\ttfamily 2009.03309}}].

\bibitem{Colberg:2008qg}
J.M.~Colberg et~al., \emph{{The Aspen--Amsterdam Void Finder Comparison
  Project}}, \href{https://doi.org/10.1111/j.1365-2966.2008.13307.x}{\emph{Mon.
  Not. Roy. Astron. Soc.} {\bfseries 387} (2008) 933}
  [\href{https://arxiv.org/abs/0803.0918}{{\ttfamily 0803.0918}}].

\bibitem{Massara:2022lng}
E.~Massara, W.J.~Percival, N.~Dalal, S.~Nadathur, S.~Radinovi\'c, H.A.~Winther
  et~al., \emph{{Velocity profiles of matter and biased tracers around voids}},
   \href{https://arxiv.org/abs/2206.14120}{{\ttfamily 2206.14120}}.

\bibitem{Neyrinck:2007gy}
M.C.~Neyrinck, \emph{{ZOBOV: a parameter-free void-finding algorithm}},
  \href{https://doi.org/10.1111/j.1365-2966.2008.13180.x}{\emph{Mon. Not. Roy.
  Astron. Soc.} {\bfseries 386} (2008) 2101}
  [\href{https://arxiv.org/abs/0712.3049}{{\ttfamily 0712.3049}}].

\bibitem{Sutter:2014haa}
P.M.~Sutter, G.~Lavaux, N.~Hamaus, A.~Pisani, B.D.~Wandelt, M.S.~Warren et~al.,
  \emph{{VIDE: The Void IDentification and Examination toolkit}},
  \href{https://doi.org/10.1016/j.ascom.2014.10.002}{\emph{Astron. Comput.}
  {\bfseries 9} (2015) 1} [\href{https://arxiv.org/abs/1406.1191}{{\ttfamily
  1406.1191}}].

\bibitem{Woodfinden:2022bhx}
A.~Woodfinden, S.~Nadathur, W.J.~Percival, S.~Radinovi\'c, E.~Massara and
  H.A.~Winther, \emph{{Measurements of cosmic expansion and growth rate of
  structure from voids in the Sloan Digital Sky Survey between redshift 0.07
  and 1.0}},  \href{https://arxiv.org/abs/2205.06258}{{\ttfamily 2205.06258}}.

\bibitem{Peebles:1994xt}
P.J.E.~Peebles, \emph{{Principles of physical cosmology}} (1994).

\bibitem{Nadathur:2015qua}
S.~Nadathur and S.~Hotchkiss, \emph{{The nature of voids \textendash{} I.
  Watershed void finders and their connection with theoretical models}},
  \href{https://doi.org/10.1093/mnras/stv2131}{\emph{Mon. Not. Roy. Astron.
  Soc.} {\bfseries 454} (2015) 2228}
  [\href{https://arxiv.org/abs/1504.06510}{{\ttfamily 1504.06510}}].

\bibitem{CUTE}
D.~{Alonso}, \emph{{CUTE solutions for two-point correlation functions from
  large cosmological datasets}}, {\emph{arXiv e-prints} (2012) arXiv:1210.1833}
  [\href{https://arxiv.org/abs/1210.1833}{{\ttfamily 1210.1833}}].

\bibitem{Wu_Jackknife}
C.F.J.~Wu, \emph{Jackknife, bootstrap and other resampling methods in
  regression analysis}, {\emph{The Annals of Statistics} {\bfseries 14} (1986)
  1261}.

\bibitem{Norberg:2008tg}
P.~Norberg, C.M.~Baugh, E.~Gaztanaga and D.J.~Croton, \emph{{Statistical
  Analysis of Galaxy Surveys - I. Robust error estimation for 2-point
  clustering statistics}},
  \href{https://doi.org/10.1111/j.1365-2966.2009.14389.x}{\emph{Mon. Not. Roy.
  Astron. Soc.} {\bfseries 396} (2009) 19}
  [\href{https://arxiv.org/abs/0810.1885}{{\ttfamily 0810.1885}}].

\bibitem{2011MNRAS.418.2435N}
P.~{Norberg}, E.~{Gazta{\~n}aga}, C.M.~{Baugh} and D.J.~{Croton},
  \emph{{Statistical analysis of galaxy surveys - IV. An objective way to
  quantify the impact of superstructures on galaxy clustering statistics}},
  \href{https://doi.org/10.1111/j.1365-2966.2011.19636.x}{\emph{"Mon. Not. Roy.
  Astron. Soc."} {\bfseries 418} (2011) 2435}
  [\href{https://arxiv.org/abs/1106.5701}{{\ttfamily 1106.5701}}].

\bibitem{Mohammad:2021aqc}
F.G.~Mohammad and W.J.~Percival, \emph{{Creating jackknife and bootstrap
  estimates of the covariance matrix for the two-point correlation function}},
  \href{https://doi.org/10.1093/mnras/stac1458}{\emph{Mon. Not. Roy. Astron.
  Soc.} {\bfseries 514} (2022) 1289}
  [\href{https://arxiv.org/abs/2109.07071}{{\ttfamily 2109.07071}}].

\bibitem{1993ApJ...412...64L}
S.D.~{Landy} and A.S.~{Szalay}, \emph{{Bias and Variance of Angular Correlation
  Functions}}, \href{https://doi.org/10.1086/172900}{\emph{\apj} {\bfseries
  412} (1993) 64}.

\bibitem{Brando:2022gvg}
G.~Brando, B.~Fiorini, K.~Koyama and H.A.~Winther, \emph{{Enabling matter power
  spectrum emulation in beyond-\ensuremath{\Lambda}CDM cosmologies with COLA}},
  \href{https://doi.org/10.1088/1475-7516/2022/09/051}{\emph{JCAP} {\bfseries
  09} (2022) 051} [\href{https://arxiv.org/abs/2203.11120}{{\ttfamily
  2203.11120}}].

\bibitem{Euclid:2020rfv}
{\scshape Euclid} collaboration, \emph{{Euclid preparation: IX. EuclidEmulator2
  \textendash{} power spectrum emulation with massive neutrinos and
  self-consistent dark energy perturbations}},
  \href{https://doi.org/10.1093/mnras/stab1366}{\emph{Mon. Not. Roy. Astron.
  Soc.} {\bfseries 505} (2021) 2840}
  [\href{https://arxiv.org/abs/2010.11288}{{\ttfamily 2010.11288}}].

\bibitem{Kaushal:2021hqv}
N.~Kaushal, F.~Villaescusa-Navarro, E.~Giusarma, Y.~Li, C.~Hawry and M.~Reyes,
  \emph{{NECOLA: Toward a Universal Field-level Cosmological Emulator}},
  \href{https://doi.org/10.3847/1538-4357/ac5c4a}{\emph{Astrophys. J.}
  {\bfseries 930} (2022) 115}
  [\href{https://arxiv.org/abs/2111.02441}{{\ttfamily 2111.02441}}].

\bibitem{DeRose:2018xdj}
J.~DeRose, R.H.~Wechsler, J.L.~Tinker, M.R.~Becker, Y.-Y.~Mao, T.~McClintock
  et~al., \emph{{The Aemulus Project I: Numerical Simulations for Precision
  Cosmology}}, \href{https://doi.org/10.3847/1538-4357/ab1085}{\emph{Astrophys.
  J.} {\bfseries 875} (2019) 69}
  [\href{https://arxiv.org/abs/1804.05865}{{\ttfamily 1804.05865}}].

\bibitem{agarwal2012}
S.~{Agarwal}, F.B.~{Abdalla}, H.A.~{Feldman}, O.~{Lahav} and S.A.~{Thomas},
  \emph{{PkANN - I. Non-linear matter power spectrum interpolation through
  artificial neural networks}},
  \href{https://doi.org/10.1111/j.1365-2966.2012.21326.x}{\emph{\mnras}
  {\bfseries 424} (2012) 1409}
  [\href{https://arxiv.org/abs/1203.1695}{{\ttfamily 1203.1695}}].

\bibitem{habib2007}
S.~Habib, K.~Heitmann, D.~Higdon, C.~Nakhleh and B.~Williams, \emph{{Cosmic
  Calibration: Constraints from the Matter Power Spectrum and the Cosmic
  Microwave Background}},
  \href{https://doi.org/10.1103/PhysRevD.76.083503}{\emph{Phys. Rev. D}
  {\bfseries 76} (2007) 083503}
  [\href{https://arxiv.org/abs/astro-ph/0702348}{{\ttfamily
  astro-ph/0702348}}].

\bibitem{angulo2009}
R.E.~Angulo and S.D.M.~White, \emph{One simulation to fit them all - changing
  the background parameters of a cosmological $n$-body simulation},
  \href{https://doi.org/10.1111/j.1365-2966.2010.16459.x}{\emph{Monthly Notices
  of the Royal Astronomical Society} (2010) }.

\bibitem{Contreras:2020kbv}
S.~Contreras, R.E.~Angulo, M.~Zennaro, G.~Aric\`o and M.~Pellejero-Iba\~nez,
  \emph{{3 per cent-accurate predictions for the clustering of dark matter,
  haloes, and subhaloes, over a wide range of cosmologies and scales}},
  \href{https://doi.org/10.1093/mnras/staa3117}{\emph{Mon. Not. Roy. Astron.
  Soc.} {\bfseries 499} (2020) 4905}
  [\href{https://arxiv.org/abs/2001.03176}{{\ttfamily 2001.03176}}].

\bibitem{pair_fixed}
R.E.~Angulo and A.~Pontzen, \emph{{Cosmological $N$-body simulations with
  suppressed variance}},
  \href{https://doi.org/10.1093/mnrasl/slw098}{\emph{Mon. Not. Roy. Astron.
  Soc.} {\bfseries 462} (2016) L1}
  [\href{https://arxiv.org/abs/1603.05253}{{\ttfamily 1603.05253}}].

\bibitem{class1}
D.~{Blas}, J.~{Lesgourgues} and T.~{Tram}, \emph{{The Cosmic Linear Anisotropy
  Solving System (CLASS). Part II: Approximation schemes}},
  \href{https://doi.org/10.1088/1475-7516/2011/07/034}{\emph{\jcap} {\bfseries
  2011} (2011) 034} [\href{https://arxiv.org/abs/1104.2933}{{\ttfamily
  1104.2933}}].

\bibitem{Dai:2018vvv}
B.~Dai, Y.~Feng and U.~Seljak, \emph{{A gradient based method for modeling
  baryons and matter in halos of fast simulations}},
  \href{https://doi.org/10.1088/1475-7516/2018/11/009}{\emph{JCAP} {\bfseries
  11} (2018) 009} [\href{https://arxiv.org/abs/1804.00671}{{\ttfamily
  1804.00671}}].

\bibitem{Lanzieri:2022zvv}
D.~Lanzieri, F.~Lanusse and J.-L.~Starck, \emph{{Hybrid Physical-Neural ODEs
  for Fast N-body Simulations}},  in \emph{{39th International Conference on
  Machine Learning Conference}}, 7, 2022
  [\href{https://arxiv.org/abs/2207.05509}{{\ttfamily 2207.05509}}].

\bibitem{Chisari:2019tus}
N.E.~Chisari et~al., \emph{{Modelling baryonic feedback for survey cosmology}},
  \href{https://doi.org/10.21105/astro.1905.06082}{\emph{Open J. Astrophys.}
  {\bfseries 2} (2019) 4} [\href{https://arxiv.org/abs/1905.06082}{{\ttfamily
  1905.06082}}].

\bibitem{Schneider:2015wta}
A.~Schneider and R.~Teyssier, \emph{{A new method to quantify the effects of
  baryons on the matter power spectrum}},
  \href{https://doi.org/10.1088/1475-7516/2015/12/049}{\emph{JCAP} {\bfseries
  12} (2015) 049} [\href{https://arxiv.org/abs/1510.06034}{{\ttfamily
  1510.06034}}].

\bibitem{Arnold:2021xtm}
C.~Arnold, B.~Li, B.~Giblin, J.~Harnois-D\'eraps and Y.-C.~Cai, \emph{{FORGE --
  the f(R) gravity cosmic emulator project I: Introduction and matter power
  spectrum emulator}},  \href{https://arxiv.org/abs/2109.04984}{{\ttfamily
  2109.04984}}.

\bibitem{Winther:2019mus}
H.~Winther, S.~Casas, M.~Baldi, K.~Koyama, B.~Li, L.~Lombriser et~al.,
  \emph{{Emulators for the nonlinear matter power spectrum beyond
  $\Lambda$CDM}},
  \href{https://doi.org/10.1103/PhysRevD.100.123540}{\emph{Phys. Rev. D}
  {\bfseries 100} (2019) 123540}
  [\href{https://arxiv.org/abs/1903.08798}{{\ttfamily 1903.08798}}].

\bibitem{LatinHypercubeSampling}
M.D.~McKay, R.J.~Beckman and W.J.~Conover, \emph{Comparison of three methods
  for selecting values of input variables in the analysis of output from a
  computer code},
  \href{https://doi.org/10.1080/00401706.1979.10489755}{\emph{Technometrics}
  {\bfseries 21} (1979) 239}
  [\href{https://arxiv.org/abs/https://doi.org/10.1080/00401706.1979.10489755}{{\ttfamily
  https://doi.org/10.1080/00401706.1979.10489755}}].

\bibitem{SavitzkyGolay}
A.~Savitzky and M.J.E.~Golay, \emph{Smoothing and differentiation of data by
  simplified least squares procedures.},
  \href{https://doi.org/10.1021/ac60214a047}{\emph{Analytical Chemistry}
  {\bfseries 36} (1964) 1627}
  [\href{https://arxiv.org/abs/https://doi.org/10.1021/ac60214a047}{{\ttfamily
  https://doi.org/10.1021/ac60214a047}}].

\bibitem{NNbook}
C.M.~Bishop, \emph{Neural Networks for Pattern Recognition}, Oxford University
  Press, Inc., USA (1995).

\bibitem{Jolliffe2002}
I.T.~Jolliffe, \emph{Principal Component Analysis}, Springer Series in
  Statistics, Springer-Verlag (2002),
  \href{https://doi.org/10.1007/b98835}{10.1007/b98835}.

\bibitem{Fletcher1988PracticalMO}
R.~Fletcher, \emph{Practical methods of optimization},  1988.

\bibitem{Donald-McCann:2021nxc}
J.~Donald-McCann, F.~Beutler, K.~Koyama and M.~Karamanis, \emph{{matryoshka:
  halo model emulator for the galaxy power spectrum}},
  \href{https://doi.org/10.1093/mnras/stac239}{\emph{Mon. Not. Roy. Astron.
  Soc.} {\bfseries 511} (2022) 3768}
  [\href{https://arxiv.org/abs/2109.15236}{{\ttfamily 2109.15236}}].

\bibitem{Donald-McCann:2022pac}
J.~Donald-McCann, K.~Koyama and F.~Beutler, \emph{{$\texttt{matryoshka}$ II:
  Accelerating Effective Field Theory Analyses of the Galaxy Power Spectrum}},
  \href{https://arxiv.org/abs/2202.07557}{{\ttfamily 2202.07557}}.

\bibitem{Nadathur:2018pjn}
S.~Nadathur, P.~Carter and W.~Percival, \emph{{A Zeldovich reconstruction
  method for measuring redshift space distortions using cosmic voids}},
  \href{https://doi.org/10.1093/mnras/sty2799}{\emph{Mon. Not. Roy. Astron.
  Soc.} {\bfseries 482} (2019) 2459}
  [\href{https://arxiv.org/abs/1805.09349}{{\ttfamily 1805.09349}}].

\bibitem{Wright:2022krq}
{\scshape LSST Dark Energy Science} collaboration, \emph{{$\texttt{Hi-COLA}$:
  Fast, approximate simulations of structure formation in Horndeski gravity}},
  \href{https://arxiv.org/abs/2209.01666}{{\ttfamily 2209.01666}}.

\end{thebibliography}\endgroup



\providecommand{\href}[2]{#2}\begingroup\raggedright\endgroup



\end{spacing}







\end{document}